\documentclass[10pt,letterpaper]{article}

\usepackage{latexsym,epsfig,multirow}
\usepackage{float,epstopdf,mathrsfs,chemarr}

\usepackage{graphicx,amsmath,amssymb}
\usepackage{relsize,longtable,multirow,multicol,rotating}
\usepackage{tikz}
\newcommand*\circled[1]{\tikz[baseline=(char.base)]{
            \node[shape=circle,draw,inner sep=2pt] (char) {#1};}}

\usepackage{helvet}
\usepackage{epstopdf}

\usepackage{changepage}

\usepackage[utf8]{inputenc}

\usepackage{textcomp,marvosym}

\usepackage{fixltx2e}
\usepackage{amsmath,amssymb}

\usepackage{cite}

\usepackage{nameref}

\usepackage[pagewise]{lineno}

\usepackage{rotating}

\usepackage[aboveskip=1pt,labelfont=bf,labelsep=period,justification=raggedright,singlelinecheck=off]{caption}

\makeatletter \renewcommand{\@biblabel}[1]{\quad#1.}  \makeatother

\date{}

\def\delt#1{\frac{\partial #1}{\partial t}} \def\tn#1{\textnormal{#1}}
 
\def\cR{{\mathcal R}}
\def\Gp{{\rm G}}
\def\Gbg{{{\rm G}_{\beta\gamma}}}
\def\Ga{{{\rm G}_{\alpha}}}
\def\Gbgm{{{\rm G}_{\beta\gamma,m}}}
\def\Gbgc{{{\rm G}_{\beta\gamma,c}}}
\def\rxn{{\rm r}}
\def\rk{{\rm k}}
\def\tr{{\rm h}}
\def\jt{{\rm j}}
\def\G2{{\rm G2}}

\begin{document}
\vspace*{0.35in}

\begin{flushleft}
{\Large \textbf\newline{A Model for Direction Sensing in {\em Dictyostelium Discoideum}:
    Ras Activity and Symmetry Breaking Driven by a G$_{\beta\gamma}$- Mediated,
    G$_{\alpha2}$-Ric8 -- Dependent Signal Transduction Network} }
\newline
\\ Yougan Cheng\textsuperscript{1,\Yinyang}, Hans
Othmer\textsuperscript{1,*\Yinyang}
\\
\bigskip
\bf{1} School of Mathematics, University of Minnesota, Minneapolis, MN, USA \\
\bigskip

\Yinyang These authors contributed equally to this work.


* othmer@umn.edu

\end{flushleft}

\section*{Abstract}

Chemotaxis is a dynamic cellular process, comprised of direction sensing,
polarization and locomotion, that leads to the directed movement of eukaryotic
cells along extracellular gradients. As a primary step in the response of an
individual cell to a spatial stimulus, direction sensing has attracted numerous
theoretical treatments aimed at explaining experimental observations in a
variety of cell types. Here we propose a new model of direction sensing based on
experiments using the free soil-living amoeba {\em Dictyostelium discoideum}
(Dicty). The model is built around a reaction-diffusion-translocation system
that involves three main component processes: a signal detection step based on
G-protein-coupled receptors (GPCR) for cyclic AMP (cAMP), a transduction step
based on a heterotrimetic G protein G$_{\alpha_{2}\beta\gamma}$ (\G2), and an
activation step of a monomeric G-protein Ras.  The model can predict the
experimentally-observed response of cells treated with latrunculinA, which
removes feedback from downstream processes, under a variety of stimulus
protocols. We show that \G2 cycling modulated by Ric8, a nonreceptor guanine
exchange factor for $\Ga$ in Dicty, drives multiple phases of Ras activation and
leads to direction sensing and signal amplification in cAMP gradients. The model
predicts that both G$_{\alpha_2}$ and G$_{\beta\gamma}$ are essential for
direction sensing, in that membrane-localized G$^*_{\alpha_2 }$, the activated
GTP-bearing form of G$_{\alpha_2}$, leads to asymmetrical recruitment of RasGEF
and Ric8, while globally-diffusing G$_{\beta\gamma}$ mediates their
activation. We show that the predicted response at the level of Ras activation
encodes sufficient 'memory' to eliminate the 'back-of-the wave' problem, and the
effects of diffusion and cell shape on direction sensing are also
investigated. In contrast with existing LEGI models of chemotaxis, the results
do not require a disparity between the diffusion coefficients of the Ras
activator GEF and the Ras inhibitor GAP. Since the signal pathways we study are
highly conserved between Dicty and mammalian leukocytes, the model can serve as
a generic one for direction sensing.

\section*{Author Summary}
Many eukaryotic cells, including {\em Dictyostelium discoideum} (Dicty),
neutrophils and other cells of the immune system, can detect and reliably orient
themselves in chemoattractant gradients. In Dicty, signal detection and
transduction involves a G-protein-coupled receptor (GPCR) through which
extracellular cAMP signals are transduced into Ras activation via an
intermediate heterotrimeric G-protein (\G2). Ras activation is the first {\em
  polarized} response to cAMP gradients in Dicty. Recent work has revealed
mutiple new characteristics of Ras activation in Dicty,  thereby providing new
insights into direction sensing mechanisms and pointing  to the need for new
models of chemotaxis.  Here we propose a novel reaction-diffusion model of Ras
activation based on three major components: one involving the GPCR, one centered on \G2,
and one involving the monomeric G protein Ras. In contrast to existing local
excitation, global inhibition (LEGI) models of direction sensing, in which a
fast-responding but slowly-diffusing activator and a slow-acting rapidly
diffusing inhibitor set up an internal gradient of activity, our model is based
on equal diffusion coefficients for all cytosolic species, and the unbalanced
local sequestration of some species leads to gradient sensing and
amplification. We show that Ric8-modulated \G2 cycling between the cytosol and
membrane can account for many of the observed responses in Dicty. including
imperfect adaptation, multiple phases of Ras activity in a cAMP gradient,
rectified directional sensing,  and cellular memory.


\section*{Introduction}

Many eukaryotic cells can detect both the magnitude and direction of
extracellular signals using receptors embedded in the cell membrane. When the
signal is spatially nonuniform they may respond by directed migration either up
or down the gradient of the signal, a process called taxis. When the
extracellular signal is an adhesion factor attached to the substrate or
extracellular matrix, the response is haptotaxis \cite{bruce2002}, and when it
is a diffusible molecule the process is called chemotaxis. Chemotaxis plays
important and diverse roles in different organisms, including mediation of
cell-cell communication \cite{christensen1998}, in organizing and re-organizing
tissue during development and wound healing \cite{schneider2010, wood2006,
  eisenbach2004}, in trafficking in the immune system \cite{kubes2013}, and in
cancer metastasis \cite{bravo2012}.

Chemotaxis can be conceptually divided into three interdependent processes:
direction sensing, polarization, and locomotion \cite{yulia2014,
  kristen2010}. In the absence of an external stimulus, cells can extend random
pseudopodia and 'diffuse' locally, which is referred as random motility
\cite{li2011}. Direction sensing refers to the molecular mechanism that detects
the gradient and generates an internal amplified response,
providing an internal compass for the cell \cite{parent1999}. Polarization involves the
establishment of an asymmetric shape with a well-defined anterior and posterior,
a semi-stable state that allows a cell to move in the same direction without an
external stimulus. These three processes are linked through interconnected
networks that govern (i) receptor-mediated transduction of an extracellular
signal into a primary intracellular signal, (ii) translation of the primary
signal into pathway-specific signals for one or more signalling pathways, and
(iii) the actin cytoskeleton and auxiliary proteins that determine polarity of
the cell. A single extracellular signal may activate numerous pathways, but our
focus herein is on the first pathway, which involves transduction of an
extracellular cAMP signal via a GPCR, and one specific pathway of the second
type, the Ras pathway, which is involved in activating the appropriate
downstream networks that govern chemotactic locomotion.

  Dicty is an amoeboid eukaryotic cell that utilizes chemotaxis during various
  stages of its life cycle. In the vegetative phase, it locates a food source by
  migrating toward folic acid secreted by bacteria or yeast. When the food
  supply is depleted Dicty undergoes a transformation from the vegetative to the
  aggregation phase, in which cells sense and migrate toward locally-secreted
  3'-5' cyclic adenosine monophosphate (cAMP), which serves as a messenger for
  control of chemotaxis and other processes \cite{othmer1998,
    kristen2010}. Dicty has served as an excellent model for studying the
  interconnected  signalling pathways governing chemotaxis due to its genetic and
  biochemical tractability \cite{charest2007, sasaki2004, nichols2015}. The
  major components of the network topology for chemotaxis have been identified
  by analyzing the effects of gene knockouts and the response of cells to
  various spatio-temporal signalling protocols \cite{kolsch2008, dawit2013,
    yulia2014}.

The first step of the chemotactic process involves signal transduction by
GPCR's, which activates G-protein  and is described in detail in the
following section. The activated G-protein can in turn activate numerous
pathways, and the pathway we analyze here involves Ras, which is a monomeric G
protein that functions as a molecular switch that activates downstream effectors
such as PI3K in its activated GTP-bound state. Activation of Ras is the earliest
measurable polarized signalling event downstream of G protein
activation\cite{sasaki2004, charest2006}. A major question from both the
experimental and the theoretical viewpoints is how the cell transduces a shallow
spatial gradient of extracellular cAMP into a steeper internal gradient of
activated Ras.  Recent experiments show that Ras activity exhibits multiple
temporal phases in cAMP gradients \cite{kortholt2013}. The first phase is
transient activation of Ras that is essentially uniform over the entire cell
boundary. In the second phase, symmetry is broken and Ras is reactivated
exclusively at the up-gradient side of the cell. The third phase is confinement,
in which the crescent of activated Ras localizes further to the region exposed
to the highest cAMP.  Other recent observations that are not incorporated in
existing models are as follows. Firstly, the Ras symmetry breaking does not
depend on the presence of the actin cytoskeleton -- treatment of cells with
latrunculin A (LatA), which leads to depolymerization of the network -- does not
destroy the symmetry-breaking \cite{kortholt2013}.  Secondly, it was found that
when two brief stimuli are applied to the same cell, the response to the second
stimulus depends on the interval between the stimuli, which indicates that there
is a refractory period \cite{huang2013}. Other experiments show that the
adaptation of Ras activation is slightly imperfect, and Ras activity is
suppressed when the chemoattractant concentration is decreasing in time, a
phenomenon called rectification\cite{nakajima2014}. Finally, it was reported
that there is a persistent memory of Ras activation, even when the cells are
treated with LatA \cite{skoge2014}.

These new results are difficult to interpret in the framework of existing
models, a number of which have been proposed
\cite{iglesias2008, iglesias2012, parent1999, levchenko2002,
  levine2006, takeda2012, tang1994, skoge2014, huang2013}.  Most of these models
are based on an activator and inhibitor mechanism called LEGI -- local
excitation, global inhibition -- to explain both direction sensing and
adaptation when the chemoattractant level is held constant
\cite{ming2014}. While these models shed some light on direction sensing, their
usefulness is limited due to the oversimplification of the signal transduction
network -- as will be elaborated later.  In particular, none of the existing
models incorporates sufficient mechanistic detail to satisfactorily explain the
spectrum of observations described above, which provides the rationale for a
more comprehensive model that enables us to test hypotheses and make predictions
concerning the expected behavior of the signal transduction pathways.

The key components in the model we develop herein are the G-protein G2, RasGEF
and RasGAP, which control rapid excitation and slower adaptation of Ras, and
Ric8, a guanine nucleotide exchange factor that activates the $\Ga$-component of
G2 \cite{kataria2013}. The model is developed for LatA-treated cells so as to
remove the feedback effect from the actin cytoskeleton on Ras, and we show that
it can replicate many of the observed characteristics of Ras activation in
Dicty.  It is known that activated Ras activates PI3K, which stimulates further
downstream steps that affect actin polymerization, but we can restrict attention
to the Ras dynamics and its upstream effectors because there is no known direct
feedback to Ras from downstream steps between Ras and the actin cytoskeleton.
We show that $\Gbg$ mediates adaptation of Ras activity in a uniform stimulus
and transient activation in a gradient. It is also shown that $G_{\alpha 2}$
contributes to the imperfect adaptation in a uniform stimulus, and that it is an
essential element for front-to-back symmetry breaking in a gradient,
highlighting the important roles of $G_{\alpha 2}$ and G2 cycling between the
bound and dissociated states. We also show that Ric8 contributes to the
amplification of Ras activity by regulating $G_{\alpha 2}$ dynamics: the
reactivation of G$_{\alpha 2}$ by Ric8 induces further asymmetry in \G2
dissociation, which in turn amplifies the Ras activity. Finally, we investigated
the effects of diffusion and cell shapes on direction sensing, and the potential
role of Ric8 in the establishment of persistent Ras activation, which is a form
of cellular memory.

\section*{Signal transduction pathways}

\begin{figure}[h!]
\centerline{\includegraphics[width=5cm]
  {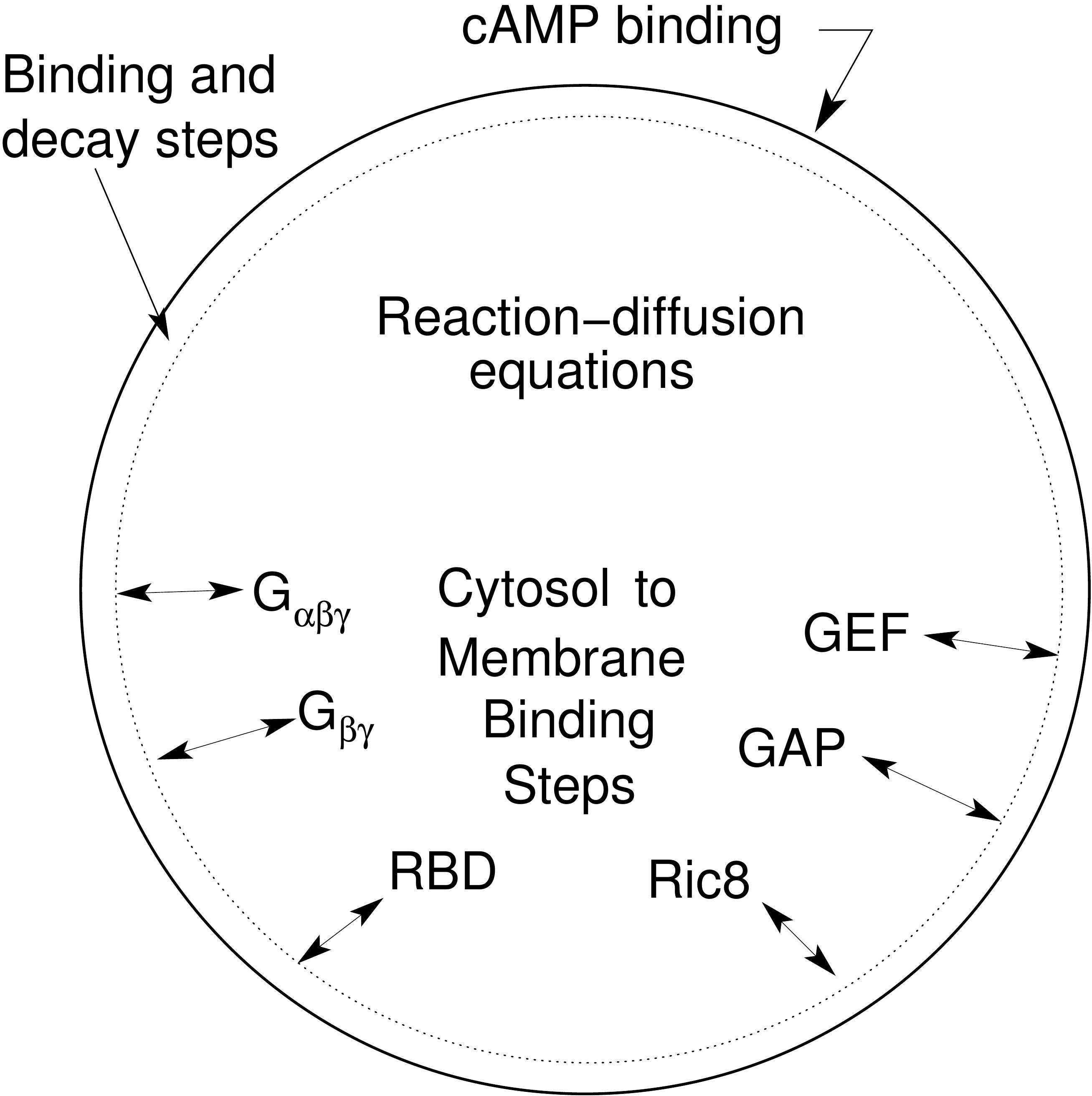}\hspace*{50pt}\includegraphics[width=6cm]
  {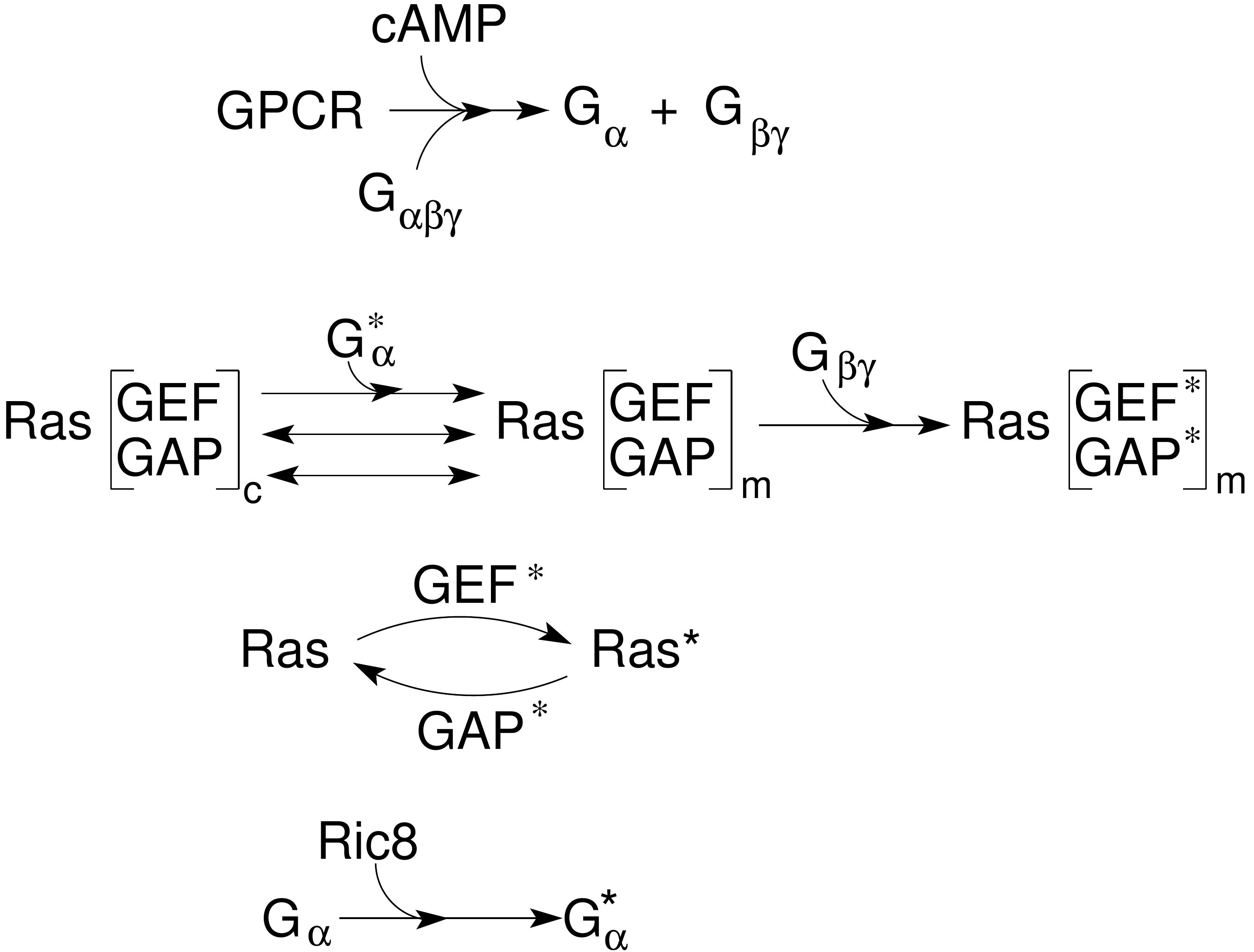} }
\vspace*{10pt}
\caption{A schematic of the major
  processes in the model  (left) and the primary steps in the network (right). }
\label{sigpath}
\end{figure}

In light of the restriction to LatA-treated cells, the backbone of the
chemotactic pathway activated in response to changes in extracellular cAMP is
$\Delta$\,cAMP $\rightarrow \Delta $\,GPCR occupation $\rightarrow \Delta$
G$_{\alpha\beta\gamma} $\,activation $\rightarrow \Delta $\,Ras activity. We
describe this pathway in terms of three modules: the GPCR surface receptors
cAR1-4, \G2 and Ras, as illustrated in Figure \ref{sigpath}.

\paragraph{The GPCR surface receptor}

The first step in Dicty chemotaxis is binding of cAMP to the G-protein coupled
receptors (GPCRs) cAR1-4.  A cAMP-bound GPCR serves as a GEF that catalyses the
exchange of GDP for GTP on the G$_\alpha$ subunit of G2, and leads to the
activated G$^*_\alpha$ subunit and the G$_{\beta\gamma}$ subunit following
dissociation.  Experimental results suggest that ~30\% of the G protein
heterotrimers exist in the cytosol.  \footnote{ It is well established in
  mammalian cells that ligand-induced phosphorylation of GPCRs leads to
  recruitment of arrestin family proteins, which uncouple receptors from
  downstream G proteins \cite{shukla2011, evron2012}. cAR1 is phosphorylated at
  multiple cytoplasmic residues upon chemoattractant stimulation
  \cite{hereld1994, vaughan1988}, which is correlated to agonist-induced loss of
  ligand binding \cite{caterina1995}. The functional consequence of receptor
  phosphorylation for chemotaxis has not been fully addressed, but it is known
  that receptor phosphorylation is not essential for chemotaxis or termination
  of G-protein-mediated responses \cite{kim1997}, and since there is no evidence
  that receptor phosphorylation affects Ras we do not include it.}  The four
receptor types, which have different affinities, are expressed sequentially
throughout the developmental transition from a unicellular to a multicellular
organism. Switching of receptor subtypes enable Dicty to response to changing
chemoattractant concentrations in a wide range and hence program morphogenesis
appropriately \cite{kim1998, johnson1992, vanhaastert1984}, but n the model we
treat only one type.  Lateral diffusion of the receptors has been suggested by
the observation of green fluorescent protein (GFP)-tagged receptors in
Dicty. The diffusion coefficient measured from the movements of individual
receptors is about 2.7 $\pm 1.1\times 10^{-10} cm^2s^{-1}$ \cite{ueda2001},
which is small at the scale of the cell size, but could be locally significant
on the scale of structures such as filipodia.

\paragraph{The  G protein module}

The heterotrimeric G proteins function as transducers of extracellular cAMP
signals for gradient sensing, since studies show that localized responses such
as Ras activation  occur upstream of PI3-kinase activity and downstream of G protein
activity \cite{charest2006}. There are 11 G$_\alpha$ subunits and a single
G$_\beta$ and G$_\gamma$ subunit in Dicty \cite{zhang2001}. This single
G$_{\beta\gamma}$ subunit is essential for chemotactic signal transduction
since  $g^-_\beta$ cells do not show any Ras activation \cite{kortholt2013} and do
not chemotact \cite{sasaki2004}. The primary $G_\alpha$ subunit in
chemotaxis is G$_{\alpha_2}$, since $g^-_{\alpha_2}$ cells lack an essential component of
the response to cAMP, as described later \cite{kortholt2013, kataria2013}.

 Ligand binding to the GPCR catalyzes the exchange of GTP for GDP on the
 G$_\alpha$ subunit, causing the dissociation of activated G$^*_\alpha$ subunits
 and G$_{\beta\gamma}$ subunits.  Hydrolysis of GTP in G$^*_\alpha$ induces
 reassociation, which reduces active G-protein subunits when the chemoattractant
 is removed \cite{ming2014, yulia2014}. By monitoring fluorescence resonance
 energy transfer (FRET) between $\alpha$ and $\beta$ subunits, the membrane
 dynamics of the heterotrimer prior to and after simulation in Dicty has been
 visualized \cite{janetopoulos, xuehua2005}, and it has been shown that G
 protein activation reaches a persistent dose-dependent steady state level during
 continuous stimulation, {\em i.e.,}  no adaptation occurs  at this level
 \cite{janetopoulos, iglesias2012}.

These and other studies show that \G2 and G$_{\beta\gamma}$ subunits cycle
between the cytosol and the plasma membrane, while the activated G$_\alpha$
probably remains membrane-bound \cite{elzie2009, dawit2013}. Moreover, although
asymmetric distributions of G$_{\beta\gamma}$ subunits are observed in highly
polarized Dicty, in LatA-treated cells $\Gbg$ is uniformly distributed along the
plasma membrane and within the cytosol in the presence of a cAMP gradient
\cite{jin2000}, which further suggests that G$_{\beta\gamma}$ is also cycling
between the membrane and the cytosol.  Finally, it is reported that Dicty
'resistant to inhibitors of cholinesterase 8' (Ric8) is a nonreceptor GEF for
$\Ga$, which converts $G_{\alpha_2} $GDP  into the activated $G_{\alpha_2}-$GTP
form \cite{kataria2013}. The regulation of Ric8 activity is currently not clear,
but its role as a GEF probably involves binding of $\Ga$ to Ric8
\cite{kataria2013}.

\paragraph{Ras GTPases}
Ras belongs to the family of small G proteins that function as molecular
switches to control a wide variety of important cellular functions.  In Dicty,
there are 5 characterized isoforms: RasS, RasD, RasB, RasC, and RasG encoded by
14 Ras family genes \cite{yulia2014}. RasC and RasG proteins appear to be
particularly important for chemotaxis, of which RasG is the key Ras protein in
the regulation of cAMP-mediated chemotaxis \cite{kortholt2013}.

In the chemotactic backbone the Ras module provides a link between G proteins
and downstream pathways. Ras proteins exist in an inactive GDP-bound state and
an active GTP-bound state, and conversion between these is regulated by RasGEFs
and GTPase activating proteins (RasGAPs). RasGEFs catalyze the exchange of GDP
for GTP, thereby activating Ras, whereas RasGAPs stimulate the GTPase activity,
converting the protein  into the inactive GDP-bound form.  Regulation by GEF
and GAP conversion of Ras includes protein-protein or protein-lipid
interactions, binding of second messengers, and post-translational modifications
that induce one or more of the following changes: translocation to a specific
compartment of the cell, release from autoinhibition, and the induction of
allosteric changes in the catalytic domain \cite{bos2007}.  Several methods have
been developed to detect Ras protein and small GTPases activation
\cite{sasaki2009}, while the dynamics of Ras are usually monitored by the
translocation of a tagged Ras-binding domain (RBD) peptide. The RBD of Raf1 only
binds to the activated Ras-GTP, which enables localized visualization of Ras
activity.  The response of activated Ras in Dicty shows near-perfect adaptation,
although some deviation from perfect adaptation can be observed
\cite{nakajima2014}.

The full set of reactions and translocation steps are given in Table
\ref{rxntab}, wherein reactions and translocations are labeled as ${\cal R}_{\rm
  s}$ and ${\cal J}_{\rm s}$, respectively, and the corresponding rate laws,
which are derived by assuming mass-action kinetics for all steps, are denoted by
$\rxn_s$ and $j_s$, respectively.  In reality the translocation of a substance
between the cytosol and the membrane takes place within a layer near the
membrane, but we treat this as a surface reaction.  Moreover, we assume that
complex formation is always fast and that a negligible amount of the factors is
in the complex form, so that the conversion rate of the substrate is
proportional to the product of regulator and substrate densities, unless
otherwise indicated.  To eliminate the effects of intrinsic polarity and
investigate the system dynamics without feedback from the cytoskeleton, we
assume that the cells are pretreated with LatA, in which case they lose polarity
and become rounded and immobile.

The applicable conservation conditions on the various species are implicit in
the evolution equations, which are given in detail in the Materials and Methods
section.  For simplicity, we model a cell as a 3D sphere centered at the origin,
of radius $5 \mu m$ \cite{jin2000}. The initial condition for the system is the
steady state in a very small concentration ($0.001 pM$) of cAMP in the extracellular space. The system is
solved numerically by a finite element discretion in space and backward
differentiation for the time stepping, implemented in the COMSOL multiphysics
package.  In the following sections, we exhibit the cell response under various
stimulation protocols, and for notational simplicity, we use $G_{\alpha}$ in
place of $G_{\alpha 2}$.  Some of the results that will be discussed are as
follows.

\begin{itemize}
\item Under uniform stimuli --
\begin{itemize}
\item The transient response

\item Imperfect adaptation

 \item The response of $g^-_\alpha$ and $ric8^-$ cells.
\end{itemize}

\item Under graded stumuli --

\begin{itemize}

\item The origin of the biphasic Ras activation and the necessity of 'activator' diffusion

\item How the magnitude of gradient amplification depends on the cAMP amplitude and
  gradient

\item The response of $g^-_\alpha$ and $ric8^-$ cells in a gradient.

 \item The 'back-of-the-wave' problem.

\end{itemize}
\end{itemize}

 \begin{minipage}{1.0\textwidth}
 \renewcommand{\thefootnote}{\thempfootnote}
 \footnotesize
    \begin{longtable}[h]{llll}
    \caption{Kinetics and rates of the reactions.}\label{Table_reaction} \\
    Label and Description & Kineti\hspace*{25pt}& Rate & \hspace*{5pt} Reference\\ \hline
    \circled{1} ligand binding & $ {\cal R}_1: cAMP + R \underset{\rk_1^-}{\overset{\rk_1^+}{\displaystyle \rightleftharpoons}} R^* $
    & $ \rk_{1}^{+} $,$ \rk_{1}^{-} $
    & \cite{ueda2001, janetopoulos}   \footnote{The affinity values of four
    receptors in Dicty have been measured in various conditions
    \cite{johnson1992} and receptors have the ability of switching their
    affinity between high affinity and low affinity \cite{vanhaastert1984}. To
    avoid modeling all four receptors, we model the binding with an averaged
binding affinity and dissociation rate.}  \\
   \circled{2} G2 cycling
    & $ {\cal J}_1: G_{\alpha\beta\gamma, m}  \underset{\tr_2}{\overset{\tr_1}{\displaystyle \rightleftharpoons}} G_{\alpha\beta\gamma, c} $
    & $ \tr_1,\tr_2  $
    & \cite{elzie2009, dawit2013} \\
    \circled{3} G2 dissociation
    & $ {\cal R}_2: G_{\alpha\beta\gamma, m}+R^* \overset{\rk_2}{\displaystyle\rightarrow} G_\alpha^*+G_{\beta\gamma, m}+R^*$
    & $ \rk_2 $
    & \cite{ming2014, yulia2014} \\
    \circled{4} G$_{\beta\gamma}$ cycling
    & $ {\cal J}_2: G_{\beta\gamma, m}  \underset{\tr_4}{\overset{\tr_3}{\displaystyle \rightleftharpoons}} G_{\beta\gamma, c} $
    & $ \tr_3,\tr_4  $
    & \cite{elzie2009, dawit2013}    \footnote{Following the dissociation, the free G$_{\beta\gamma}$ could diffuse away from the
membrane and enter the cytosol.}\\
    \circled{5} GTPase of G$_\alpha^*$
    & $ {\cal R}_3: G_{\alpha}^*\overset{\rk_3}{\displaystyle\rightarrow} G_\alpha$
    & $ \rk_3 $
    & \cite{ming2014, yulia2014}
    \footnote{The intrinsic guanosine triphosphatase (GTPase) activity of the
activated $\alpha$ hydrolyses the bound GTP on the plasma membrane, whose rate
can be varying depending on the influence of regulator of G protein signalling
(RGS) proteins \cite{vries2000, berman1998}}\\
    \circled{6} Ric8 cycling
    & $ {\cal J}_3: Ric8_{m}  \underset{\tr_6}{\overset{\tr_5}{\displaystyle \rightleftharpoons}} Ric8_{c} $
    & $ \tr_5, \tr_6 $
    & \cite{kataria2013}    \footnote{The regulation of Ric8 activity is still not clear. We assume a translocation-activation mechanism here. The possibility of translocation-only mechanism is investigated.}\\
     \circled{7} Promoted Ric8 cycling
    & $ {\cal J}_4: Ric8_c+G_{\alpha}^*\overset{\tr_7}{\displaystyle\rightarrow} Ric8_m+G_\alpha^*$
    & $ \tr_7 $
    & Assumed
    \footnote{We assume that Ric8 translocation can be promoted by G$_\alpha^*$. The scenarios of G$_\alpha$ promotion and no promotion (g$\alpha$-null) are also investigated.}\\
     \circled{8} Ric8 activation
    & $ {\cal R}_4: Ric8_m+G_{\beta\gamma, m}\overset{\rk_4}{\displaystyle\rightarrow} Ric8^*+G_{\beta\gamma, m}$
    & $ \rk_4 $
    & Assumed
    \footnote{We assume Ric8 is activated by G$_{\beta\gamma, m}$. The scenario
      of translocation-only is investigated, in which case $Ric8_m$
      converts G$_\alpha$ directly into G$_\alpha^*$. The simulations suggest that
      this activation is not essential to induce symmetry breaking.}\\
     \circled{9} G$_\alpha$ reactivation
    & $ {\cal R}_5: Ric8^*+G_{\alpha}\overset{\rk_5}{\displaystyle\rightarrow} Ric8^*+G_{\alpha}^*$
    & $ \rk_5 $
    & \cite{kataria2013}\\
     \circled{10} Ric8 inactivation
    & $ {\cal R}_6: Ric8^*\overset{\rk_6}{\displaystyle\rightarrow} Ric8_m$
    & $ \rk_6 $
    & Assumed
    \footnote{An inactivation is introduced to balance  the Ric8 activation
      step. In the translocation-only scenario, this step is eliminated.}\\
     \circled{11} G2 reassociation
    & $ {\cal R}_7: G_{\alpha}+G_{\beta\gamma,m}\overset{\rk_7}{\displaystyle\rightarrow} G_{\alpha\beta\gamma,m}$
    & $ \rk_7 $
    & \cite{ming2014, yulia2014}\\
    \circled{12} RasGEF cycling
    & $ {\cal J}_5: RasGEF_{m}  \underset{\tr_9}{\overset{\tr_8}{\displaystyle \rightleftharpoons}} RasGEF_{c} $
    & $ \tr_8, \tr_9 $
    & \cite{wilkins2005}\\
    \circled{13} Promoted RasGEF cycling
    & $ {\cal J}_6: RasGEF_c+G_{\alpha}^*\overset{\tr_{10}}{\displaystyle\rightarrow} RasGEF_m+G_\alpha^*$
    & $ \tr_{10} $
    & \cite{fuku2001, hart1998, jackson2001, zheng2001}\\
    \circled{14} RasGAP cycling
    & $ {\cal J}_7: RasGAP_{m}  \underset{\tr_{12}}{\overset{\tr_{11}}{\displaystyle \rightleftharpoons}} RasGAP_{c} $
    & $ \tr_{11}, \tr_{12} $
    & \cite{bos2007}\\
    \circled{15} RasGEF activation
    & $ {\cal R}_8: RasGEF_m+G_{\beta\gamma,m}\overset{\rk_{8}}{\displaystyle\rightarrow} RasGEF^*+G_{\beta\gamma,m}$
    & $ \rk_{8} $
    & \cite{takeda2012, kortholt2013}\\
    \circled{16} RasGEF inactivation
    & $ {\cal R}_9: RasGEF^*\overset{\rk_{9}}{\displaystyle\rightarrow} RasGEF_m$
    & $ \rk_{9} $
    & \cite{takeda2012, kortholt2013}\\
    \circled{17} RasGAP activation
    & $ {\cal R}_{10}: RasGAP_m+G_{\beta\gamma,m}\overset{\rk_{10}}{\displaystyle\rightarrow} RasGAP^*+G_{\beta\gamma,m}$
    & $ \rk_{10} $
    & \cite{takeda2012, kortholt2013}\\
    \circled{18} RasGAP inactivation
    & $ {\cal R}_{11}: RasGAP^*\overset{\rk_{11}}{\displaystyle\rightarrow} RasGAP_m$
    & $ \rk_{11} $
    & \cite{takeda2012, kortholt2013}\\
    \circled{19} Ras activation
    & $ {\cal R}_{12}: RasGEF^*+Ras\overset{\rk_{12}}{\displaystyle\rightarrow} RasGEF^*+Ras^*$
    & $ \rk_{12} $
    & \cite{takeda2012, kortholt2013}\\
    \circled{20} Ras inactivation
    & $ {\cal R}_{13}: RasGAP^*+Ras^*\overset{\rk_{13}}{\displaystyle\rightarrow} RasGAP^*+Ras$
    & $ \rk_{13} $
    & \cite{takeda2012, kortholt2013}\\
    \circled{21} Spontaneous Ras activation
    & $ {\cal R}_{14}: Ras\overset{\rk_{14}}{\displaystyle\rightarrow} Ras^*$
    & $ \rk_{14} $
    & \cite{takeda2012, kortholt2013}\\
    \circled{22} Spontaneous Ras inactivation
    & $ {\cal R}_{15}: Ras^*\overset{\rk_{15}}{\displaystyle\rightarrow} Ras$
    & $ \rk_{15} $
    & \cite{takeda2012, kortholt2013}\\
    \circled{23} RBD cycling
    & $ {\cal J}_8: RBD_{m}  \underset{\tr_{14}}{\overset{\tr_{13}}{\displaystyle \rightleftharpoons}} RBD_{c} $
    & $ \tr_{13},  \tr_{14} $
    & \cite{takeda2012, kortholt2013}\\
    \circled{24} Promoted RBD cycling
    & $ {\cal J}_9: RBD_c+Ras^*\overset{\tr_{15}}{\displaystyle\rightarrow} RBD_m+Ras^*$
    & $ \tr_{15} $
    & \cite{takeda2012, kortholt2013}\\
\label{rxntab}
\end{longtable}
\end{minipage}

\section*{Results}

\subsection*{The response under a uniform stimulus}

\begin{figure}[h!]
\centerline{\includegraphics[width=6cm]
  {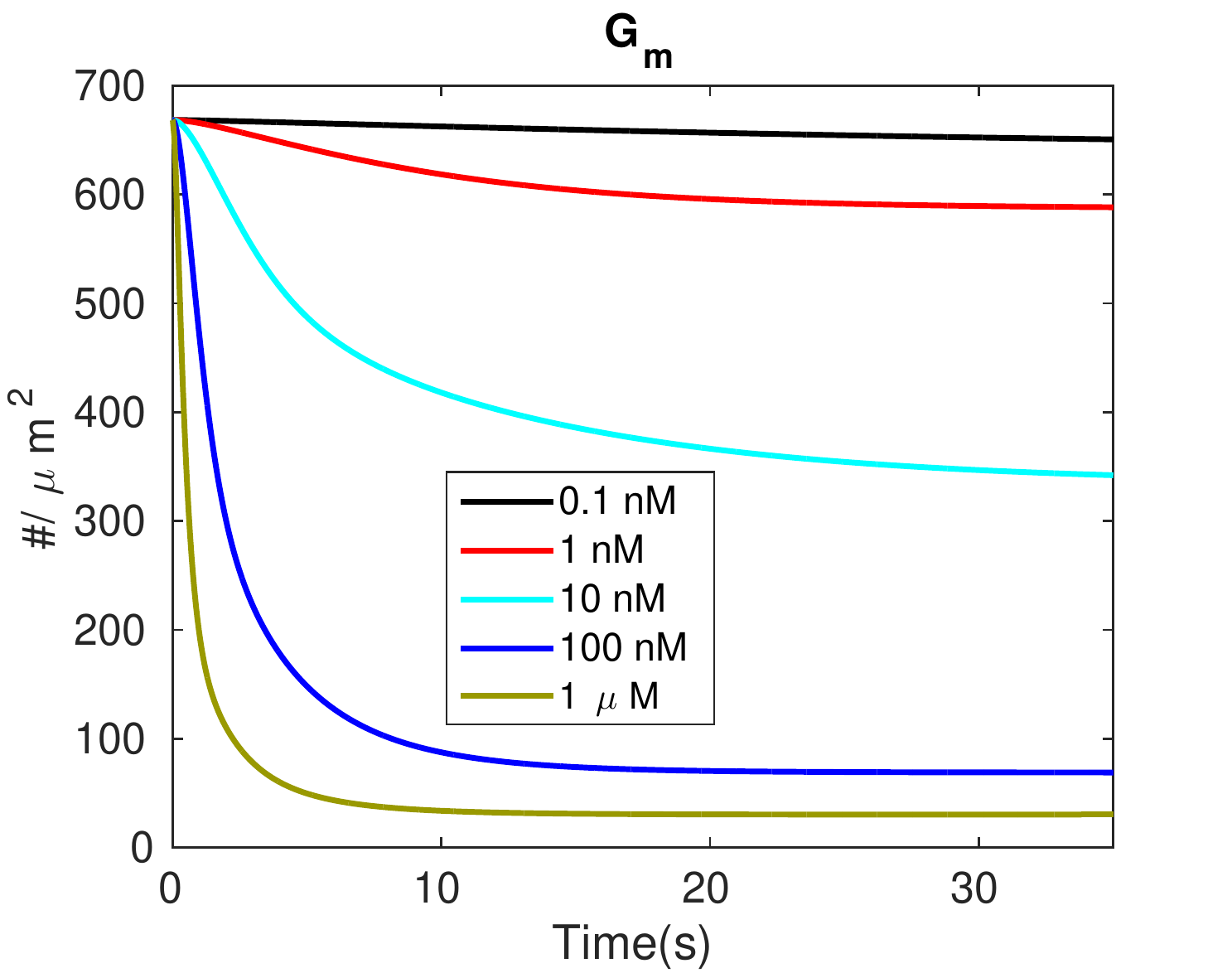}\includegraphics[width=6cm]{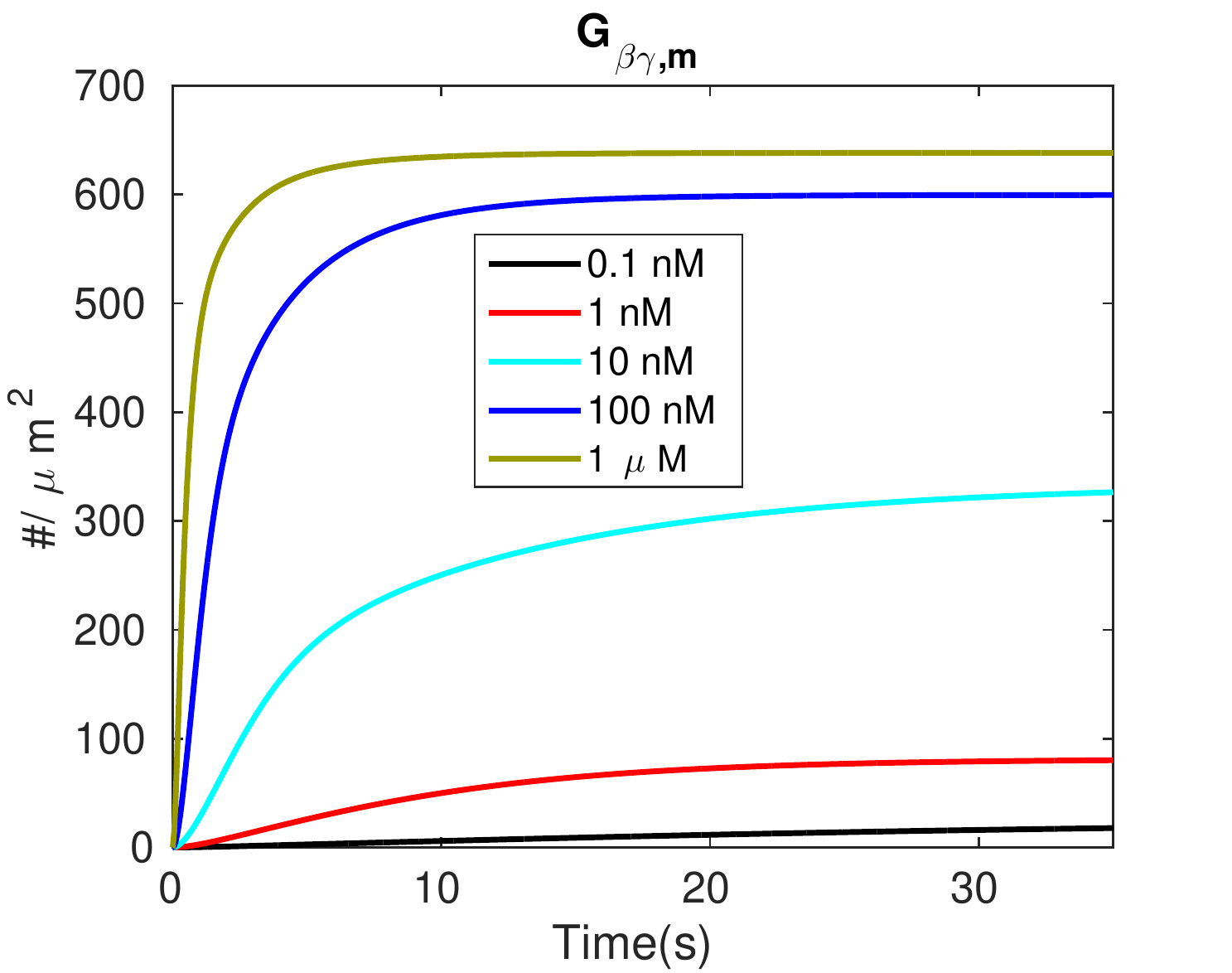}}
\centerline{\includegraphics[width=6cm]{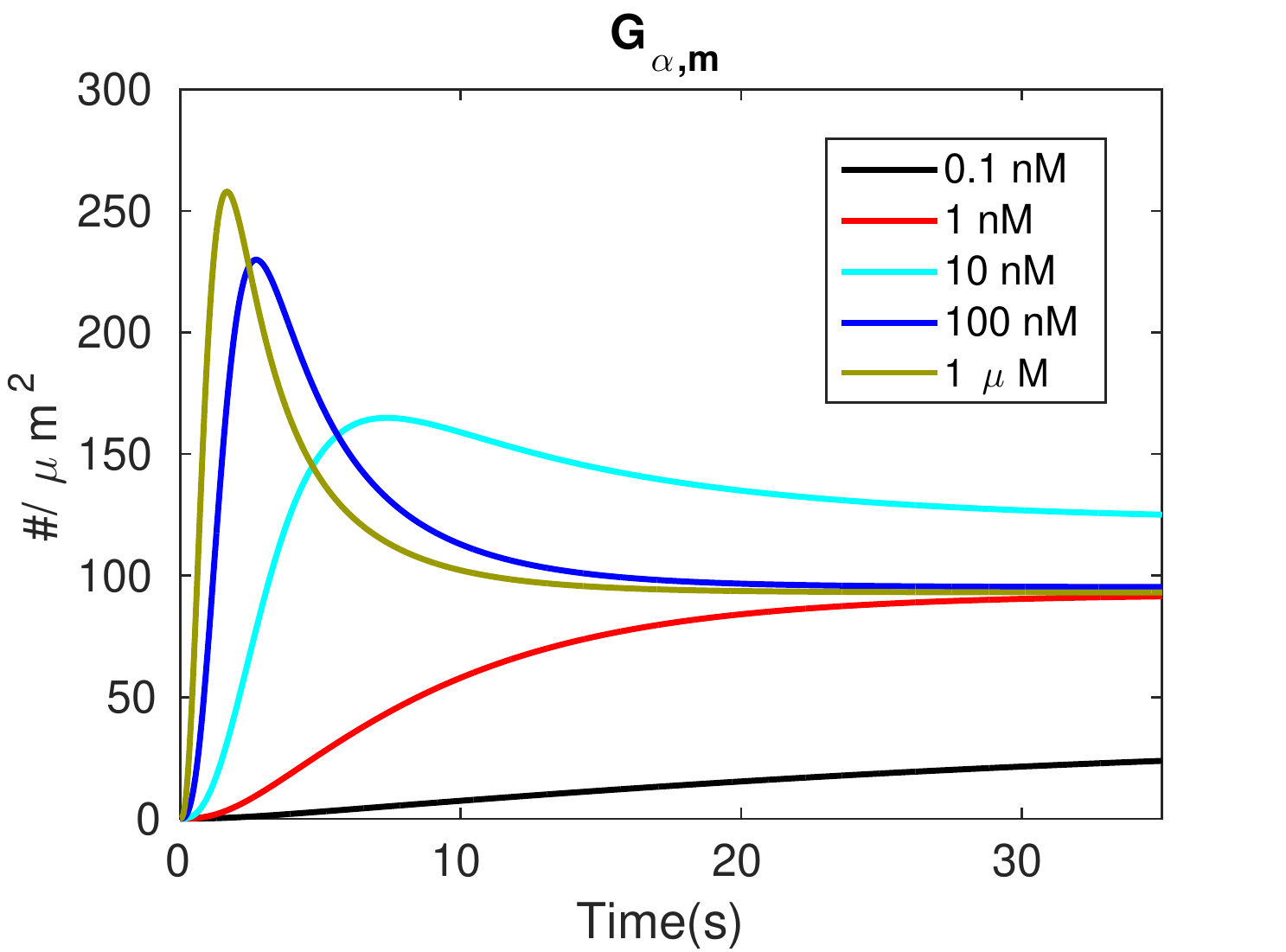}\includegraphics[width=6cm]{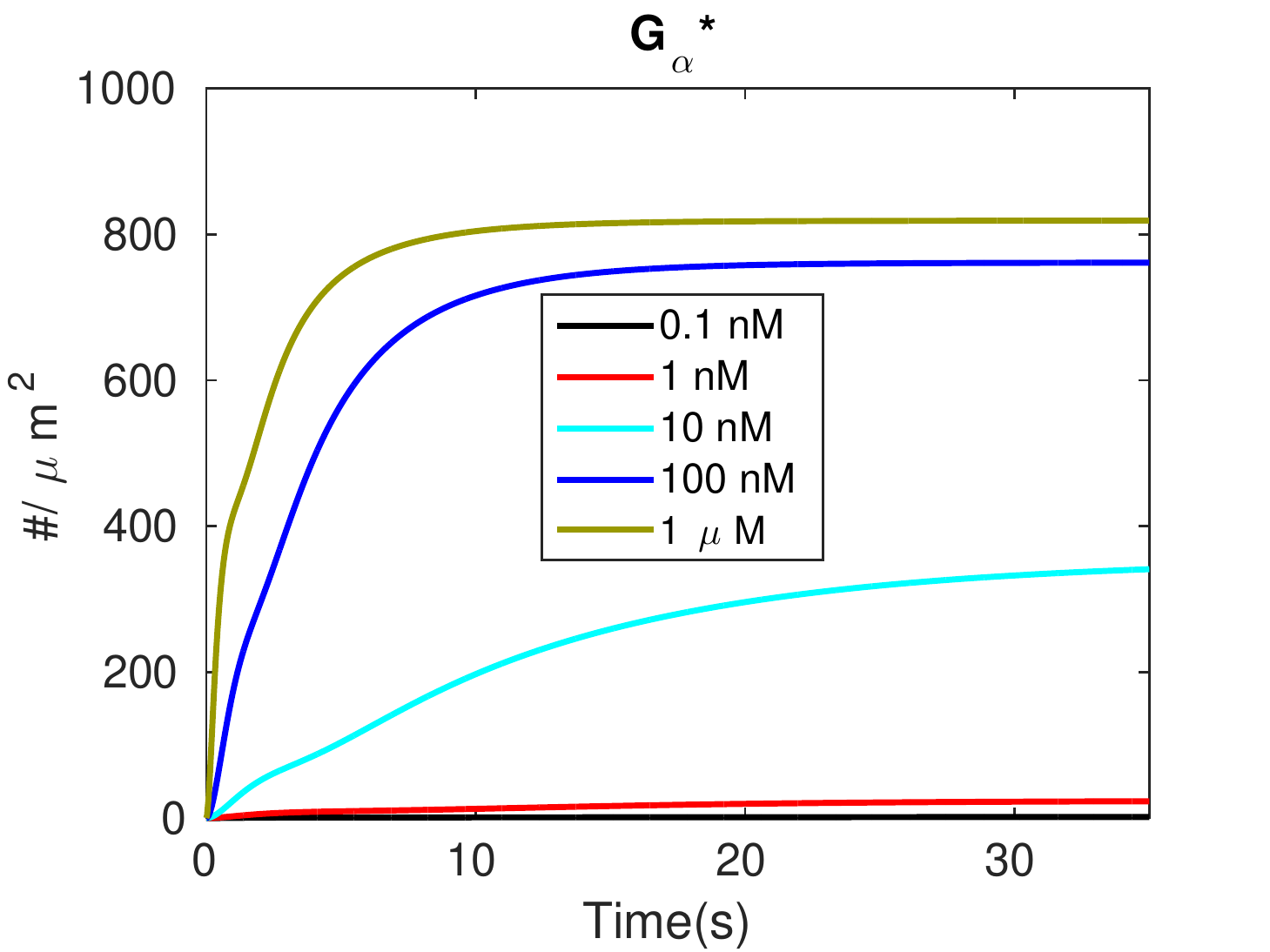}}
\centerline{\includegraphics[width=6cm] {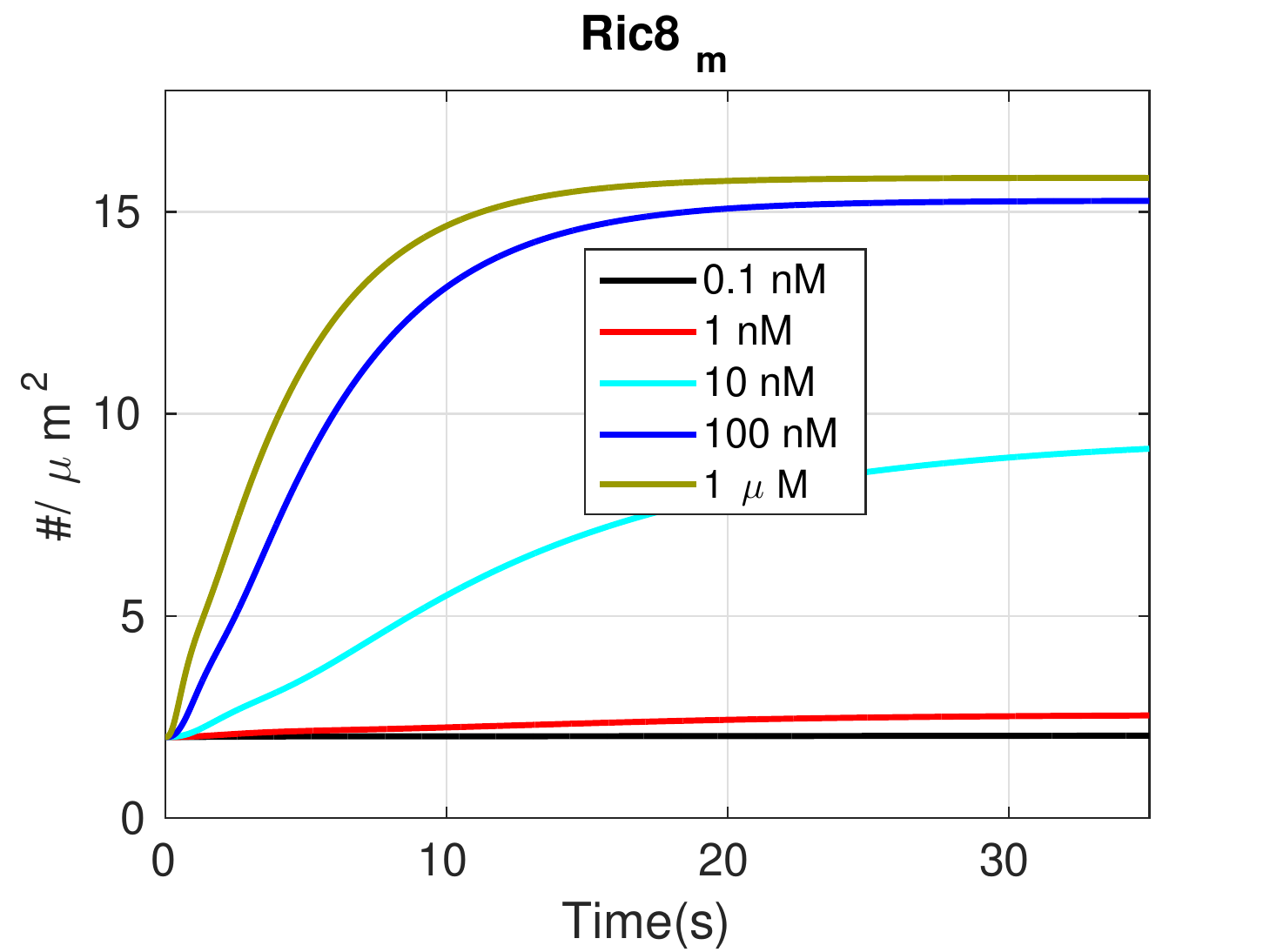}\includegraphics[width=6cm] {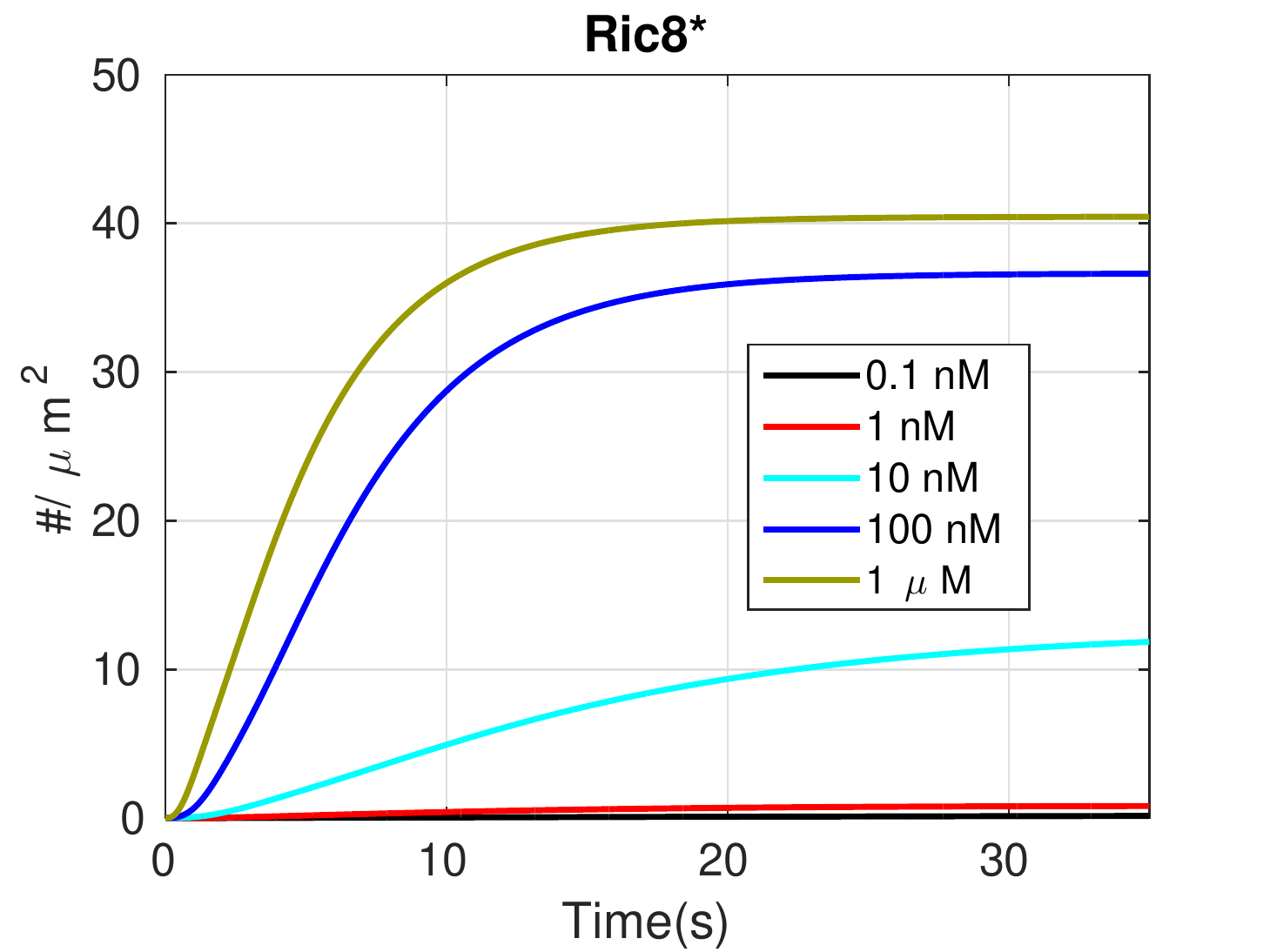}}
\centerline{\includegraphics[width=6cm] {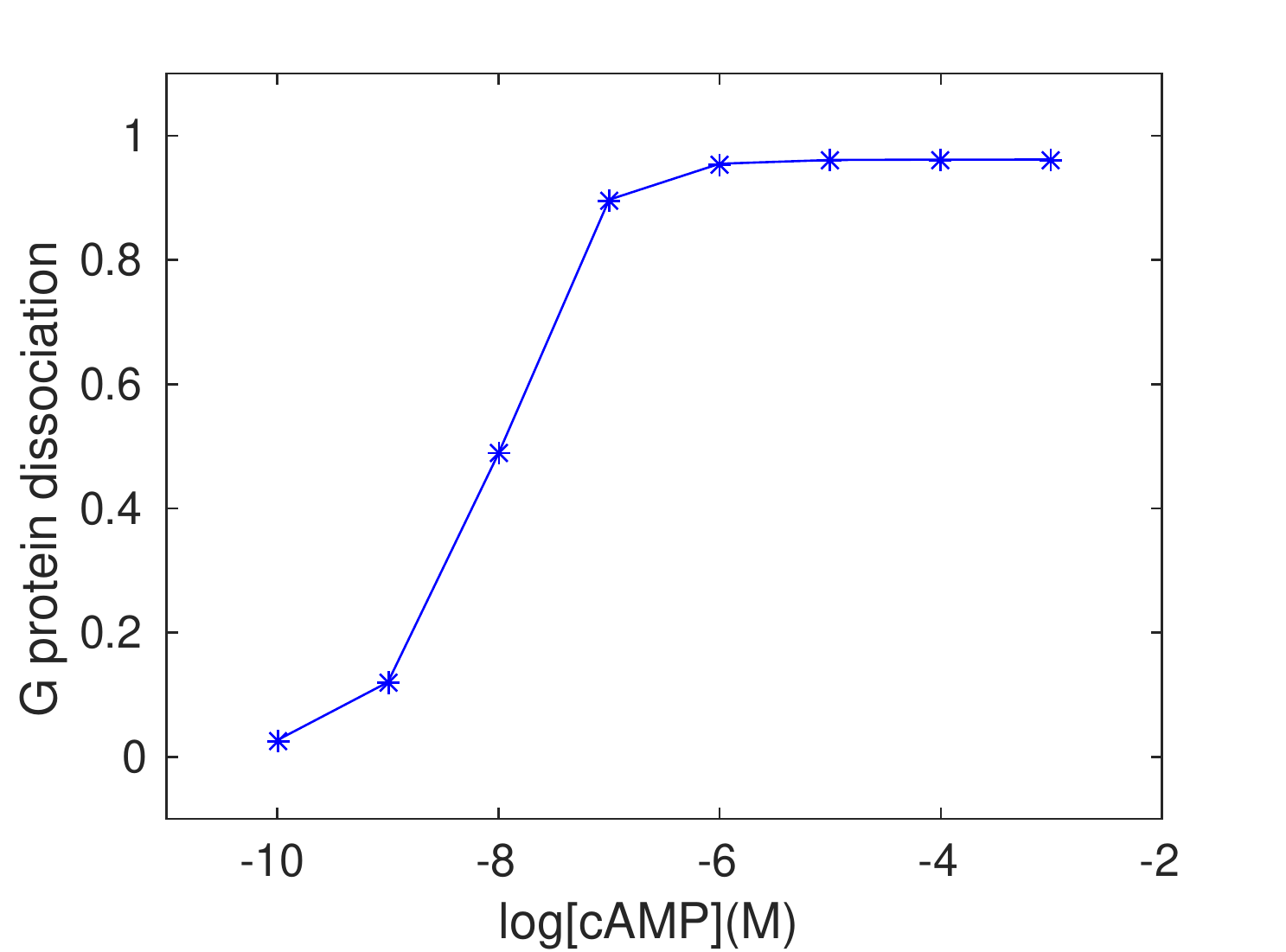}\includegraphics[width=6cm] {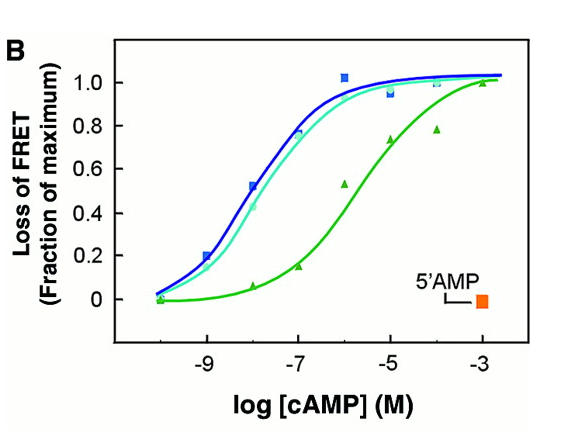}}
\caption{\label{GandRic8} The time course of various components
under different levels of uniform stimuli: \emph{First row} -- \G2 and
$\Gbg$: \emph{Second row} -- $G_{\alpha}$ and
$G^*_\alpha$ ; \emph{Third row}: --  Ric8 and $Ric8^*$ .  \emph{Fourth row} --
The dose dependent
dissociation at steady state. Left: model prediction; \emph{Right}:  Dose-response
curves for cAMP (dark blue), 2'-dcAMP(light blue), and 8-Br cAMP (green), and 5'
AMP (orange) from \cite{janetopoulos}.}
\end{figure}

\paragraph{$G_{\alpha}$ dynamics}\hspace*{-8pt}\footnote{Hereafter we use
  $G_{\alpha}$ for $G_{\alpha_2}$ where possible to
  simplify the notation.}  As previously noted, G2 dissociates rapidly upon
addition of chemoattractant and G$^*_{\alpha}$ and $\Gbg$ reach a dose-dependent
steady-state level during continuous stimulation, even though downstream
responses subside \cite{janetopoulos}. The computed dose-dependent time
evolutions of \G2 and $\Gbg$ are shown in the first row of Fig.~\ref{GandRic8}.
Under a spatially-uniform stimulus the concentration of \G2 decreases due to
dissociation induced by cAMP-bound cAR, the concentration of $\Gbg$ subunits
increases, and the steady state level of each is dose-dependent.  The time to
reach a steady state level decreases as the cAMP increases, and at $1 \mu M$
cAMP the dissociation is stabilized within 5 seconds of activation, which is
consistent with results in  \cite{janetopoulos}.

The dynamics of the $G_{\alpha}$ subunits are shown in the second row of
Fig.~\ref{GandRic8}. As shown in the right panel, $G_{\alpha}$ is activated in a
dose-dependent persistent manner similar to $\Gbg$, but $G^*_{\alpha}$ reaches
steady state  more slowly than $\Gbg$ and the steady state concentration
is higher at a given cAMP stimulus,  because both forms of $G_{\alpha}$ remain
membrane-bound. Surprisingly, the simulation shows that $G_{\alpha}$ exhibits a
biphasic response when the cAMP concentration is above a certain threshold. When
the cAMP concentration is lower than 1 nM the $G_{\alpha}$ concentration
increases to the steady state monotonically, but if the cAMP concentration is
greater than 10 nM the $G_{\alpha}$ concentration shows an initial overshoot and
then decreases to the steady state, which illustrates the kinetic diversity of G
protein signalling \cite{vladimir2007}. Furthermore, unlike the response of
$\Gbg$ and $G^*_{\alpha}$, for which a higher concentration of cAMP produces a
higher steady state levels of subunits, for $G_{\alpha}$ there is an optimal
cAMP concentration at which the steady state level of $G_{\alpha}$ is
maximized.

In light of our assumption that Ric8 is localized on the membrane by
$G^*_{\alpha}$ and activated by $\Gbg$, it follows that the model predicts that
Ric8 activation is also nonadaptative, as demonstrated in the third row of
Fig.~\ref{GandRic8}.  In the fourth row of Fig.~\ref{GandRic8} we show the
comparison of dose-dependent \G2 dissociation between the observations in
\cite{janetopoulos} and our model prediction. One sees that the predictions
matches the experimental data and both show that dissociation of G2 is saturated
at 1 $\mu M$ cAMP.

\paragraph{Imperfect adaptation at the level of Ras}

\begin{figure}[H]
\centerline{\includegraphics[width=6.5cm]
{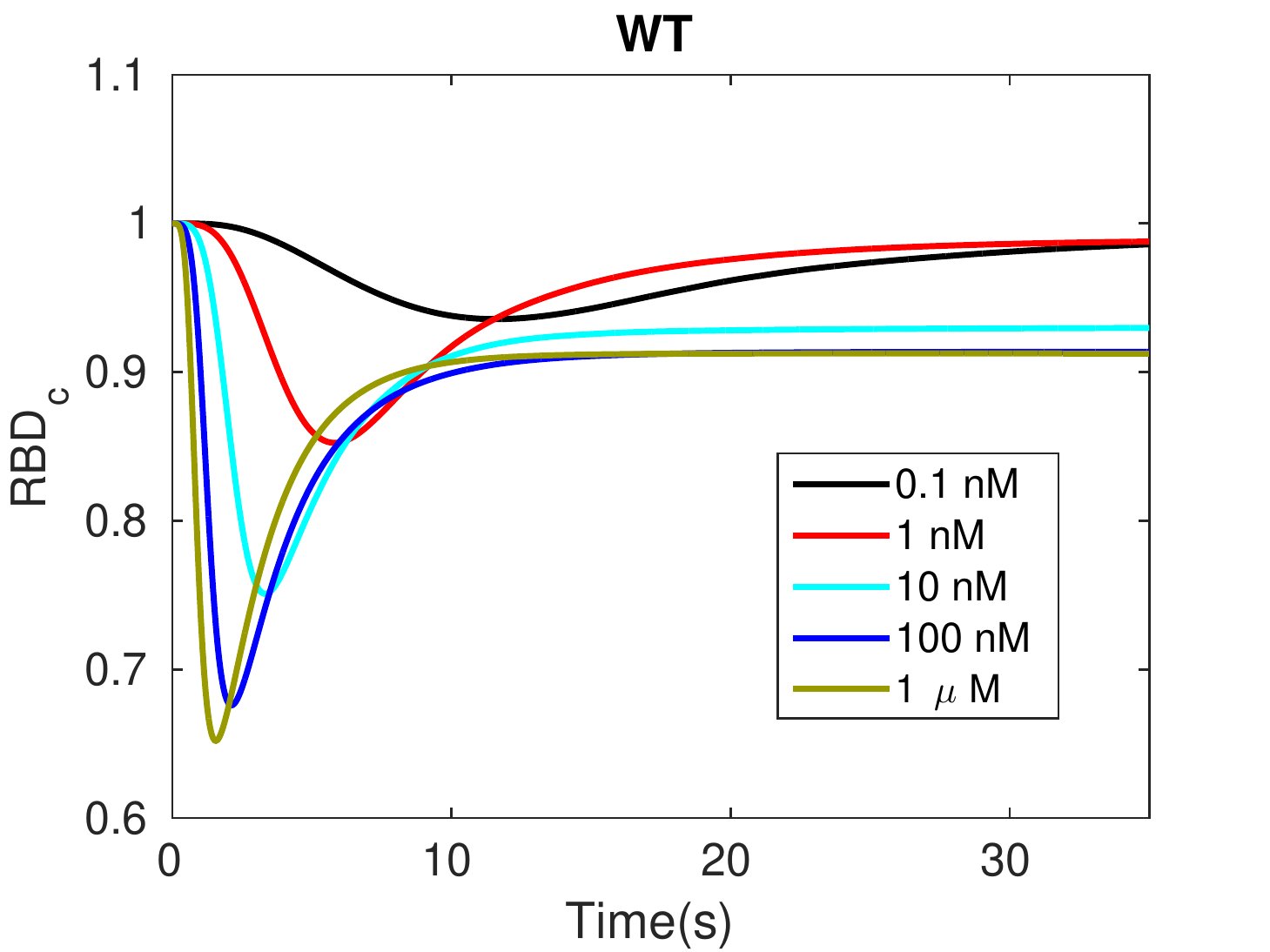}\includegraphics[width=6cm]
{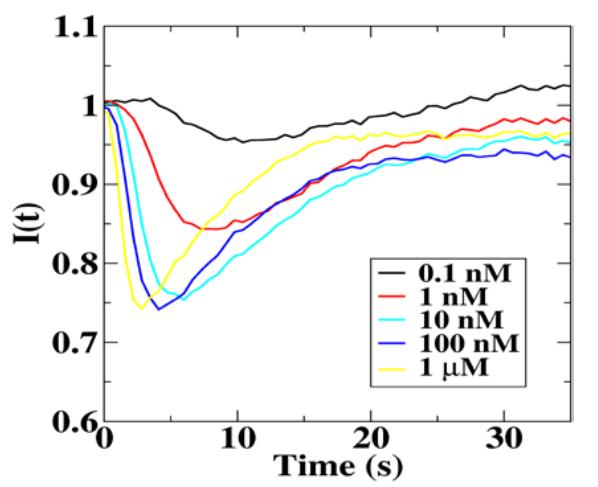}}
\centerline{\hspace*{10pt}\includegraphics[width=6.5cm]
  {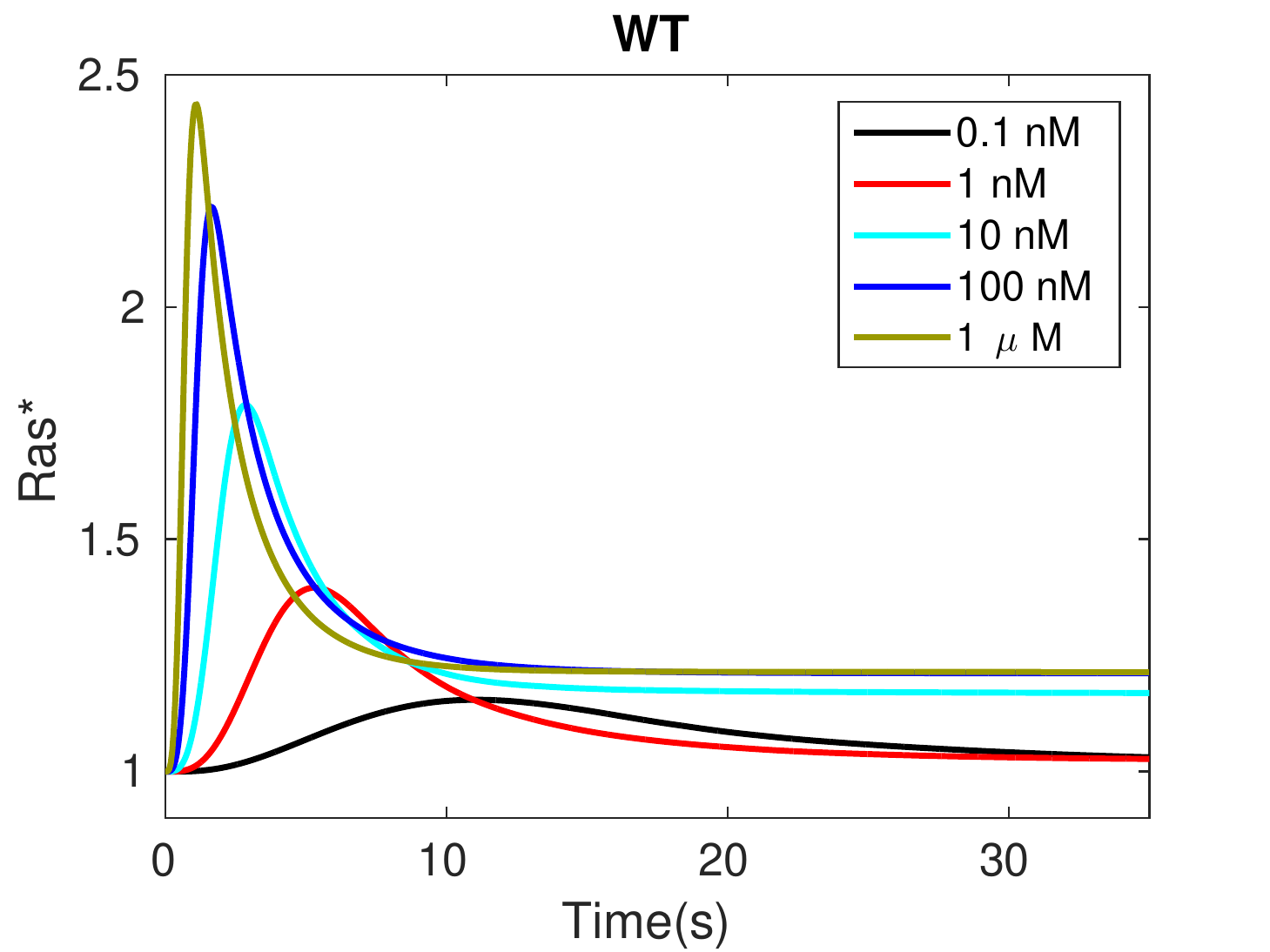}\includegraphics[width=6.5cm]
  {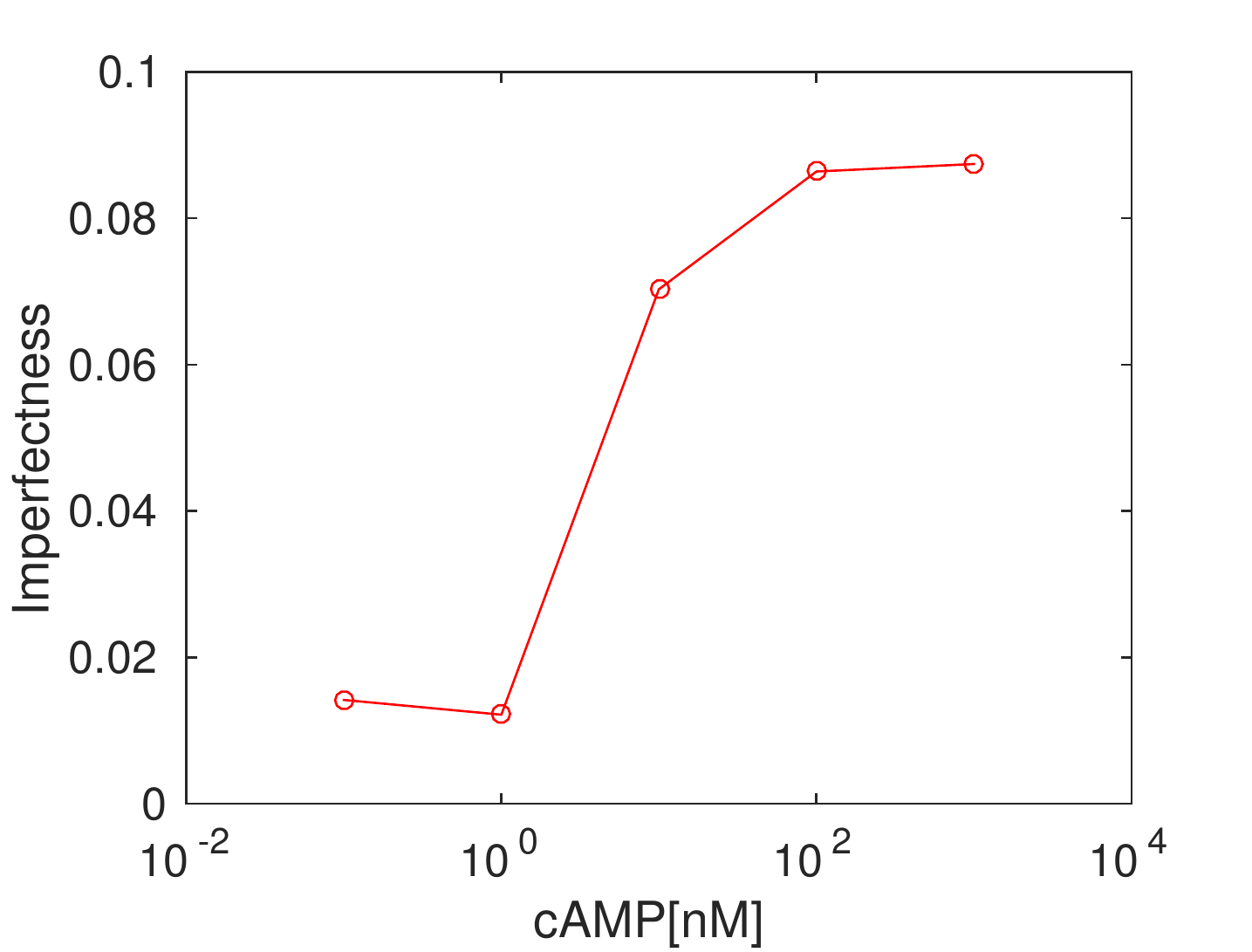}}
\caption{\emph{Top}: uniform stimulation causes a transient decrease
  in the average cytosolic concentration of RBD. WT signifies a  wild type
  cell. Left: simulation; \emph{Right}: experimental measurements from \cite{takeda2012}. \emph{Bottom}:
  transient Ras activation and imperfection of Ras activation, computed as the
  relative difference between the steady state Ras level under stimulus and
  without.}
\label{ras}
\end{figure}

It is suggested in \cite{takeda2012} that adaptation of Ras activity is due to
incoherent feedforward control via activation and inactivation of Ras by RasGEF
and RasGAP, resp. Ras activation is monitored via membrane localization of RBD,
which diffuses freely in the cytosol and is localized at the membrane by binding
to active Ras. The comparison between the experimental results for LatA-treated
cells and the model predictions are shown in the top row of Fig.~\ref{ras}. One
sees that the model captures several basic aspects seen in  the observed Ras
activation.

\begin{itemize}
\item After an increase in cAMP,  RBD rapidly translocates  to the
  membrane and binds  to $Ras^*$ --  whose dynamics are  shown in bottom left of
  Fig.~\ref{ras} -- reaching a maximum in a few seconds. This is followed by a
  more gradual return to the cytosol, where RBD  returns  to approximately its
  basal level.
\item The maximum response increases with  increasing concentrations and saturates
  at about  1 $\mu M$ cAMP.
\item The time  to the   peak of the $Ras^*$  response decreases with increasing cAMP concentration.
\end{itemize}

While perfect adaptation has been confirmed in bacterial gradient sensing
\cite{alon1999}, the experimental evidence in eukaryotes is mixed and sometimes
suggests that only partial adaptation takes place\cite{arrieumerlou2005,
  servant2000, postma2004}. Although it was claimed that the adaptation is
near-perfect in Dicty \cite{takeda2012}, the experimental results in the top
right panel of Fig.~\ref{ras} show that it is not. Imperfect activation is also
reported in \cite{nakajima2014}, and the degree of imperfection is quantified at
various cAMP stimulus levels there. The model also predicts imperfect
adaptation, as shown in the top left panel of Fig.~\ref{ras}, and the deviation
from perfect adaptation is shown in the bottom right panel of that figure.
Both the simulations and experimental measurements show that the deviation from
perfect adaptation increases with the level of stimulation and saturates at
about 100 nM, and in both cases  the relative deviation from
perfect adaptation does not exceed 0.1.

It is suggested in \cite{takeda2012} that the local activator and global
inhibitor of a LEGI model are RasGEF and RasGAP, respectively, and that only
RasGAP diffuses in the cytosol. Our model differs from this at the level of Ras
activation by incorporating a diffusion-translocation-activation mechanism for
both RasGEF and RasGAP. In other words, RasGEF and RasGAP are both globally
supplied through diffusion -- with the same diffusion coefficients -- while only
localization of RasGEF is increased by the locally constrained $G^*_{\alpha}$,
resulting in stronger persistent RasGEF activation. Consequently, RasGAP
activation cannot offset this, even under spatially-uniform stimuli, thereby
inducing imperfect adaptation (see Supporting Information for a theoretical
analysis).

\paragraph{Refractoriness induced by subtle temporal regulation of RasGEF and RasGAP activation}

\begin{figure}[H]
\centerline{\includegraphics[width=10cm,height=5cm]{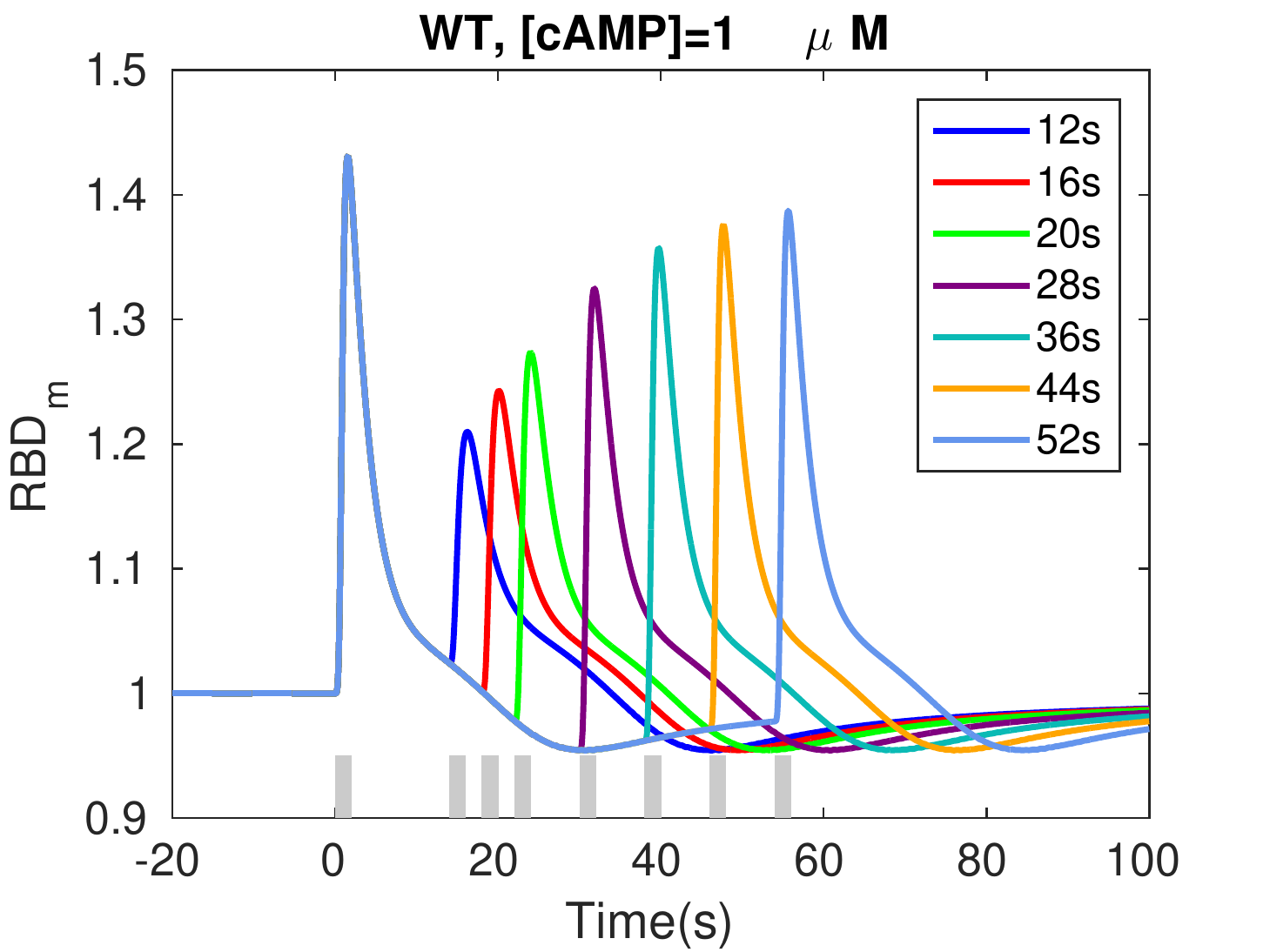}\includegraphics[width=10cm]  {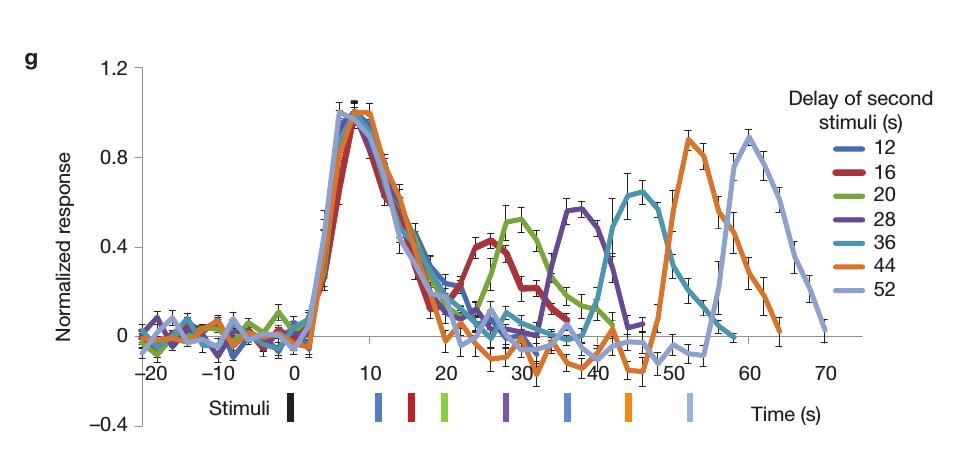}}
\caption{ Refractoriness under
  uniform stimululation. Left: Simulation. The gray bar indicates the duration
  of the stimulus; \emph{Right}: experimental results from \cite{huang2013}.The black
  bar indicates the first stimulus. The other bars are color-coded to show the
  delay. All values are normalized to the peak of the first response.}
\label{refractory}%
\end{figure}

Refractoriness, which is a characteristic of excitable systems, has been
reported for Dicty in \cite{huang2013}.  When two brief large stimuli are
applied to the same cell, the response to the second stimulus depends on the
interval between it and the first, as shown in Figure \ref{refractory} (right),
which suggests the existence of a refractory period. We repeated this experiment
computationally by applying 1 $\mu M$ cAMP stimuli for 2 sec \footnote{The model exhibits a maximal response to this short saturating stimuli, see Supporting Information} separated by
increasing intervals. As shown in Figure \ref{refractory}, refractoriness is
observed and the decrease in the second response decreases as the separation
time increases, consistent with the experimental observations. Moreover, the
peak response with a 52 delay is still weaker than the first response, both in
simulation and experimental measurements, probably due to the fact that Ras does not
adapt perfectly.

\begin{figure}[h!]
\centerline{\includegraphics[width=6cm]
  {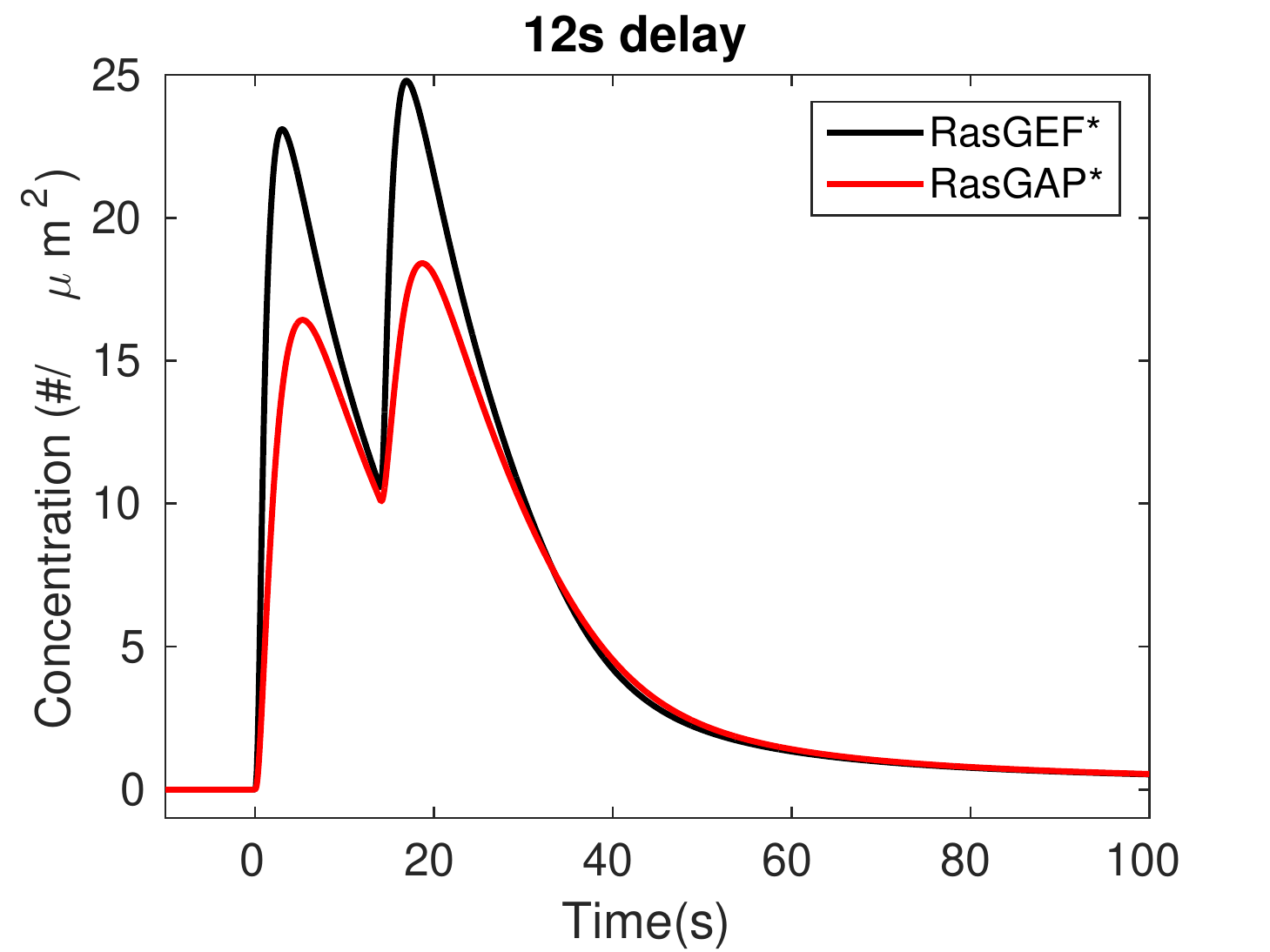}\includegraphics[width=6cm]
  {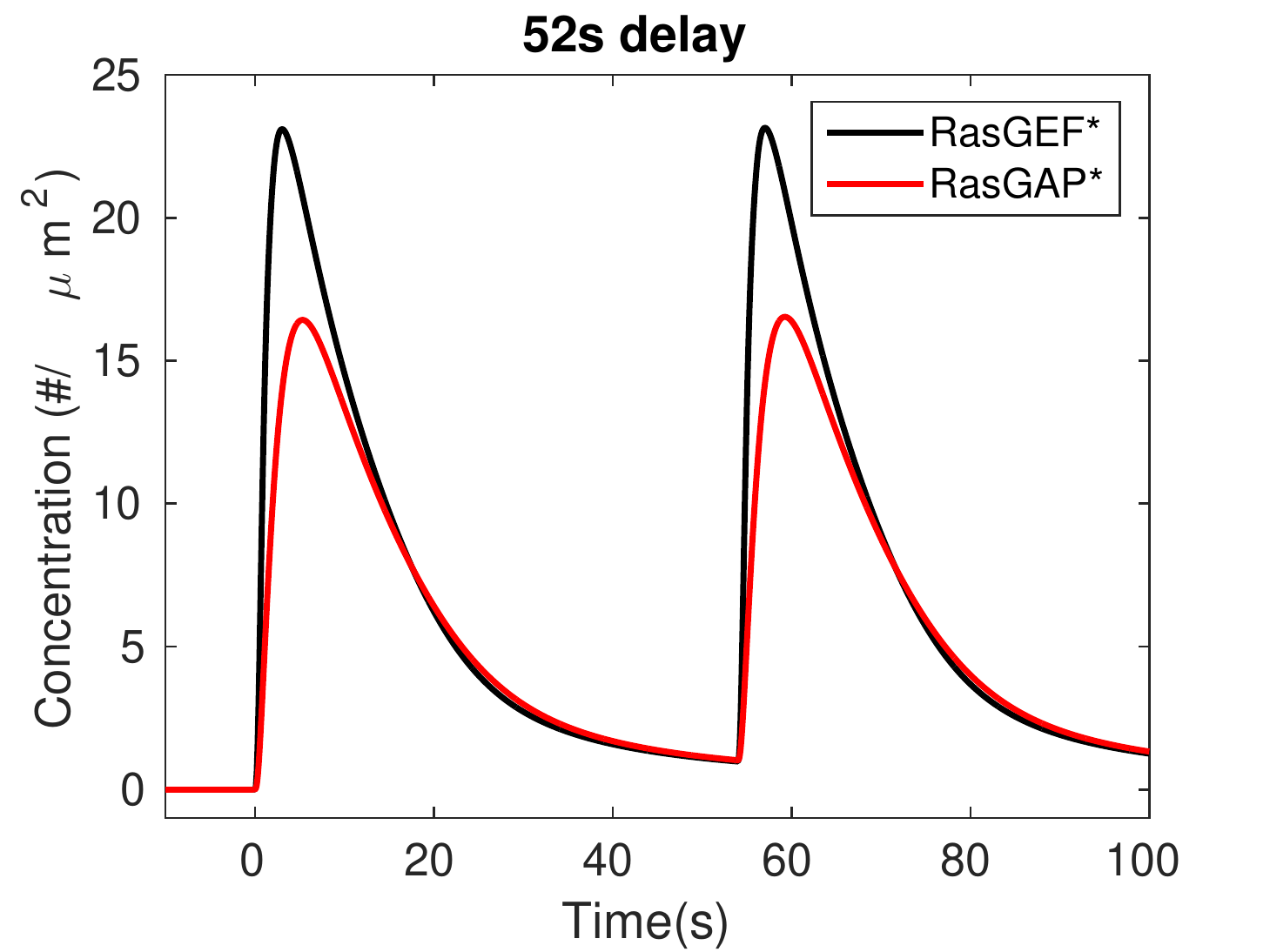}\includegraphics[width=6cm]
  {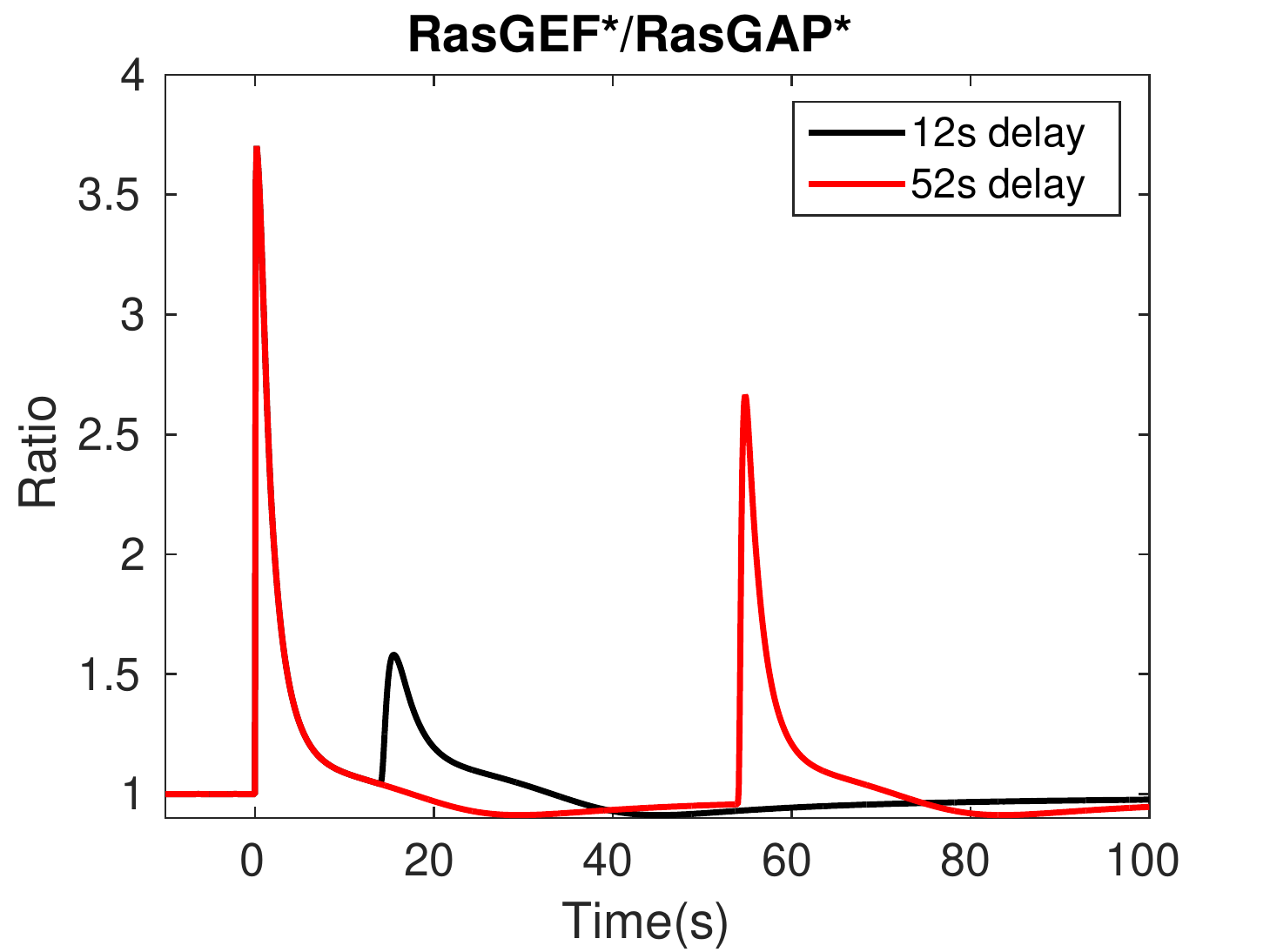}}
\caption{The time courses of $RasGEF^*$ and $RasGAP^*$ for a 12s delay (left)
  and a 52s delay (center).  \emph{Right}: The time course of the
  $RasGEF^*$ /$RasGAP^*$ ratio, which  reaches a peak before the
  two   factors reach their peaks.}
\label{gefgap}
\end{figure}

As to the refractory period, note that under large stimuli large fractions of
RasGEF and RasGAP are activated, and when the duration between the stimuli is
too short, neither RasGEF nor RasGAP can return to prestimulus levels, as shown
by comparison of the left and center panels of Fig.~\ref{gefgap}.  As a result,
the peak ratio of activated RasGEF and RasGAP decreases for short inter-stimulus
intervals as compared with long intervals, as shown in the right panel of
Fig.~\ref{gefgap}. Note that the ratio for a 12 sec interval in
Fig. \ref{gefgap} differs from the corresponding RBD ratio in
Fig.~\ref{refractory} because there is a basal, unstimulated translocation of
RBD to the membrane.\footnote{The refractory periods for non-saturating cAMP
  stimuli are reported in Supporting Information.}

\begin{figure}[H]
\centerline{\includegraphics[width=6cm] {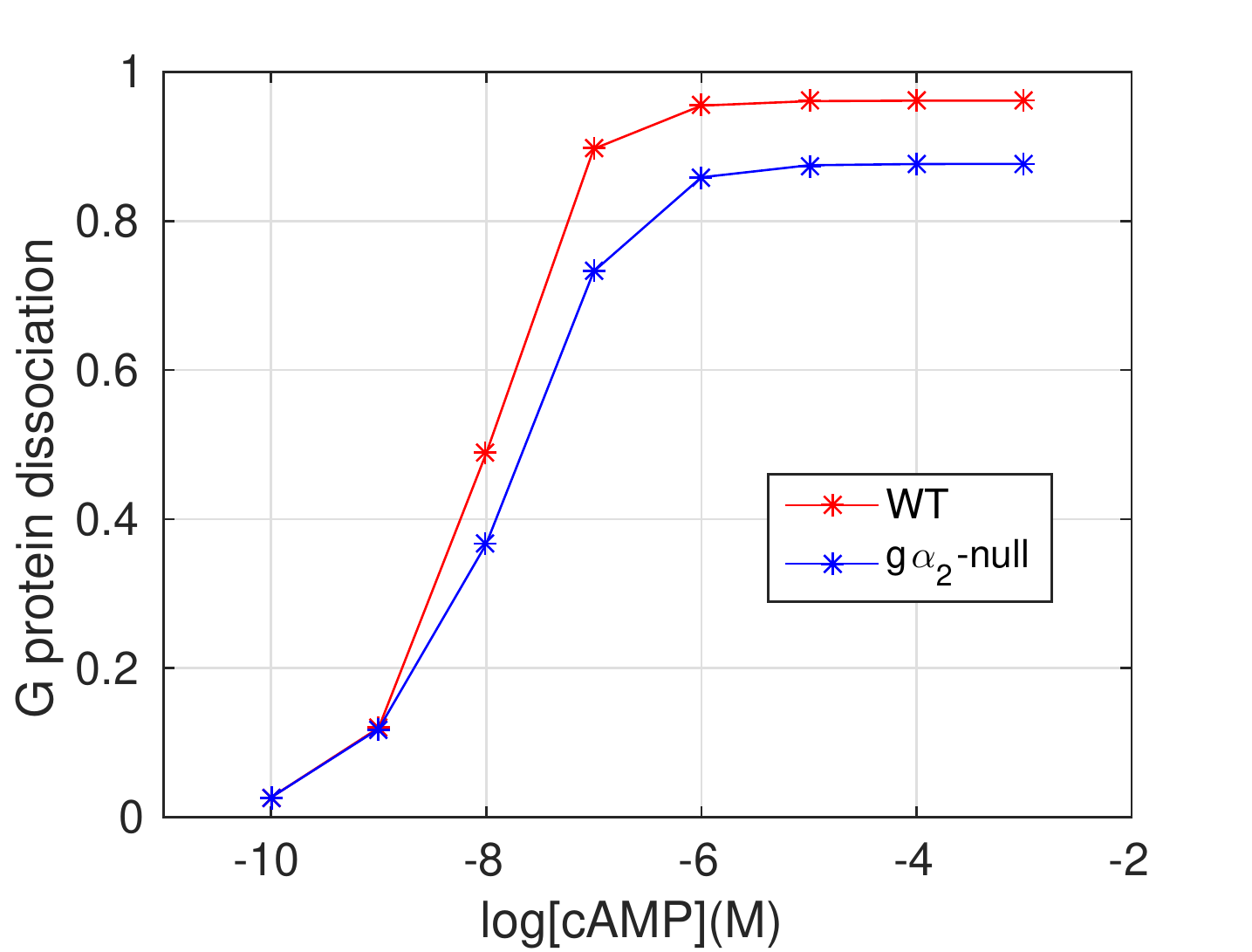}\includegraphics[width=6cm] {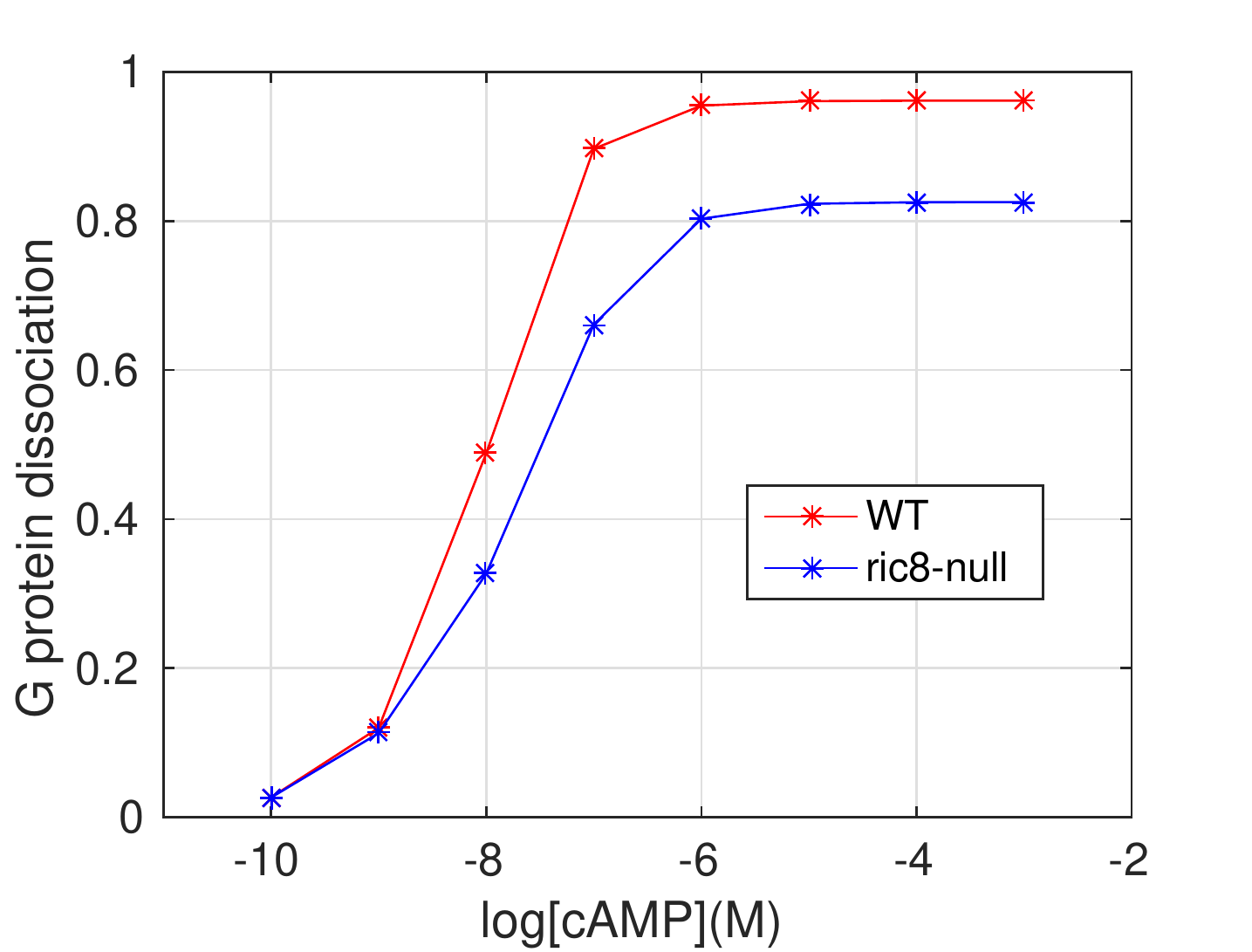}}
\centerline{\includegraphics[width=6cm] {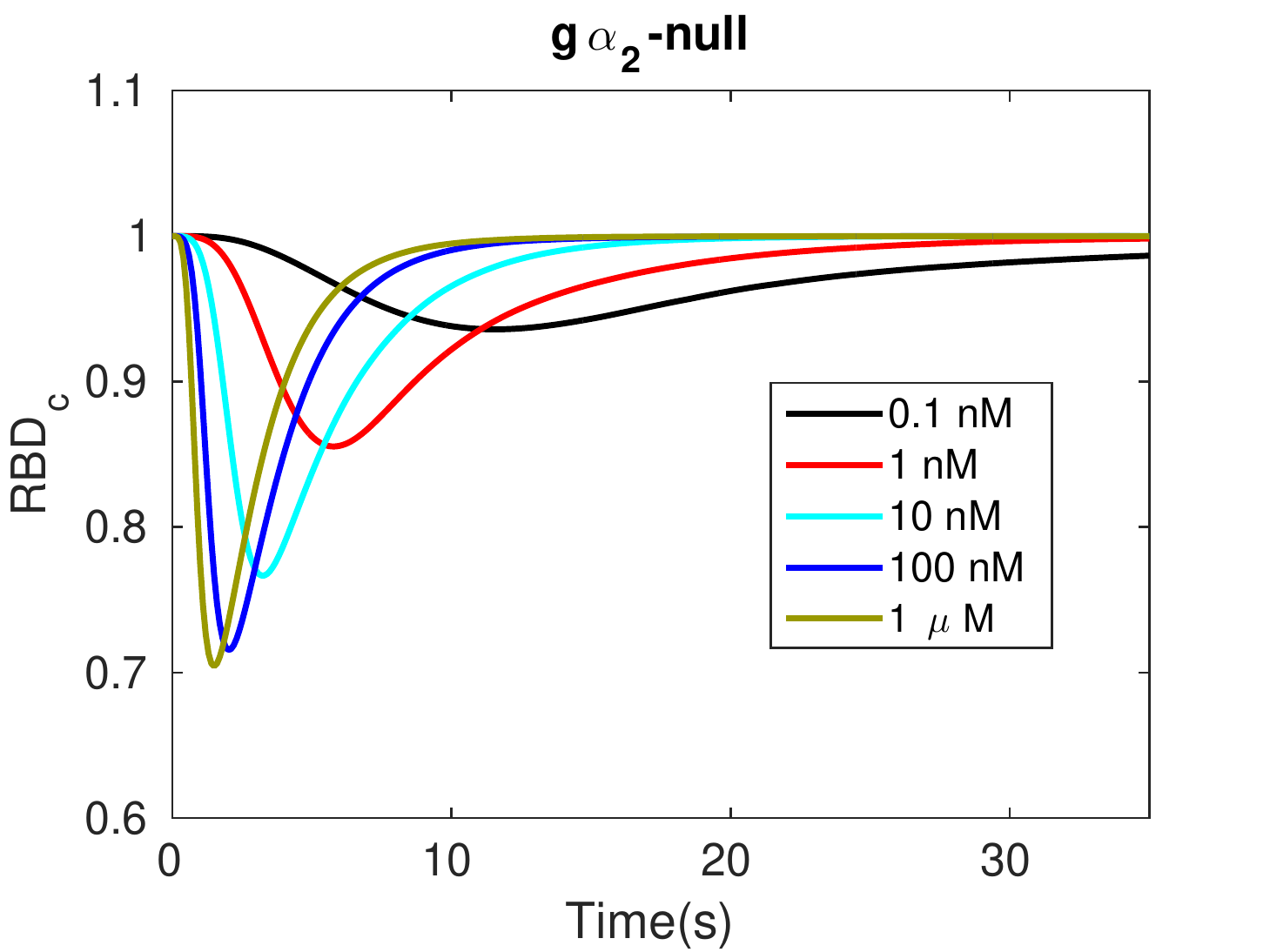}\includegraphics[width=6cm] {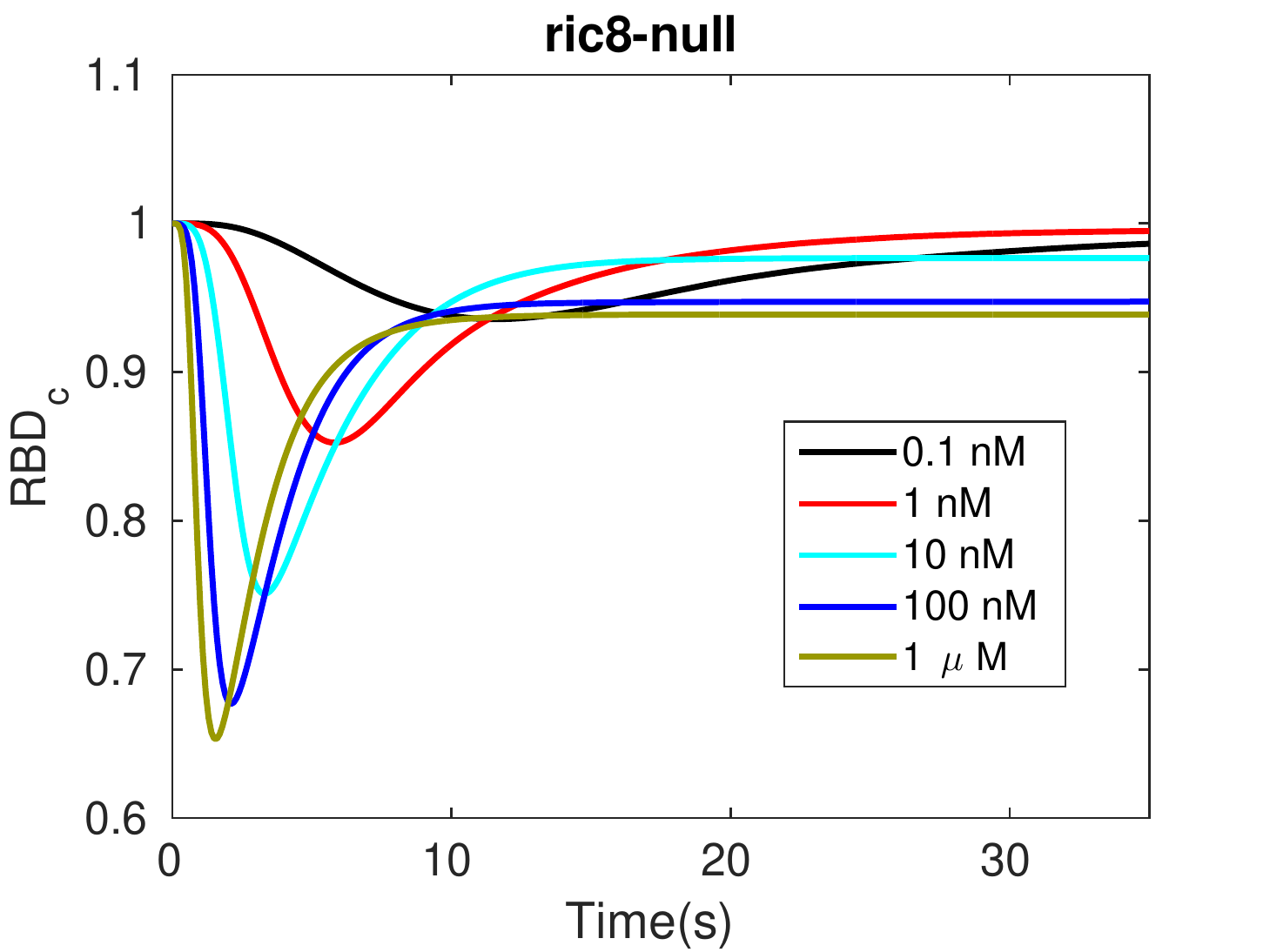}}
\centerline{\includegraphics[width=6cm] {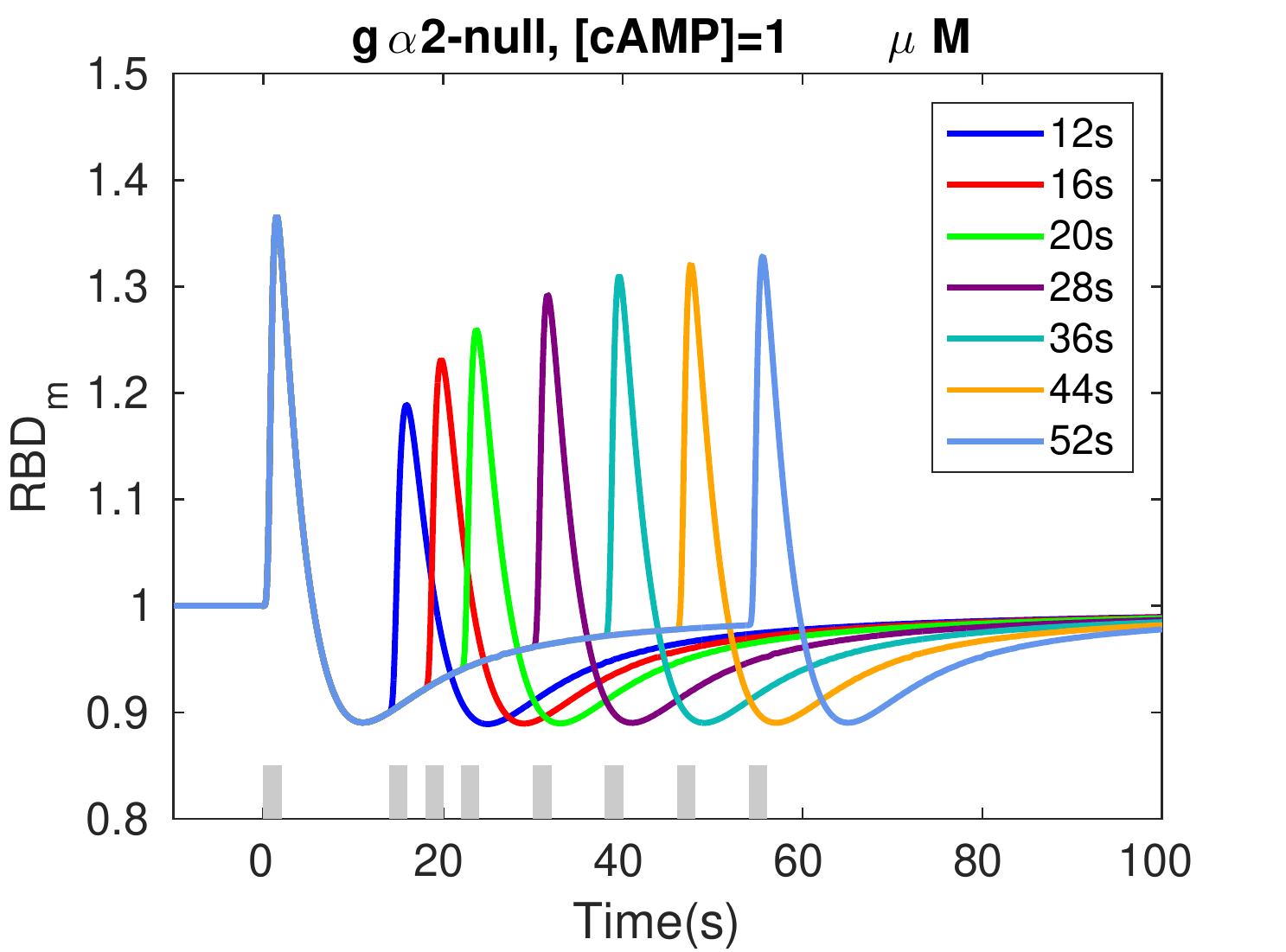}\includegraphics[width=5cm] {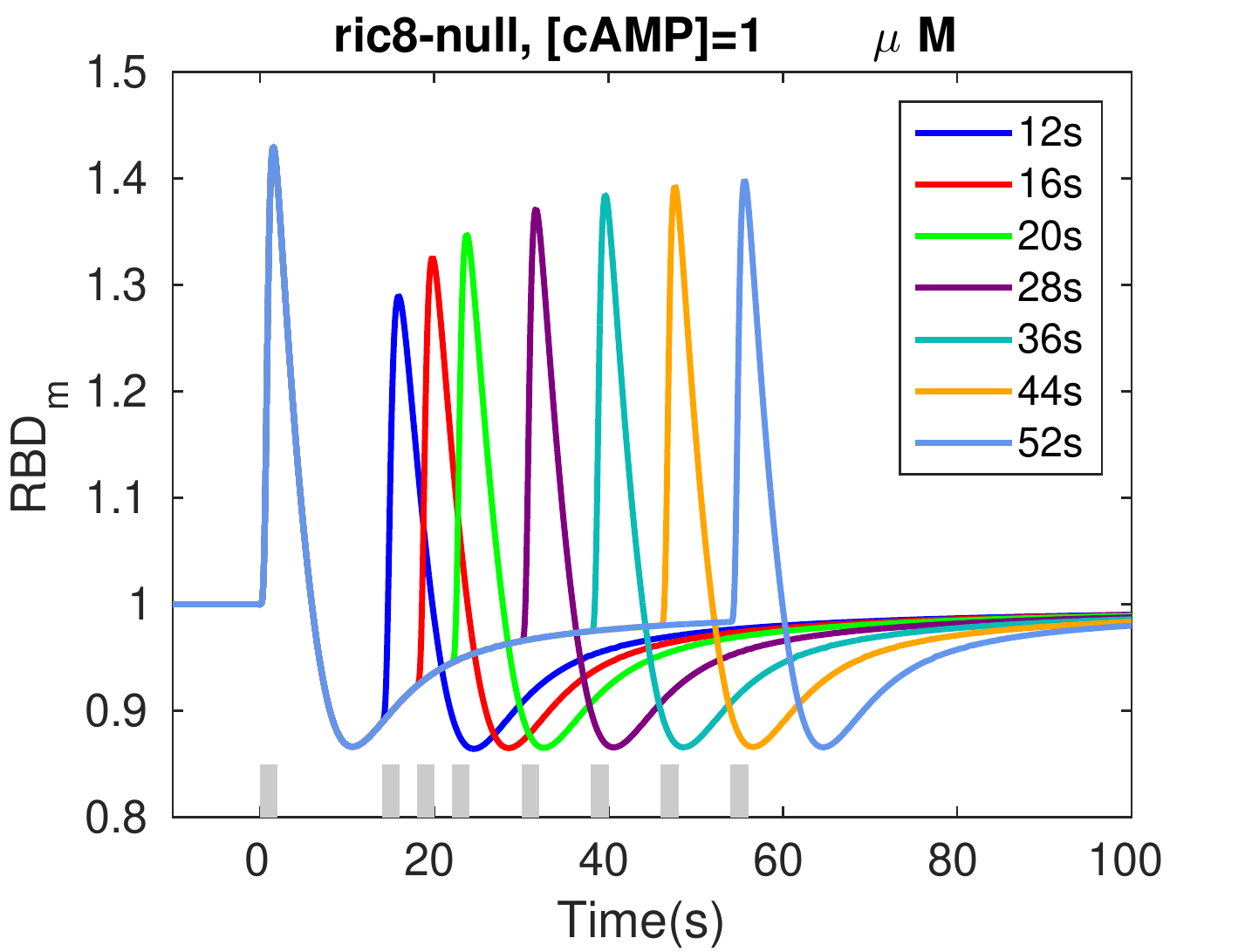}}
\caption{\label{nullcells} \emph{Top}: Dose dependent dissociation at steady state in
  the simulated $g_{\alpha_22}$-null cell and ric8-null cells. \emph{Middle}:
  Time course of RBD dynamics in g$\alpha_2$-null and ric8-null
  cells. \emph{Bottom:} refractoriness in g$\alpha_2$-null and ric8-null cells.}
\end{figure}

\paragraph{$g_{\alpha_2}$-null and ric8-null cells}
To investigate the role of G$_{\alpha_2}$ and Ric8, we simulated
$g_{\alpha_2}$-null cells and ric8-null cells by blocking the $G_{\alpha_2}$-
and Ric8-related pathways, respectively\footnote{In wild type cells, most of the
  G$_{\beta\gamma}$ comes from $G_{\alpha_2\beta\gamma}$, but G$_{\beta\gamma}$
  can also be released from other G proteins \cite{kortholt2013}. To simulate
  the responses in $g_{\alpha_2}$-null cells, we assume for simplicity that
  G$_{\beta\gamma}$ is released from $G_{\alpha_x\beta\gamma}$ and that the
  total amount of $G_{\alpha _x\beta\gamma}$ is the same as
  $G_{\alpha_2\beta\gamma}$ in WT cells. Moreover, we assume that the dynamics
  of $G_{\alpha_x\beta\gamma}$ are the same as for $G_{\alpha_2\beta\gamma}$,
  but that the released G$_{\alpha_x}^*$ subunits do not promote RasGEF and Ric8
  localization. More precisely, we assume that when cAMP binds to the receptor,
  $G_{\alpha_x\beta\gamma}$ dissociates at the same rate as in WT cells, and
  that Ric8 regulates G$_{\alpha_x}^*$ hydrolysis through spontaneous membrane
  localization and G$_{\beta\gamma}$-mediated activation.  G$_{\alpha_x}^*$ and
  G$_{\alpha_x}$ only affect G protein cycling and no other components in the
  network.}.  As shown in the first row of Fig.~\ref{nullcells}, G2 dissociation
decreases in both $g_{\alpha_2}$-null cell and ric8-null cells. Note that since
Ric8 translocation is not enhanced in $g_{\alpha_2}$-null cells, $G_{\alpha_x}$
is reactivated at a lower rate $G_{\alpha_2}$ in wild type cells. Consequently,
$G_{\alpha_x\beta\gamma}$ cycling dynamics is altered and
$G_{\alpha_x\beta\gamma}$ dissociation decreases. Similarly, $ric8$-null cells
also show decreased \G2 dissociation because there is no Ric8 binding to $\Gbg$.

The RBD responses are shown in the second row of
Fig.~\ref{nullcells}. Adaptation is perfect for any physiologically-reasonable
cAMP stimulus in $g_{\alpha_2}$-null cells, and the rate of Ras activation is
initially the same as in WT cells, but the RBD response is less pronounced ({\em
  cf.} Fig.~\ref{ras}). RasGEF activation is weaker in $g_{\alpha_2}$-null cells due
to the absence of G$^*_{\alpha_2}$-promoted RasGEF recruitment, and the
incoherent feedforward circuit in the model guarantees that the activation of
RasGEF and RasGAP are perfectly balanced. Hence perfect adaptation occurs and
the maximum response is reduced compared to that in WT cells, which agrees with
the results in \cite{kataria2013}. For ric8-null cells, one sees that
$ric8$-null cells still exhibit imperfect adaptation, since
$G_{\alpha_2}^*$-promoted RasGEF translocation still occurs, but the
imperfectness is reduced due to the fact that there is no Ric8 available to
reactivate $G_{\alpha_2}$. Simulations show that $ric8$-null cells with a
reduced $G_{\alpha_2}^*$-GTP hydrolysis rate approximately resemble the WT
behaviors (not shown).

The bottom row shows that the refractory response is still observable in both
mutant cells, but the dependence on the time interval is less sensitive compared
with WT cells. For $g_{\alpha_2}$-null cells, the change in the RBD response is
less than 10\% (from $\sim$ 1.11 to $\sim$ 1.2) when the interval ranges from 12
to 52 seconds, compared with a 20\% change (from $\sim$ 1.2 to $\sim$ 1.4) in WT
cells (Fig.~\ref{refractory}). There is less than a 10\%
difference in maximum response between a 12s interval and a 52s interval ( right
panel, from $\sim $ 1.3 to $\sim$ 1.38) for ric8-null cells.

\subsection*{The response under a graded stimulus}

Next  we investigate how cells respond to a linear cAMP gradient along the
x-axis, which we define as follows.
$$
C(x,y,z)= \dfrac{\Delta C}{10}\cdot ( x-x_r)  + C_r
$$
where $C(x,y,z)$ is the cAMP concentration on the membrane at $(x,y,z) \in
S^2_5$ (a sphere of radius 5),  $\Delta C \equiv C_f -C_r$, and subscripts $f$ and $r$ denote the
points (5,0,0) (the 'front') and (-5,0,0) (the 'rear').

\paragraph{Biphasic Ras activation in LatA-treated cells}
It was shown in \cite{kortholt2013} that spatially-localized stimuli lead to
three phases of Ras activation.  In the first, which is transient, Ras is
activated on the entire membrane, and this phase requires $\Gbg$ and exists in
$g_{\alpha_2}$-null cells. The second phase is symmetry breaking, in that Ras is
only activated at the side of the cell facing the higher cAMP concentration, and
this phase requires $G_{\alpha_2}$. The third phase is confinement, wherein the
crescent of activated Ras at the front half of the cell localizes to a small
area around the high point of the gradient. The first two phases are observed in
LatA-treated cells, but the third phase requires actin polymerization.  Since
the model is based on LatA- pretreated cells, we only test whether it exhibits
the first two phases of Ras activation.
\begin{figure}[H]
\centerline{\includegraphics[width=7cm] {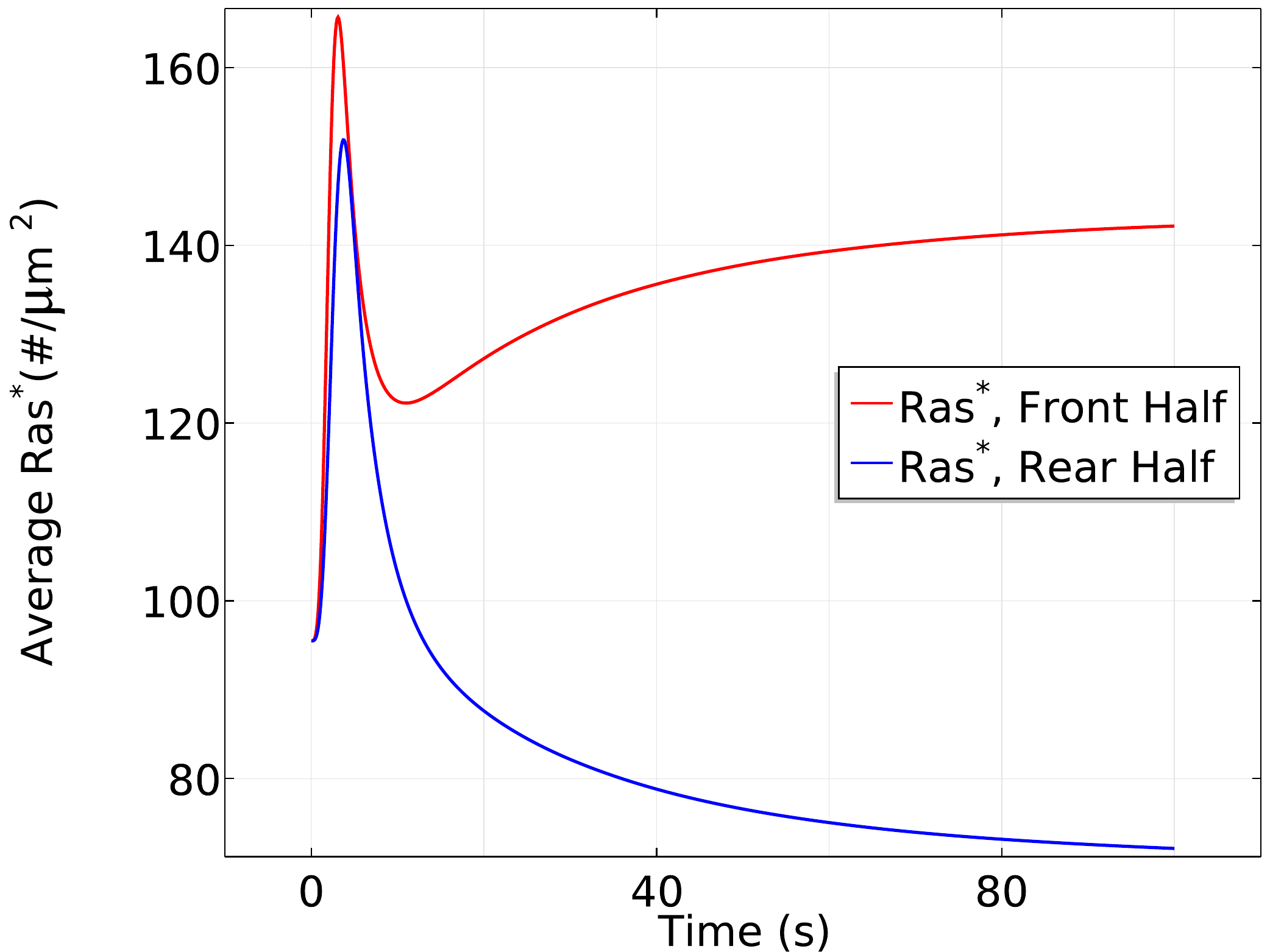}\hspace*{15pt} \includegraphics[width=7cm]{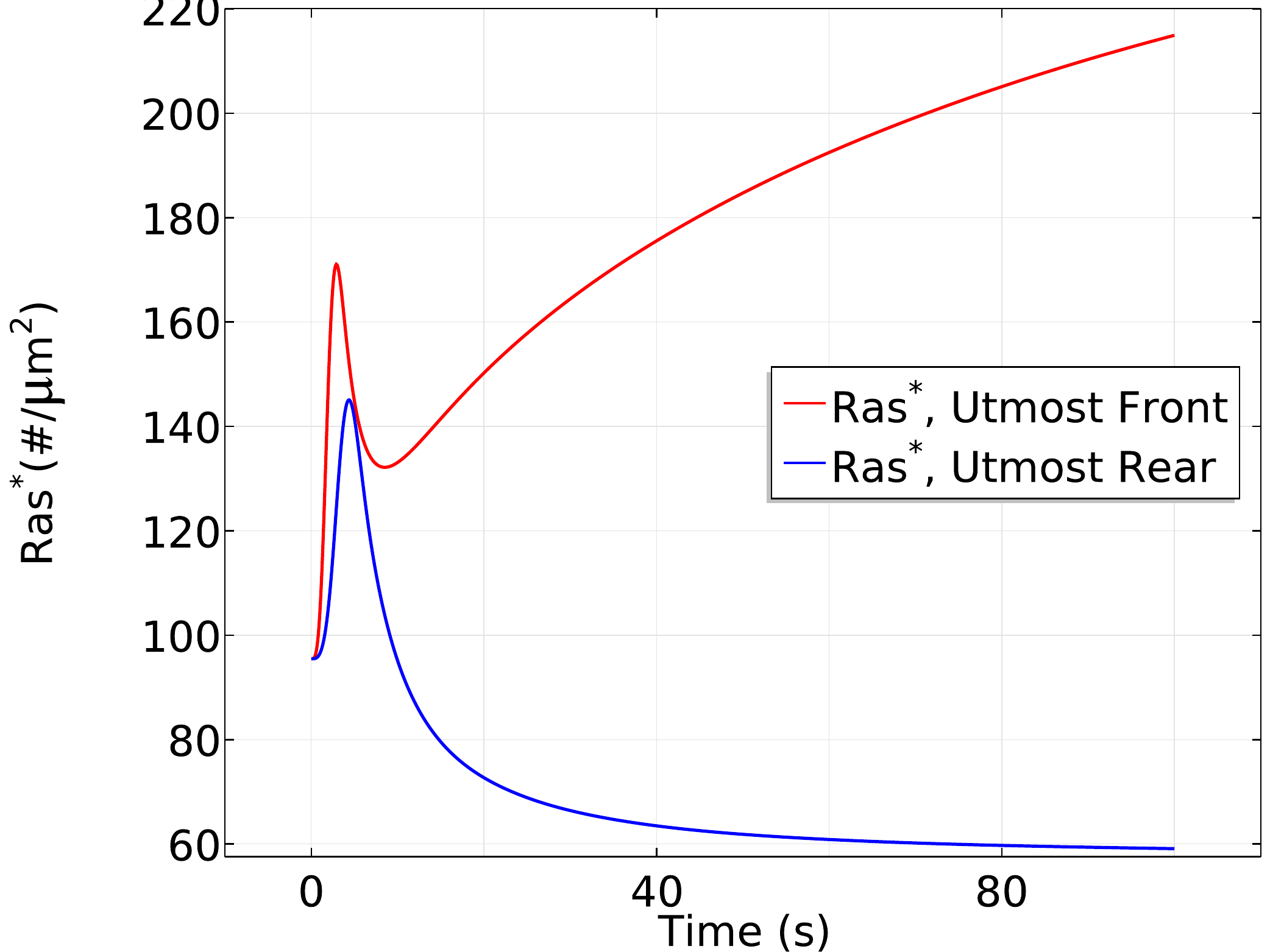}}
\caption{ \emph{Left}: The time course of average $Ras^*$ activity
  in a cAMP gradient defined by $C_f = 10$ nM and $C_r$ = 1nM.
  \emph{Right}: The  $Ras^*$ activity in the same gradient at $x_f$ and $x_r$.}
\label{symmetrybreaking}
\end{figure}
Fig.~\ref{symmetrybreaking} (left) shows that the initial response is transient
activation of Ras on the entire boundary, which is completed in $\sim\!10$\,s,
followed by a pronounced asymmetric activation pattern\footnote{Here and
  hereafter we display the average of various species at  the front and rear
  halves  of a cell because this is how experimental results are reported.}.  In the second phase
 Ras is reactivated exclusively at the front half of the cell,
where the peak Ras$^*$ activation is roughly twice that at the rear, which
reflects the difference in receptor occupancy and G protein activation.  Thus
symmetry breaking occurs in this phase, which is stabilized at around
$t=100$\,s. The biphasic  behavior in a cAMP gradient is even more pronounced in a
time plot of $Ras^*$ at the antipodal points of the gradient, as shown in the
right panel of Fig.~\ref{symmetrybreaking}.

The critical components that give rise to the biphasic response are several
globally diffusing molecules ($\Gp$, $\Gbg$, $Ric8$, $RasGEF$ and $RasGAP$) and
localized  G$^*_{\alpha_2}$.  The sequence of events following application of the
graded stimulus is as follows.

\begin{itemize}
\item [(i)]   \G2 dissociation is
  higher at the front, resulting in more G$_{\beta\gamma}$ there initially
  (Fig.~\ref{GEFGAP} (left)), but $\Gbg$ can diffuse in the cytosol, which  reduces
  the spatial difference. A similar difference applies to $G_{{\alpha}_2}^*$, but it
  remains membrane-bound.

\item [(ii)] G$_{\beta\gamma}$ activates RasGEF faster than RasGAP everywhere
  (Fig.~\ref{GEFGAP} (right) for $0<t\leq 10s$) which favors the activation of
  Ras. Because the dissociation of G$_{\alpha\beta\gamma}$ is higher at the
  front $Ras^*$ increases faster there and induces a higher maximum.
\begin{figure}[H]
\centerline{\includegraphics[width=7cm]
  {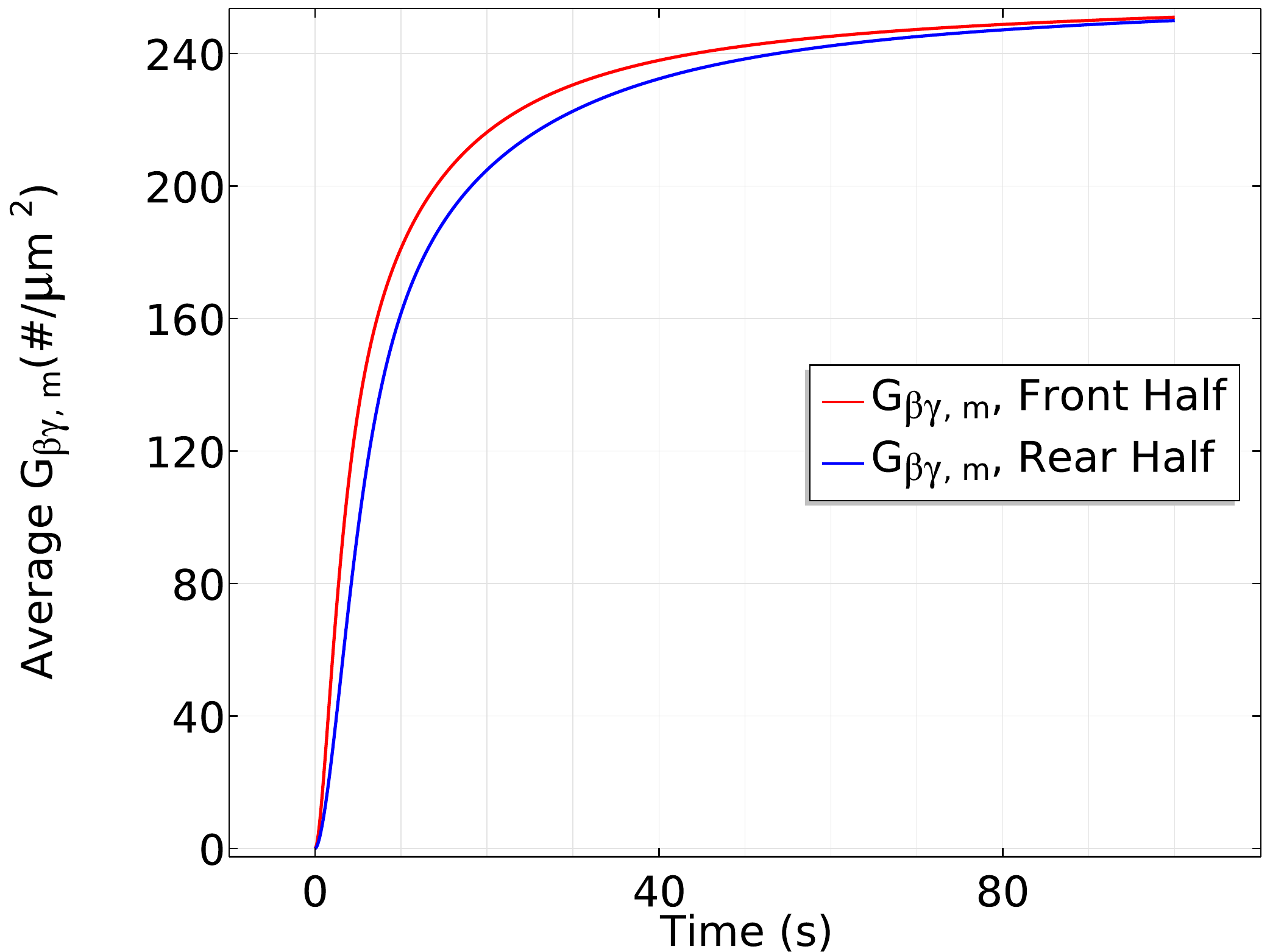}\hspace*{15pt}\includegraphics[width=7cm]
  {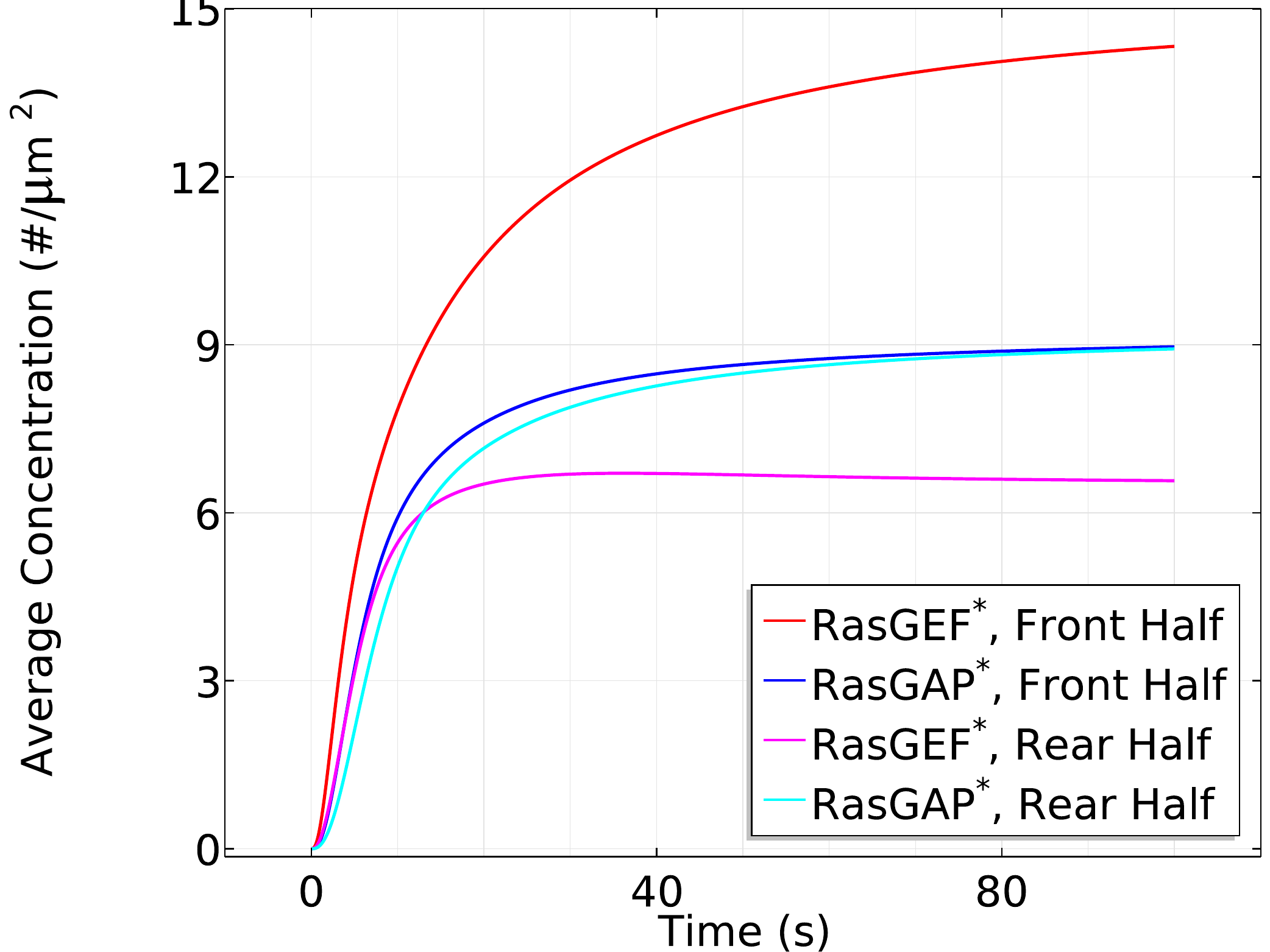}}
\caption{Left: The time course of membrane $\Gbg$ at the front and rear halves
  of the cell in the cAMP gradient used in Fig.~\ref{symmetrybreaking}.  \emph{Right}: The time cource of
  $RasGEF^*$ and $RasGAP^*$ activity in the same cAMP gradient at the front half
  and rear half.}
\label{GEFGAP}
\end{figure}
\item [(iii)]   $RasGAP^*$ activation increases  on a slower time scale,
  resulting in a decrease of $Ras^*$ everywhere. However, the localization of
  $G_{{\alpha}_2}^*$ at the membrane enhances translocation of RasGEF from the
  cytosol to the membrane, and this is higher at the front than at the rear
  (Fig.~\ref{Gaa} (left)). This leads to higher RasGEF activation at the front
  (Fig.~\ref{GEFGAP} (right)), which offsets the Ras deactivation due to
  $RasGAP^*$, and reactivation of Ras occurs.

\item [(iv)] At the same time, the nonuniform distribution of $G_{{\alpha}_2}^*$
  on the membrane induces a nonuniform localization of Ric8.  Although diffusion
  of $\Gbg$ tends to equalize Ric8 activation, this is offset by the difference
  in the distribution of $G_{{\alpha}_2}^*$ (Fig.~\ref{Gaa}
  (right)). Consequently, $G_{{\alpha}_2}$ is reactivated at the front of the
  cell, which further promotes RasGEF localization at the front. Moreover, the
  asymmetrical $G_{{\alpha}_2}$ reactivation generates an asymmetrical \G2
  reassociation profile -- less reassociation at the front and more at the
  rear. As a result, diffusion of $G_{\alpha\beta\gamma}$ that re-associated at
  the rear provides a source of $G_{\alpha\beta\gamma}$ needed at the front,
  which further contributes to symmetry breaking.
\begin{figure}[H]
\centerline{\includegraphics[width=7cm] {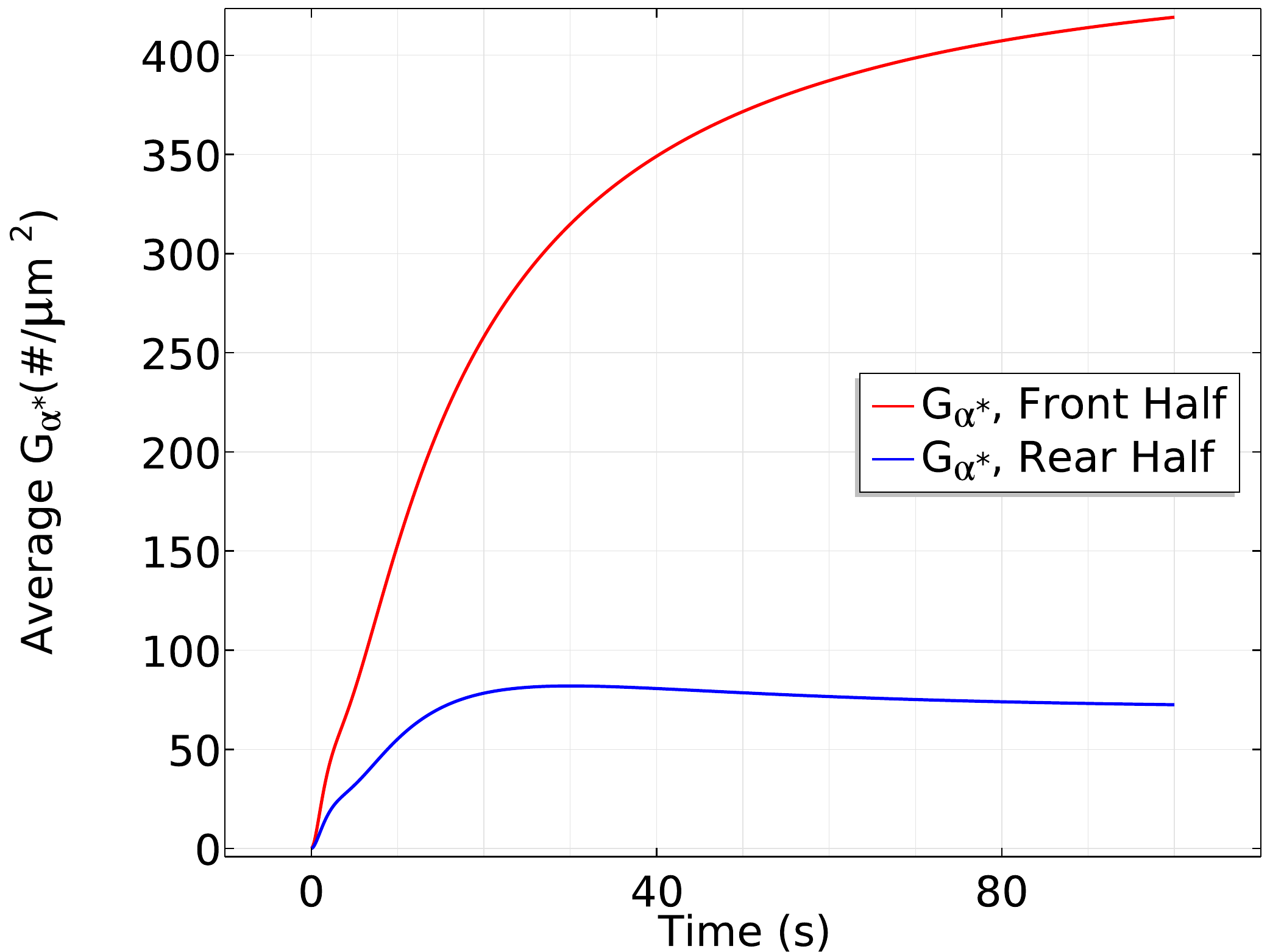}\hspace*{15pt}\includegraphics[width=7cm] {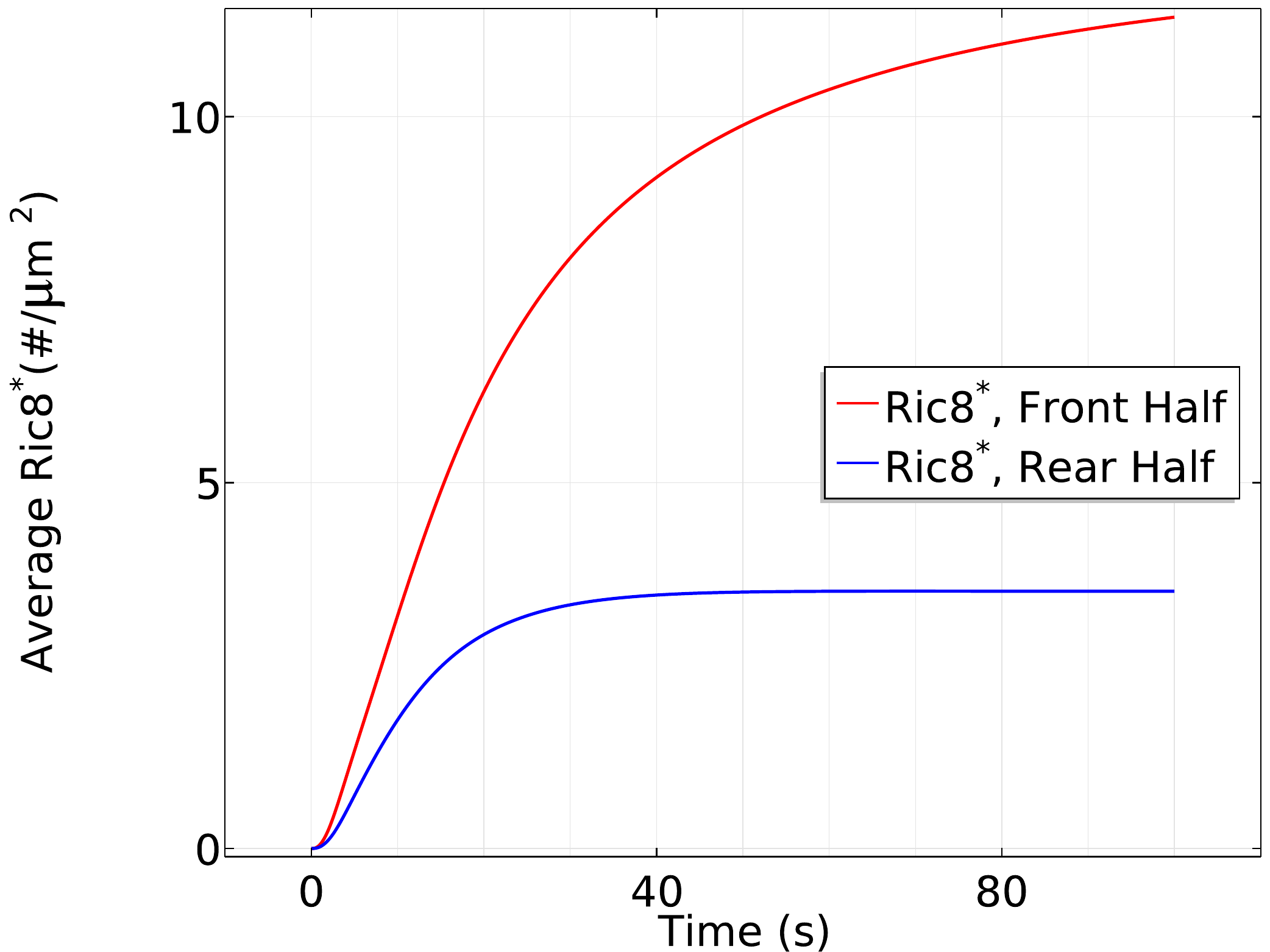}}
\caption{ Left: The time course of membrane $G^*_{\alpha}$ at the front half and
  rear half in the cAMP gradient used in Fig.~\ref{symmetrybreaking}. The
  gradient generates a difference of $G^*_{\alpha}$ concentrations, which is
  amplified further by Ric8. \emph{Right}: The time course of membrane $Ric8^*$ at the
  front half and rear half in the same cAMP gradient.}
\label{Gaa}
\end{figure}

\item [(v)] Note that the cAMP gradient introduces a larger sink of
  G$_{\alpha\beta\gamma}$ and a larger G$_{\beta\gamma}$ concentration at the
  front initially, but the diffusion of G$_{\alpha\beta\gamma}$ guarantees the
  continuous supply at the membrane as long as saturation is not
  reached. Moreover, the distribution of G$_{\beta\gamma}$ is essentially
  uniform on the membrane and within the cytosol (Fig.~\ref{GEFGAP} (left))
  after $\sim\!100s$, as was reported in \cite{jin2000}. This eventually leads
  to a uniform distribution of $RasGAP^*$ at the entire cell boundary, but
  $RasGEF^*$ is higher at the front due to the asymmetrical recruitment of
  RasGEF from the cytosol. Ras activity at the rear of the cell decreases below
  the prestimulus level because the $RasGAP^*$ activity offsets the $RasGEF^*$
  activity there.
\end{itemize}
In summary, the fast time scale of $\Gbg$-mediated RasGEF and RasGAP activation
induces the first transient Ras activation on the entire membrane, while the
slow time scale of overall equilibration (redistributions due to diffusion and
membrane localization) induces the delayed secondary response that produces the
symmetry breaking.

\paragraph{The effects of diffusion}

The results in the previous section suggest that diffusion plays an important
role in inducing the biphasic response. To investigate this, we do simulations
in which the diffusion coefficients of $\Gbg$, RasGEF/GAP, G2, and Ric8, all of
which are present in the cytosol and diffuse, are individually set to $0.003 \mu
m^2/s$ ($10^{-5}$ of the normal value) and compare the Ras response with that in
WT cells.
\begin{itemize}
\item Slow G$_{\beta\gamma}$ diffusion

  In the absence of apparent $\Gbg$ diffusion after dissociation, localized
  $\Gbg$ leads to highly polarized activation of RasGEF
  (Fig.~\ref{nogbdiffusionGEFGAP} (left)). Correspondingly, in the transient
  activation phase the peak value of $Ras^*$ at the rear half is the same as in
  WT cells, but the peak at the front half increases from $\sim 165$ $\#/\mu
  m^2$ to $\sim 172$ $\#\mu m^2$ (cf.~the right panel of
  Fig.~\ref{nogbdiffusionGEFGAP} and the left panel of
  Fig.~\ref{symmetrybreaking}). Moreover, RasGAP activity is polarized (cf.~the
  left panels of Fig.~\ref{nogbdiffusionGEFGAP} and Fig.~\ref{GEFGAP}), causing
  a stronger $Ras^*$ deactivation at the front. Hence we observe a slightly
  reduced steady state response ($\sim\!140$ $\#/\mu m^2$ v.s. $\sim\!143$ $\#/\mu
  m^2$) in the front half during the symmetry breaking phase of Ras
  activation. It is not surprising that the reduced  $\Gbg$ diffusion  still
  captures the biphasic behavior in the sense that $G_{{\alpha}_2}^*$ is still
  polarized and its downstream pathways are minimally affected. Although
  $RasGAP^*$ varies along the cell perimeter, it is counterbalanced by a
  stronger polarized $RasGEF^*$ (Note that both $RasGEF^*$ and $RasGAP^*$ at the
  front in the left panel of Fig.~\ref{nogbdiffusionGEFGAP} are much larger than
  the ones in the right panel of Fig.~\ref{GEFGAP}).
\begin{figure}[H]
\centerline{\includegraphics[width=7cm]
{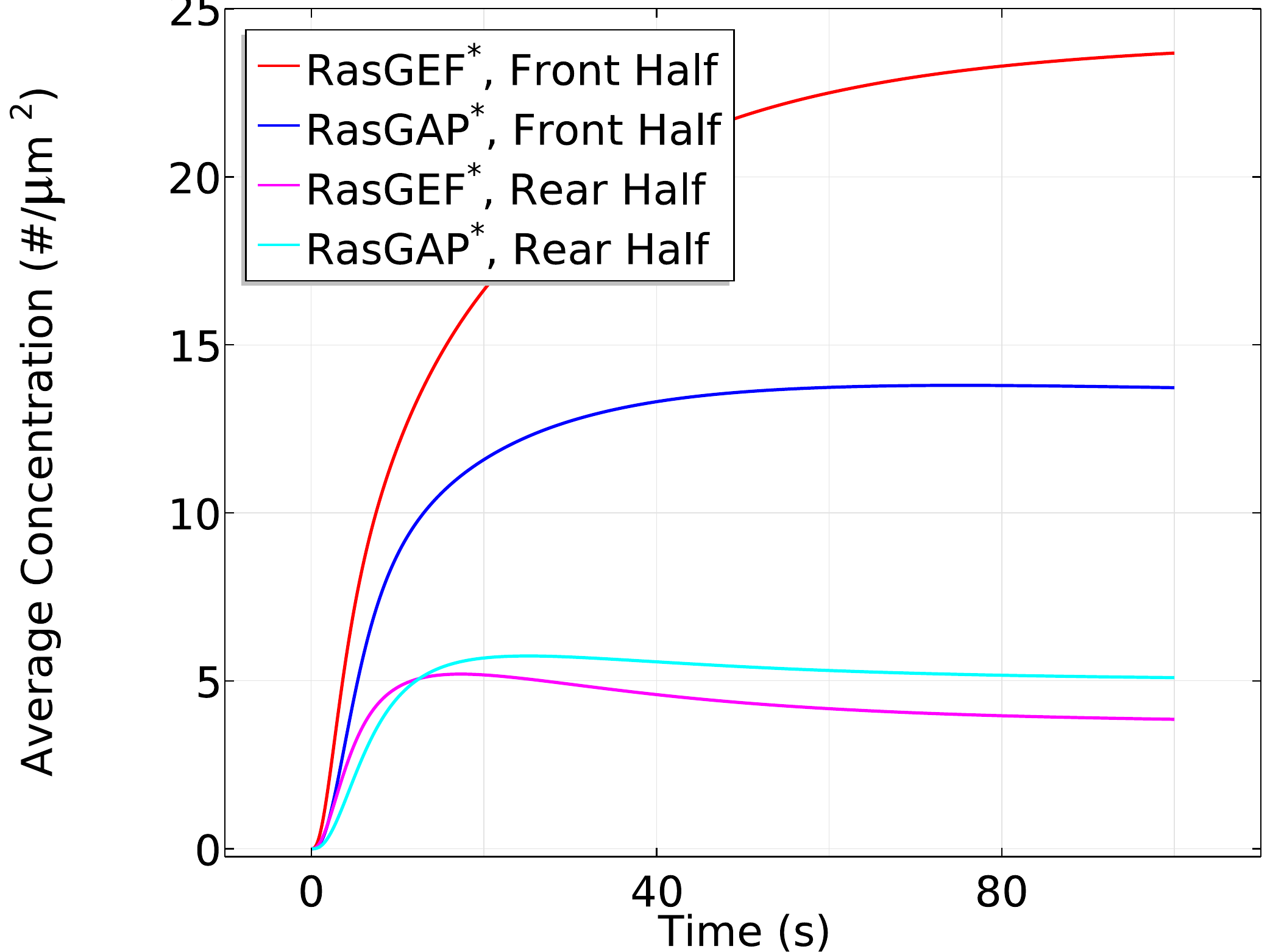}\hspace*{15pt}\includegraphics[width=7cm]
{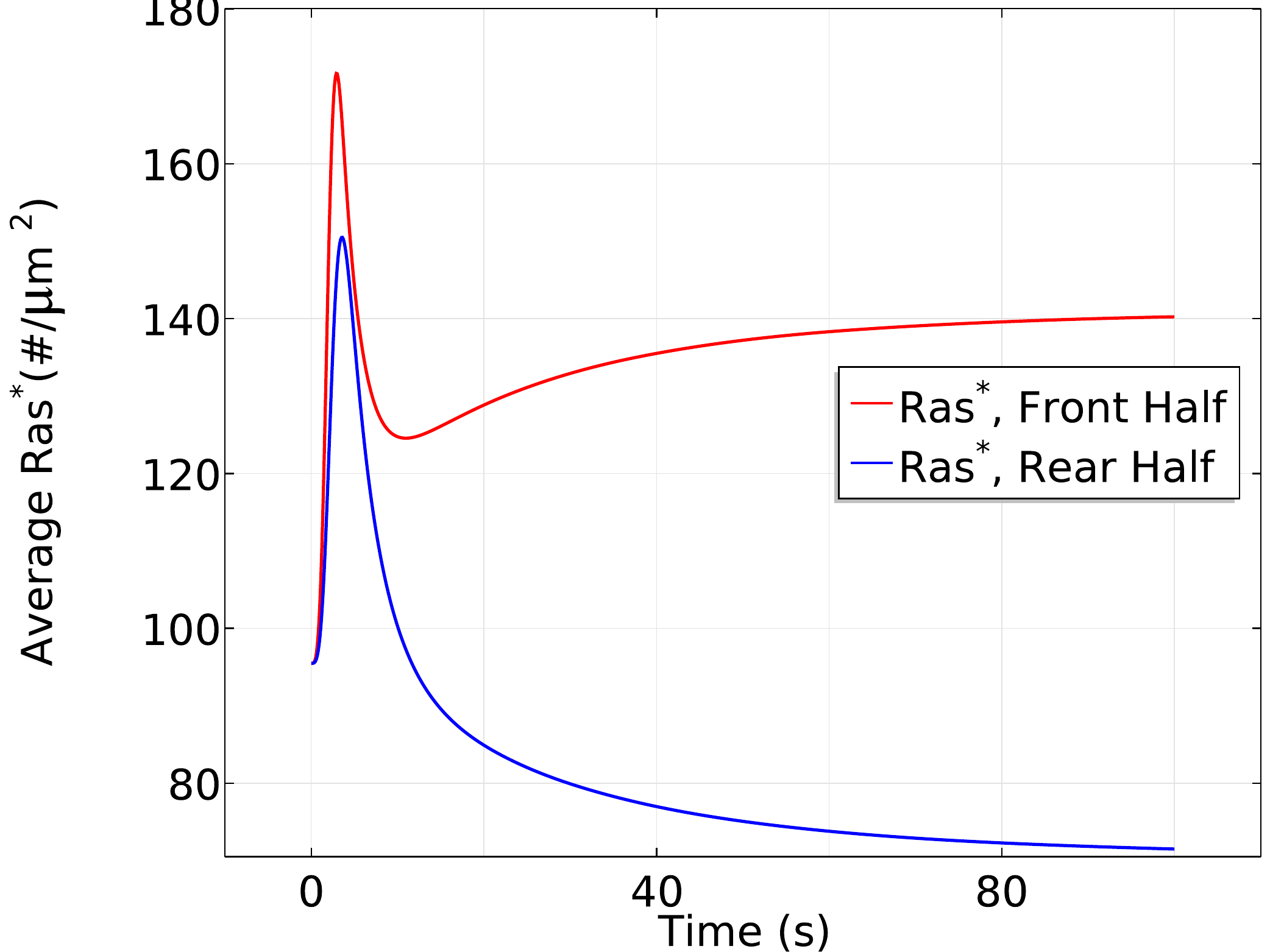}}
\caption{\emph{Left}: The time course of $RasGEF^*$ and $RasGAP^*$ activity at the front and rear halves in the absence of apparent G$_{\beta\gamma}$ diffusion in the same gradient as previously used. \emph{Right}: The time course of average $Ras^*$ activity in the absence of apparent G$_{\beta\gamma}$ diffusion in that gradient.}  \label{nogbdiffusionGEFGAP}
\end{figure}
%

\item Slow RasGEF diffusion: the necessity of `activator' diffusion

  The supply of RasGEF is localized on the membrane when RasGEF
  diffuses slowly, since $G_{{\alpha}_2}^*$ can only attract very limited RasGEF
  from the cytosol very close to the membrane. Moreover, diffusion of $\Gbg$
  ensures an almost uniform RasGAP and RasGEF activity at the front and the rear
  at steady state. Consequently, we observe that both the front and rear half of
  the cell adapts to the cAMP gradient and there is no Ras reactivation at the
  front due to limited availability of RasGEF, as shown in
  Fig.~\ref{nogefdiffusionGEFGAP}. The front settles down at a slightly higher
  level of $Ras^*$ comparing to the rear due to a slightly stronger RasGEF
  activity.
\begin{figure}[H]
\centerline{\includegraphics[width=7cm]
  {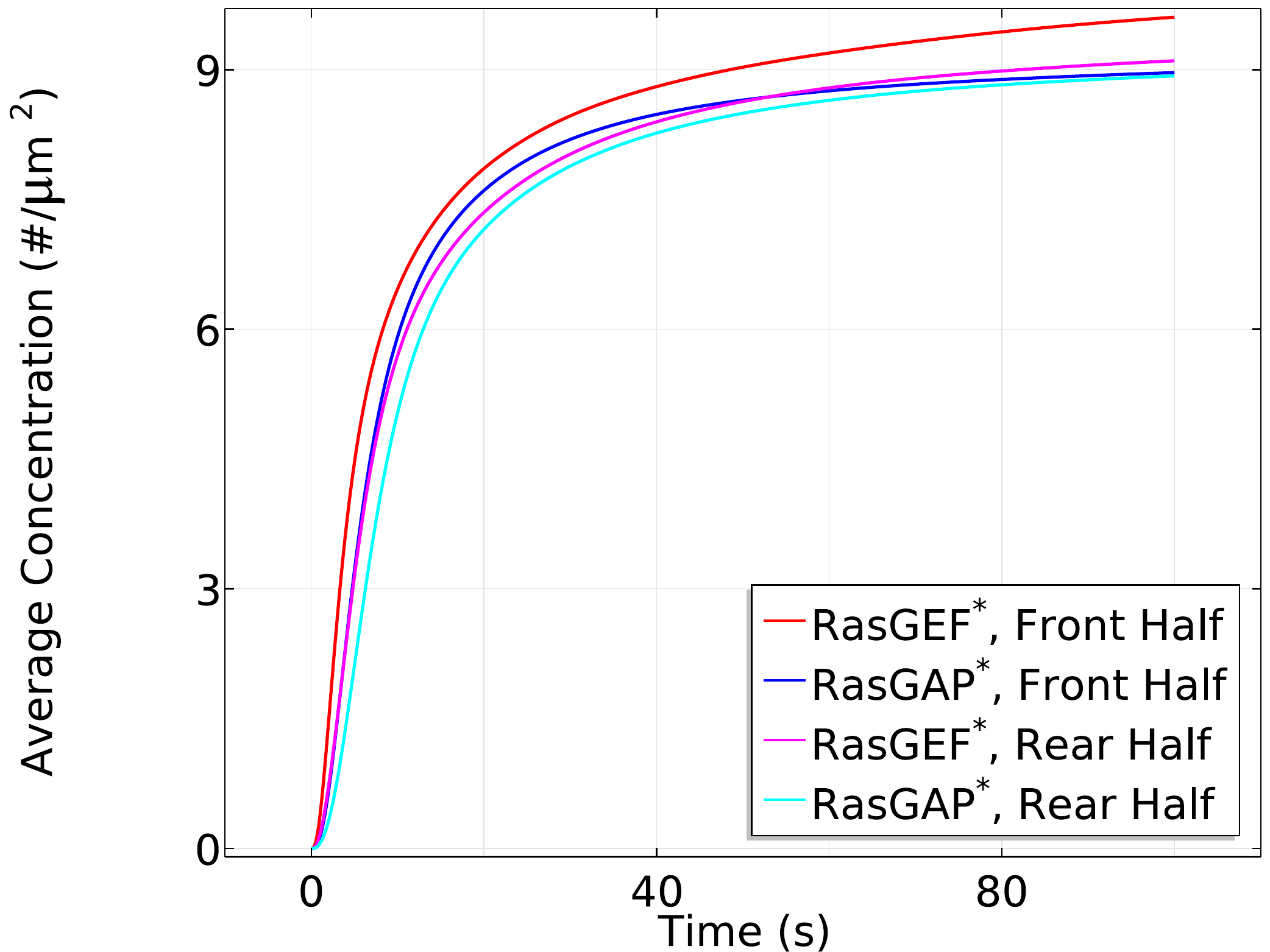}\hspace*{15pt}\includegraphics[width=7cm]
  {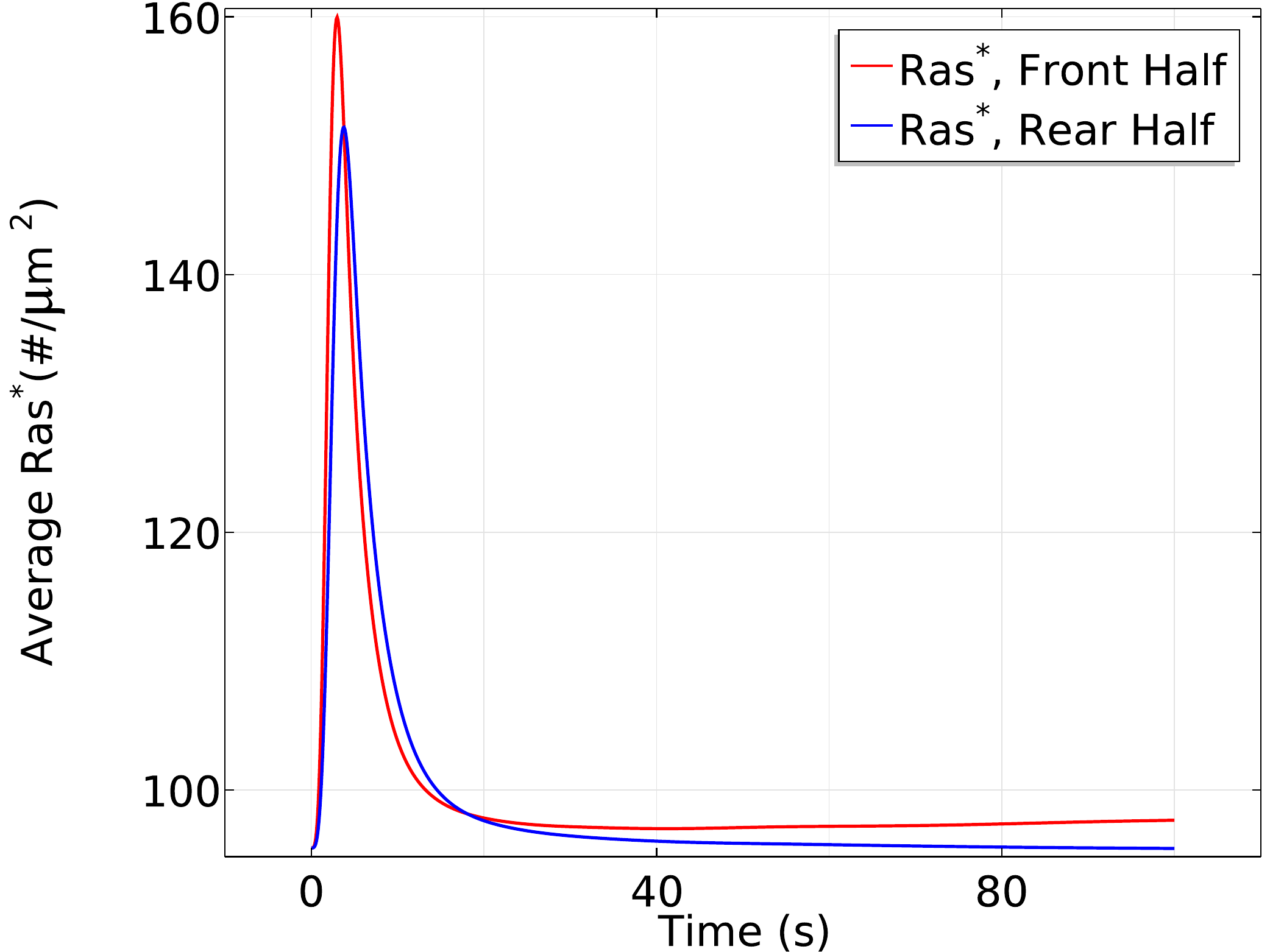}}
\caption{\label{nogefdiffusionGEFGAP}\emph{Left}: The time course of average $RasGEF^*$ and $RasGAP^*$ at the front and rear halves in the absence of apparent RasGEF
  diffusion in the same gradient as previously used. \emph{Right}: The time course of
  average $Ras^*$ in the absence of apparent RasGEF diffusion in that gradient at the front and rear halves.}
\end{figure}
%

\item Slow RasGAP diffusion: `inhibitor' diffusion is not necessary

  Although the supply of RasGAP is also primarily restricted to the membrane
  when RasGAP diffuses slowly, there is enough RasGAP  on the membrane due
  to the relatively small mean cAMP concentration (5.5 nM) in the gradient
  used. As a result, the biphasic behavior is not affected, as shown  in
  Fig.~\ref{nogapdiffusionGEFGAP}.
\begin{figure}[H]
  \centerline{\includegraphics[width=7cm]
    {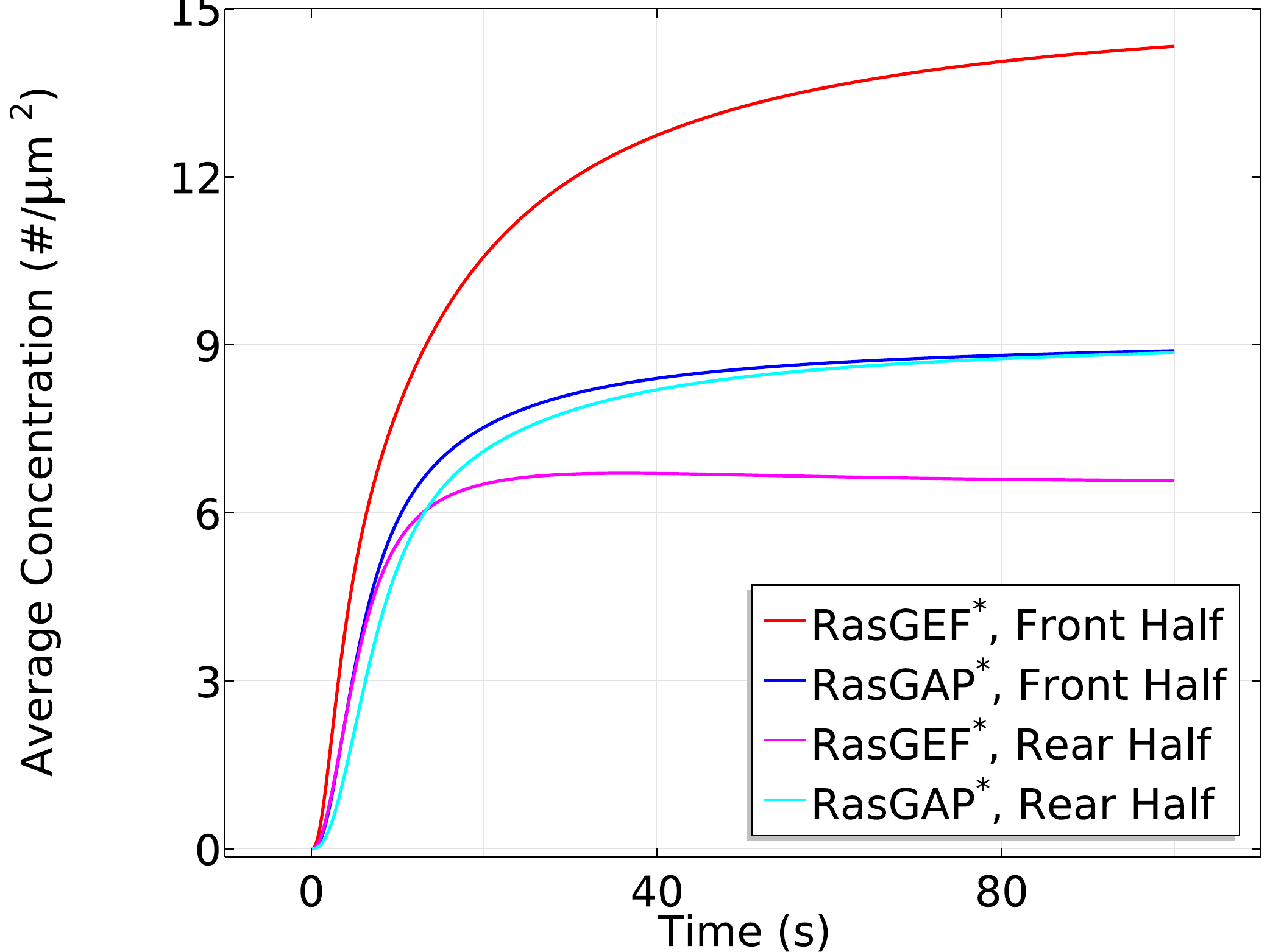}\includegraphics[width=7cm]
    {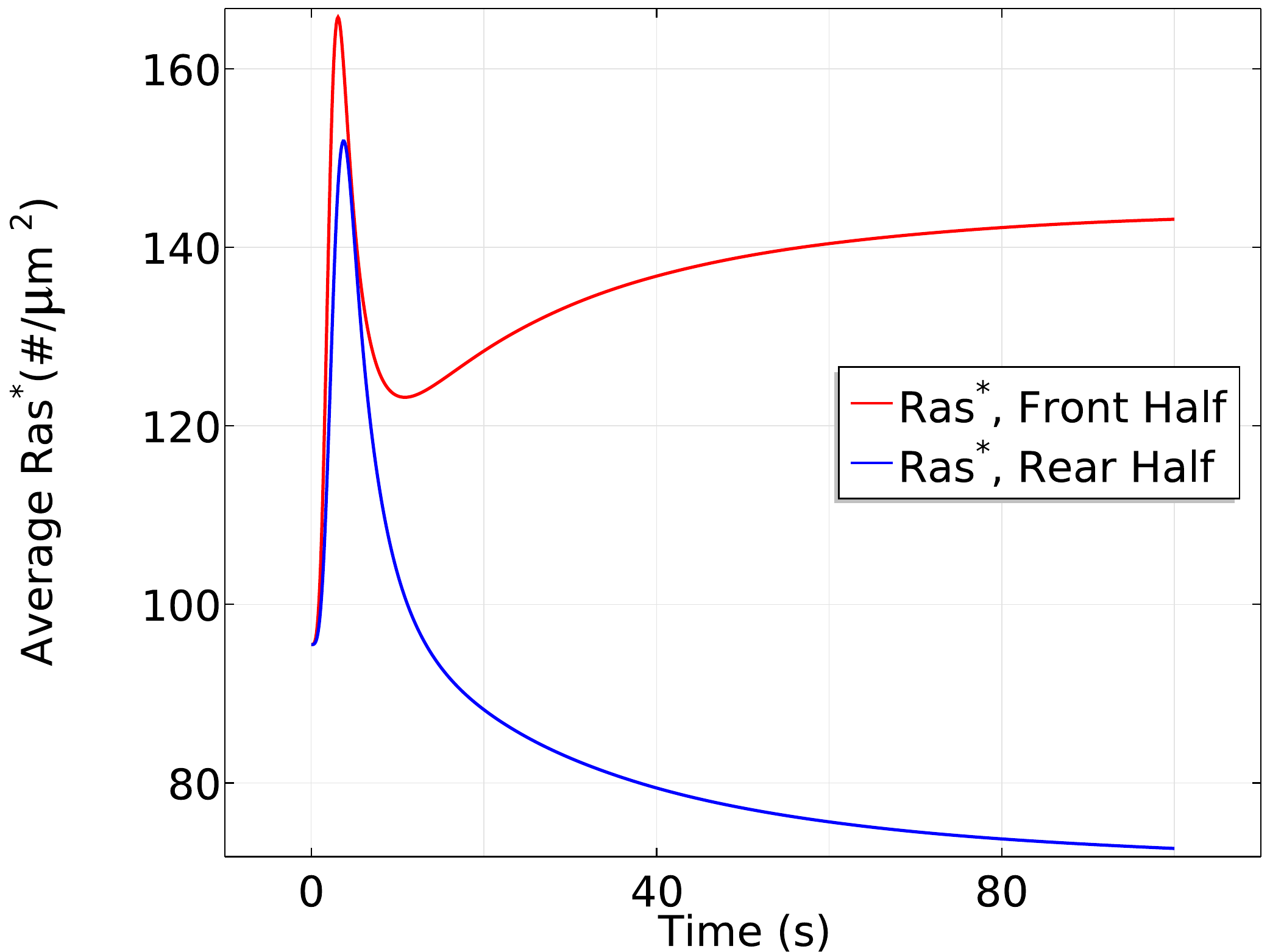}}
\caption{\label{nogapdiffusionGEFGAP}\emph{Left}: The time course of the average $RasGEF^*$ and $RasGAP^*$  at the front  and rear halves in the absence of apparent
  RasGAP diffusion using the previous gradient. \emph{Right}: The time course of
  average $Ras^*$ in the absence of apparent
  RasGAP diffusion in that gradient.}
\end{figure}
In LEGI based models, the global diffusion of inhibitor is essential for
inducing symmetry breaking. It is proposed \cite{takeda2012} that the inhibitor
might be RasGAP, but our model predicts that the diffusion of RasGAP is not a
key component as long as there is sufficient amount of RasGAP on the
membrane. Instead, a diffusible activator RasGEF becomes essential to induce
symmetry breaking. In a LEGI scheme, the diffusion of inhibitor creates a
uniform inhibitor distribution and the gradient induced nonuniform activator
activity generates the symmetry breaking. In our model, the incoherent
activation of both activator (RasGEF) and inhibitor (RasGAP) are induced through
diffusing $\Gbg$. Hence the Ras activity induced by $\Gbg$ alone is balanced
along the cell. In other words, the cell can not develop a sensitive gradient
sensing from a diffusing $\Gbg$. Alternatively, $G_{{\alpha}_2}^*$ facilitated
pathways are the critical elements.

We also tested scenarios in which both G2 and Ric8 diffuse slowly, and when both
G$_{\beta\gamma}$ and RasGEF diffuse slowly (see Supporting Information). In
summary, various Ras activity patterns can be realized by controlling only the
diffusion rates, thus revealing a potential role for diffusion in explaining the
observed diverse sensitivities of genetically identical Dicty species in response to cAMP
\cite{wang2012diverse}.
\end{itemize}

\paragraph{The dependence of Ras activation on the magintude of the gradient and
 the mean concentration }

\begin{figure}[H]
\centerline{\includegraphics[width=7cm]
  {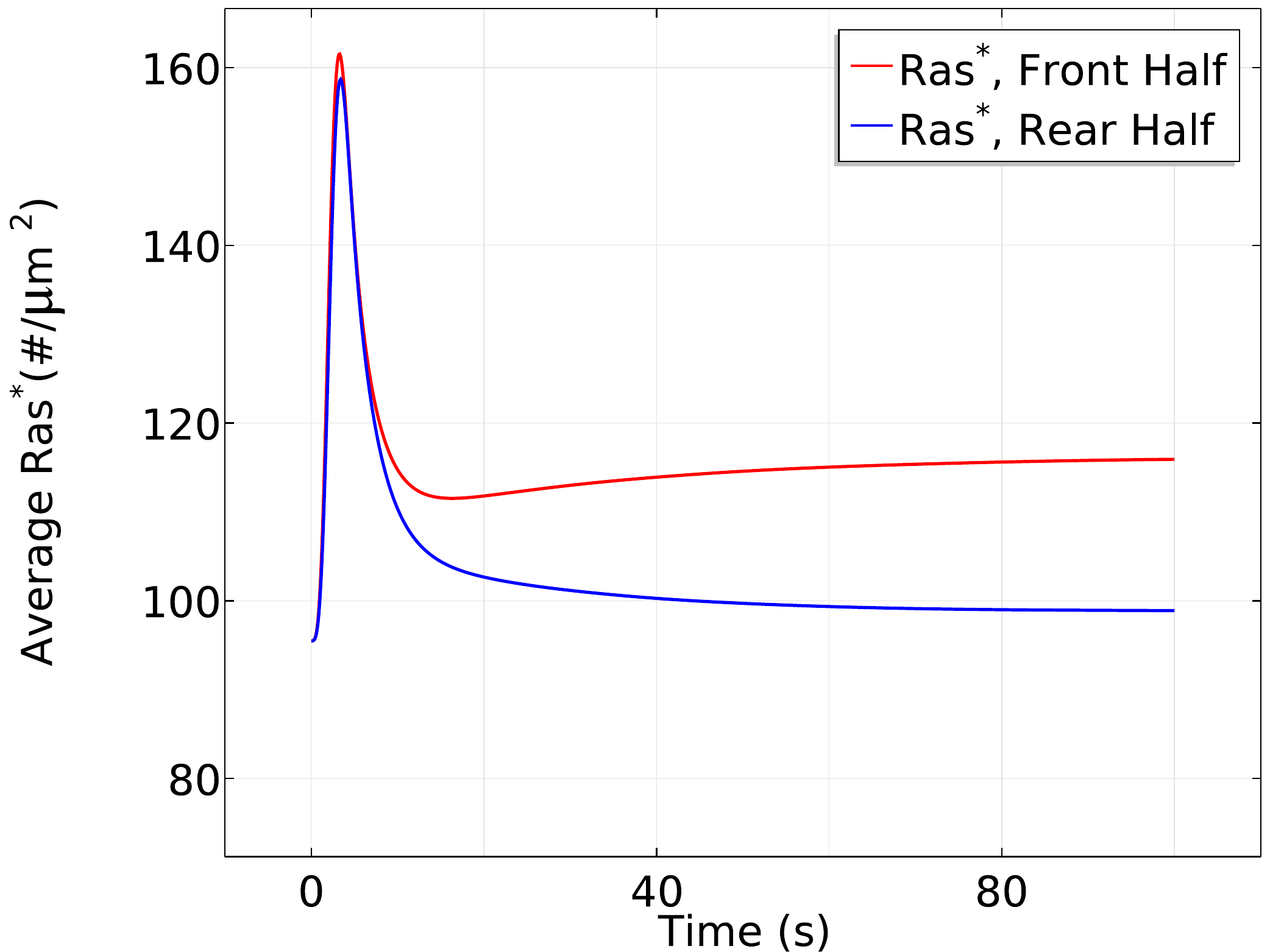}\includegraphics[width=7cm]
  {Gradient900pMRasaHalfHalf-eps-converted-to.pdf}}
\caption{\emph{Left}: The time course of average $Ras^*$ at the front and rear halves
  using $c_f = 6.5$ nM and $c_r = 4.5$ nM.  \emph{Right}: The time course using $c_f=10$
  nM and $c_r=1$ nM.}
\label{gradientdependent}
\end{figure}

\begin{figure}[H]
\centerline{\includegraphics[width=7cm] {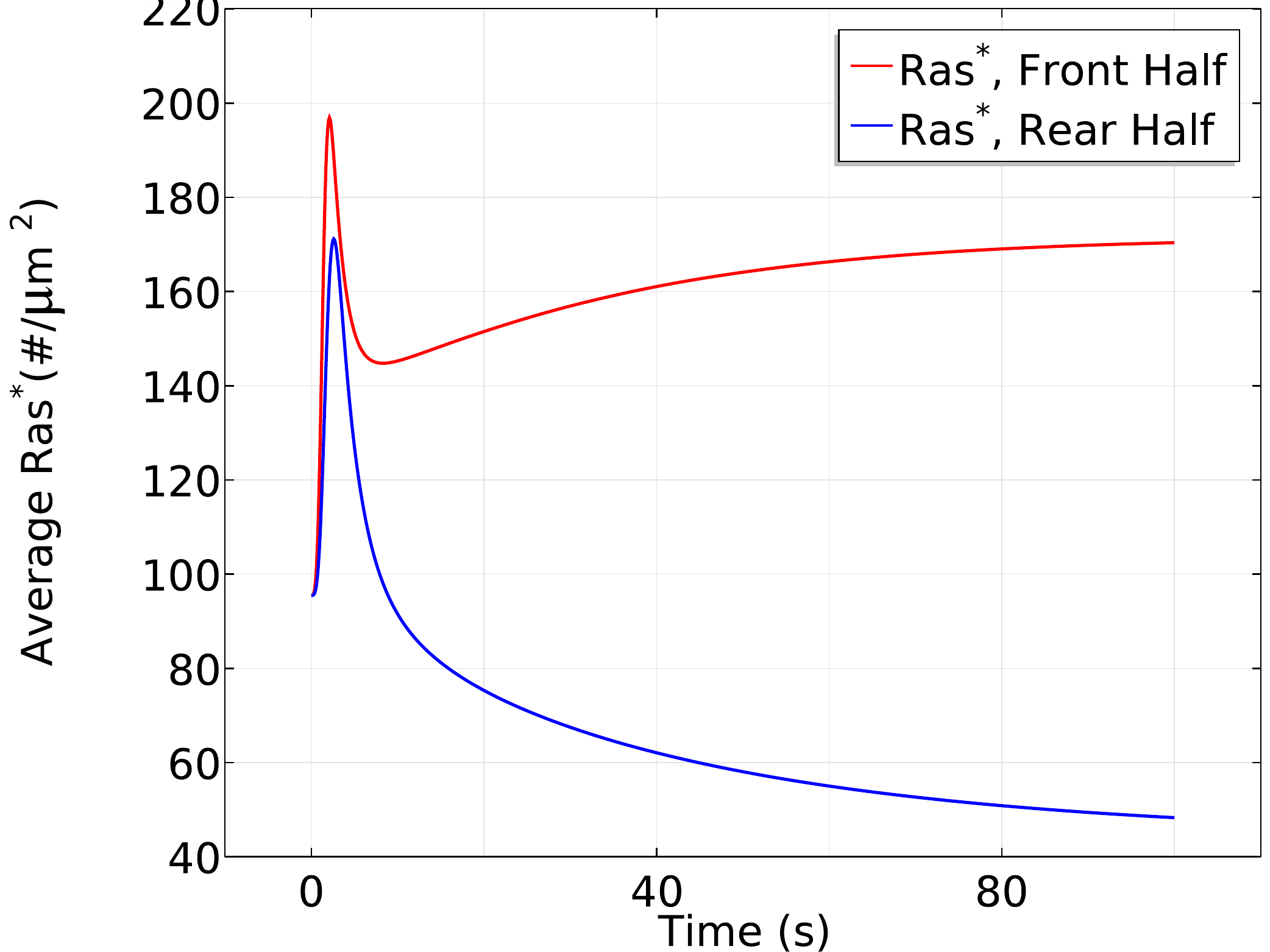}\includegraphics[width=7cm] {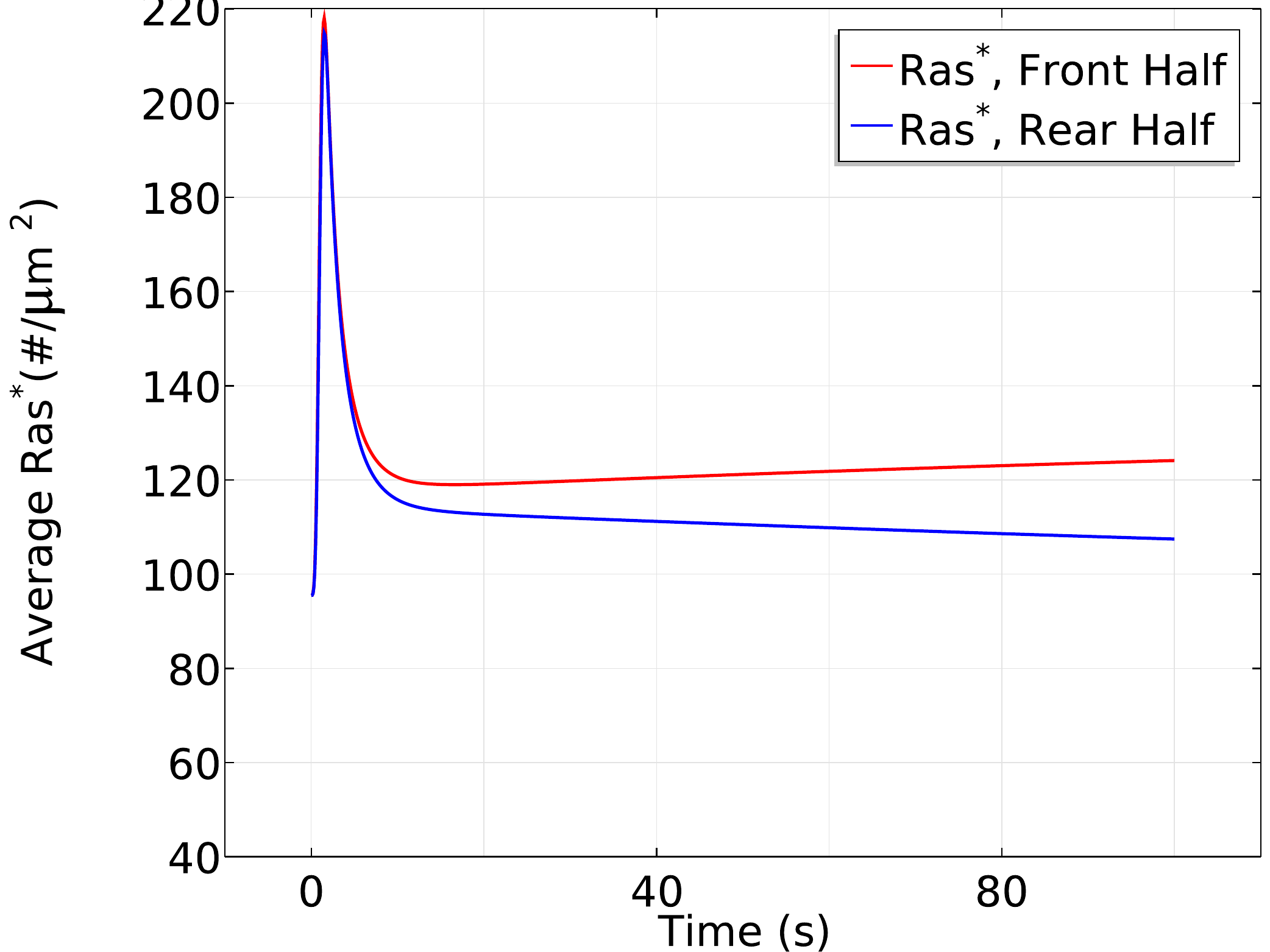}}
\caption{\label{backdependent}\emph{Left}: The time course of $Ras^*$ at the front and rear halves
  using $c_f = 50$ nM and $c_r = 0$ nM. \emph{Right}: The time course using $c_f = 175$ nM and $c_r = 125$ nM.}
\end{figure}

To determine how the front-to-back gradient affects the activation of Ras, we
stimulate the cell using two gradients: a shallow one with $c_f = 6.5$ nM and
$c_r = 4.5$ nM, and the previously-used gradient with $c_f=10$ nM and $c_r=1$nM,
both at the same mean cAMP concentration of 5.5 nM. The cell responses are shown
in Fig.~\ref{gradientdependent}. Ras activation is qualitatively similar in both
a shallow gradient and a steep gradient, but smaller in magnitude in both phases
for a shallow gradient. This is not surprising, since a steep gradient produces
more G$_{\beta\gamma}$ locally, which accounts for the slightly higher initial
response, and a steeper $G^*_{\alpha}$ gradient that initiates the second
phase. Note that the front-rear difference in a steep gradient is around 70
$\#/\mu m^2$ while the front-rear difference in shallow gradient is around $17
\#/\mu m^2$, giving a ratio of $\sim 4$, which is roughly the ratio of the
front-rear difference between the steep gradient (9 nM across the cell) and the
shallow gradient (2 nM across the cell).  Our model predicts results similar to
those reported in \cite{kortholt2013}, where gradient-dependent activation of
Ras is observed.
Next we test whether the cell responds differently in the same large gradient
($5 nM/\mu m$) with different mean concentrations. As shown in
Fig.~\ref{backdependent}, in a steep gradient at a mean concentration of 25 nM,
the front and back halves respond differently in the first phase of Ras
activation --- the front half reaches a maximum of $200 \#/\mu m^2$ while the
rear half only reaches a maximum of $170 \#/\mu m^2$.  Ras is reactivated at the
front when the average $Ras^*$ drops to $150 \#/\mu m^2$ and symmetry breaking
is well established after 100 seconds of cAMP stimulation, resulting in a 3.5
fold difference ($120 \#/\mu m^2$) between the front half and rear
halves. Surprisingly, we observe different response when the cell is exposed to
the steep gradient at a higher mean concentration of 150 nM. In the first phase
of Ras activation, the front and the rear responses almost exactly the same --
both increase to a maximum of $\sim 220 \#/\mu m^2$ -- which is followed by a
decrease to $\sim 120 \#/\mu m^2$. Then Ras is slowly reactivates at the front
and the front-rear difference reaches less than $20 \#/\mu m^2$ after 100
seconds of stimulation.

\begin{figure}[H]
\centerline{\includegraphics[width=7cm] {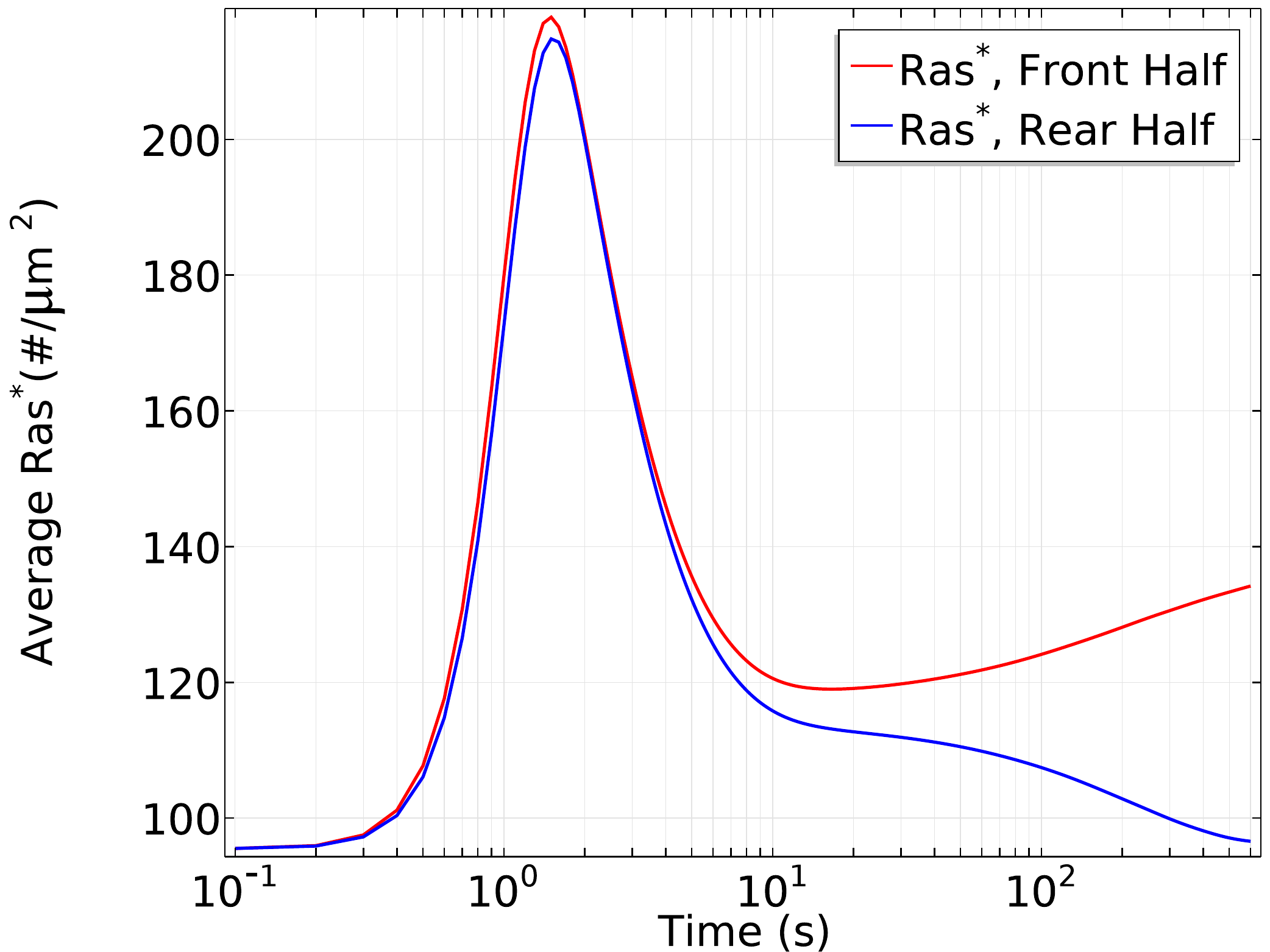}\includegraphics[width=7cm] {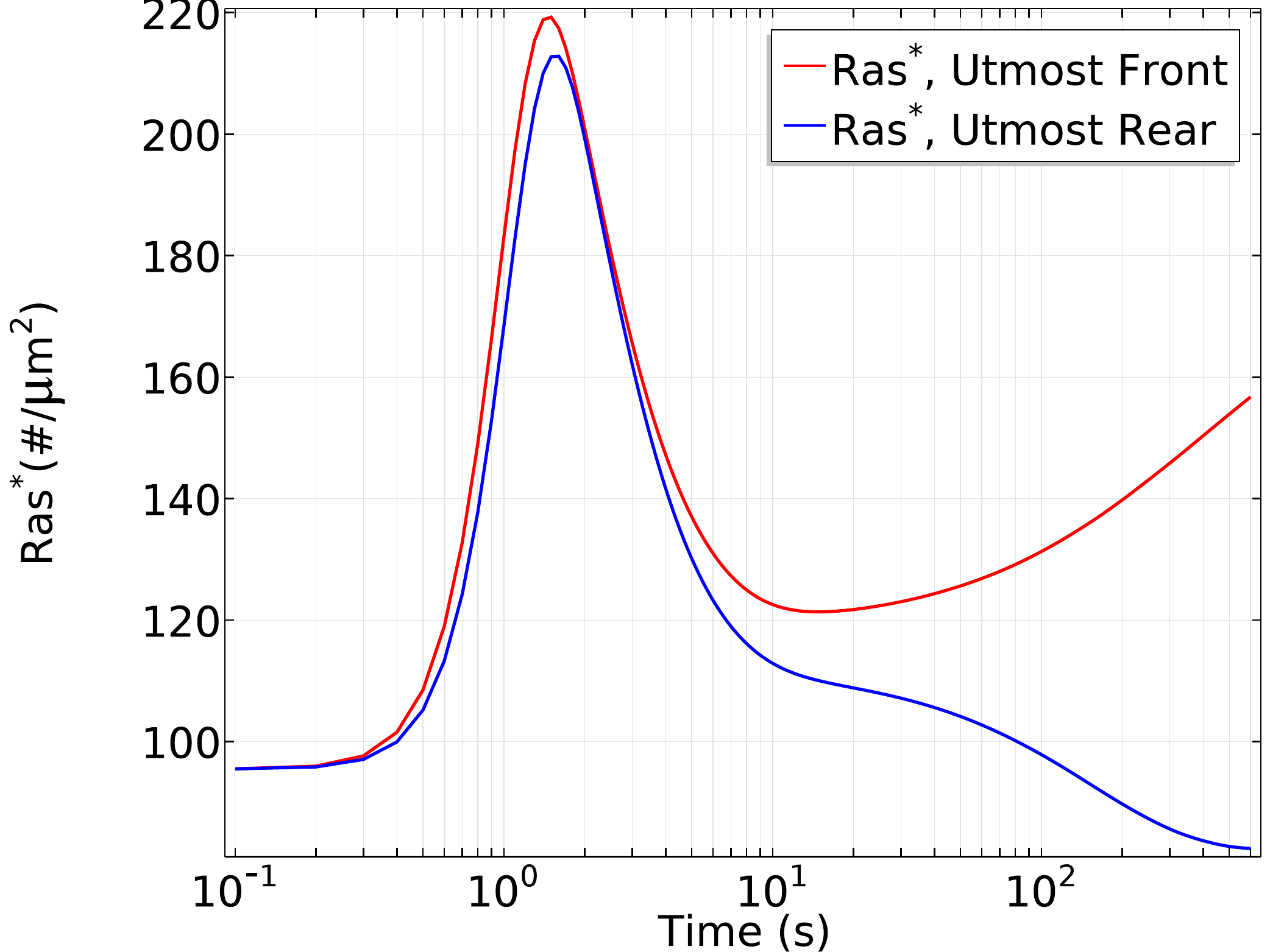}}
\caption{\label{backhigh2}\emph{Left}: The log scale time course of average $Ras^*$ at the front and rear halves
  using $c_f = 175$ nM and $c_r = 125$ nM. \emph{Right}: The $Ras^*$ activity in the same gradient at $x_f$ and $x_r$.}
\end{figure}

It is tempting to say that symmetry breaking is strongly reduced when the mean
concentration increases to a saturation level, but strong symmetry breaking
appears and the steady state difference between front and rear halves reaches
approximate 1.3 fold if we observe the cell for a longer time, as shown in
Fig.~\ref{backhigh2}. This shows  that a higher mean concentration induces a
more 'uniform' initial transient activation followed by much slower symmetry
breaking.


\paragraph{No symmetry breaking in $g_{\alpha 2}$-null cells}

\begin{figure}[H]
\centerline{\includegraphics[width=6cm] {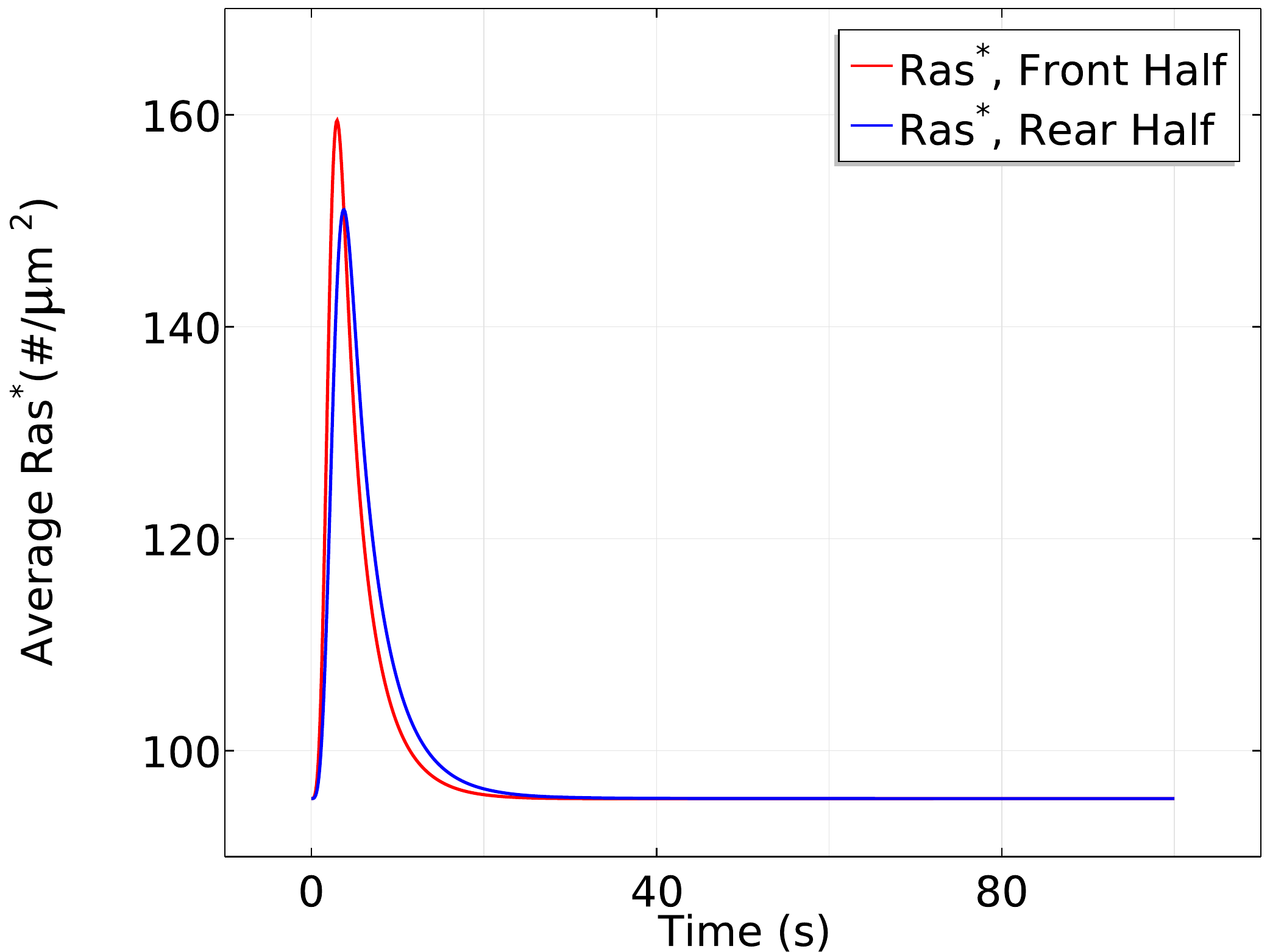}\includegraphics[width=6cm] {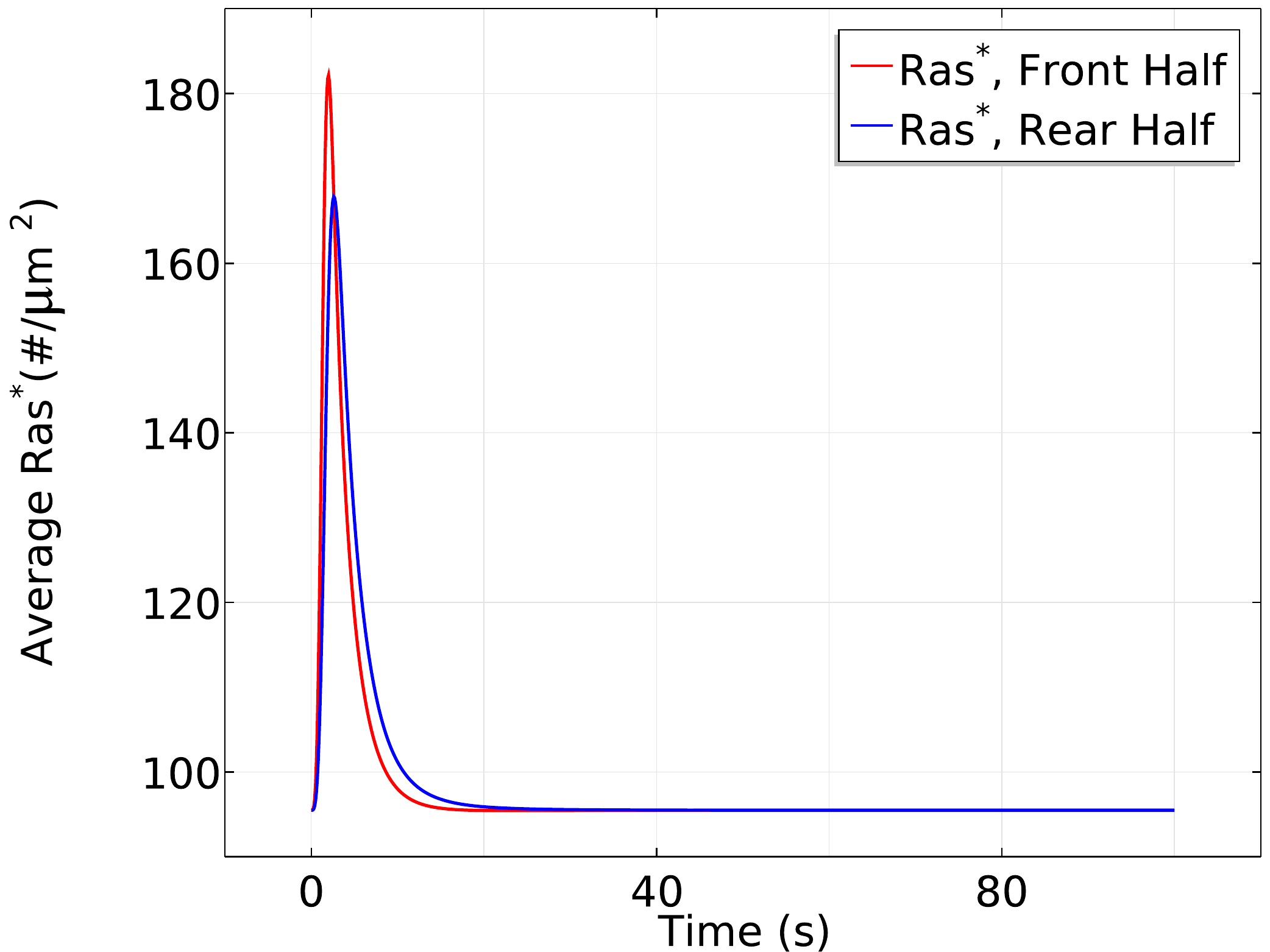}\includegraphics[width=6cm] {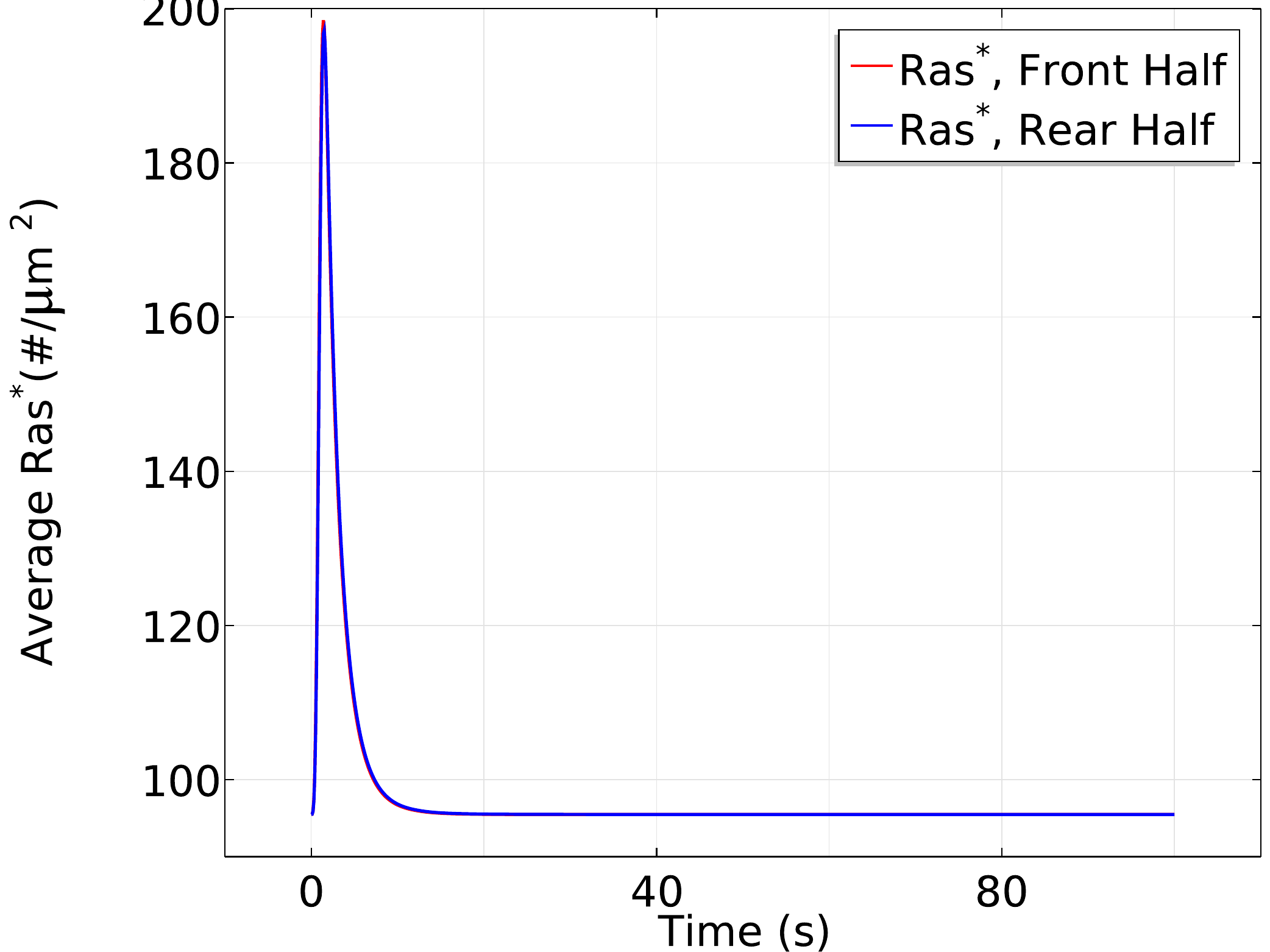}}
\caption{\label{ganullgradient}\emph{Left}: The time course of $Ras^*$ at the front and rear halves
  using $c_f = 10$ nM and $c_r = 1$ nM in $g\alpha 2$-null cells. \emph{Center}: The $Ras^*$ activity using $c_f = 50$ nM and $c_r = 0$ nM in $g\alpha 2$-null cells. \emph{Right}: The $Ras^*$ activity using $c_f = 175$ nM and $c_r = 125$ nM in $g\alpha 2$-null cells.}
\end{figure}

It is reported \cite{kortholt2013} that in $g_{\alpha 2}$-null cells, the cAMP
gradient induces a short transient uniform Ras activation but the specific
upgradient Ras reactivation never occurs. We test our model for $g_{\alpha 2}$-null
cells by blocking the $G_{{\alpha}_2}^*$ promoted RasGEF and Ric8 localization, and
the simulation results are illustrated in Fig.~\ref{ganullgradient} for
different gradients and same gradient with different mean concentrations. In all
three gradients we tested, $g\alpha 2$-null cells only exhibit the initial
transient activation of Ras in consistent with the experimental findings. The
cell settles down at the same level of $Ras^*$ at both the front and rear of the
cell, suggesting the failure of direction sensing. Both the experimental
measurements and computational simulation reveal the essential role of
$G_{{\alpha}_2}^*$ in generation of direction sensing.

\paragraph{No direction sensing when ric8-null cells are exposed to a shallow
  gradient or a steep gradient with high mean concentration}

\begin{figure}[H]
\centerline{\includegraphics[width=7cm] {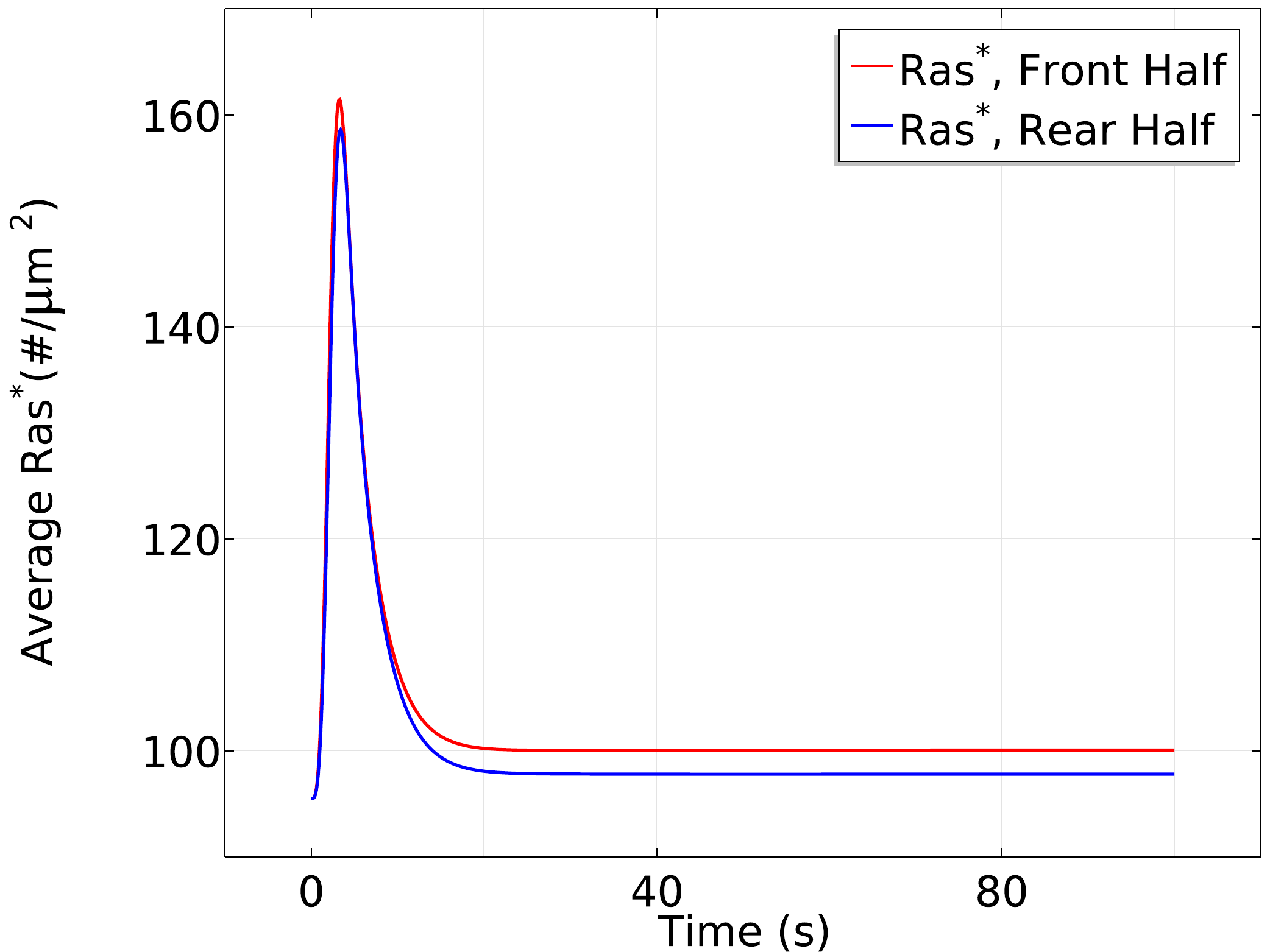}\includegraphics[width=7cm] {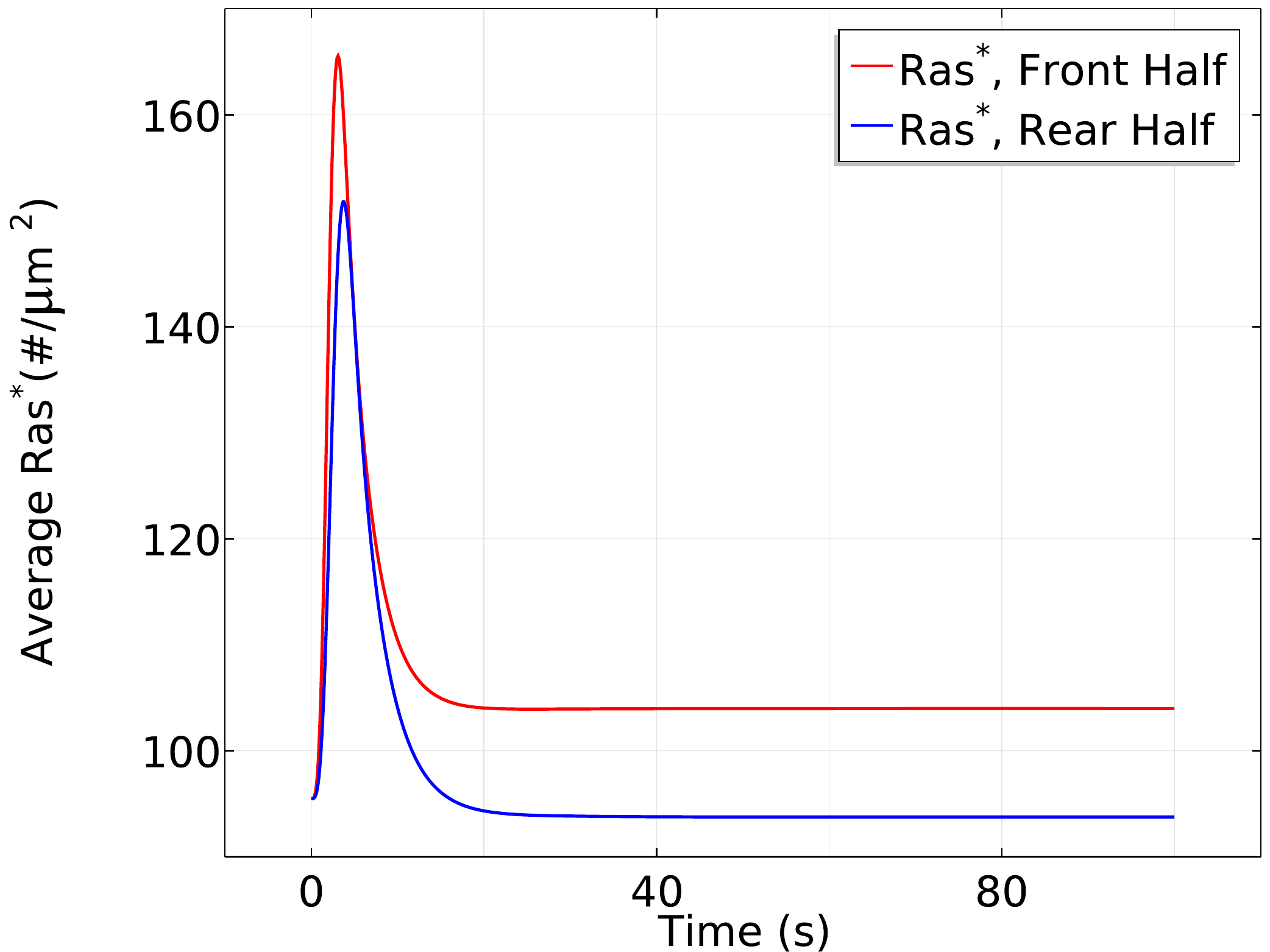}}
\caption{\label{ric8null1}The time course of average $Ras^*$  at the front
  and rear halves in ric8-null cells.  \emph{Left}:  The gradient set by using $c_f =
  6.5$ nM and $c_r = 4.5$ nM.   \emph{Right}: The gradient set by using $c_f=10$
  nM and $c_r=1$ nM.}
\end{figure}

Recall that ric8-null cells have a decreased \G2 dissociation at the steady state compared with WT
cells in uniform stimulus, and here we test whether ric8-null cells are able to
sense directions effectively in a cAMP gradient. Ras activation is illustrated
in Fig.~\ref{ric8null1} when ric8-nulls are exposed to gradients of the same mean
concentrations with different steepness. Comparing with the plot in the left
panel of Fig.~\ref{gradientdependent}, the average front-rear difference is
reduced 8 fold for the shallow gradient (from $\sim 15 \#/\mu m^2$ in WT cells
to $\sim 2 \#/\mu m^2$). Consistent with experimental findings
\cite{skoge2014}, the almost identical $Ras^*$ activity at the front and rear
suggests failure of direction sensing when ric8-null cells are exposed to a
shallow gradient. The plot in the right panel suggests that the cell is still able to sense direction when the gradient is large enough, but the biphasic responses disappear.

\begin{figure}[H]
\centerline{\includegraphics[width=7cm]
  {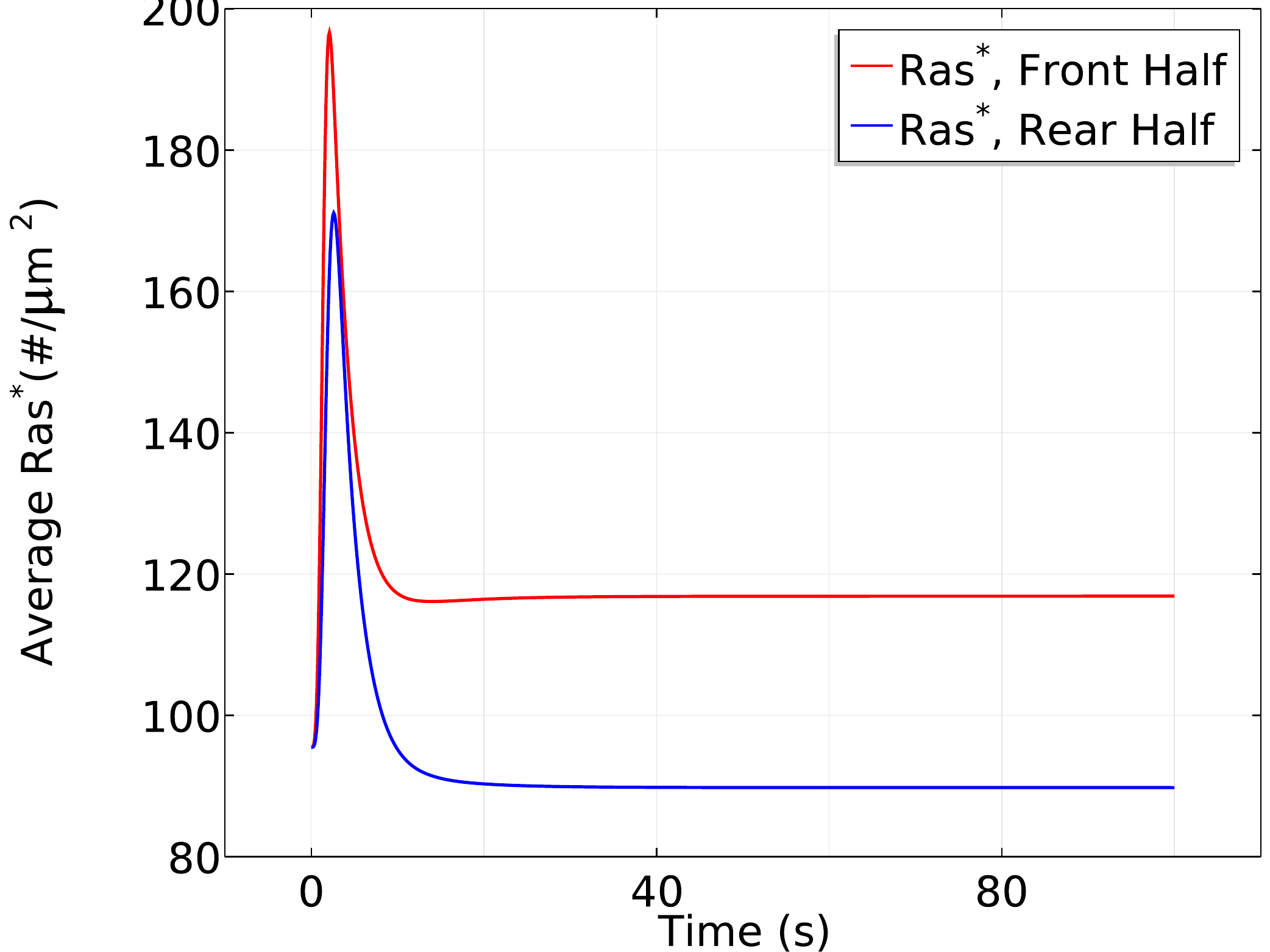}
\includegraphics[width=7cm]  {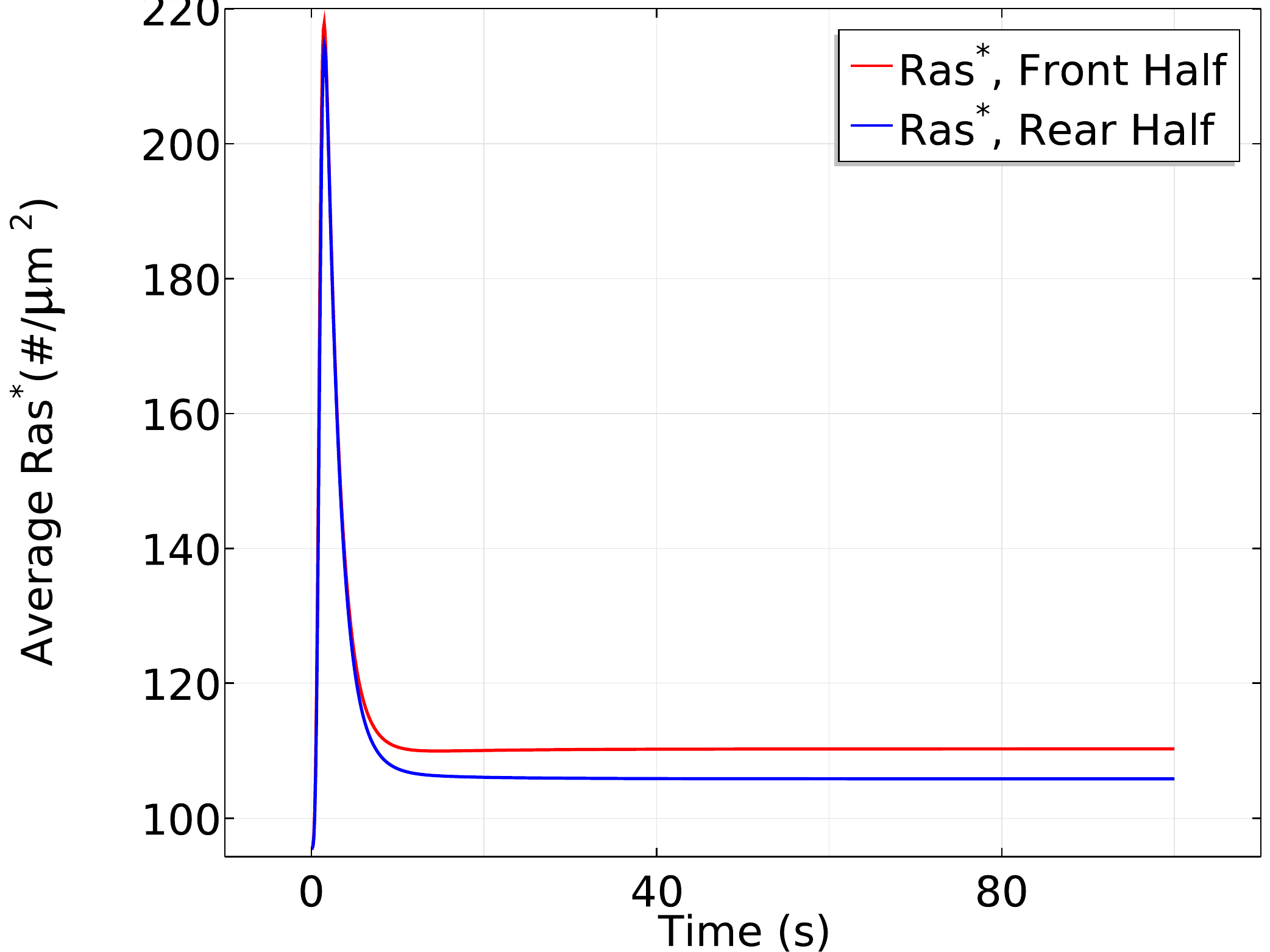}}
\caption{\label{ric8null2} The time course of average $Ras^*$  at the front
  and rear halves in ric8-null cells in steep gradients.\emph{Left}: The gradient set by using $c_f =
  25$ nM and $c_r = 0$ nM. \emph{Right}: The gradient set by using $c_f =
  175$ nM and $c_r = 125$ nM.}
\end{figure}

It has been shown  that ric8-null cells migrate with an
efficiency similar to that of wild-type cells when cells are exposed to a steep
gradient of cAMP ($>10 nM/\mu m$) \cite{skoge2014}. We tested our model with a gradient of $5nM
/\mu m$ with different mean concentrations, and the results are shown in
Fig.~\ref{ric8null2}. As shown in the left figure, ric8-null cells still sense
direction by creating an asymmetrical distribution of $Ras^*$. However, the
asymmetry is strongly reduced comparing to WT cells (left
panel of Fig.~\ref{backdependent}). Moreover, ric8-null cells do not exhibit a
biphasic response. Instead, the front and rear half of
the cell settle at different levels after initial transient
activation. Surprisingly, when the mean concentration is elevated to 150 nM,
ric8-null cells lose the  ability to sense direction, as shown in the right
panel of Fig.~\ref{ric8null2} (front rear difference is less that $5 \#/\mu
m^2$). Hence our model predicts that Ric8 is essential for chemotaxis in both
shallow gradients of cAMP and steep gradients  with high mean
concentration. In the range of cAMP gradients where ric8-null cells can sense
direction, our model predicts that there is  no biphasic Ras activation and little
amplification.

\subsection*{A solution to  the back-of-the-wave problem}

\begin{figure}[H]
\centerline{\includegraphics[width=5cm] {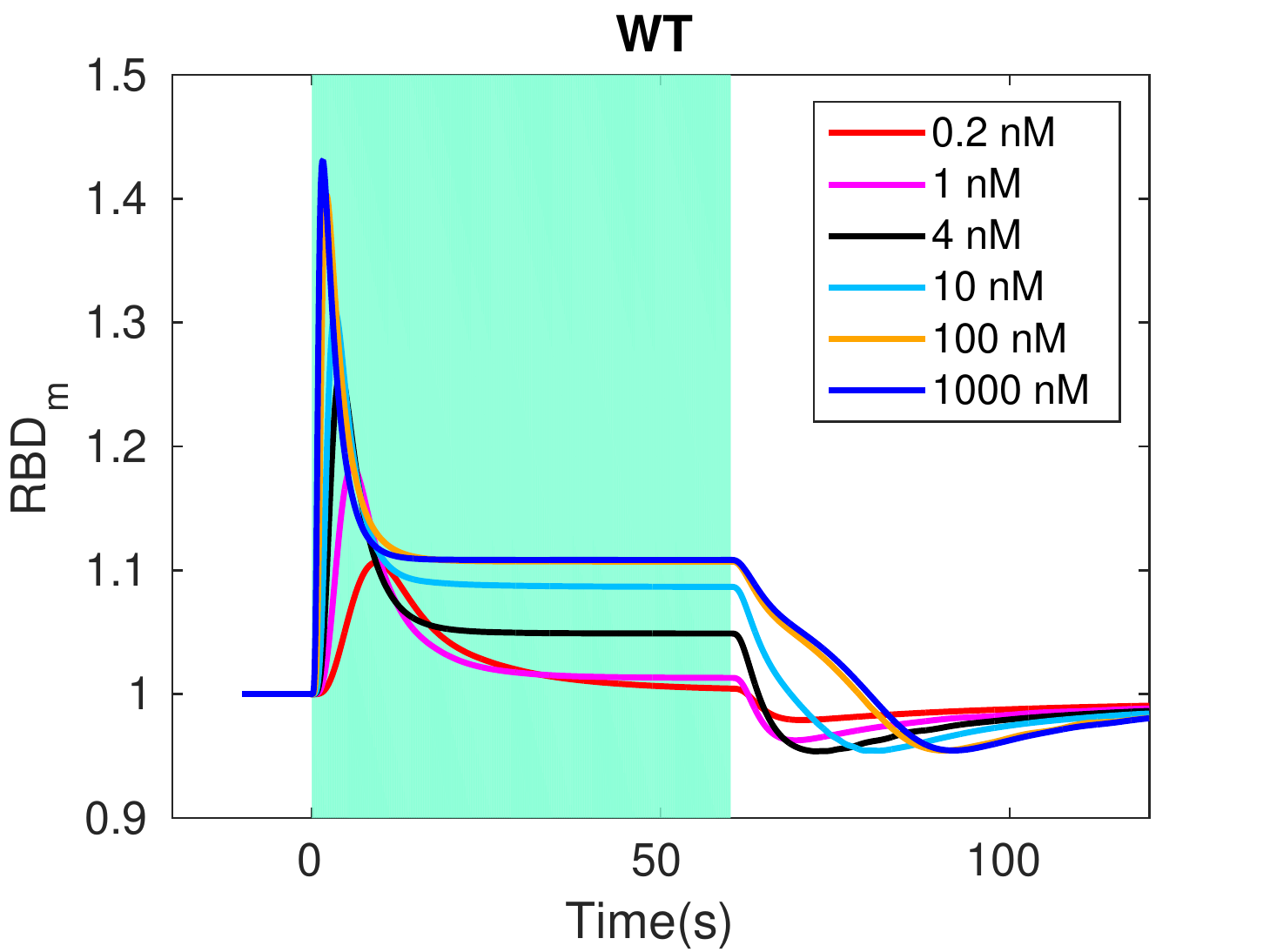}\hspace*{.25in}\includegraphics[width=5cm] {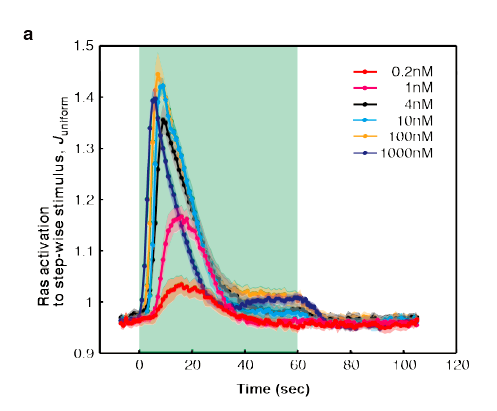}\hspace*{.25in}\includegraphics[width=5cm] {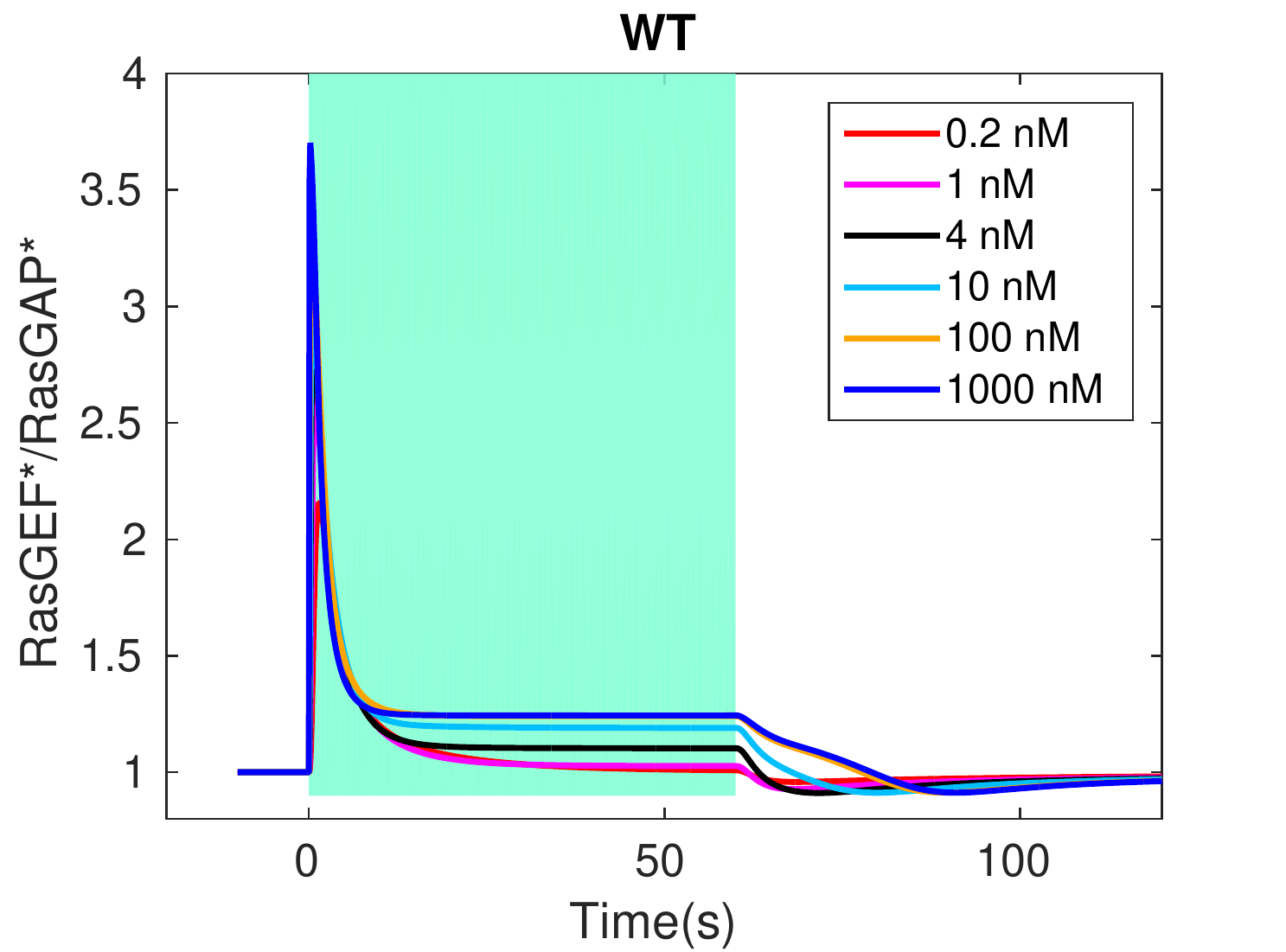}}
\caption{\label{rectify}Rectification. \emph{Left}: The time course of membrane RBD under uniform stimuli of various concentrations. \emph{Middle}: experimental
  measurements extracted from \cite{nakajima2014}. \emph{Right}: The time course of the ratio of $RasGEF^*$ to
  $RasGAP^*$.}
\end{figure}

In the context of Dicty aggregation, the `back-of-the-wave' problem refers to
the fact that cells do not turn to follow the cAMP gradient after the wave has
passed, despite the fact that the spatial gradient reverses as the wave passes
over a cell \cite{huang2014cell, nichols2015}.  This requires some level of
persistence of 'orientation' of a cell, but there is as yet no agreed-upon
mechanistic solution for this problem, since polarization and other factors may
play a role. Under uniform stimuli, cells are said to show
rectification if there is an asymmetry in the amplitude and evolution of the
response to a step increase in cAMP compared with the response following removal
of the stimulus \cite{nakajima2014}. To test whether the proposed network
exhibits rectification in this sense, we apply a uniform stimulus of various
concentrations for 60 seconds and then remove it, as was done experimentally in
fully aggregation-competent cells \cite{nakajima2014}.  Fig.~\ref{rectify} (left
and center) show the simulation and the experimental results, resp. In both
cases the concentration of cAMP is increased from 0 M to the concentrations
indicated for 60 seconds (green shaded area), followed by a decrease to 0 M, and
in both cases one sees a much larger and faster change in RBD following
application of the stimulus than on removal.  We also applied the same stimuli
as used above to $g_\alpha$-null cells and ric8-null cells. Results given in the
Supporting Information show that Ric8 plays a significant role in the
rectification, as will also be seen later in the traveling wave analysis.
Some insight into this behavior can be gained from simple models of excitation
and adaptation, such as the cartoon description defined by the system of
equations
\begin{equation}
\frac{dy_1}{dt} = \frac{S(t) - (y_1 + y_2)}{t_e}, \qquad \qquad
\qquad \qquad
\frac{dy_2}{dt} = \frac{S(t) -  y_2}{t_a}.
\label{adapteq}
\end{equation}
Here $S(t)$ represents the signal and the magnitudes of $t_e$ and $t_a$ reflect
the time scale for excitation and adaptation, resp., and one see that $y_1$
adapts perfectly to a constant stimulus whereas $y_2$ compensates for the
stimulus. However, the temporal responses to increasing and decreasing stimuli
are symmetric, and therefore such a simple model cannot explain the observed
response.  Nakajima et al. \cite{nakajima2014} suggest that a single-layered
incoherent feedforward circuit with zero-order ultrasensitivity \cite{koshland}
is necessary to generate rectification, but our model does not include an
ultrasensitive circuit. Instead, rectification is induced solely by the balanced
regulation of RasGEF and RasGAP activity. The ratio of $RasGEF^*$ to $RasGAP^*$
increases 2-4 fold very rapidly in response to a step increase in the cAMP
concentration, but when the stimulus is removed this ratio does not drop
significantly, as shown in the right panel of Fig.~\ref{rectify}. Thus Ras
activation persists because the ratio equilibrates rapidly while the absolute
levels of the factors decrease more slowly.

\begin{figure}[H]
\centerline{\includegraphics[width=6.75cm]
  {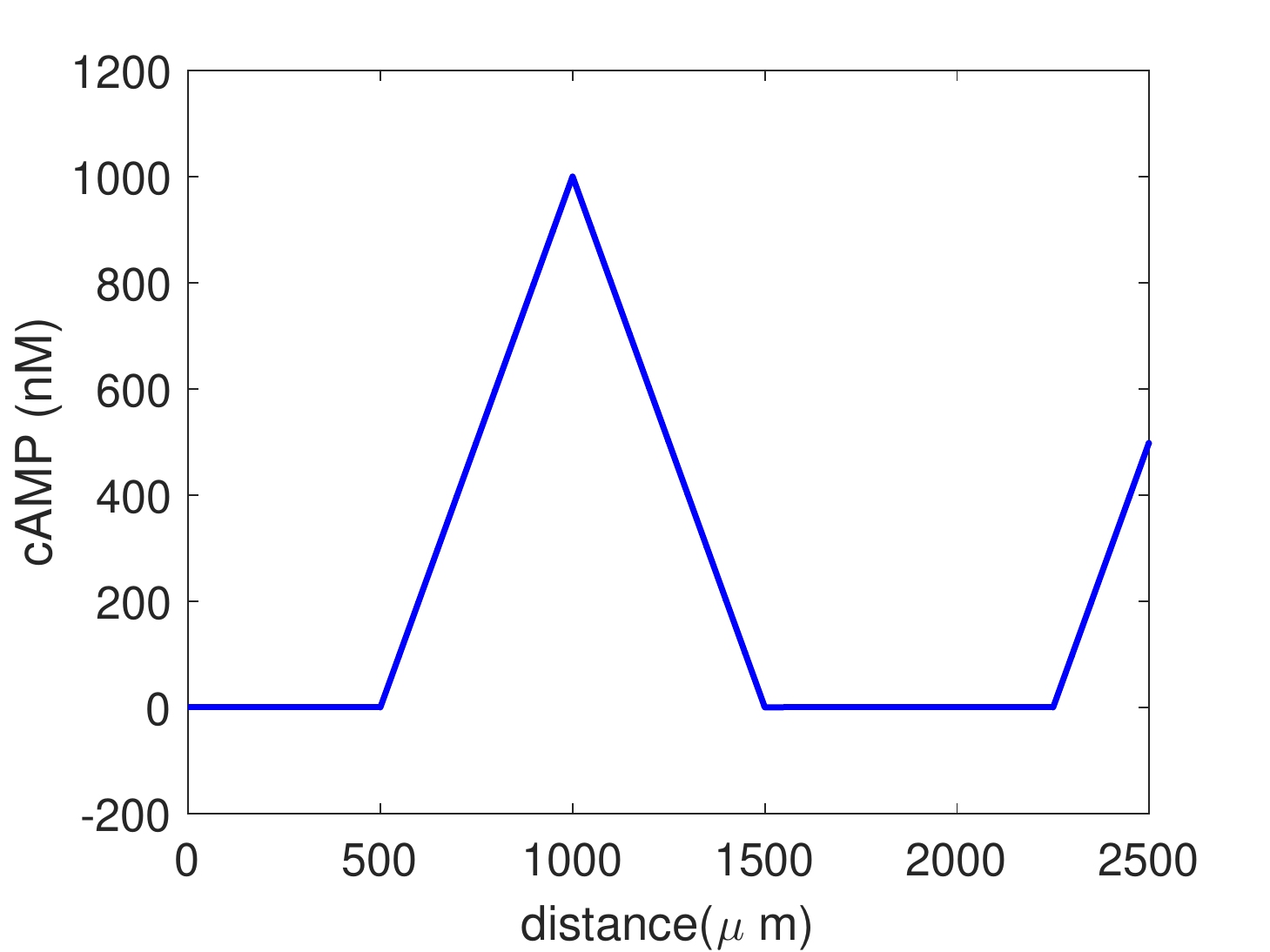}\hspace*{.25in}\includegraphics[width=7cm] {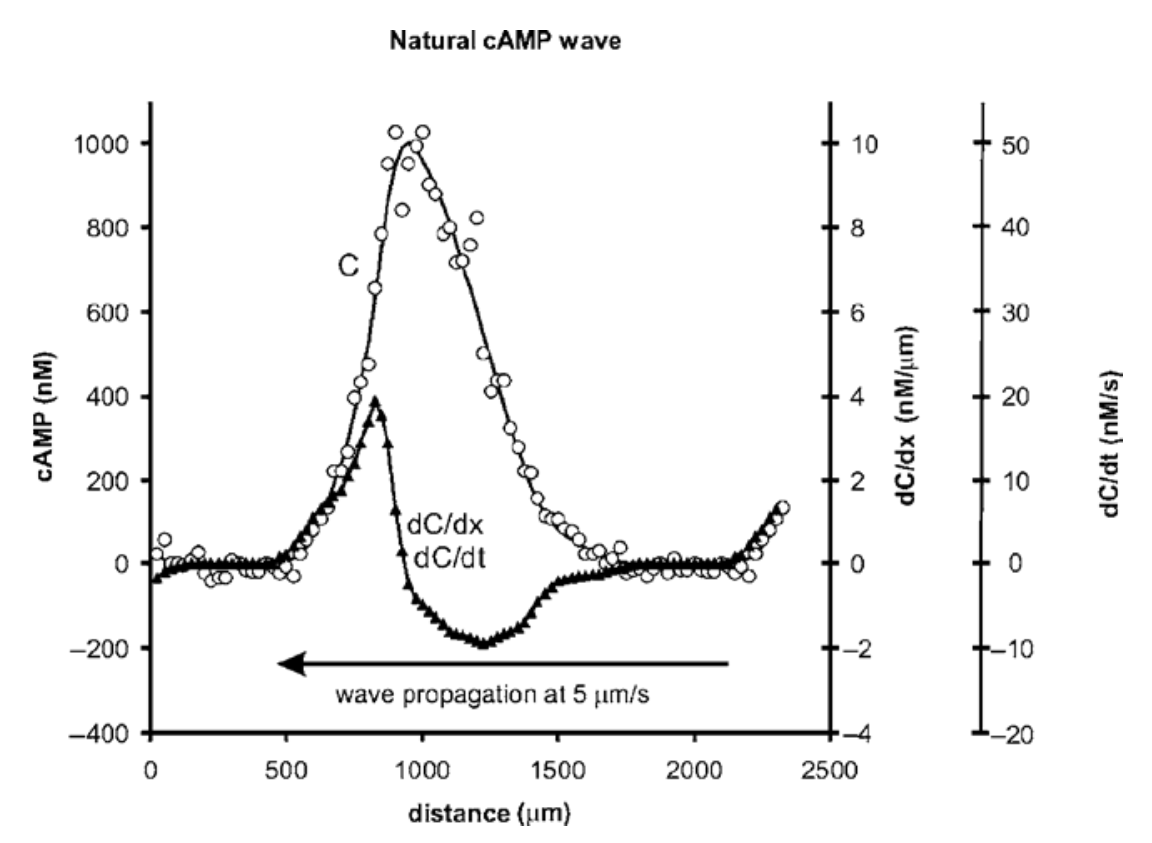}}
\caption{\emph{Left}: The triangle wave. \emph{Right}: A natural
  wave -- from \protect\cite{Postma:2009:MEG} with permission.}
\label{trianglewave}
\end{figure}

To study how cells would respond in wave-like spatially-graded stimuli, we first
generate a simple trianglular wave that approximates a natural cAMP wave.  Let
$W(x,y,z,t)$ denote the cAMP concentration at $(x,y,z)$ of the cell at
time $t$, and  specify it as
\[
W(x,y,z,t)=\left\{ \begin{array}{ll}
0, & 0+350k\leq t\leq\frac{x+5}{v}+350k\\
10(t-\frac{x+5}{v}-350k), & \frac{x+5}{v}+350k<t\leq\frac{x+5}{v}+100+350k\\
-10(t-\frac{x+5}{v}-350k)+2000, & \frac{x+5}{v}+100+350k<t\leq\frac{x+5}{v}+200+350k\\
0, & \frac{x+5}{v}+200+350k<t\leq350(1+k)
\end{array}\right.,
\]
where $v$ is the wave speed and $-5\leq x, y, z \leq 5, k=0,1,\cdots$. This wave
resembles a natural wave when we choose the natural wave speed $v=5\mu m /s$, as
shown in Fig.~\ref{trianglewave}. The wave length is $1000 \mu$m, and at the
natural speed any point on a cell is subject to an increasing stimulus for 100
sec on the upstroke of the wave and a decreasing stimulus for 100 sec on the
downstroke.

\begin{figure}[H]
\centerline{\includegraphics[width=7cm]
  {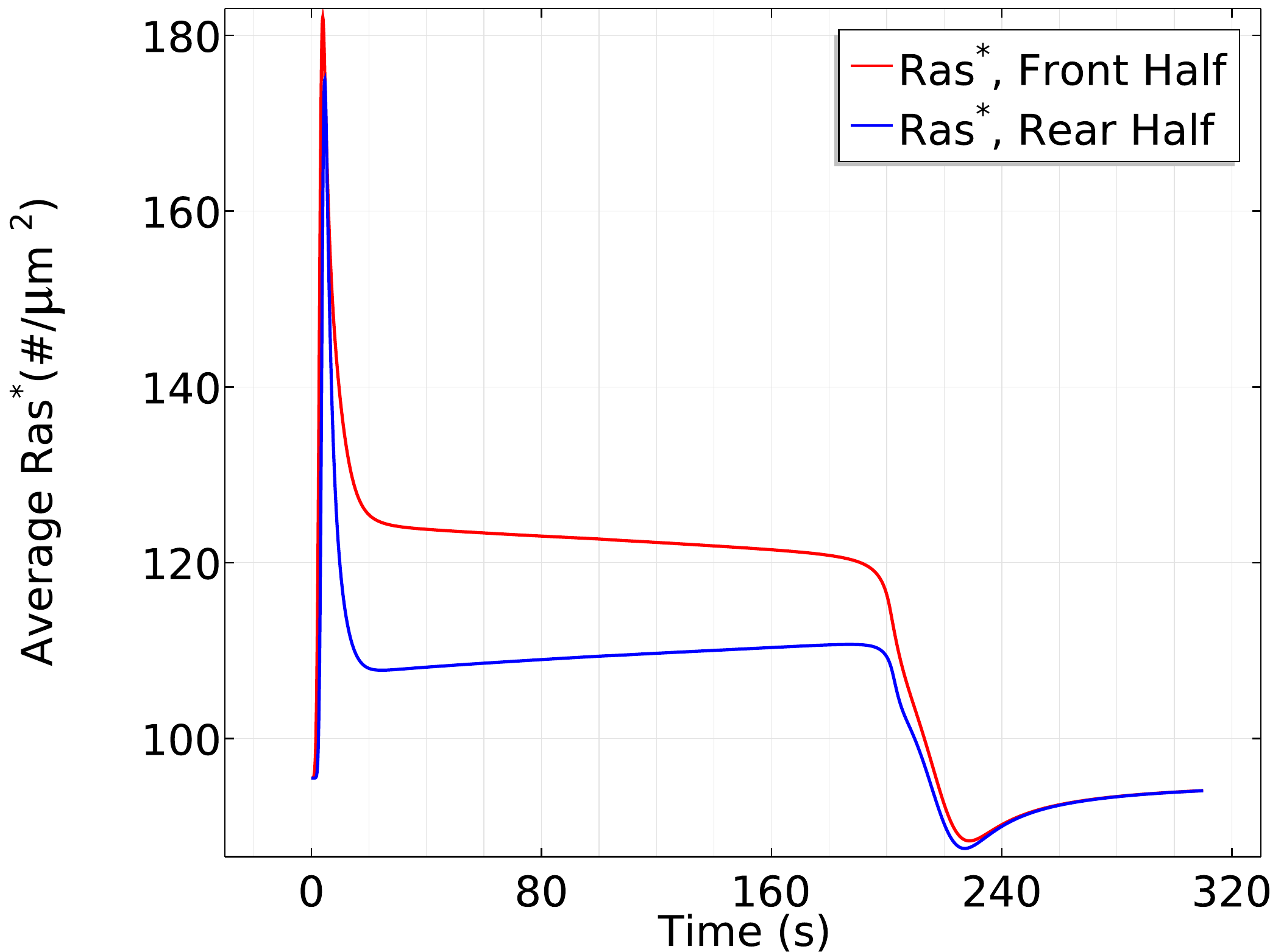}\hspace*{.25in}\includegraphics[width=7cm]
  {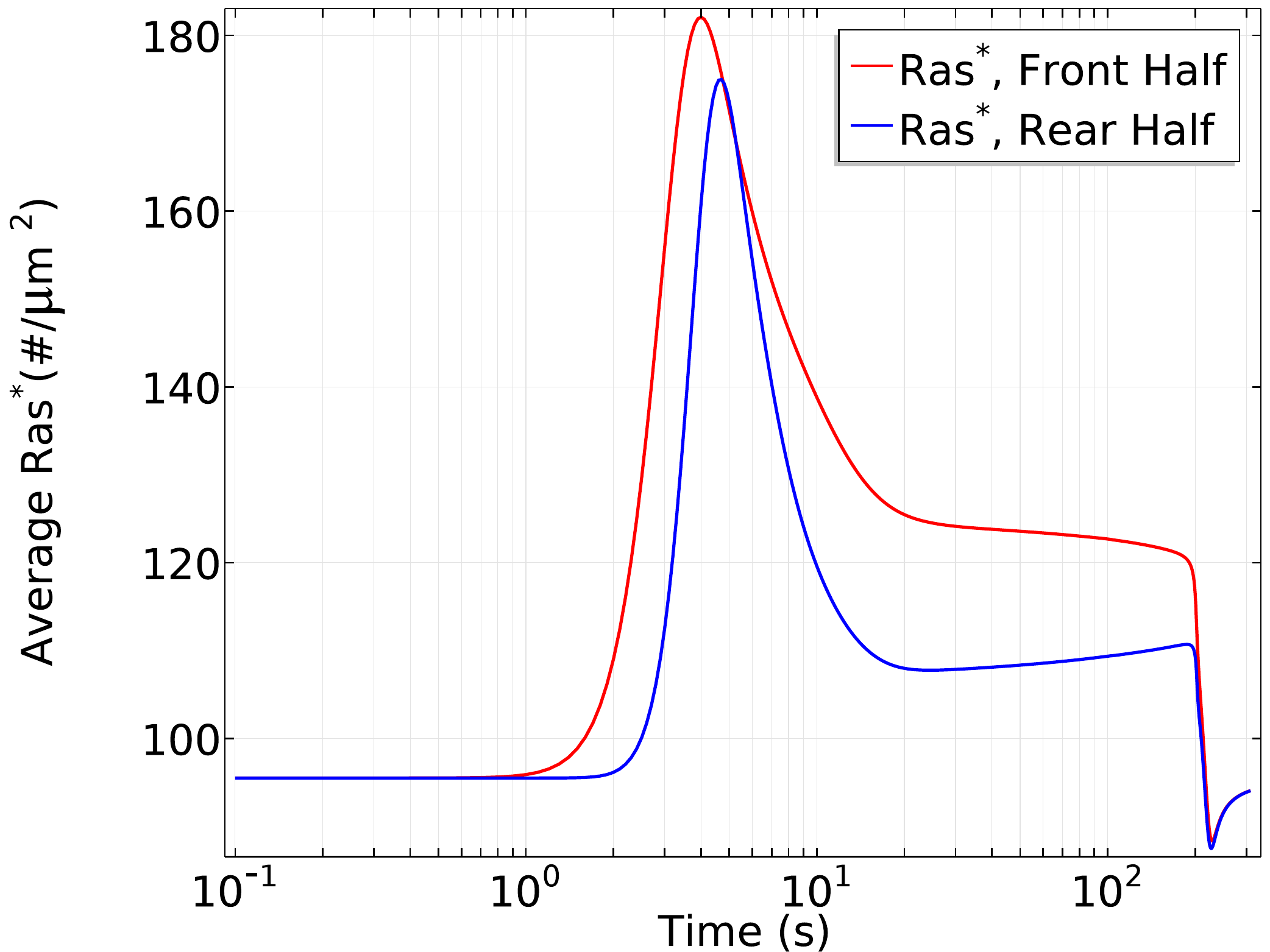}}
\caption{\label{normalwave}\emph{Left}: The time course of activated Ras at the
  front and rear halves   when the triangle wave passes over a WT cell at $v=5\mu
  m/s$. \emph{Right}: A log
  plot of time to show the delay at the rear of the cell.}
\end{figure}

\begin{figure}[H]
\centerline{\includegraphics[width=7cm]
  {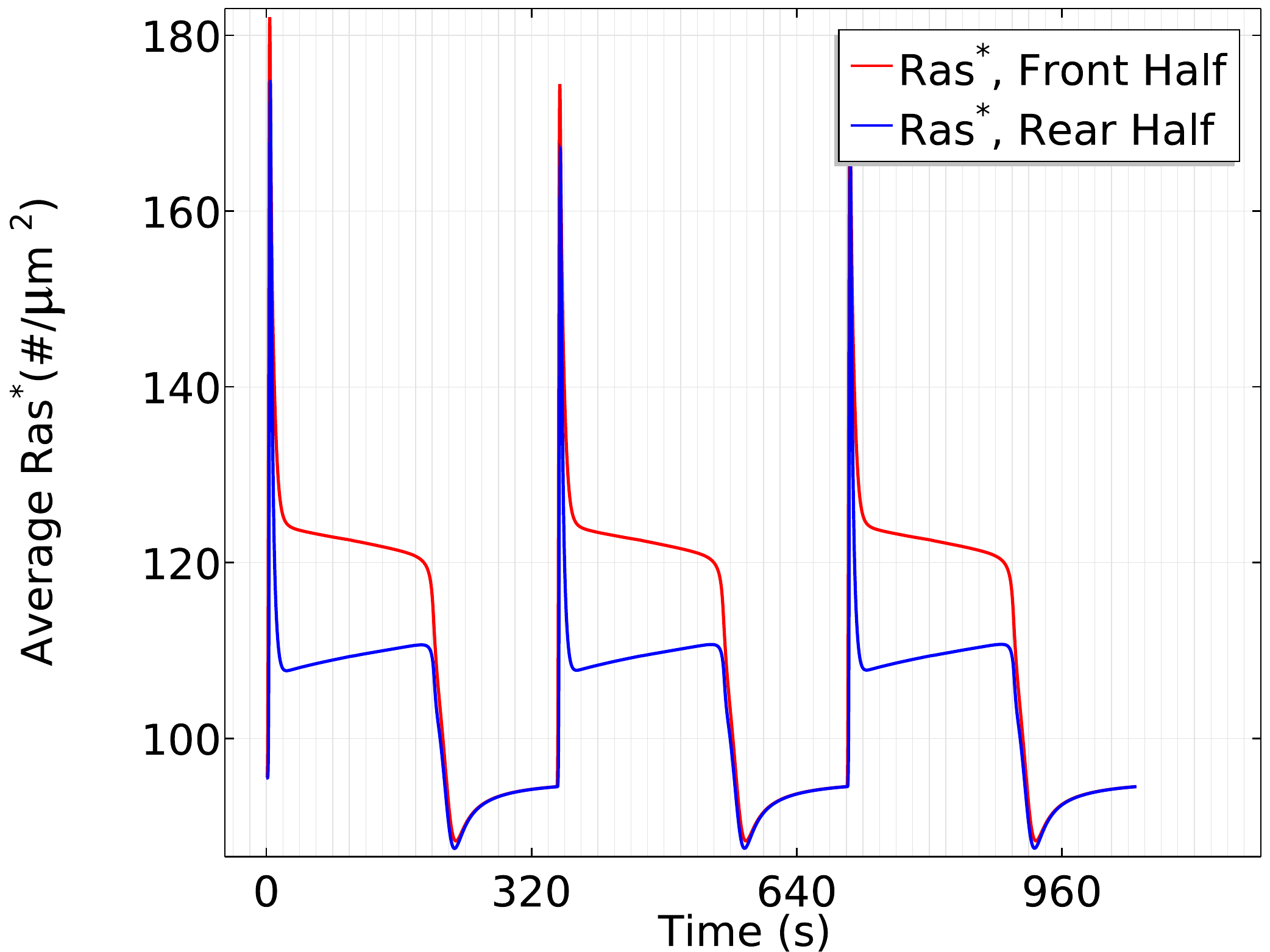}\hspace*{.25in}\includegraphics[width=7cm]
  {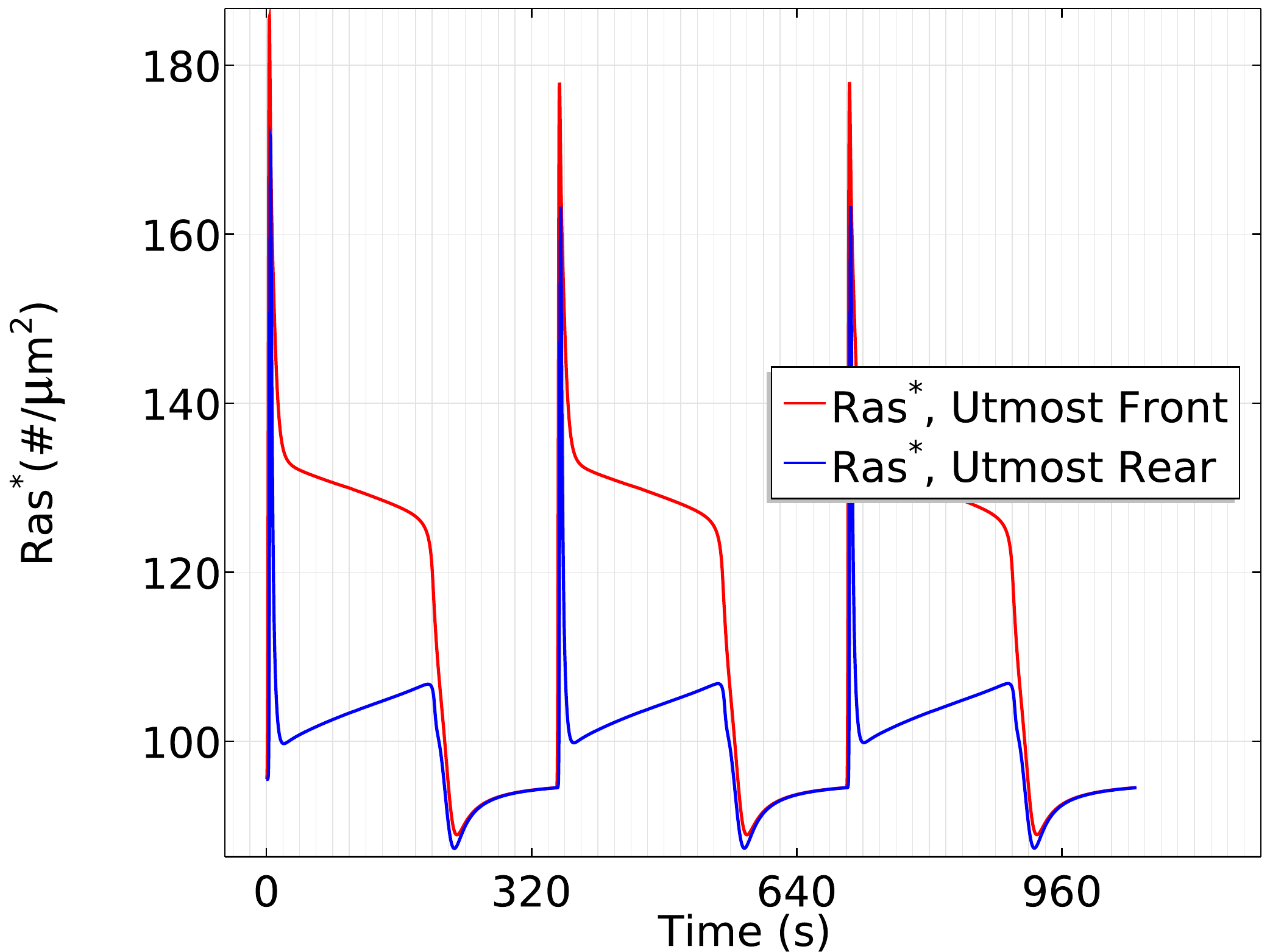}}
\caption{\emph{Left}: The time course of the front and rear halfes when three
  waves pass the WT cell at $v=5\mu m/s$. \emph{Right}:  The time course of
  $Ras^*$ activity at the antipodal points.}
\label{threeperiods}
\end{figure}

As shown in Fig.~\ref{normalwave}, Ras is activated everywhere as the wave
passes, but Ras activation is delayed about 1 sec in the rear half
(Fig.~\ref{normalwave} -right) for a wave traveling at the natural wave
speed. Ras activation is higher at the front of the cell than at the rear
throughout passage of the wave, thereby providing persistent directionality in
Ras activation and the potential for persistent orientation as the wave passes.
It should be emphasized that we are simulating the rounded LatA-treated cells
that have no intrinsic polarity, which suggests that polarity is not necessary
for the persistence of direction sensing at the natural wave speed, even at the
level of Ras activity.  By comparing  Fig.~\ref{rectify} and Fig.~\ref{normalwave}, one  sees a
similar pattern in Ras activation. In fact, due to the rectification characteristic observed in
uniform stimuli, $Ras^*$ activity does not drop significantly in a wave, and therefore the
front is able to maintain a higher $Ras^*$. To determine whether the cell is able to respond after
the first wave passes, we applied the same wave for three periods, and one sees in
Fig.~\ref{threeperiods} that the cell responses are almost identical for three
successive passages of a wave.

\begin{figure}[h!]
\centerline{\includegraphics[width=7cm]  {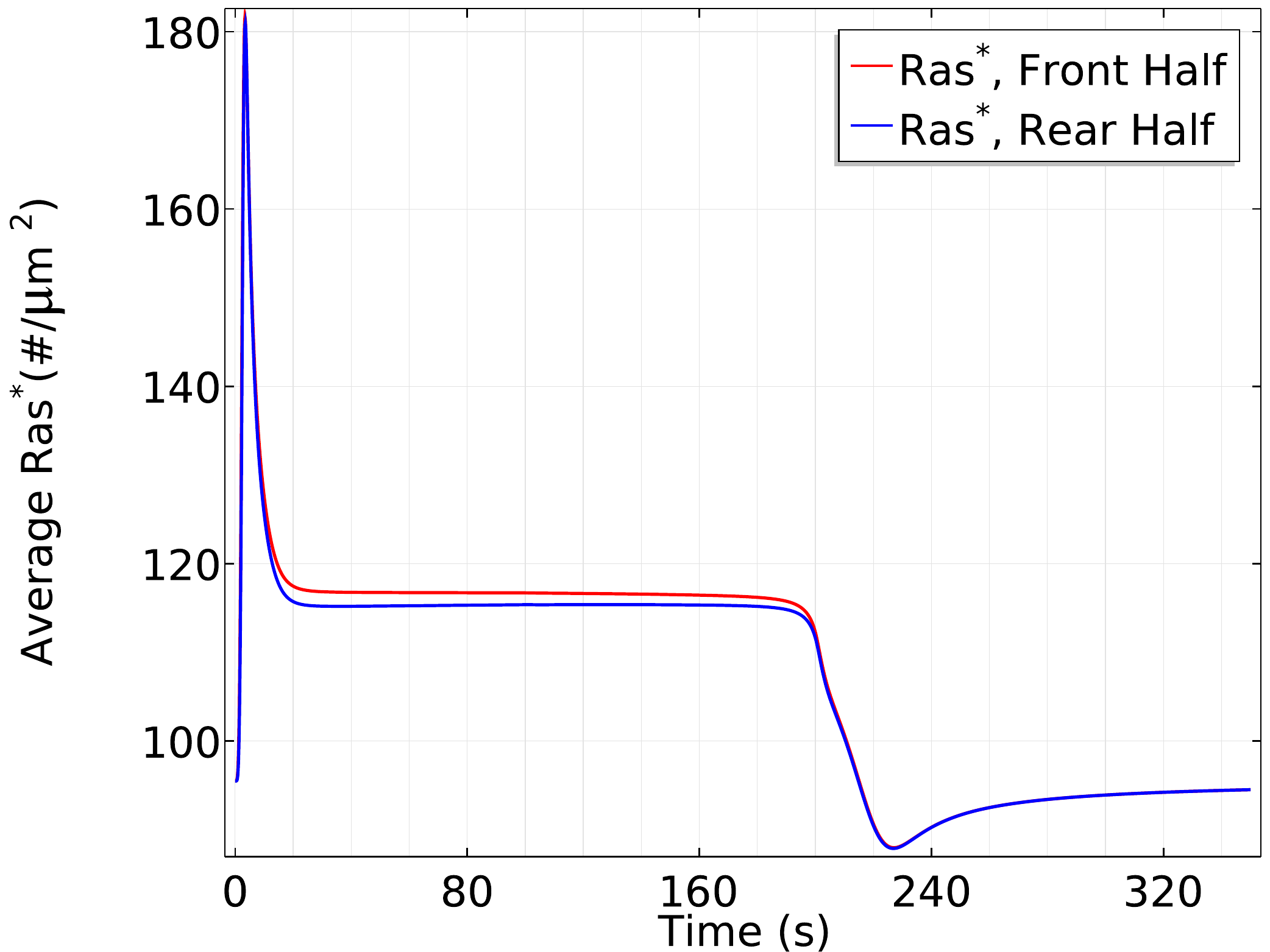}\hspace*{.25in}\includegraphics[width=7cm]
  {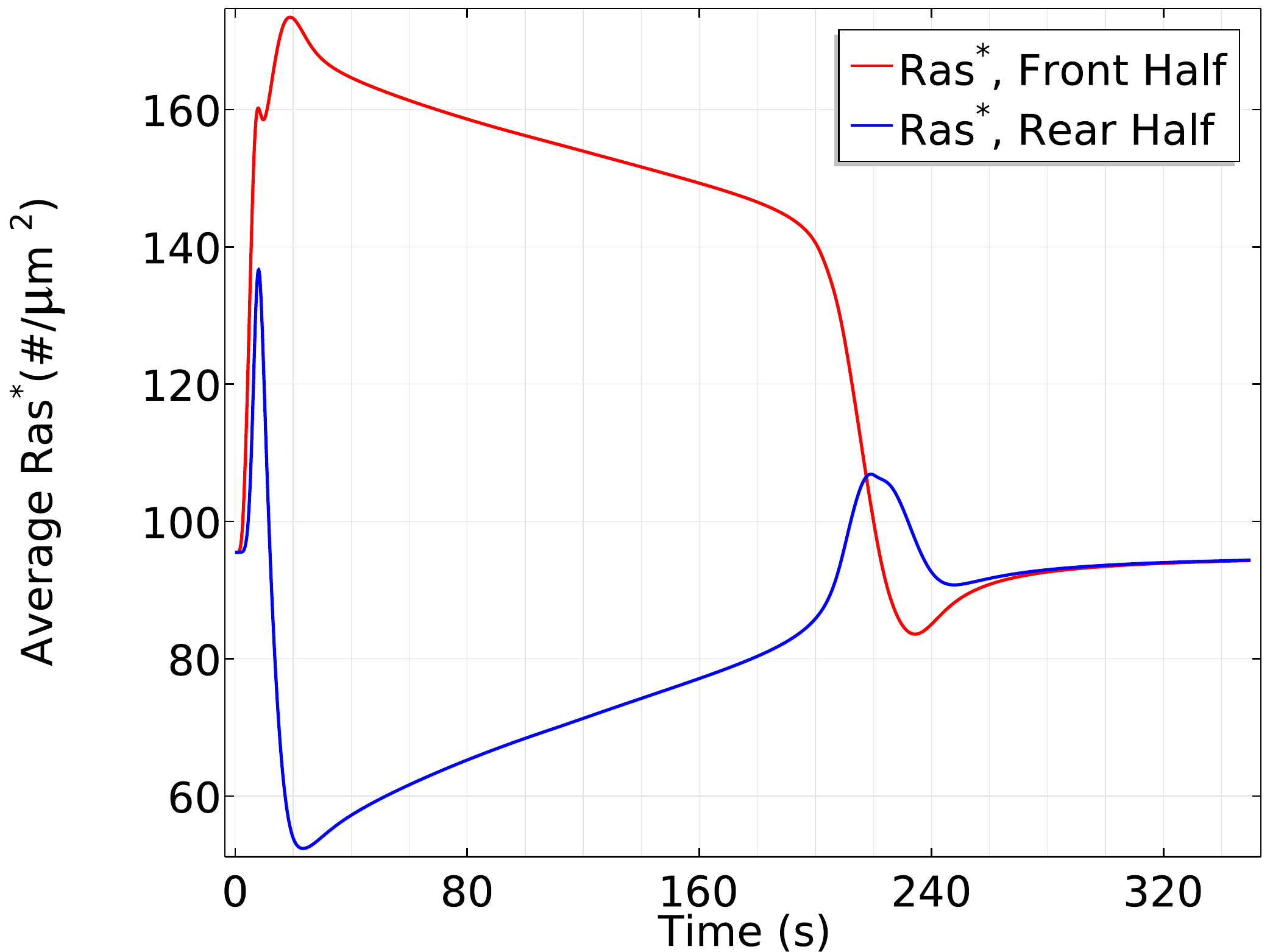}}
\caption{The time course of Ras activation at the front and rear
  half for a wave speed  $v=50\mu m/s$ (left), and  $v=0.5\mu m/s$ (right).}
\label{fastandslowwave}
\end{figure}

It is also known that wave speeds affect the spatial pattern of Ras activity
over a cell \cite{nakajima2014},  in that Ras is
activated uniformly for a fast wave,  and  activated at  both the wavefront and
waveback for slow waves. To test the effects of the wave speed, we apply a fast
wave ($50 \mu m/s$) and a slow wave ($0.5 \mu m/s$) to the rounded LatA-treated
cells. The results are shown in Fig.~\ref{fastandslowwave}.  At a wave speed of
$50 \mu m /s$, Ras activation is uniform along the cell periphery, as is
observed in the experiments, but at  $0.5 \mu m/s$ we see a
significant Ras reactivation at the rear of the cell and the $Ras^*$ distribution
reverses at the back of the wave.

\begin{figure}[H]
\centerline{\includegraphics[width=7cm]
  {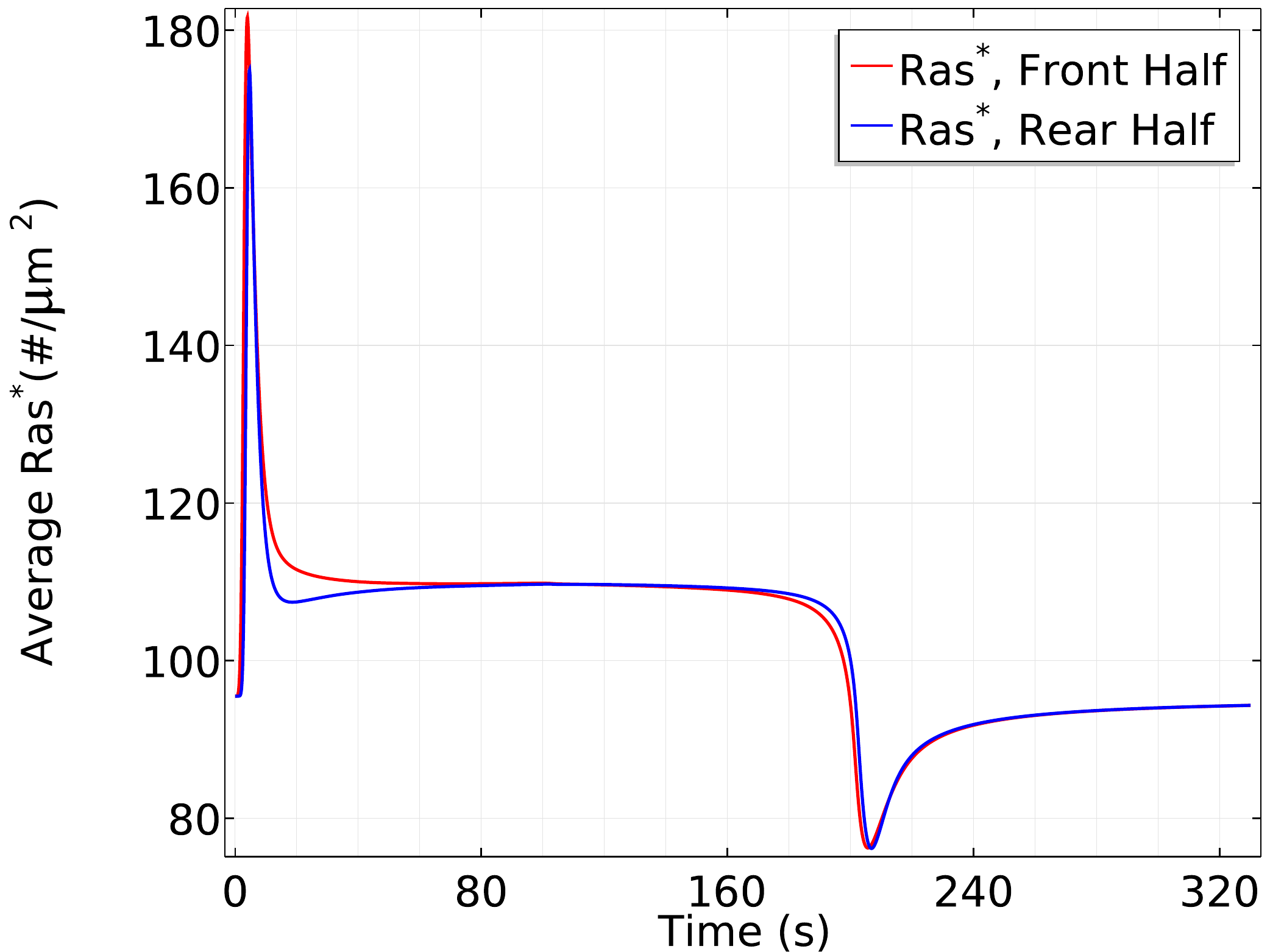}\hspace*{.25in}\includegraphics[width=7cm]
  {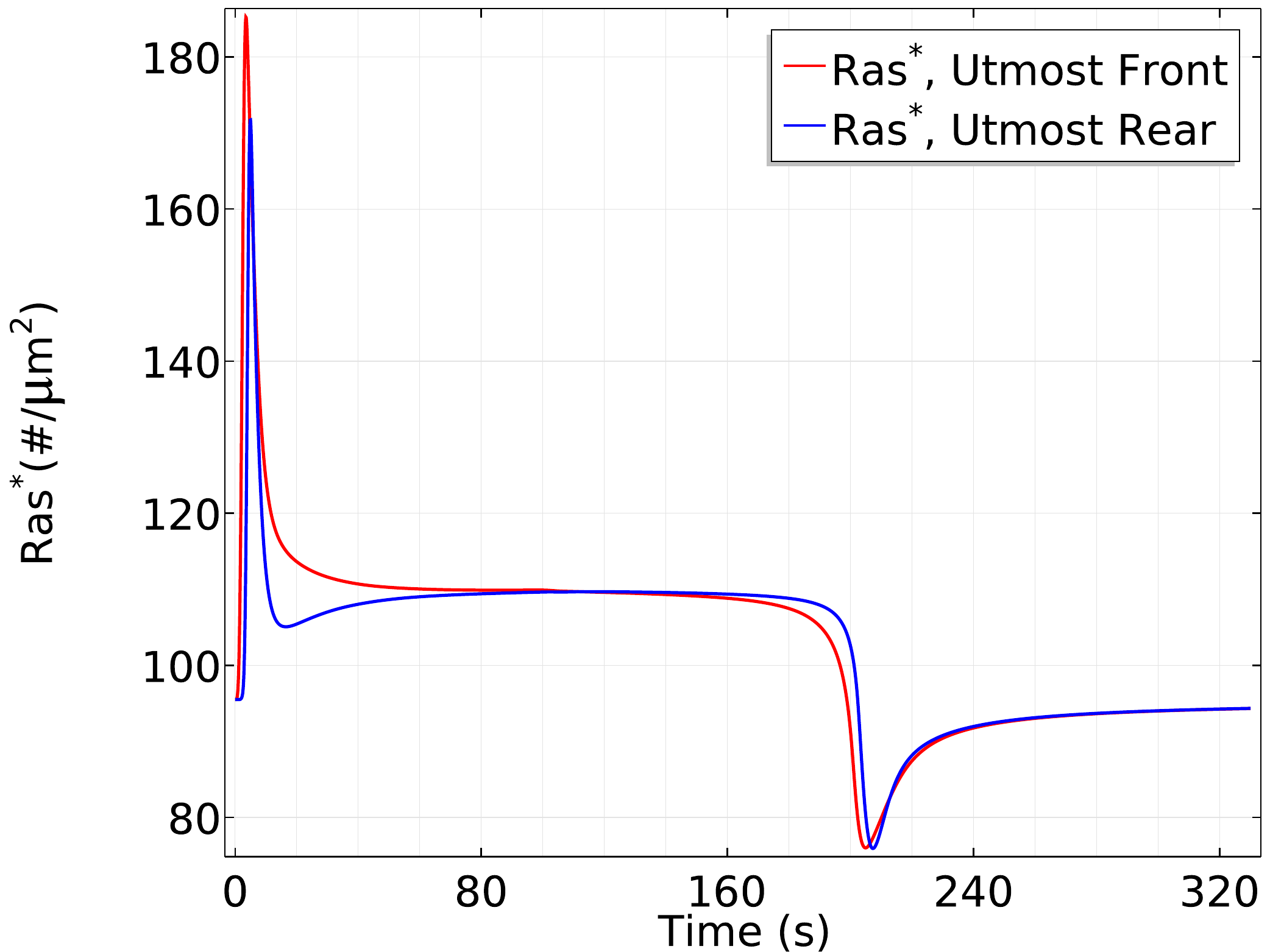}}
\caption{\emph{Left}: Time course of the front and rear half when the triangle
  wave passes the ric8-null cell at $v=5\mu m/s$. \emph{Right}: Time course of
  the point $Ras^*$ activity.}
\label{waveric8null}
\end{figure}

As was pointed out earlier, Ric8 plays an essential role in rectification under
uniform stimuli, and to further emphasize that the back of the wave problem is
closely connected with the disparity in the response to increasing {\em vs.}
decreasing stimuli, we applied the same wave used previously to a ric8-null
cell. The $Ras^*$ activity is shown in Fig.~\ref{waveric8null}, where one sees
that the persistence of directional information is essentially lost. It is not
surprising to see that $Ras^*$ at the front becomes smaller than the rear, which
indicates a reversal in the $Ras^*$ distribution, further reinforcing the
importance of the asymmetric response to increasing {\em vs} decreasing stimuli
in solving the back of the wave problem.

\subsection*{The trade-off between persistence of directionality and the ability to reorient}

\begin{figure}[H]
\centerline{\includegraphics[height=4.5cm] {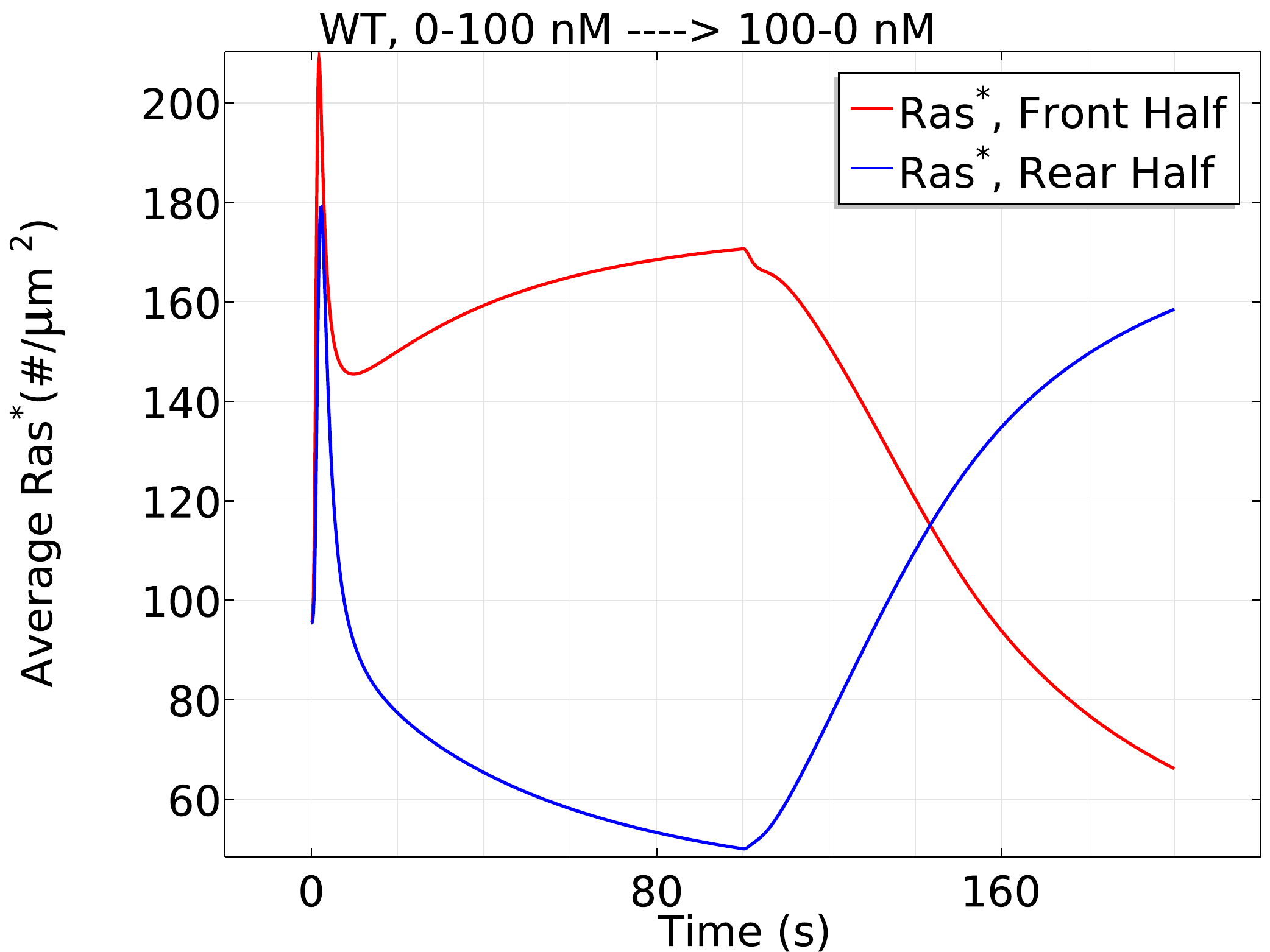}\includegraphics[height=4.5cm] {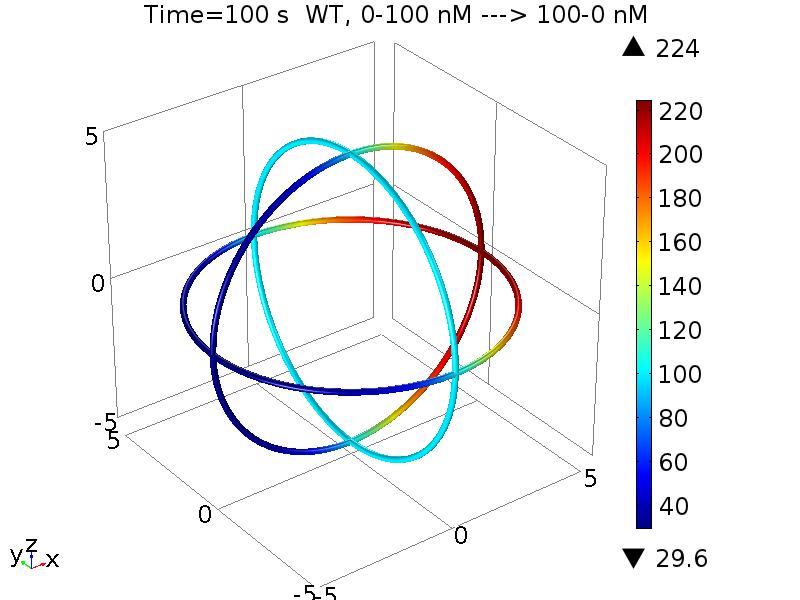}\includegraphics[height=4.5cm] {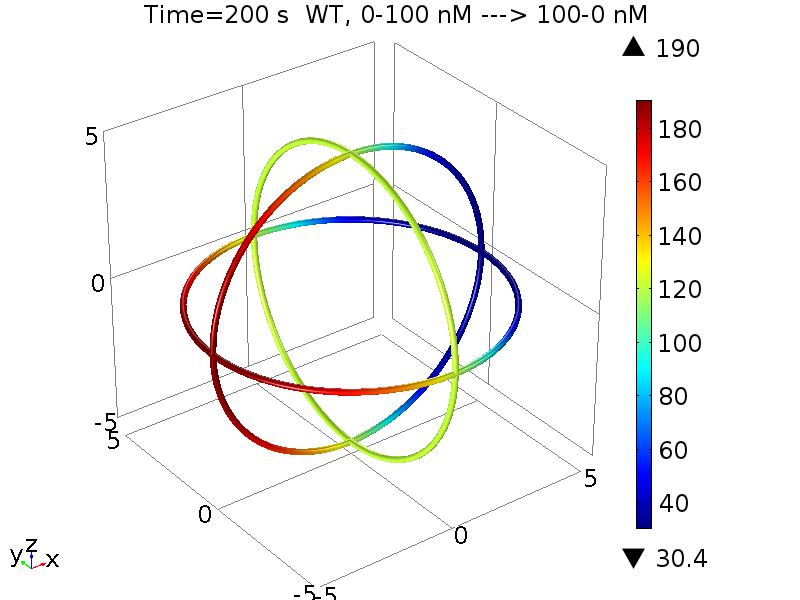}}
\caption{The response to gradient reversal.  A linear gradient of $10 nM/\mu m$ with
  mid point 50 nM (0-100 nM) is applied at $t=0$\,s and reversed at
  $t=100$\,s. The time course of average $Ras^*$ at the front and rear
  halves of WT cells (left) and the spatial profile of $Ras^*$ at three cutting
  lines of the sphere at $t=100$\,s (center) and   at $t=200$\,s (right). }
\label{reverse1}
\end{figure}

\begin{figure}[H]
\centerline{\includegraphics[width=7cm] {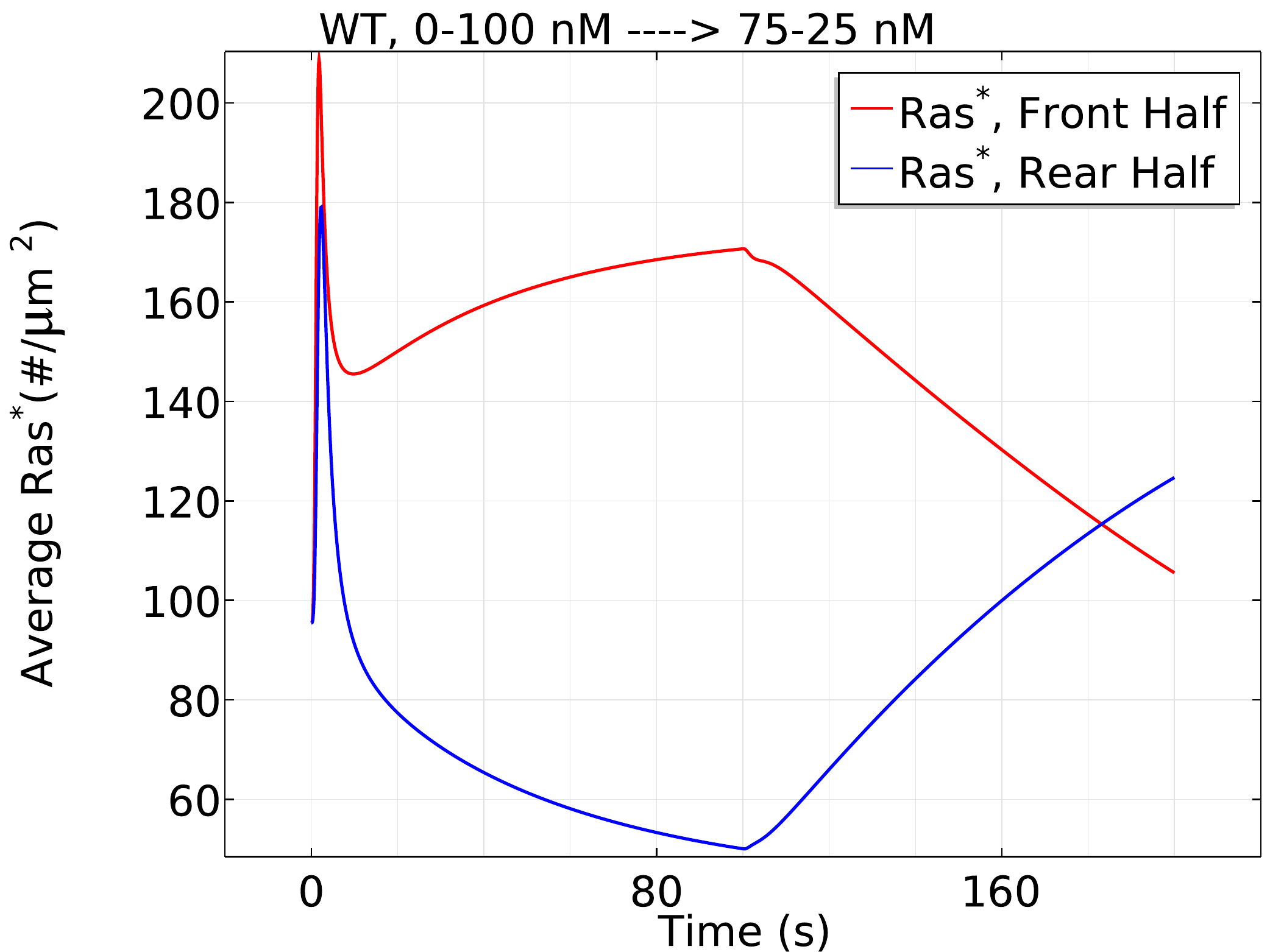}\hspace*{.25in}\includegraphics[width=7cm] {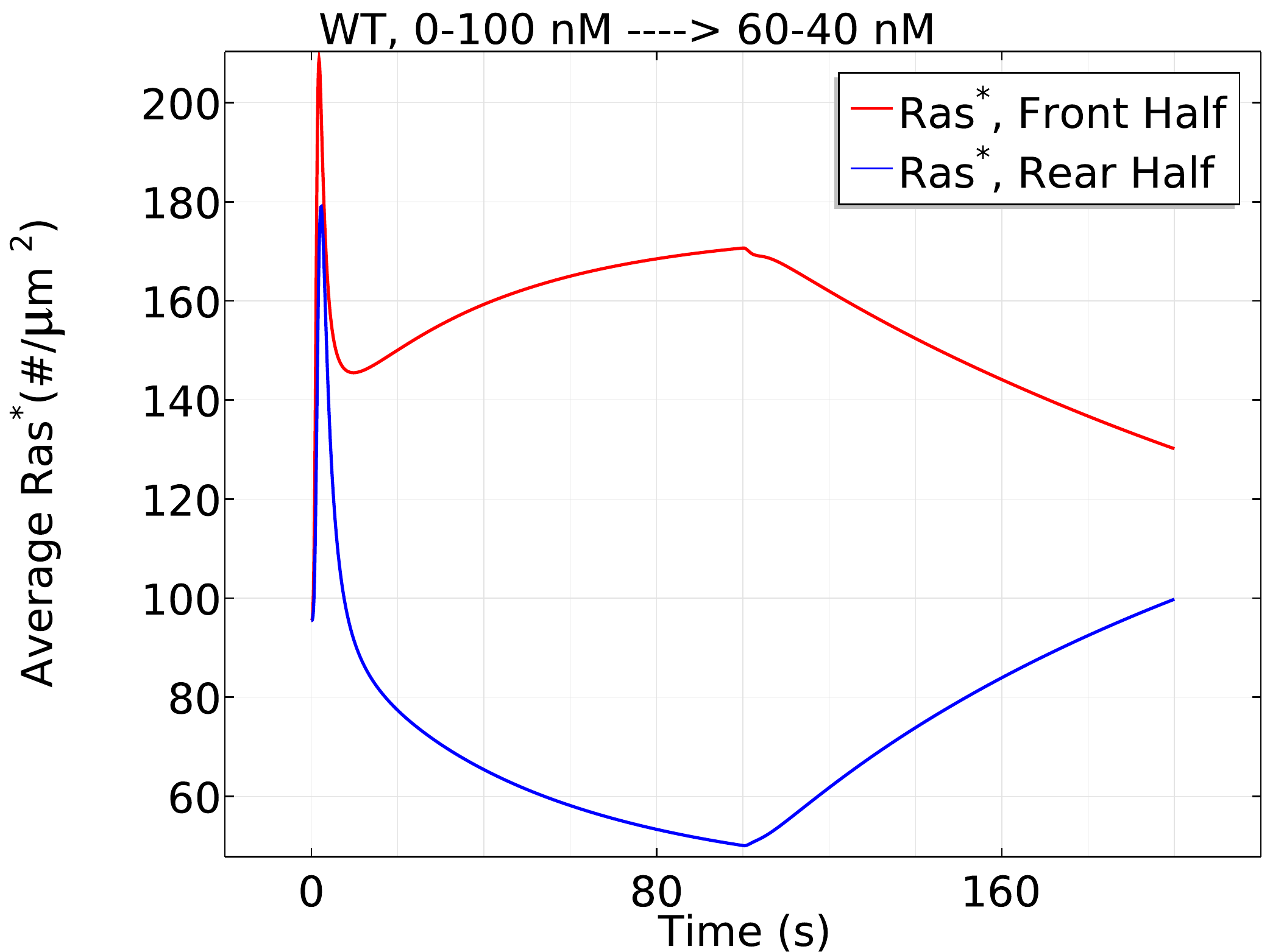}}
\caption{The response when the reversed gradient is shallower. \emph{Left}:  75-25 nM
  after 100 secs. \emph{Right}: 60-40 nM after 100 secs.}
\label{reverse2}
\end{figure}

Clearly there is a trade-off between the persistence of directionality in Ras
activation and the ability of cells to respond to new gradients. To investigate
whether the Ric8-induced rectification has an adverse effect on reorientation in
response to a reversed gradient, we subject cells in a 0-100 nM gradient to
reversals to increasingly weaker gradients. In each case we keep the mean
concentration experienced by the cell fixed to eliminate the mean concentration
effect (see. Fig.~\ref{backdependent}). For an equally strong reverse gradient
(100-0 nM), the directional persistence of $Ras^*$ is reversed within 100
seconds of gradient reversal, as shown in Fig.~\ref{reverse1}. The spatial
profile also indicates that $Ras^*$ distribution is strongly reversed after
switching to equally strong reversed gradients, (Fig.~\ref{reverse1} --center
and right). It is observed in Dicty that all cells ($20/20$) reversed their
direction of migration under this protocol \cite{skoge2014}. For intermediate
gradients (75-25 nM), $Ras^*$ is slightly reversed (Fig.~\ref{reverse2} --left)
in the same time window (0-200\,s). The spatial plot of $Ras^*$ indicates a
fluctuation along the cell periphery at the end of time window $t=200$\,s, (not
shown) suggesting uncertainties in $Ras^*$ redistribution. Consistently,
experiments show that a fraction of the cells ($5/17$) did not reverse their
migration direction. For weak gradients (60-40 nM) a difference in Ras
activation is still maintained at the end of the time window ($t=200$\,s)
(Fig.~\ref{reverse2} (right)), consistent with the observation that that all
cells continued moving in their original direction in this case
\cite{skoge2014}. These simulations suggest that Ric8-induced rectification does
not harm cells' reorientation in response to large amplitude reversals of the
gradient, but it delays the reorientation in a weak reversed gradient.

\subsection*{Variants of the model}
\paragraph{Robustness of the G$_\alpha$-G$_\beta\gamma$-Ric8 triangle}

In the current signal transduction  mechanism  Ric8  cycles
between a cytosolic pool and the membrane, where  it is activated by G$_{\beta\gamma}$
and it in turn  reactivates  G$_\alpha$. There is some evidence in other systems
that Ric8 may not require an activation step on the membrane \cite{Gong:2014:RHP,
  Van-Eps:2015:GNE}, and here we investigate the robustness of the
G$_\alpha$-G$_{\beta\gamma}$-Ric8 triangle by considering other
possibilities.  For  convenience in comparing schemes, we call the  current translocation-activation
mechanism  \emph{Mode 1}, and consider two alternative schemes.

\begin{itemize}
\item \emph{Mode 2}: Translocation-only mechanism. Reaction \circled{8} and
  \circled{10} in Table~\ref{Table_reaction} are eliminated. \circled{9} is
  modified so that $Ric8_m$ reactivates G$_\alpha$ directly.

\item \emph{Mode 3}: Alternative translocation-only mechanism. We remove the
  activation steps as in Mode 2, and G$_\alpha$ is assumed to be the membrane
  recruitment promoter in reaction  \circled{7}.
\end{itemize}

The simulations demonstrate that \emph{Mode 2} still captures the basic
characteristics of Ras activation, very similar to the results for \emph{Mode
  1}, except that the magnitudes are slightly changed (see Supporting
Information for plots). This  suggests that G$_{\beta\gamma}$ activation (Reaction
\circled{8} in Table~\ref{Table_reaction}) is not an essential step.

As for \emph{Mode 3}, it is shown that the cell is still able to sense direction
and exhibit biphasic responses under various cAMP gradients (see Supporting
Information for plots). They differ from the results in \emph{Mode 1} and
\emph{Mode 2} in that the point Ras activity equilibrates more rapidly and the
magnitudes of the front-back differences are smaller.

These results demonstrate  the robustness of the the G$_\alpha$-G$_\beta\gamma$-Ric8
triangle in the signal transduction pathways, providing flexibility in modeling
this triangle.

\paragraph{Amplification at the level of Ras}

It has been reported that the gradient of
active Ras across the cell is substantial in an imposed  cAMP  gradient \cite{zhang2008}. Recent quantitative
analysis also suggests that amplification may occur at the level of Ras
\cite{kortholt2013}. We test the magnitude of amplification by calculating the
amplification factor \cite{koshland1982, iglesias2002modeling}
$$
\sigma = \frac{(Ras^*_f-Ras^*_r)/Ras^*_m}{(cAMP_f-cAMP_r)/cAMP_m},
$$
where $X_m$ is the mean value of $X$. $X_f$ and $X_r$ are the concentrations of $C$
at the point on the cell surface exposed to the highest and lowest concentration
of stimulus, respectively. If $\sigma>1$, $Ras^*$ the signal is amplified.

The amplification indices are summarized in Table \ref{amplify}. As one sees
in the table, the signal is amplified at the level of Ras in both Mode 1 and
Mode 2,  but the signal amplification indices for Mode 3 are smaller than 1,
which indicates that the signal is not amplified.
\begin{table}[H]
\begin{center}
\caption{{\bf Amplification indices under various modes and gradients} \label{amplify}}
\begin{tabular}{llll}
  & 1-10 nM & 0-50 nM & 125-175 nM \\ \hline
Mode 1 & 1.3 & 1.7 & 2.7 \\
Mode 2 & 1.6 & 2.0 & 1.6 \\
Mode 3 & 0.6 & 0.7 & 0.7 \\
\hline
\end{tabular}
\end{center}
\end{table}

There are two sources of amplification in the proposed network. Firstly, the
higher concentration of $G_{{\alpha}_2}^*$ on the membrane at the front of the
cell induces a higher localization and activation of Ric8, which reactivates
$G_{{\alpha}_2}$ and further promotes RasGEF localization at the front.
Secondly, faster \G2 reassociation at the back due to higher $G_{{\alpha}_2}^*$
hydrolysis induces a faster \G2 cycling, providing more \G2 at the back. As a
result, the faster reassociated $G_{\alpha\beta\gamma}$ at the back can provides
a source of $G_{\alpha\beta\gamma}$ needed at the front by diffusion, which
creates an imbalanced sequestration of $G_{\alpha\beta\gamma}$ between the front
and the back. These two positive feedback loops are built into Mode 1 and Mode
2, but not into Mode 3.

\paragraph{The effect of cell shape}

\begin{figure}[H]
  \centerline{\includegraphics[width=7cm]
    {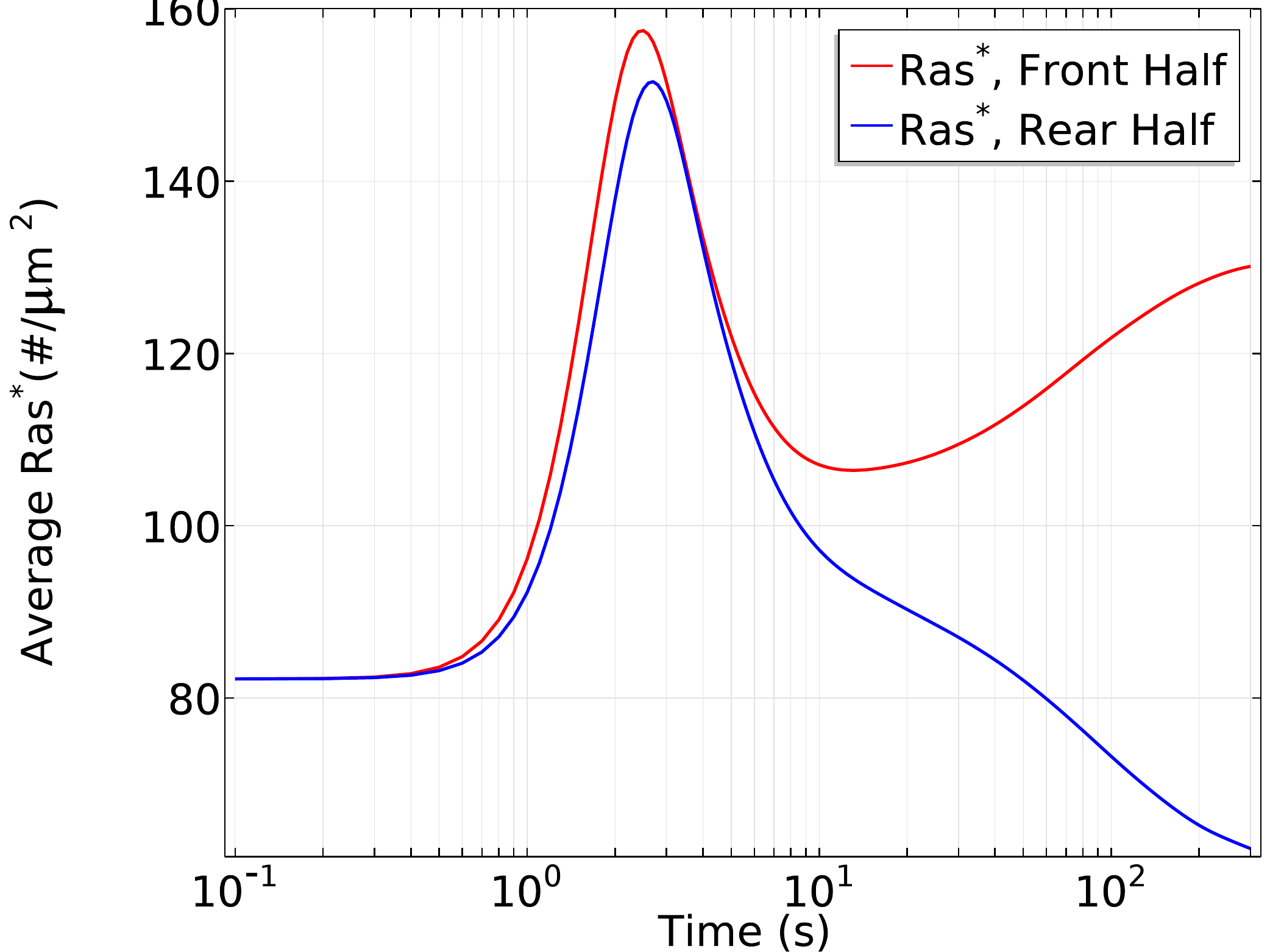}\hspace*{.25in}\includegraphics[width=7cm]
    {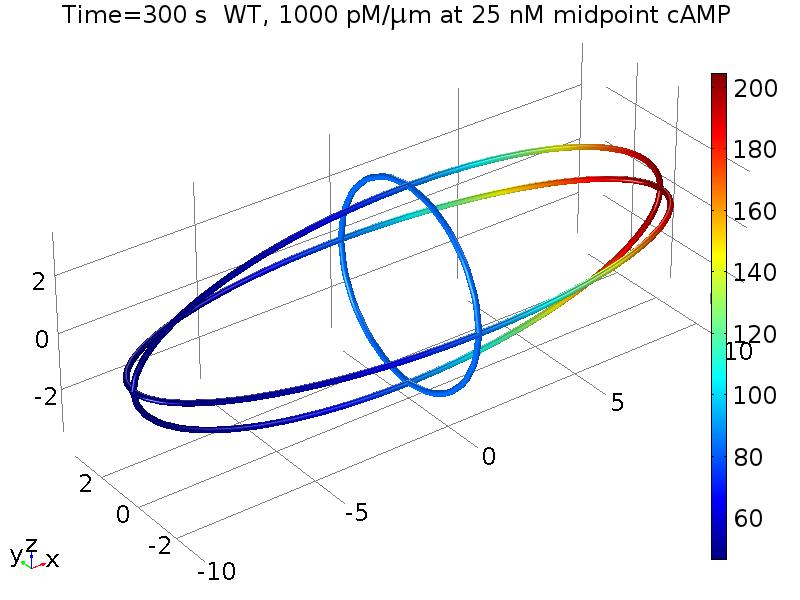}}
  \centerline{\includegraphics[width=7cm]
    {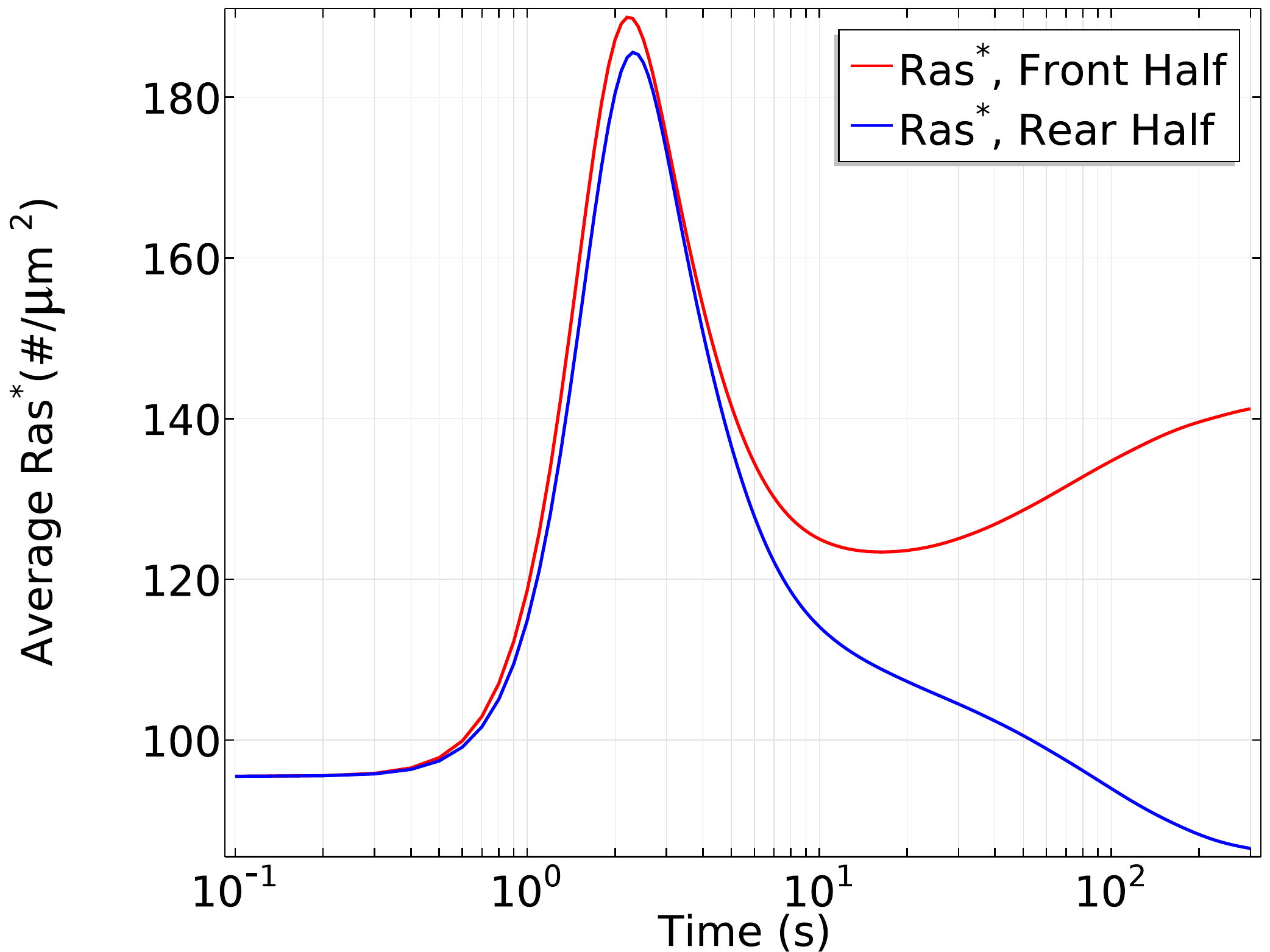}\hspace*{.25in}\includegraphics[width=7cm]
    {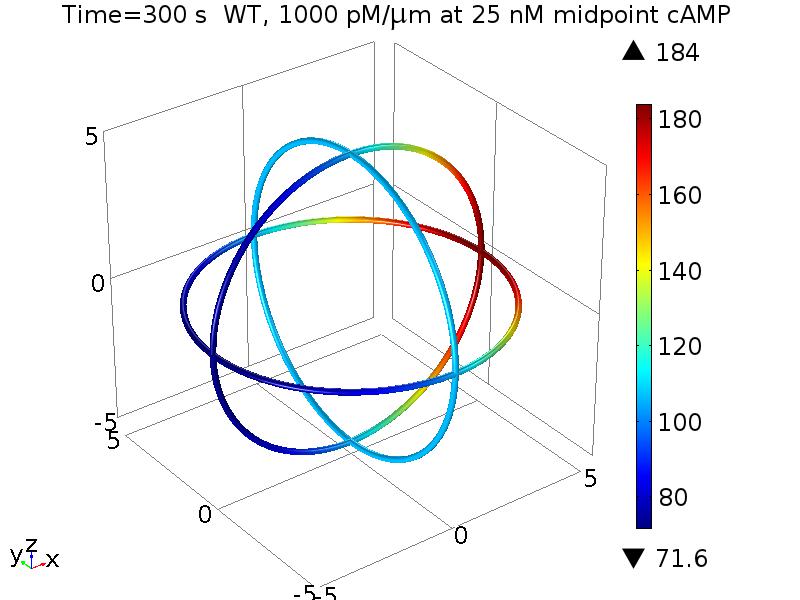}}
\caption{\emph{Top}: Ras activity for a ellipsoidal cell.  \emph{Left}: Average $Ras^*$ at front and rear half of WT
  cells; \emph{Right}: The spatial profile of $Ras^*$ at $t=300s$; \emph{Bottom}: Ras activity for a sphere cell.  \emph{Left}: Average $Ras^*$ at front and rear half of WT
  cells; \emph{Right}: The spatial profile of $Ras^*$ at $t=300s$.}
\label{changeshape}
\end{figure}

Heretofore  we have assumed that  the cell is pretreated with LatA, hence the cell
is  spherical with radius $r=5 \mu m$. To investigate how cell shape
may alter the $Ras^*$ dynamics, we construct an ellipsoid with the same volume
as that of the standard cell. By assuming that the ellipsoid is prolate, we have
$$
a=10 \mu m, b=c=3.5 \mu m.
$$
To test the effect of this shape change, we applied a cAMP gradient of 1000
$pM/\mu m$ with a 25 $\mu M$ midpoint, and the resulting responses are shown in
Fig.~\ref{changeshape}. The basic characteristics of Ras activation are still
maintained for an ellipsoidal cell: the cell first experiences a transient
activation both at the front and rear; then Ras is reactivation at the front and
a clear symmetry breaking emerges.

Fig.~\ref{changeshape} illustrates how cell shapes affect Ras activity. On one
hand, the density of molecules is reduced when the cell is changed from a sphere
to an ellipsoid with the same volume. Hence we see that the peak of first phase
for an ellipsoid is smaller than for a sphere due to lower availabilities of
molecules, although the endpoint cAMP sensed by a cell is increased from a 10 nM
difference (20-30 nM) to a 20 nM (15-35 nM) difference. On the other hand,
although the point $Ras^*$ at the frontal point for a ellipsoid cell is higher
than a sphere cell (see right panels of Fig.~\ref{changeshape}), the average
$Ras^*$ at the front half of the ellipsoid cell is still smaller than for the sphere
cell, suggesting that the larger gradident does not compensate for the smaller
molecular densities.

\section*{Discussion}

Chemotaxis is a dynamic spatio-temporal process that involves direction sensing,
polarization, and cell movement, and direction sensing is the first essential
step in this process, becuase it defines the cell's compass. A growing body of
evidence suggests that Ras is an ideal candidate within the chemotactic
signalling cascade to play an essential role in direction sensing
\cite{zhang2008, kataria2013}. In this article, we developed a novel modular
model of direction sensing at the level of Ras activation. The model
incorporates biochemical interactions in Dicty and captures many aspects of its
response. The model consists of the cAMP receptor, the G-protein \G2, and a Ras
GTPase module in which both adaptation and amplification occur. Utilizing a
rounded cell pretreated by LatA as was done in experiments, we investigated Ras
activation patterns in various cAMP stimuli. Simulations of this model give
insights into how the signal transduction network determines Ras activation
characteristics in wild type cells, how an altered network in mutant cells
changes Ras activation, and how the spatial profile and persistence of Ras
activation can lead to directional persistence.

We proposed an experimentally-based kinetic model of \G2 signaling in which
the intact \G2 and the $\Gbg$ subunit can cycle between the membrane and the cytosol,
while the $G_{{\alpha}_2}$ subunit remains membrane-bound.  Moreover,
$G_{{\alpha}_2}$ can be reactivated by the only known (to date)  GEF for $G_{{\alpha}_2}$, Ric8. The
regulation of Ric8 is not well-defined, but we assume that it is also cycles
between the cytosol and the membrane, and that its recruitment to the membrane
is promoted by $G_{{\alpha}_2}^*$. The model replicates the persistent
$G_{\alpha\beta\gamma}$ dissociation in the presence of cAMP, and  also
demonstrates that $\Gbg$ and $G_{{\alpha}_2}^*$ are produced in a dose-dependent
manner. Interestingly, the  model reveals that $G_{{\alpha}_2}$ exhibits
dose-dependent kinetic diversities. The variety of $G_{\alpha_2}$ dynamics
revealed here may have important implications in direction sensing because in
neutrophils $G_{\alpha_2}$-GDP accumulates at the leading edge and is involved
in regulating directionality \cite{kamakura2013}, although it has not been
demonstrated that Ric8 is involved there.

Adaptation of Ras activity is controlled by a balance between RasGEF and RasGAP,
both of which can cycle between the membrane and the cytosol. This component of
the network involves incoherent feed-forward, and becuase both can cycle between
membrane and cytosol, can give rise to spatial asymmetry in Ras activation. Both
RasGEF and RasGAP are activated at the membrane by free $\Gbg$, but the
translocation of RasGEF from the cytosol is enhanced by $G_{{\alpha}_2}^*$. The
proposed translocation-activation topology is able to capture the dose-dependent
Ras activation and various patterns such rectification and refractoriness under
uniform stimuli. It also predicts that imperfect adaptation is inevitable in
wild type cells due to the asymmetrical translocation of RasGEF.  Takeda {\em et
  al.\,}\cite{takeda2012} proposed an incoherent feedforward activation model to
explain adaptation of Ras activity in which RasGEF is assumed to be confined to
the membrane and RasGAP diffuses in the cytosol. In our model, both RasGEF and
RasGAP can diffuse in the cytosol at equal rates, and both can be recruited to
the membrane and activated by $\Gbg$.

Direction sensing, biphasic Ras activation and signal amplification are achieved
by complex  interactions between the modules. The
incoherent-feedforward-activation by  globally-diffusing $\Gbg$ contributes to
a transient activation along the entire cell perimeter. The activation at the front of
the cell (facing the higher cAMP concentration) is initially faster and stronger
due to the cAMP gradient, but it provides no symmetry breaking or signal
amplification since diffusion eliminates the initial $\Gbg$ concentration
gradient. This means that G$_{\beta\gamma}$ does not reflect the external
stimulus gradient and provides no basis for direction sensing in LatA-treated
cells, although it is essential for RasGEF and RasGAP activation. It is the Ric8
regulated, membrane-bound $G_{{\alpha}_2}^*$ that determines the symmetry
breaking and signal amplification. $G_{{\alpha}_2}^*$ creates an asymmetrical
recruitment of RasGEF in a cAMP gradient, which in turn induces asymmetrical
RasGEF activation, providing a basis for symmetry breaking. More importantly,
Ric8 recruitment to the membrane is elevated by $G_{{\alpha}_2}^*$, while
activated Ric8 reactivates $G_{{\alpha}_2}$, forming a positive feedback
loop. In addition, faster \G2 reassociation at the back of the cell due to less
reactivation of $G_{{\alpha}_2}$ there induces faster \G2 cycling. Since \G2 diffuses
in the cytosol, this provides a potential redistribution of \G2 from the back to
the front, which in turn results in more $G_{{\alpha}_2}^*$ at the front,
thereby forming another positive feedback loop. These two positive feedback
loops generate the symmetry breaking and signal amplification of Ras activation
in a cAMP gradient.

In contrast to LEGI-type models, the global diffusing G$_{\beta\gamma}$ does not
act as an inhibitor directly in our model -- instead, it induces both activation
and inhibition by activating RasGEF and RasGAP respectively. G$_{\beta\gamma}$
also serves as a `global' activator for the pool of $RasGEF^*$ and as a `global'
inhibitor by creating a uniform inhibition pool of $RasGAP^*$.  Asymmetry in
their localization at the membrane arises from the fact that membrane-bound
G$_{{\alpha}_2}^*$ recruits RasGEF from the cytosol, thereby creating an
asymmetrical pool of $RasGEF^*$. Hence, our model can be regarded as a
local-global transitions of both excitation and inhibition with a delayed local
sequestrations of excitation model, in the sense that initially both activation
and inhibition go through a local-global transition due to diffusion of $\Gbg$
while a delayed localized translocation by $G_{{\alpha}_2}^*$ contributes to a
local excitation. Direction sensing is results from the $\Gbg$- mediated,
$G_{{\alpha}_2}$-Ric8 dependent signal transduction network.

The genome of Dictyostelium
contains 25 genes encoding for RasGEFs and 17 genes encoding for RasGAPs that
potentially activate and inactivate Ras, respectively \cite{wilkins2005,
weeks2005}. Various RasGEF and RasGAP could be utilized at different stages of
chemotaxis \cite{kortholt2013}. Our model suggests that both refractoriness and
rectification are managed by subtle temporal regulation of RasGEF and RasGAP
activity, highlighting the importance of RasGEF and RasGAP activation.

We also studied cell responses to $g_{\alpha 2}$ and ric8 mutations extensively. It
is predicted in numerical simulations that in the presence of uniform stimulus,
adaptation of Ras activity is perfect and the maximum cytosolic RBD depletion is
reduced in $g_{\alpha 2}$-null cells. In a cAMP gradient, $g_{\alpha 2}$-null cells
fail to sense directions and there is only an initial transient Ras
activation. Adaptation of Ras activity is still imperfect in ric8-null cells,
but the magnitude of imperfectness is reduced comparing to wild type
cells. Moreover, simulations suggest that ric8-null cells fail to sense
direction when they are exposed to a shallow gradient or a steep gradient with
high mean concentration, highlight the importance of Ric8 in regulating Ras
activation.

There are still missing links in our proposed model. Firstly, there are no
direct evidence in Dicty that $G_{{\alpha}_2}^*$ is connected to RasGEF
translocation. Secondly, there is no Ric8 activation mechanism established in
Dicty. To establish the missing link, it is important to study molecules that
interact with $G_{{\alpha}_2}^*$ and Ric8 in experiments, where our model could
provide guidance in experimental designs.

Although the model is based on cAMP induced Ras activation in Dicty, GPCR
mediated Ras activation is remarkable conserved between Dicty and mammalian
leukocytes \cite{yulia2014}. GEF translocation through interacting with an
upstream GTP-bound G protein is a principle conserved in evolution
\cite{bos2007} and G$_\alpha$'s role in GPCR mediated signalling has been
emphasized in other systems \cite{moore2012, fuku2001} and in drug discovery
\cite{adam2011}. Therefore, our model could serve a generic framework for GPCR
mediated Ras activation in other systems.

\section*{Materials and Methods}

\subsection*{The evolution equations for the reaction-diffusion model }

We first  formulate  the reaction-diffusion system of signal transduction
in general terms and then list the specific equations for the model.

Consider a bounded three dimensional domain $\Omega \subset R^3$ representing a
cell, and denote $\partial \Omega$ as the plasma membrane. Then the reaction
diffusion equation for a cytosolic  species A is
\begin{equation}
\label{reactiondiffusion}
\frac{\partial C}{\partial t} = \nabla\cdot (D\nabla C) +\sum_{i}s^i\rxn^i,
\end{equation}
in which $C=C(t,x)$ represents the concentration of A at time $t$ at $x\in
\Omega$ and $D$ is the diffusion coefficient of A. The summation is a reaction
term indicating $A$ participates in cytosolic reactions which either depletes it
or produces it. The $i$th reaction produces $s^i$ molecules of A, or consumes
$-s^i >0$ molecules of A with a reaction rate $\rxn^i=\rxn^i(t,x)$. In the
signal transduction network considered in this article, $s^i = 0, 1$.

The boundary conditions involve reactions on the boundary and binding and
release of molecules at the membrane. We assume that the volume density $C$ (the
concentration in the cytosol) for A has the units $\mu M$ and that the surface
density (the concentration on the membrane), $C_m$, has the units $\#/\mu
m^2$. We also assume that the binding reactions at the membrane take place
within a layer of thickness $\delta (nm)$ at the membrane. Then the net flux to
the boundary, which can be positive or negative,  can be written as
\begin{equation}
\label{boundary}
-{\vec n} \cdot D\nabla  C = - D\frac{\partial C}{\partial n} = k^{+}\cdot
\delta \cdot C -k^{-}\cdot C_m  \equiv j^+ - j^-,
\end{equation}
where $\vec n $ is the exterior unit normal to $\partial
\Omega$, $k^{\pm}$ are the on and off rate of binding to the membrane,  and
$\kappa = 602$ relates  the units of volume density and surface density scaled
by Avogadro's constant.

For the membrane form of species A  we have the
translocation-reaction-diffusion  equation,
\begin{equation}
\label{transreaction}
\frac{\partial C_m}{\partial t} = \nabla\cdot (D_m\nabla C)+\kappa
(j^{+}-j^{-})+\sum s_m^i\rxn_m^i,
\end{equation}
where $C_m = C_m(t,x)$ denotes the concentration on the membrane and $D_m$ is
the surface diffusion coefficient \cite{rama2009, pff1995}. The first term represents the diffusion
on the membrane, which we ignore throughout, and  the second represents
transolcation between cytosol and membrane, which could be absent if A is
confined on the membrane, such as $Ras$, $Ras^*$.

There may also be conservation laws for certain substances. If the
substances are confined to the  membrane we write
\begin{equation}
\int_{\partial \Omega} \sum_{i=1}^{n}A_n dS = A^{tot},
\end{equation}
where $A_i$s are the concentrations of different forms and $A^{tot}$ represents
the total amount in the cell. If the substances are present both in the cytosol
and on the membrane, we write
\begin{equation}
\int_{\Omega} \sum_{i=1}^{k}A_i^cdx + \int_{\partial \Omega} \sum_{j=1}^{n}A_j^m
dS = A^{tot},
\end{equation}
where $A_i^c$s are the concentrations of different forms in the cytosol and
$A_i^m$s are the concentrations of different forms on the membrane.

We are now ready to assemble the system of equations that constitute the full
kinetic model in a given geometry $\Omega$. We have to account for  6 cytosolic species
in the system: $G_{\alpha\beta\gamma},c$, $\Gbgc $, $RasGEF_c$, $RasGAP_c$,
$Ric8_c$ and $RBD_c$. The evolution can be described by a system of
diffusion-translocation equations
\addtolength{\jot}{5pt}
\begin{alignat*}{1}
\delt{  G_{\alpha\beta\gamma,c} } &=\nabla \cdot (D_{G_{\alpha\beta\gamma},c}\nabla
      G_{\alpha\beta\gamma} ) \\
\delt{ \Gbgc } &= \nabla \cdot (D_{\Gbgc }\nabla  \Gbg ) \\
\delt{ RasGEF_c } &= \nabla \cdot (D_{RasGEF_c} \nabla   RasGEF_c ) \\
\delt{ RasGAP_c } &= \nabla  \cdot (D_{RasGAP_c} \nabla  RasGAP_c )\\
 \delt{ Ric8_c } &= \nabla \cdot (D_{Ric8_c} \nabla Ric8_c )\\
\delt{ RBD_c } &= \nabla    \cdot (D_{RBD_c} \nabla  RBD_c )
\end{alignat*}
with the following conditions on $\partial\Omega$,
\begin{align*}
D_{G_{\alpha\beta\gamma},c}\frac{\partial  G_{\alpha\beta\gamma},c }{\partial    n}&=\jt_1\\
D_{\Gbgc }\frac{\partial  \Gbgc  }{\partial    n}&=\jt_2\\
D_{RasGEF_c}\frac{\partial  RasGEF_c }{\partial
    n}&=\jt_5-\jt_6\\
D_{RasGAP_c}\frac{\partial  RasGAP_c }{\partial  n}&=\jt_7\\
D_{Ric8_c}\frac{\partial  Ric8_c }{\partial    n}&=\jt_3-\jt_4\\
D_{RBD_c}\frac{\partial  RBD_c }{\partial n}&=\jt_8-\jt_9.\\
\end{align*}

The species that evolve on the membrane are: $R^*$, $G_{\alpha\beta\gamma,m}$,
$\Gbgm$, $G^*_{\alpha}$, $G_{\alpha}$, $Ric8_m$, $Ric8^*$, $RasGEF_m$,
$RasGAP_m$, $RasGEF^*$, $RasGAP^*$, $Ras$, $Ras^*$ and $RBD_m$. The evolution
equations for these  are given by
\begin{align*}
\delt{ R^* } &=  \rxn_1 &\quad  \\
\delt{ G_{\alpha\beta\gamma,m} } &=  -\kappa\jt_1-\rxn_2+\rxn_7 \\
\delt{ \Gbgm }&=  -\kappa \jt_2+\rxn_2-\rxn_7 \\
\delt{ G^*_{\alpha} } &=  \rxn_2-\rxn_3+\rxn_5\\
\delt{ G_{\alpha} } &= \rxn_3-\rxn_5-\rxn_7 \\
\delt{ Ric8_m }&=  -\kappa \jt_3 +\kappa \jt_4-\rxn_4+\rxn_6 \\
\delt{ Ric8^* } &=  \rxn_4-\rxn_6 \\
\delt{ RasGEF_m } &=  -\kappa \jt_5+\kappa\jt_6-\rxn_8+\rxn_9 \\
\delt{ RasGAP_m } &= -\kappa \jt_7-\rxn_{10}+\rxn_{11}\\
\delt{ RasGEF^* }&=  \rxn_8-\rxn_9\\
\delt{ RasGAP^* }&=  \rxn_{10}-\rxn_{11}\\
\delt{ Ras^* }&= \rxn_{12}-\rxn_{13}+\rxn_{14}-\rxn_{15}\\
\delt{ Ras }&= -\rxn_{12}+\rxn_{13}-\rxn_{14}+\rxn_{15}\\
\delt{ RBD_m }&=  -\kappa \jt_8+\kappa \jt_9
\end{align*}
The following conservation laws are also imposed:
\begin{equation}
\int_{\partial \Omega} (R+R^*) ds = R^{t},
\end{equation}
where $R^t$ is the total amount of receptors.
\begin{equation}
\int_{\Omega} \left(G_{\alpha,c}
+\Gbgc +G_{\alpha\beta\gamma},c+G^*_{\alpha}\right)dx+\int_{\partial
  \Omega}\left(G_{\alpha} +\Gbgm+G_{\alpha\beta\gamma,m}\right)ds =
G_{\alpha\beta\gamma}^t,
\end{equation}
where $G_{\alpha\beta\gamma}^t$ is the total amount of heterotrimetric G protein,
indicating the cell does not produce additional heterotrimetric G protein.
\begin{equation}
\int_{\Omega} RasGEF_c dx +\int_{\partial \Omega}
\left(RasGEF_m+RasGEF^*\right)ds=RasGEF^{t}.
\end{equation}
Similarly, for RasGAP
\begin{equation}
\int_{\Omega} RasGAP_c dx +\int_{\partial \Omega}
\left(RasGAP_m+RasGAP^*\right)ds=RasGAP^{t}.
\end{equation}
For Ras, we have
\begin{equation}
\int_{\partial \Omega} \left(Ras+Ras^*\right)ds = Ras^{t}.
\end{equation}
\subsection*{Parameters}

{\bf We have to adjust the parameters to incorporate $\kappa$ and $\delta$ where
  needed.}

The parameters involved in the Receptor module are taken from the literature. We
estimated the parameters in the heterotrimeric G protein module from steady
state analysis (SSA) of the spatially lumped model averaged from the spatially
distributed model, see \nameref{parameter}. The parameters in the Ras module are
also estimated from steady state analysis and characteristics of Ras
activation. The detailed estimation scheme is described in the appendix. We
summarize the parameters in the following table.  \small
\addtolength{\tabcolsep}{-2pt}
\begin{table}[H]
\caption{{\bf Parameter values used in the model of Ras activation
    pathway.} \label{PIPparamTable}}
\begin{tabular}{llll}
\label{params}
 Parameter & Value & Description & References \\ \hline
$r$ & 5 $\mu m$ & Cell radius & \cite{jin2000} \\
$\delta$ & 10 $nm$ & Effective  length for membrane reactions & \cite{xu2010} \\ [5pt]
 \hline
$RasGEF^t$ & 80000  \#/\tn{cell} & Total RasGEF molecules & \cite{martin2006, aoki2007}\\
$RasGAP^t$ & 80000 \#/\tn{cell} & Total RasGAP molecules & \cite{martin2006} \\
$Ras^t$ & 300000 \#/\tn{cell} & Total Ras molecules on the membrane &\cite{martin2006} \\
$G_{\alpha\beta\gamma}^t$ & 300000 \#/\tn{cell} & Total  heterotrimeric G protein
molecules & \cite{martin2006, xu2010} \\
$R^t$ &  80000\#/\tn{cell} & Total receptors on the membrane & \cite{johnson1992,
   ueda2007} \\[5pt]
 \hline
$D_{RasGEF_c}$ & 30 $\mu m^2/s$ & Diffusion constant of
 RasGEF & \cite{postma2001} \\ $D_{RasGAP_c}$ & 30 $\mu m^2/s$ & Diffusion
 constant of RasGAP & \cite{postma2001} \\ $D_{G_{\alpha\beta\gamma},c}$ & 30 $\mu
 m^2/s$ & Diffusion constant of G$_{\alpha\beta\gamma}$ & \cite{postma2001}
 \\ $D_{\beta\gamma,c}$ & 30 $\mu m^2/s$ & Diffusion constant of G$_{\beta\gamma}$
 & \cite{postma2001} \\ $D_{RBD_c}$ & 30 $\mu m^2/s$ & Diffusion constant of
 $RBD_c$ & \cite{postma2001} \\ $D_{Ric8_c}$ & 30 $\mu m^2/s$ & Diffusion
 constant of $Ric8_c$ & \cite{postma2001} \\[5pt]
\hline
$\rk_1^{+}$ & 5.6 $(\mu
 M)^{-1}s^{-1}$ & Average binding rate of cAMP to GPCR & \cite{ueda2001,
   janetopoulos} \\ $\rk_{1}^{-}$ & 1 $s^{-1}$ & Average unbind rate of cAMP-bound
 GPCR & \cite{ueda2001, ueda2007} \\ [5pt]
 \hline
$\rk_2$ & 0.02 $(\#/\mu
 m^2)^{-1}s^{-1}$ & G$_{\alpha\beta\gamma}$ dissociation rate by $R^*$ &
 \tn{Estimated from SSA} and \cite{janetopoulos} \\ $\rk_3$ & 1 $s^{-1}$ &
 $G^*_{\alpha}$ GTPase rate & \cite{xu2010} \\ $\rk_4$ & 0.004 $(\#/\mu
 m^2)^{-1}s^{-1}$ & Ric8 activation rate on the membrane & \\ $\rk_5$ & 0.2
 $(\#/\mu m^2)^{-1}s^{-1}$ & $G_{\alpha}$ reactivation rate by $Ric8^*$ & \\ $\rk_6$
 & 1 $s^{-1}$ & $Ric8^*$ deactivation rate & \\ $\rk_7$ & 0.0070 $(\#/\mu
 m^2)^{-1}s^{-1}$ & Reassociation rate of G$_{\alpha}$ and G$_{\beta\gamma,m}$ &
 \tn{Estimated from SSA} and \cite{janetopoulos} \\ $\tr_1$ & 1 $s^{-1}$ &
 Off rate of G$_{\alpha\beta\gamma,m}$ & \cite{elzie2009}\\ $\tr_2$ &
 $3.9\times 10^2 s^{-1}$ & Translocation rate of G$_{\alpha\beta\gamma},c$ &
 \tn{Estimated from SSA} \\ $\tr_3$ & 1 $ s^{-1}$ & Off rate of
 G$_{\beta\gamma,m}$ & \tn{Set the same as }$G_{\alpha\beta\gamma}$ \\ $\tr_4$ &
 $3.9\times 10^2s^{-1}$ & Translocation rate of G$_{\beta\gamma,c}$ & \tn{Estimated
   from SSA} \\ $\tr_5$ & 1 $ s^{-1}$ & Off rate of $Ric8_m$ & \tn{Set the
   same as }$G_{\alpha\beta\gamma}$ \\ $\tr_6$ & $1.6667s^{-1}$ & Translocation
 rate of $Ric8_c$ & \tn{Estimated from SSA} \\ $\tr_7$ & $0.02(\#/\mu
 m^2)^{-1}s^{-1}$ & Translocation rate of $Ric8_c$ facilitated by $G^*_{\alpha}$ &\\[5pt]
 \hline
 $\rk_8$ & 0.0004 $(\#/\mu m^2)^{-1}s^{-1}$ & RasGEF activation rate by
 G$_{\beta\gamma,m}$ & \\ $\rk_9$ & $2s^{-1}$ & $RasGEF^*$ deactivation rate &
 \\ $\rk_{10}$ & 0.0001 $(\#/\mu m^2)^{-1}s^{-1}$ & RasGAP activation rate by
 G$_{\beta\gamma,m}$ & \\ $\rk_{11}$ & 0.5 $s^{-1}$ & $RasGAP^*$ deactivation rate &
 \\ $\rk_{12}$ & 0.11 $(\#/\mu m^2)^{-1}s^{-1}$ & Ras activation rate by
 $RasGEF^a$ & \\ $\rk_{13}$ & 1 $(\#/\mu m^2)^{-1}s^{-1}$ & $Ras^*$ deactivation
 rate by $RasGAP^a$ & \\ $\rk_{14}$ & $1.1\times 10^{-7}s^{-1}$ & Spontaneous Ras
 activation rate & \\ $\rk_{15}$ & $10^{-6}s^{-1}$ & Spontaneous $Ras^*$
 deactivation rate & \\ $\tr_8$ & $1s^{-1}$& Off rate of $RasGEF_m$& \tn{Set
   the same as }PTEN \cite{vazquez2006}\\ $\tr_9$ & $444.4s^{-1}$&
 Translocation rate of $RasGEF_c$& \tn{Estimated from SSA}\\ $\tr_{10}$ &$2
 (\#/\mu m^2)^{-1}s^{-1}$ & Translcation rate of $RasGEF_c$ facilitated by
 $G^*_{\alpha}$& \\ $\tr_{11}$ & $1s^{-1}$& Off rate of $RasGAP_m$& \tn{Set the
   same as }PTEN \cite{vazquez2006}\\ $\tr_{12}$ & $444.4s^{-1}$&
 Translocation rate of $RasGAP_c$& \tn{Estimated from SSA}\\ \hline
\end{tabular}
\end{table}
\addtolength{\tabcolsep}{2pt}
\normalsize

%
%



\section*{Acknowledgments} This work was supported by NSF Grant DMS \#s 9517884 and 131974 to H.G. Othmer, and by the Newton Institute and the Simons Foundation.

\section*{Supporting Information}
\subsection*{Other characteristics under uniform stimuli}
\begin{figure}[H]
\centerline{\includegraphics[width=6cm] {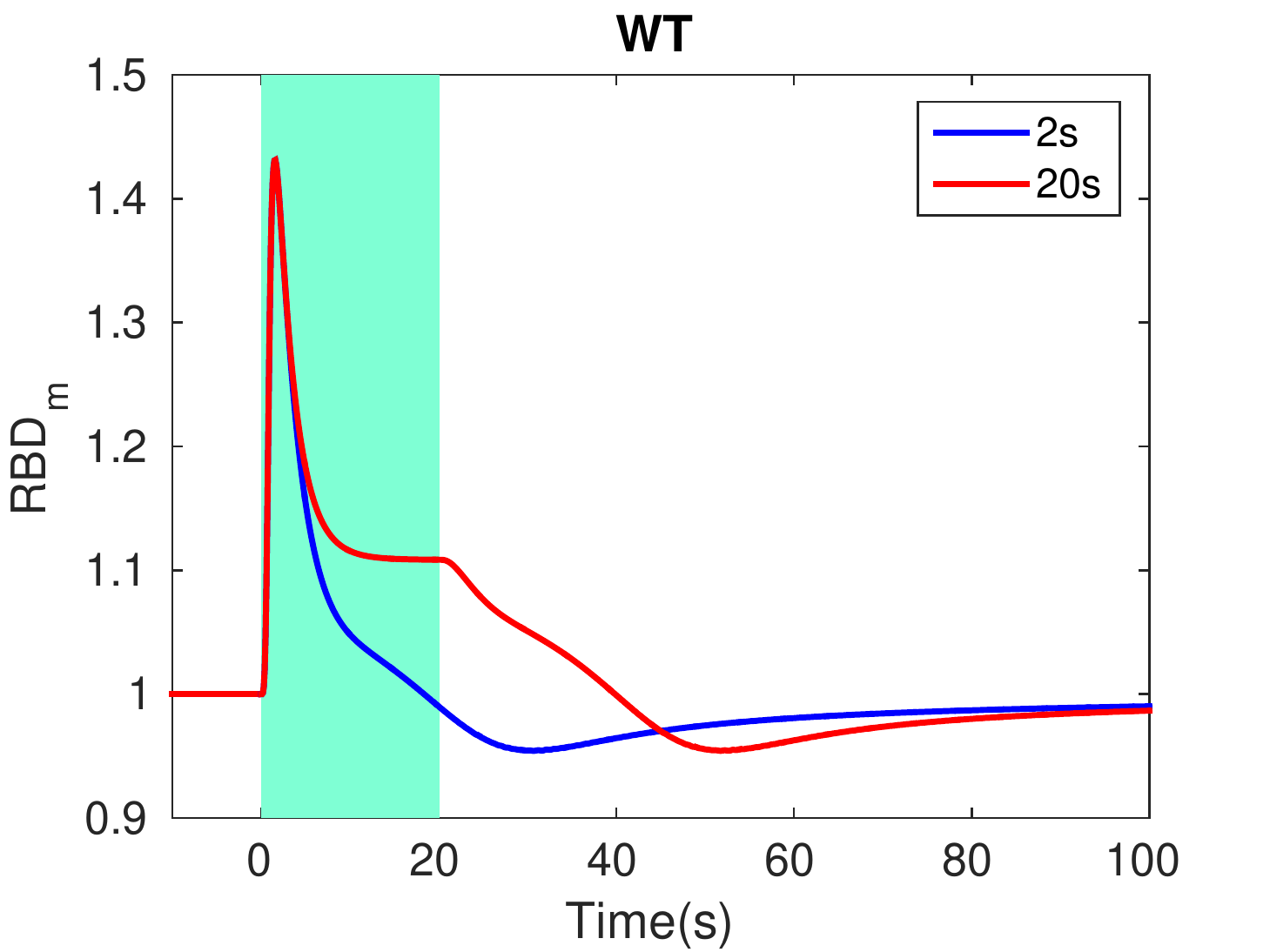}\includegraphics[width=7cm] {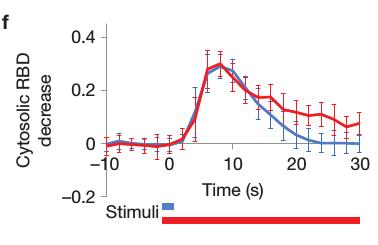}}
\caption{ \textbf{Short stimulus vs long stimulus.}\\ \emph{Left}:
  Simulation; \emph{Right}: Results reported in \cite{huang2013}. Blue indicates a 2s
  stimulus and red indicates a 20s stimulus.}
\label{shortvslong}
\end{figure}

\begin{figure}[H]
\centerline{\includegraphics[width=7cm] {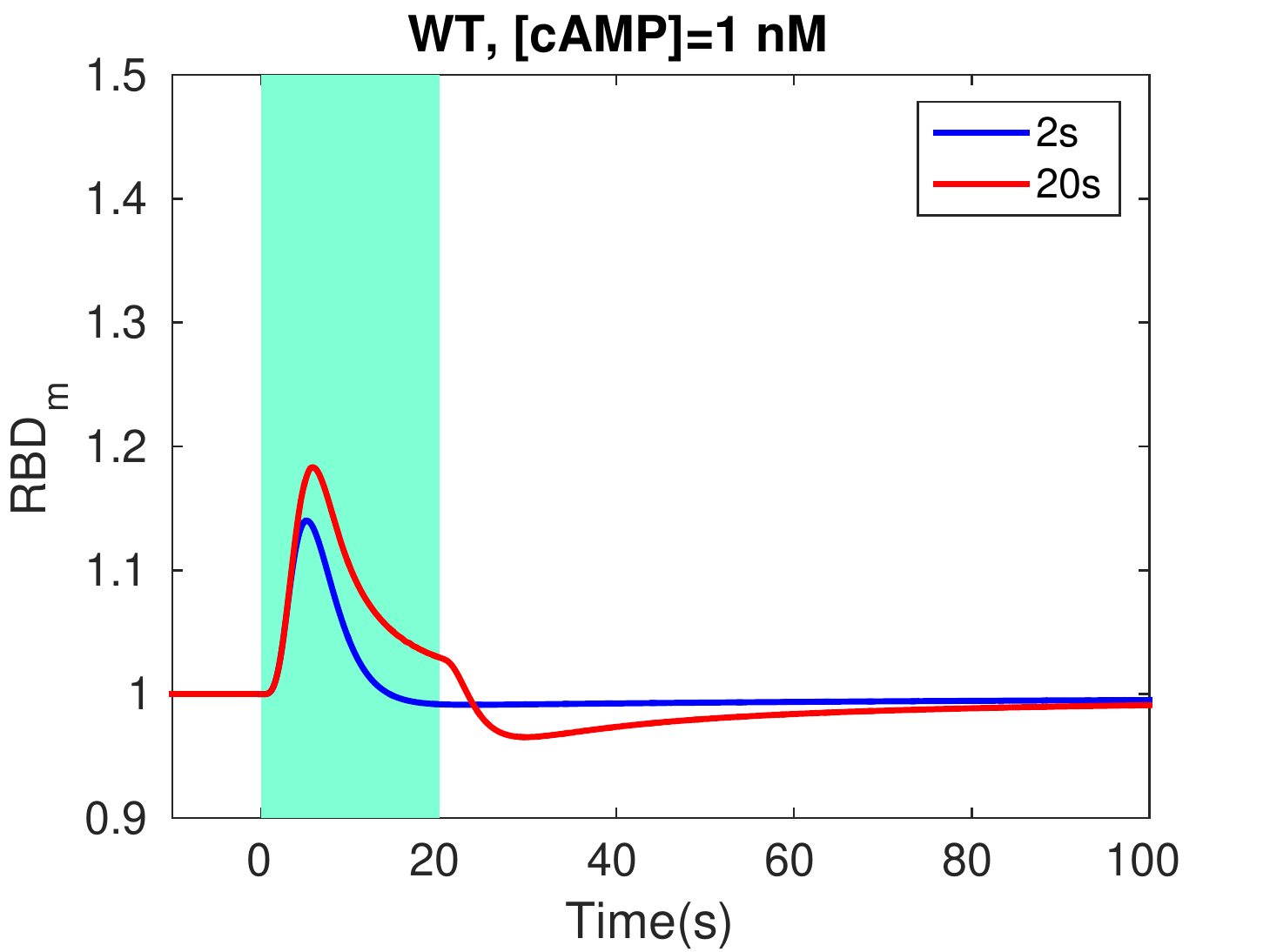}\includegraphics[width=7cm] {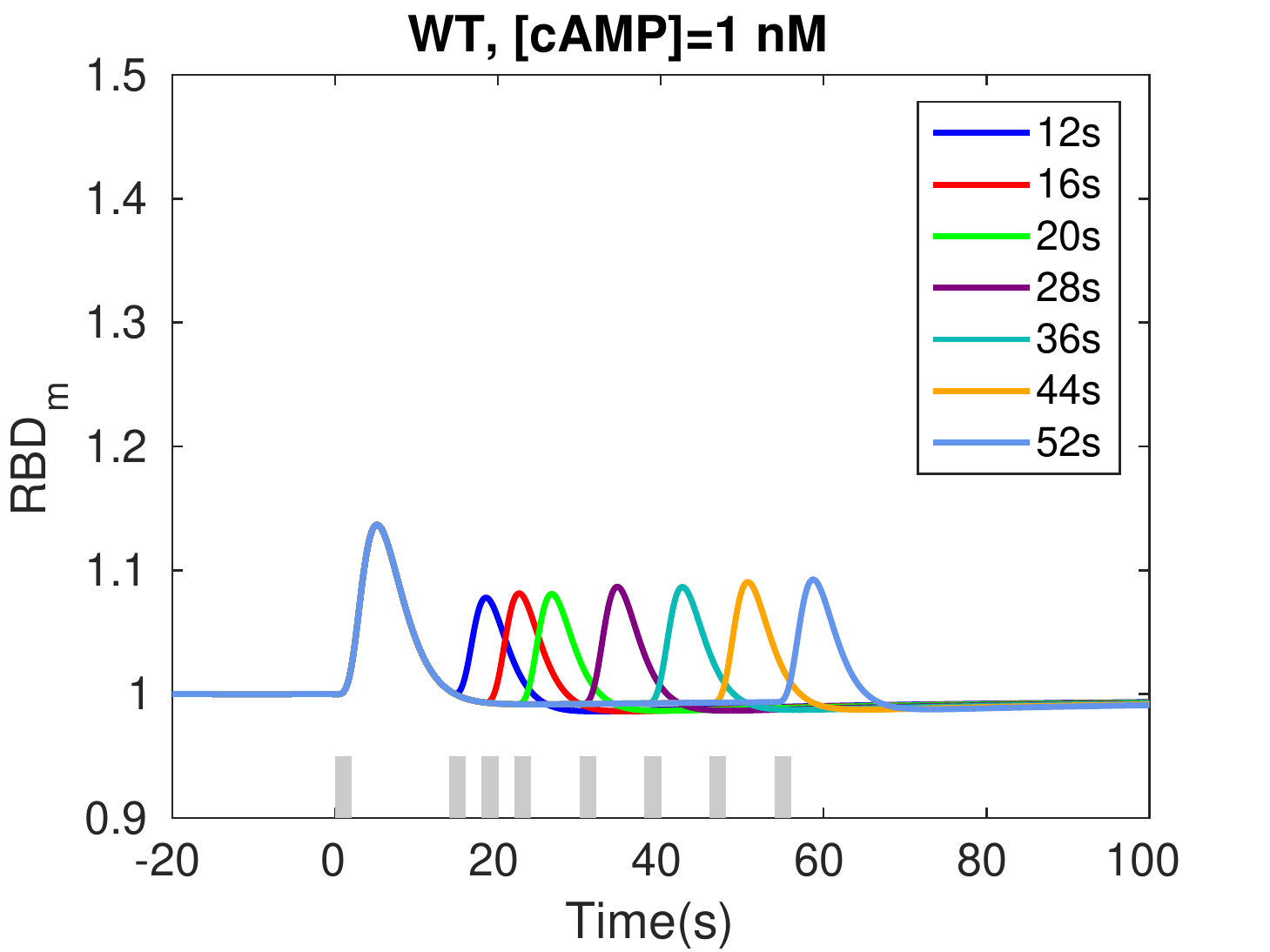}}
\caption{\textbf{Cell responses in low non-saturating cAMP stimulus.}\\ \emph{Left}: time course of $RBD_m$ when the cell is applied a
  1 nM short stimulus and long stimulus; \emph{Right}: time course of $RBD_m$ to two 2s
  cAMP stimuli of 1 nM separated by increasing duration.}
  \label{lowcAMP}
\end{figure}

\begin{figure}[H]
\centerline{\includegraphics[width=6cm] {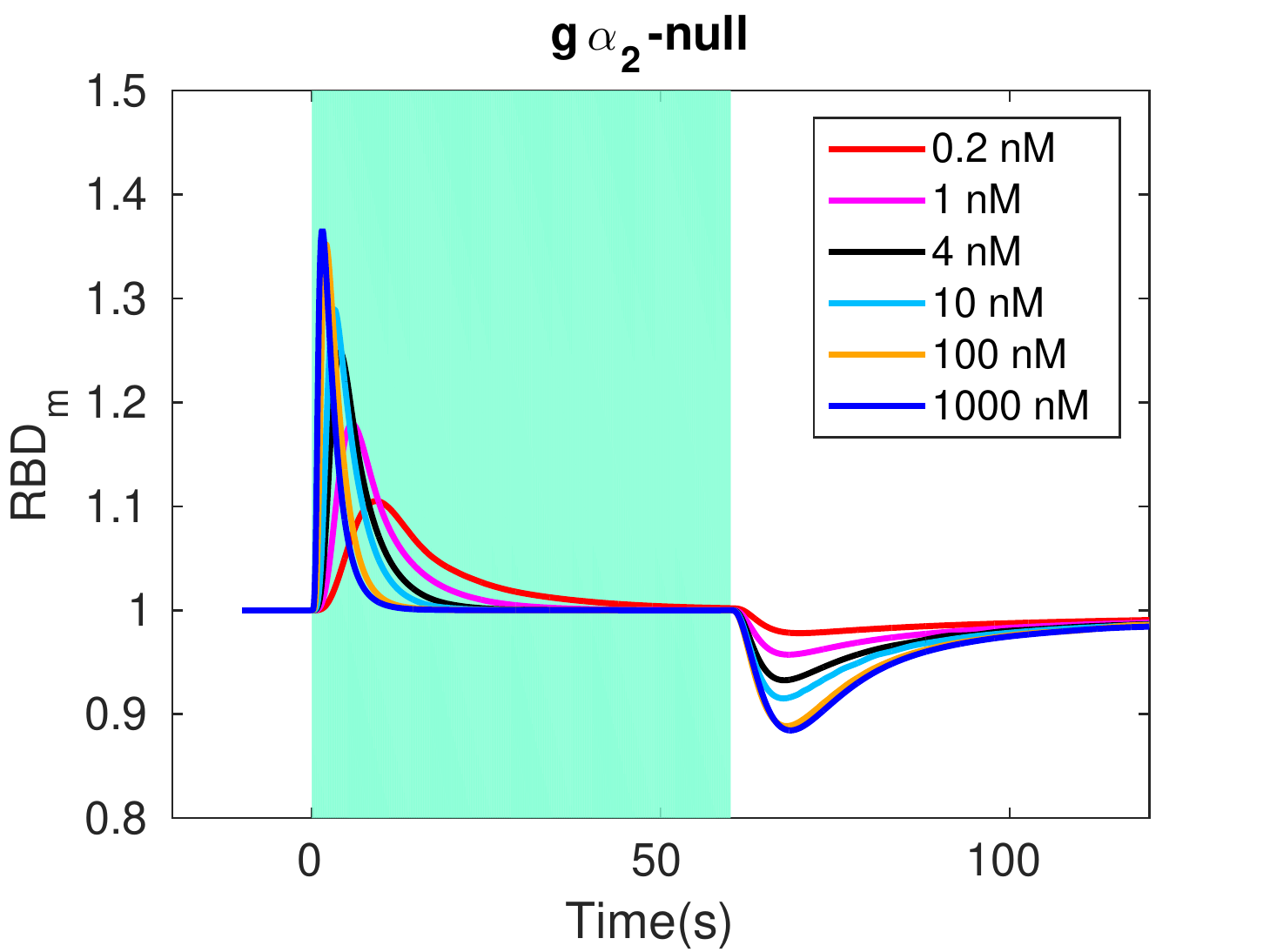}\hspace*{15pt}\includegraphics[width=6cm] {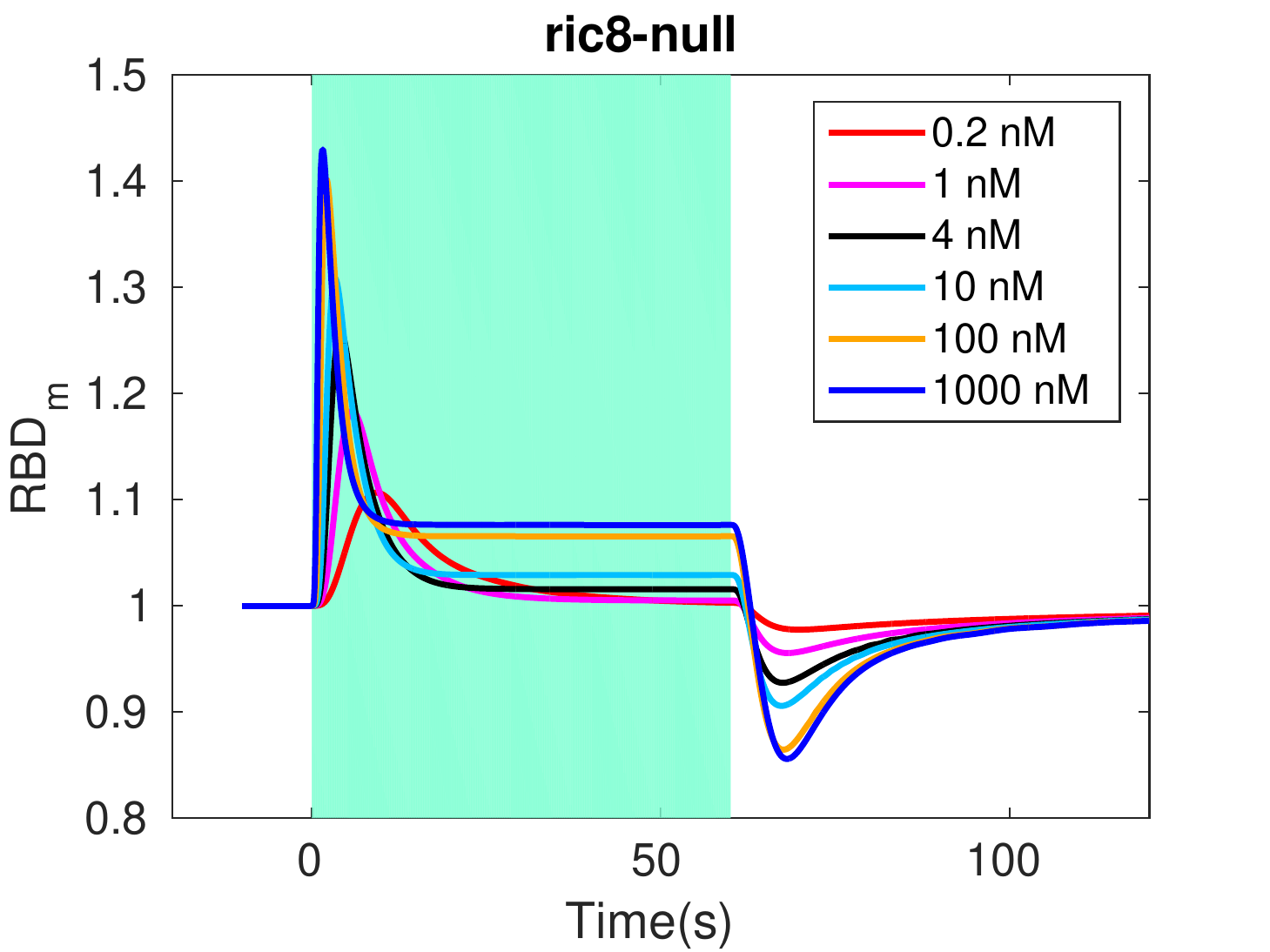}
 }
\caption{\textbf{Rectification in $g\alpha_2$-null cells(left) and ric8-null cells (right).}}
\label{rectifynull}
\end{figure}

\begin{figure}[H]
\centerline{\includegraphics[width=6cm] {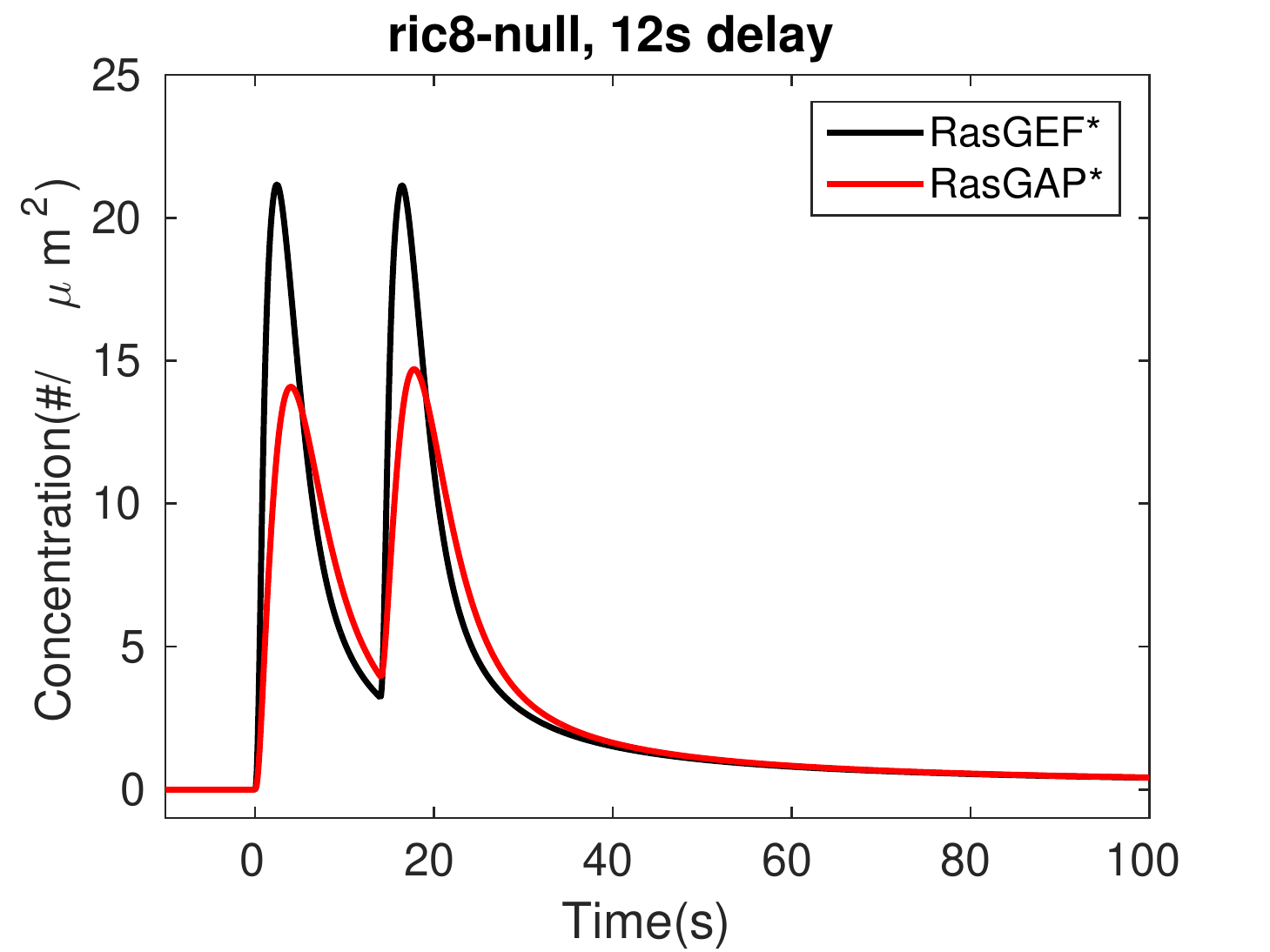}
  \includegraphics[width=6cm] {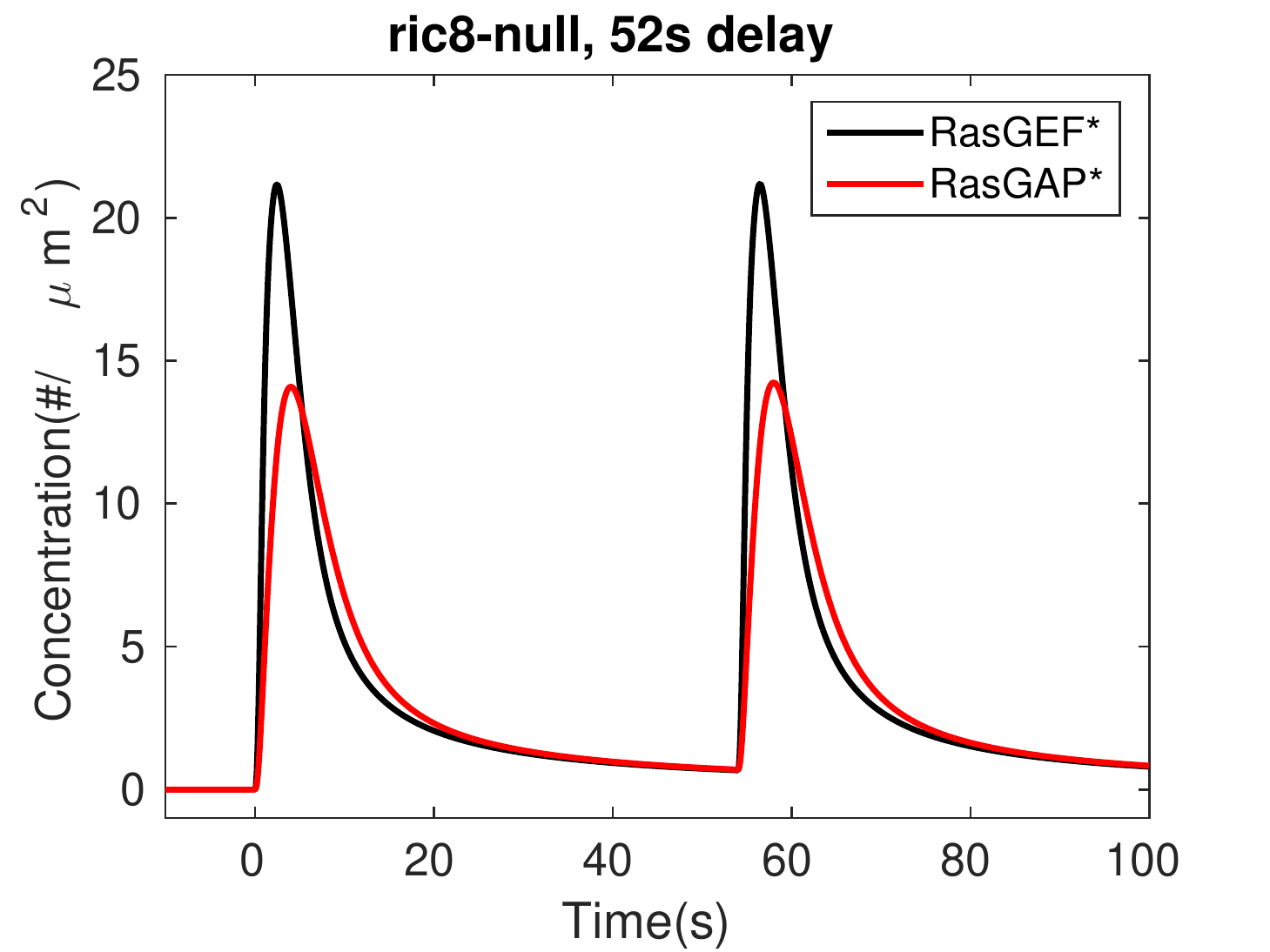}
  \includegraphics[width=6cm] {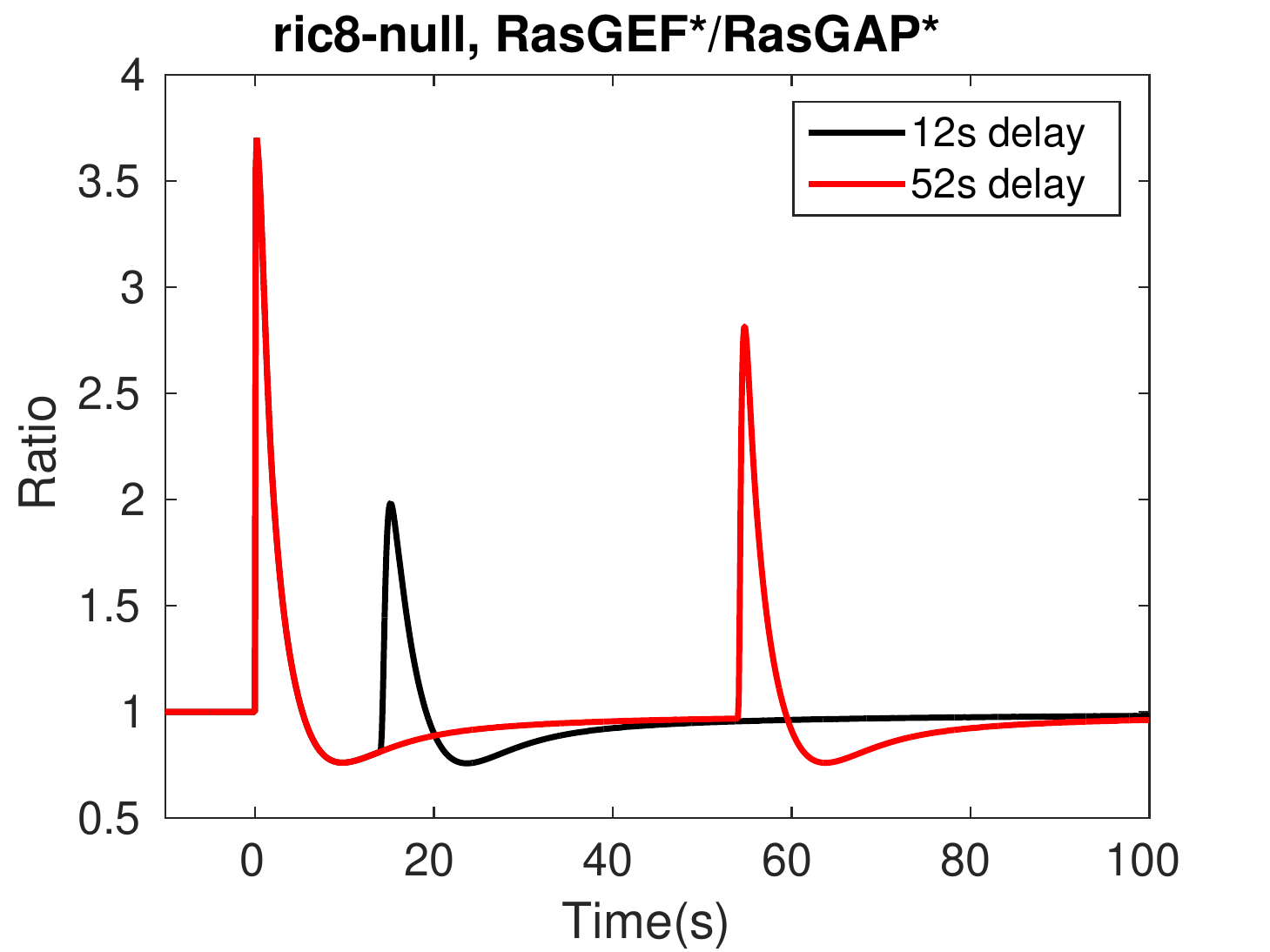}
}
\caption{\textbf{The time course of $RasGEF^*$, $RasGAP^*$ and $RasGEF^*/RasGAP^*$ activities in
  $ric8$-null cells.}}
  \label{gefgapric8null}
\end{figure}

\paragraph{Short vs Long saturating stimuli}
Our proposed network exhibits a maximal response to short saturating stimuli. RBD translocation is illustrated in Figure~\ref{shortvslong} when the cell is applied a short (2s) and a long (20s) saturating stimuli (1 $\mu M$). Short and long stimuli lead to a response with the same rise time, the same peak and the same initial decline, although there is a additional slowly-declining phase for the longer stimulus. Similar experiments are reported in \cite{huang2013}, in which it is suggested that this slowly
declining phase could be related to the start of the secondary responses usually
seen during continued stimulation \cite{chen2003, postma2003}.
 In the model the ratio of $RasGEF^*$ and $RasGAP^*$ determines the
Ras activation, and for saturating cAMP stimuli the ratio $RasGEF^*/RasGAP^*$
rises instantaneously and arrives at a maximum within 2 seconds, whereas it
takes longer to achieve maximum RasGEF and RasGAP activation separately. Hence
we observe a response with the same rise time and the same peak for short and
long stimuli. In the latter case there is an additional slowly-declining phase
due to the higher peak of $RasGEF^*$ and $RasGAP^*$. This indicates that subtle regulation of RasGEF and RasGAP
activation at saturating cAMP is essential for the observed characteristics.
\paragraph{Cell responses under non-saturating cAMP stimuli}
We predict that the cell loses the ability to induce full responses by short stimuli and
the existence of a refractory period greater than 12 seconds (which was the smallest time interval between stimuli used in \cite{huang2013}) at low cAMP level since temporal dynamics
of RasGEF and RasGAP are much weaker, which is confirmed by simulation results
shown in Fig.~\ref{lowcAMP} where the cAMP level is reduced from $1\mu M$ to $1
nM$. Indeed, longer stimulus induces higher peak of $RBD_m$ and duration between
stimuli no longer affects second response.
\paragraph{Rectification in $g\alpha_2$-null cells and ric8-null cells}
As seen from Figure~\ref{rectifynull}, the simulated $g_{\alpha_2}$-null cells show a much larger response to
termination of the stimulus, as shown in the middle panel. At 1 $\mu M$ cAMP,
RBD drops $\sim$15\% below the prestimulus level, compared to $<$ 5\% in
WT cells. Surprisingly, rectification is significantly reduced compared to that in WT
cells, and even compared to $g_{\alpha_2}$-null cells. To
understand these behaviors, recall that \G2 re-association is increased in
$ric8$-null cells because $G_{\alpha_2}$ activation is absent. As a result, the
time dynamics of RasGEF and RasGAP are altered correspondingly (see Figure~\ref{gefgapric8null}), and consequently, Ras
activation patterns are changed.
%
\subsection*{Other characteristics under a graded stimulus}
\paragraph{Effects of diffusion}
\begin{itemize}
\item Slow G2 and Ric8 diffusion

The cell responds similarly when there is no apparent $G_{\alpha\beta\gamma}$ and Ric8 diffusion, as demonstrated in Fig.~\ref{nogcdiffusionGEFGAP} and
Fig.~\ref{noricdiffusionGEFGAP}. In both cases, Ras activation first occur at
both ends, to reach a maximum, and then to decline to reach different
steady-state levels in distinct parts of the cell. Only a shallow reactivation
of Ras can be observed after it declines to a minimum, suggesting that symmetry
breaking is strongly severed in the absence of $G_{\alpha\beta\gamma}$ and Ric8
diffusion.

These two simulations suggest that the two sources of signal amplification are
equally important: The imbalanced sequestration of $G_{\alpha\beta\gamma}$ is
sabotaged in the absence of $G_{\alpha\beta\gamma}$ diffusion and asymmetrical
recruitments of Ric8 is destroyed in the absence of Ric8 diffusion. Therefore,
the symmetry breaking phase collapses in either cases due to deficiency of
signal amplification.

Our model reveals the importance of Ric8 in amplifying the signal at the level
of Ras by regulating \G2 cycling: on one hand, it amplifies RasGEF activation at
the front by reactivating $G\alpha2$; on the other hand, it
amplifies $G_{{\alpha}_2}^*$ activation by redistributing \G2 between the front and
the rear of the cell.
\begin{figure}[H]
\centerline{\includegraphics[width=7cm] {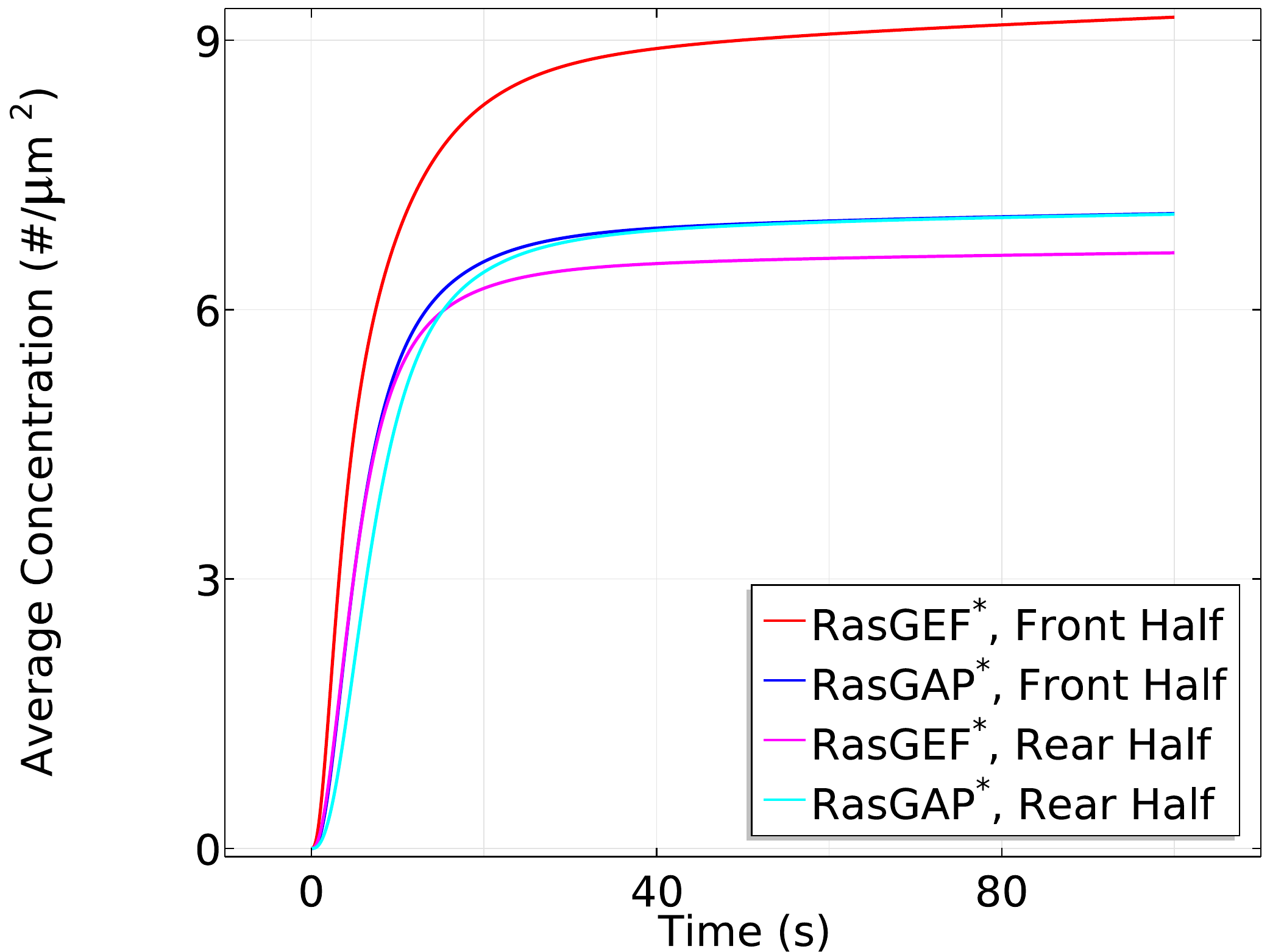}\includegraphics[width=7cm] {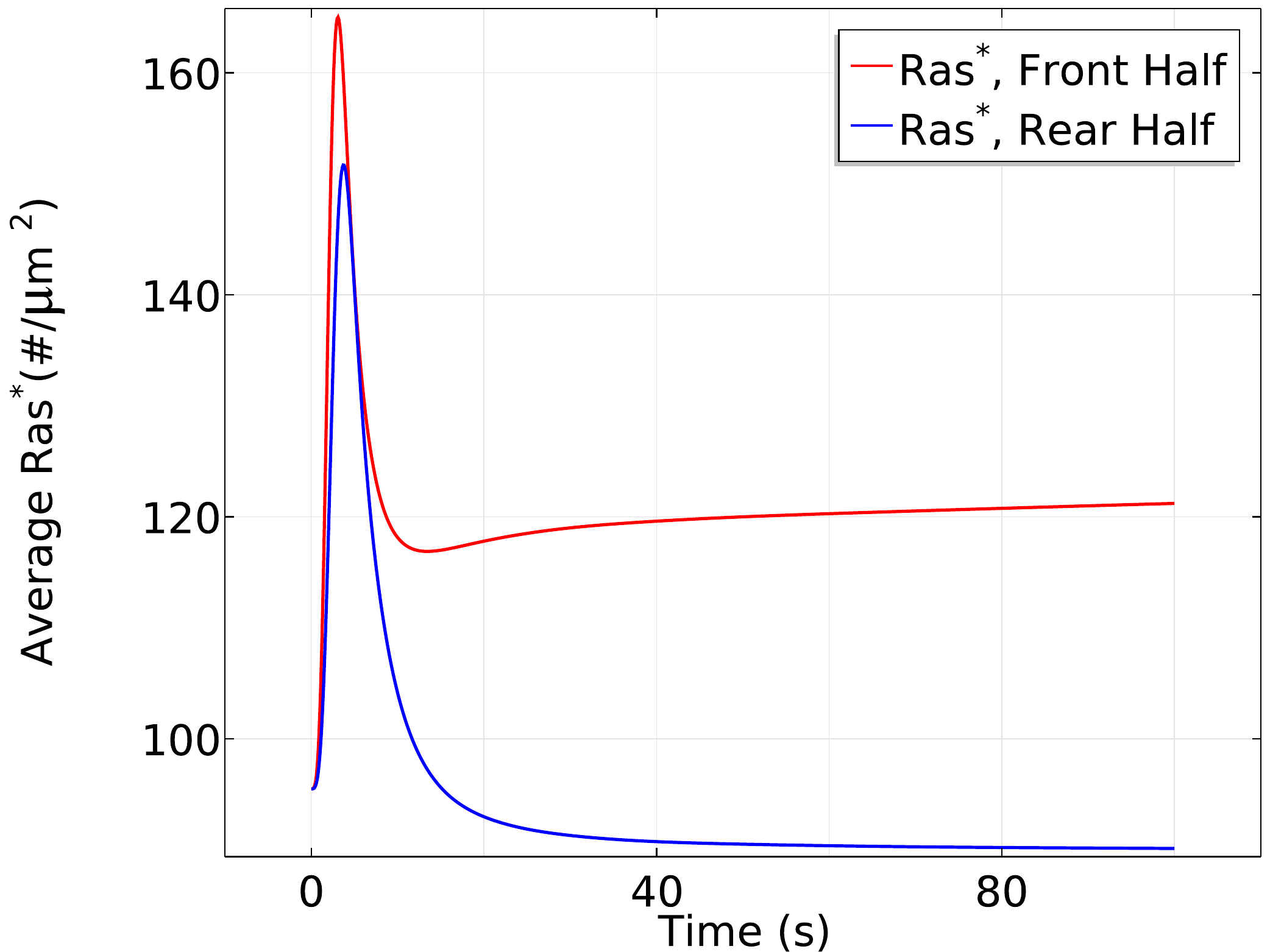}}
\caption{ \textbf{Time course of average $RasGEF^*$ and $RasGAP^*$ (left), and $Ras^*$ (right) in a cAMP gradient defined by $C_f = 10$ nM and $C_r$ = 1nM in the absence of apparent
  $G_{\alpha\beta\gamma}$ diffusion.}}
\label{nogcdiffusionGEFGAP}
\end{figure}
%
%
\begin{figure}[H]
\centerline{\includegraphics[width=7cm] {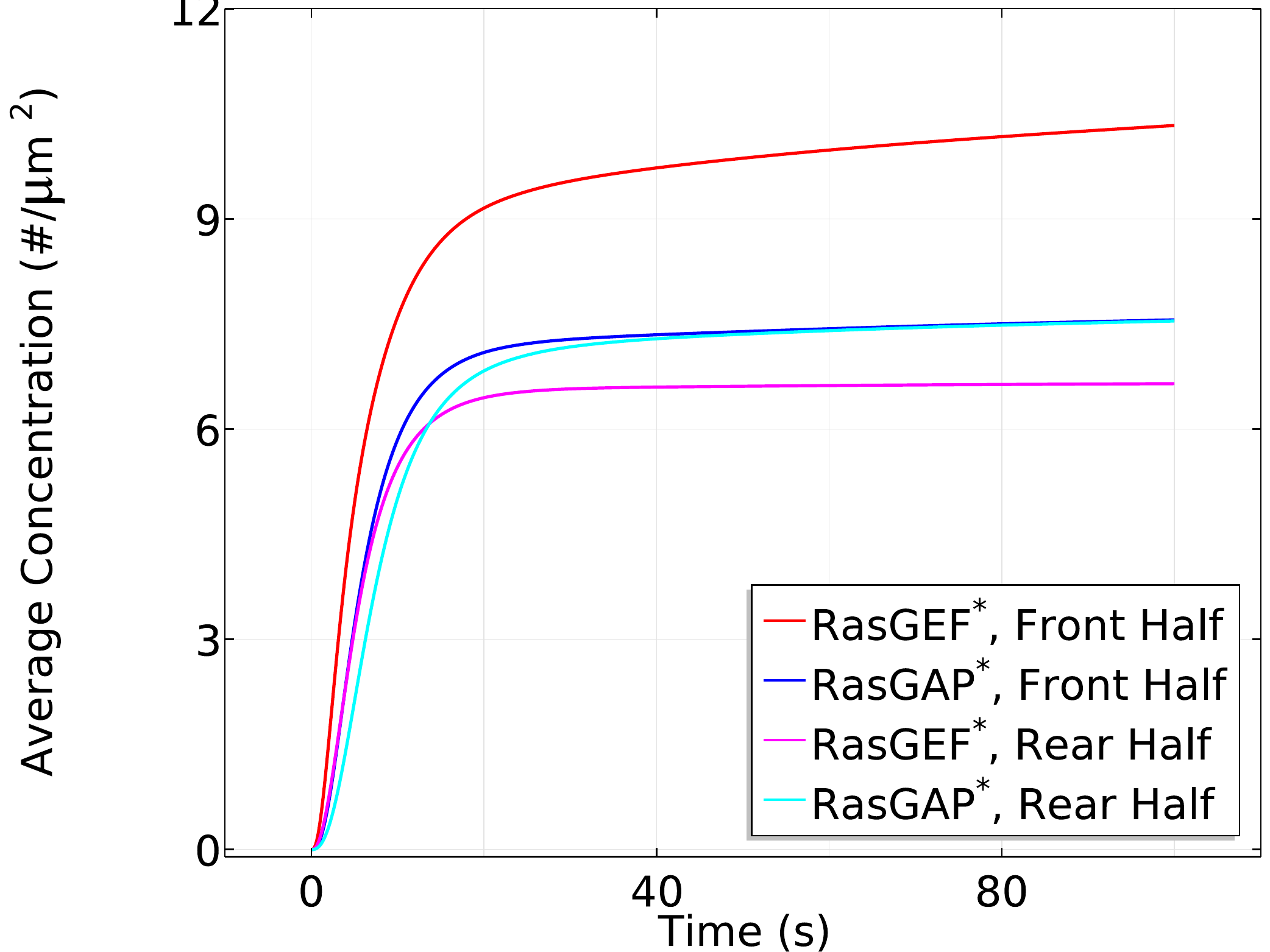}\includegraphics[width=7cm] {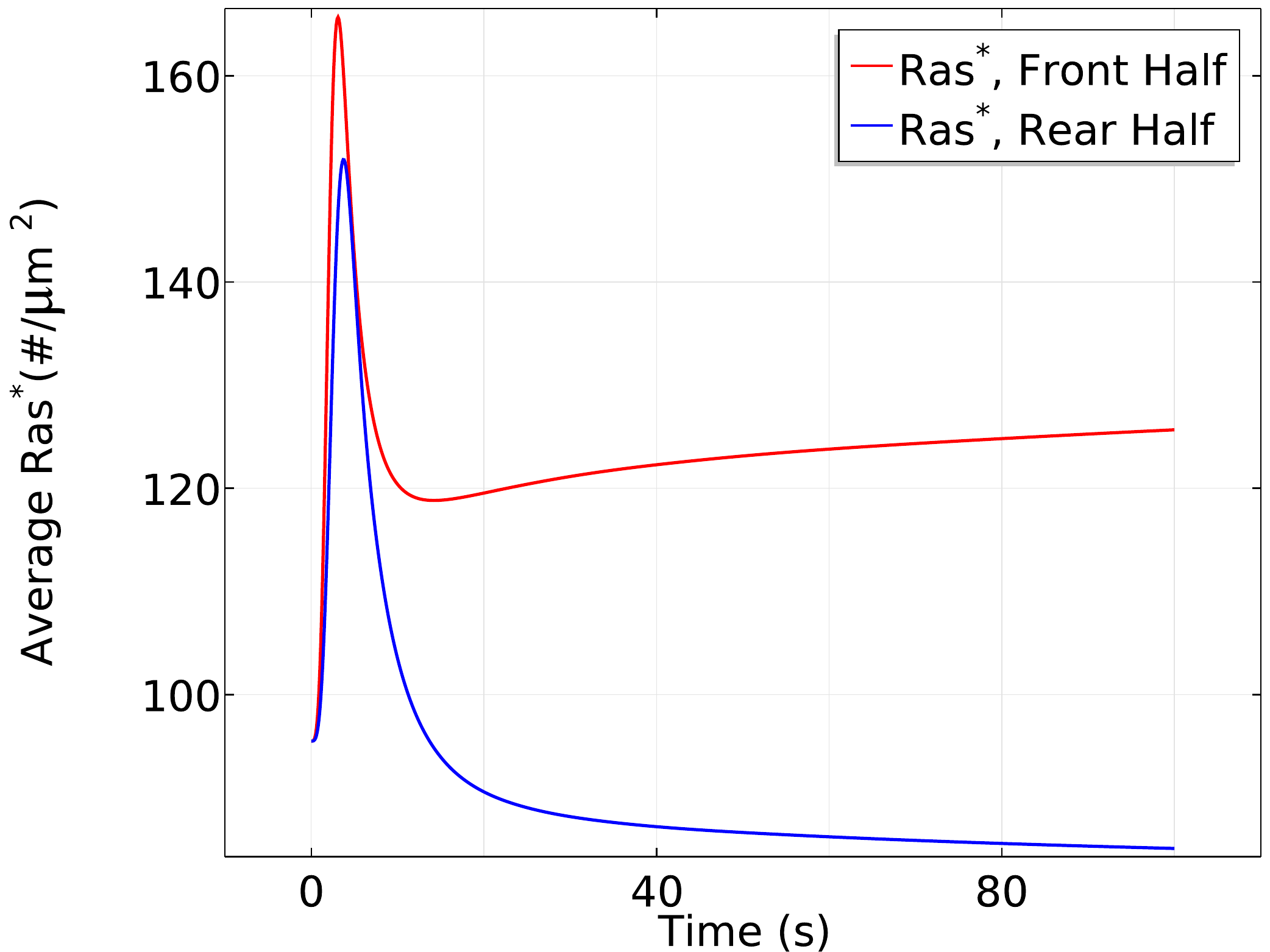}}
\caption{\textbf{Time course of average $RasGEF^*$ and $RasGAP^*$ (left), and $Ras^*$ (right) in a cAMP gradient defined by $C_f = 10$ nM and $C_r$ = 1nM in the absence of apparent Ric8 diffusion.}}
\label{noricdiffusionGEFGAP}
\end{figure}
%
\item Slow G$\beta\gamma$ and RasGEF diffusion

We also tested the cell response when both $\Gbg$ and RasGEF diffusion
are absent (see Fig.~\ref{nogbgefdiffusionGEFGAP}). Because there is no
$\Gbg$ diffusion, the activity of $RasGAP^*$ is stronger at the front of
the cell. Meanwhile, the supply of RasGEF and $G_{{\alpha}_2}^*$ facilitated
asymmetrical recruitment of RasGEF is limited since RasGEF diffusion is
absent. Therefore, we observe a higher $Ras^*$ activity at the rear of the cell
even though the cAMP gradient is opposite.
\begin{figure}[H]
\centerline{\includegraphics[width=7cm] {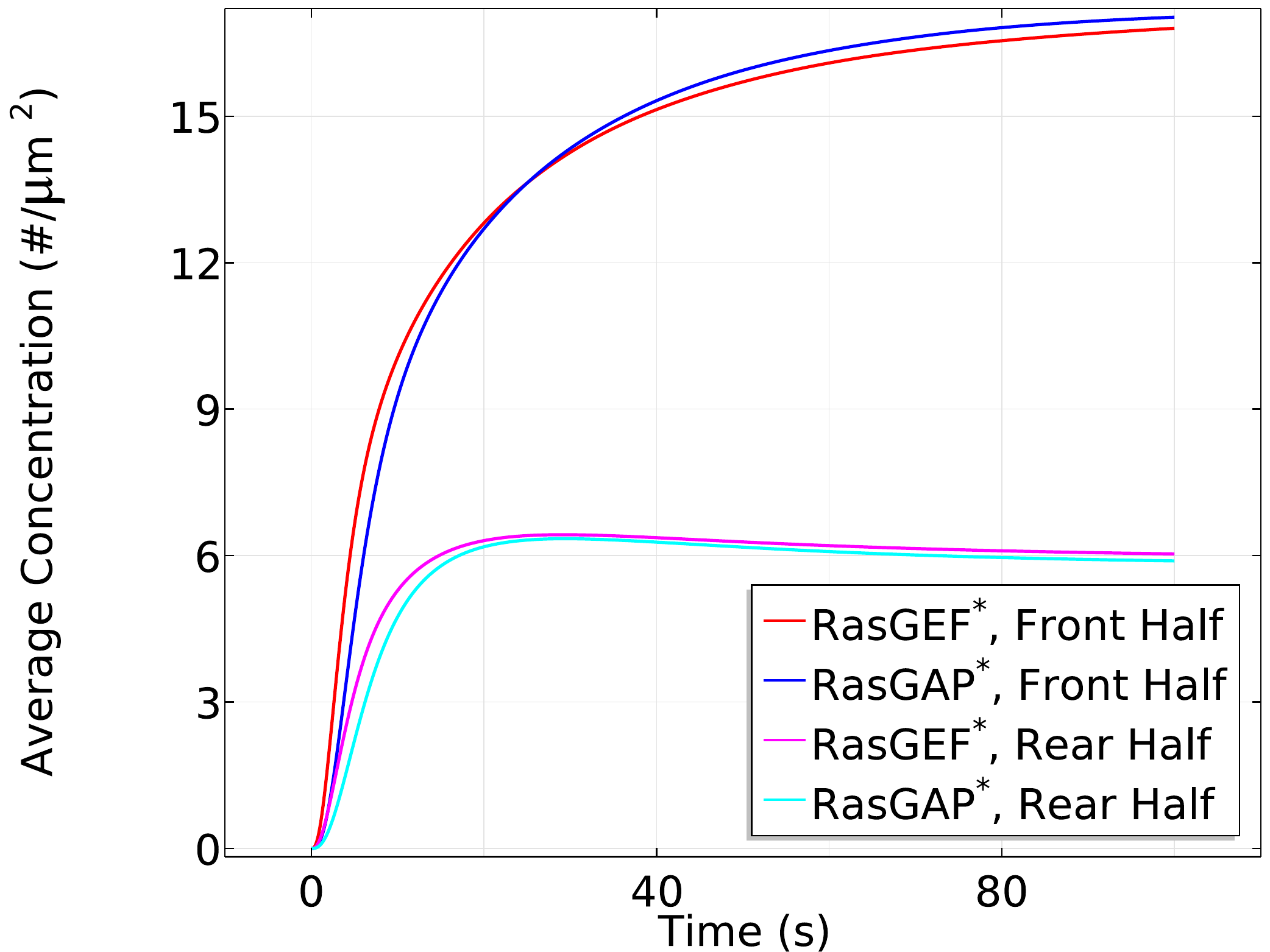}\includegraphics[width=7cm] {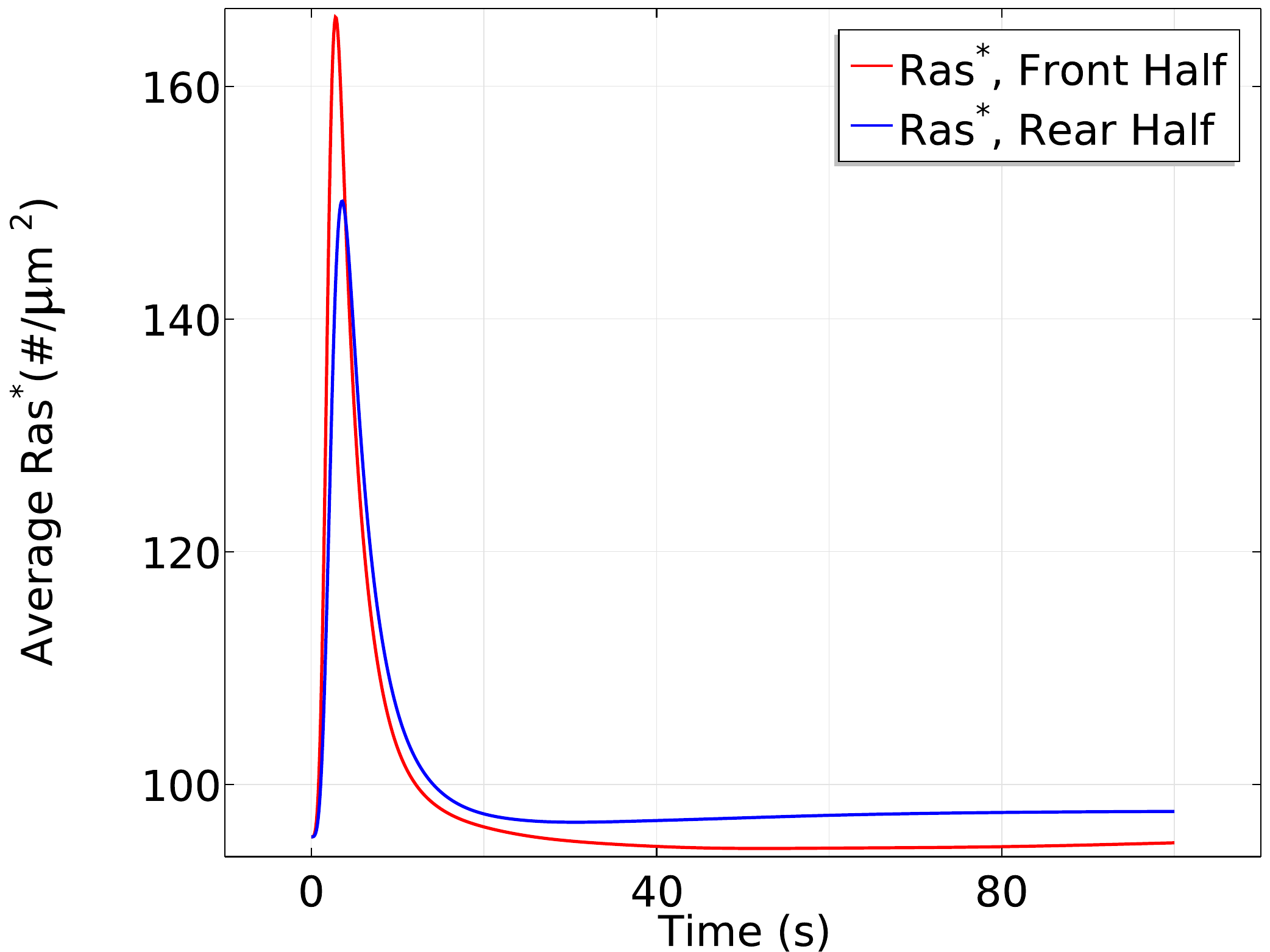}}
\caption{\textbf{Time course of average $RasGEF^*$ and $RasGAP^*$ (left), and $Ras^*$ (right) in a cAMP gradient defined by $C_f = 10$ nM and $C_r$ = 1nM in the absence of apparent G$_{\beta\gamma}$ and RasGEF diffusion.}}
  \label{nogbgefdiffusionGEFGAP}
\end{figure}
\end{itemize}

\paragraph{Robustness of the G$\alpha_2$-G$\beta\gamma$-Ric8 triangle}

\begin{figure}[H]
\centerline{\includegraphics[width=20cm]{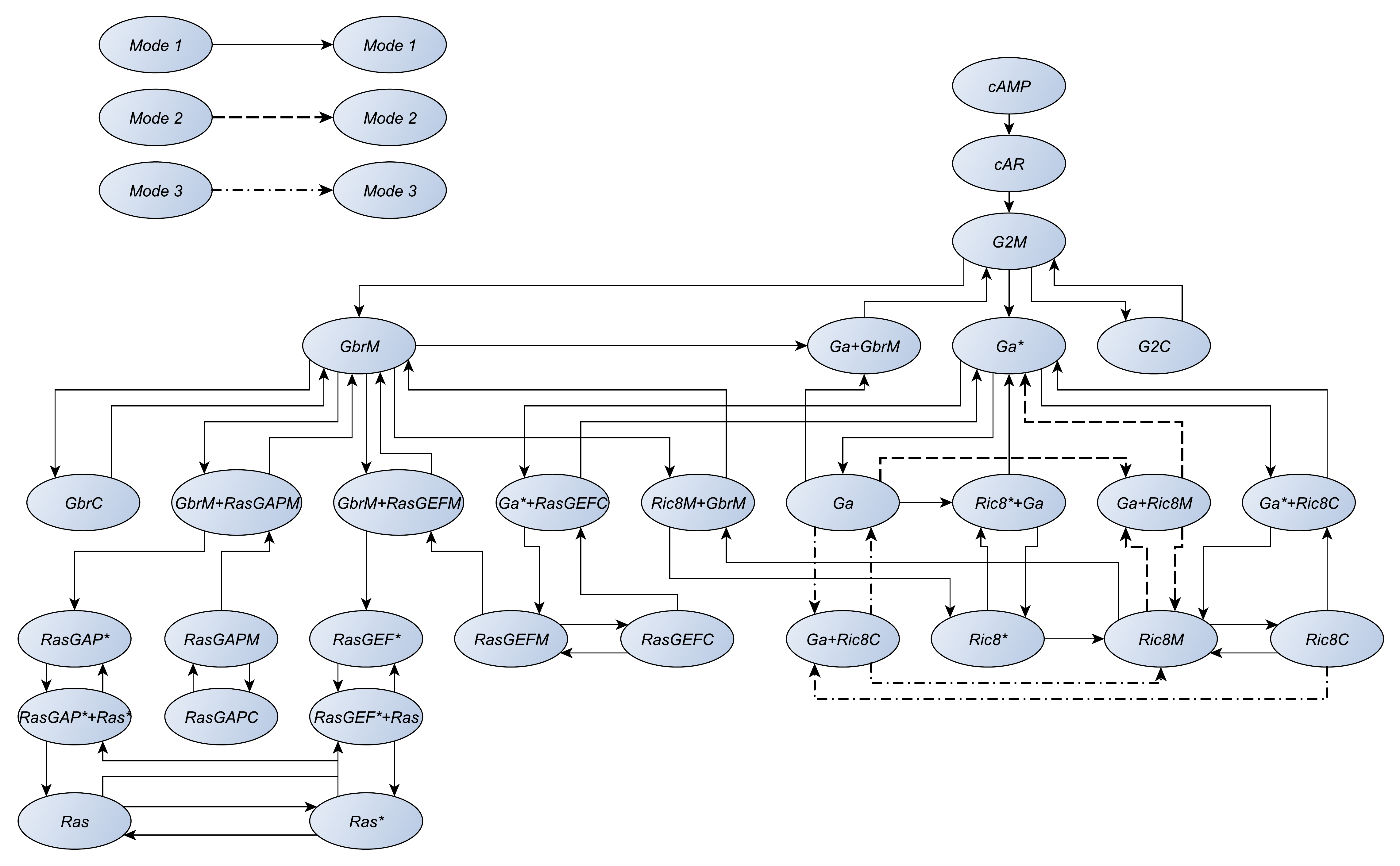}}
\caption{ \textbf{A flow chart schematics of the detailed network topology.}\\
Components end with M represents membrane species and components end with C represents cytosolic species. Gbr: G$_{\beta\gamma}$; Ga: G$_\alpha$.}
\label{flowchart}
\end{figure}

A schematics of the different modes is shown in Fig.~\ref{flowchart}, which illustrates the importance of G$_{\beta\gamma}$ amd G$_\alpha$.

\begin{figure}[H]
\centerline{\includegraphics[width=6cm] {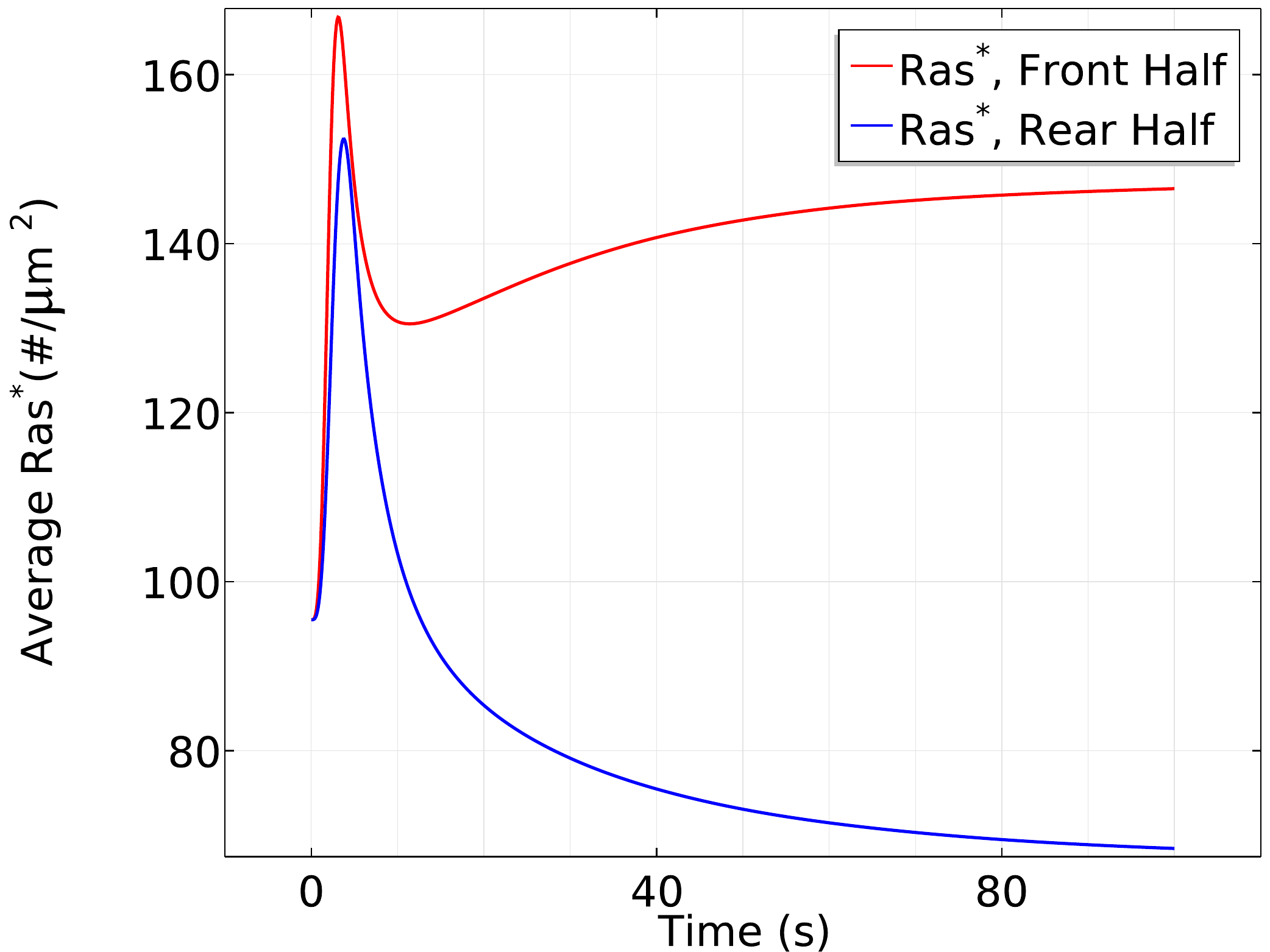}\includegraphics[width=6cm] {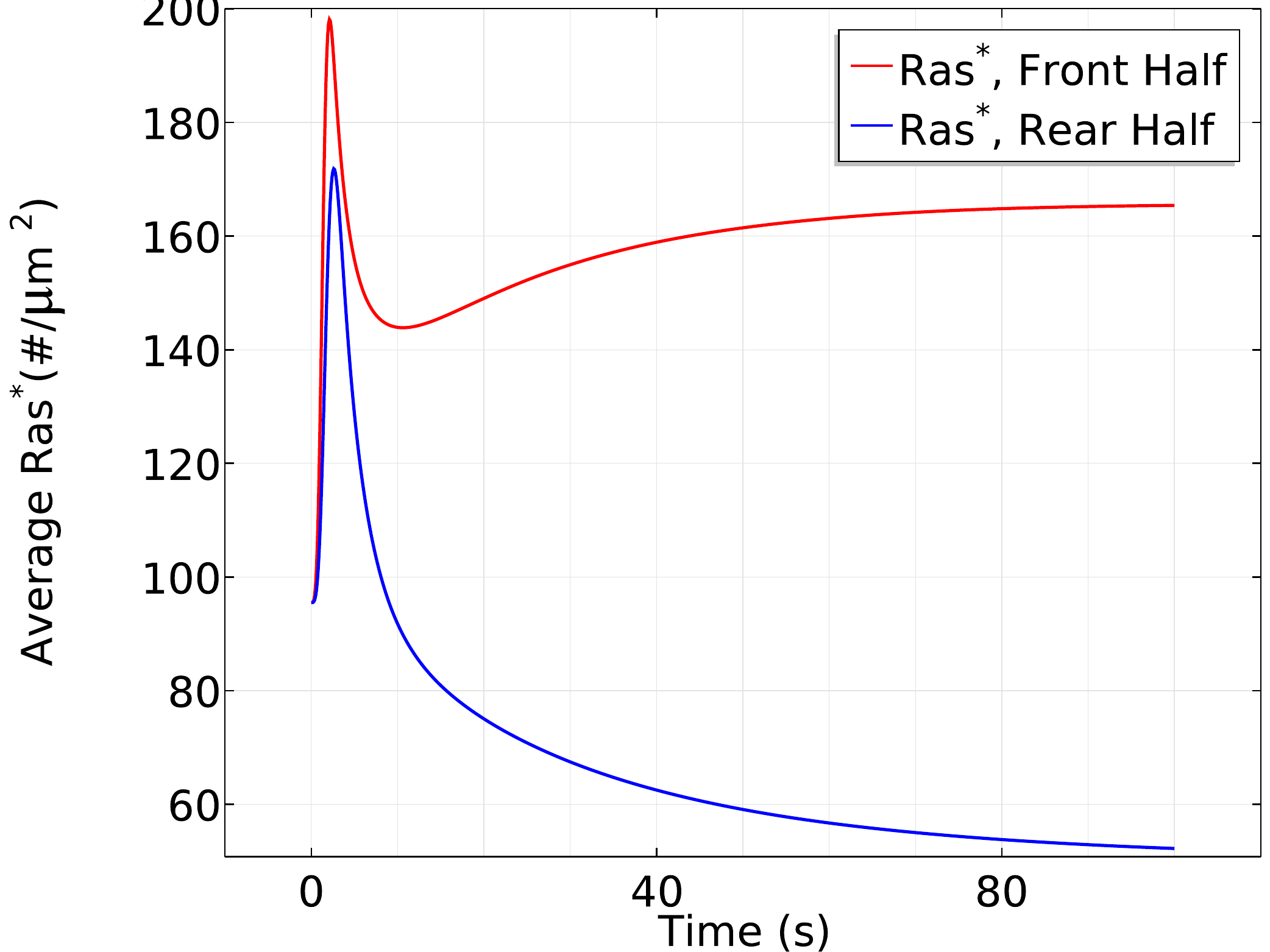}\includegraphics[width=6cm] {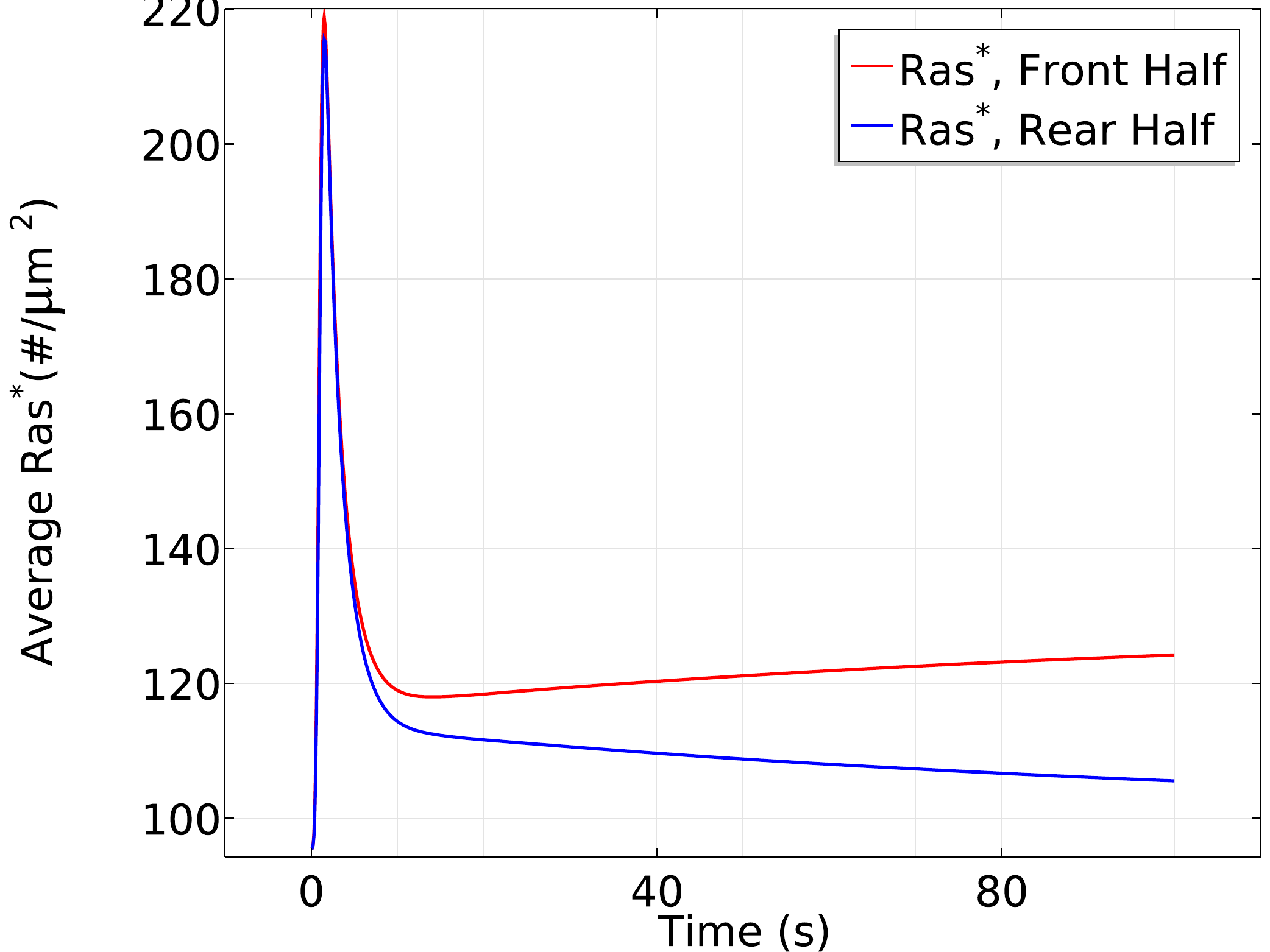}}
\centerline{\includegraphics[width=6cm] {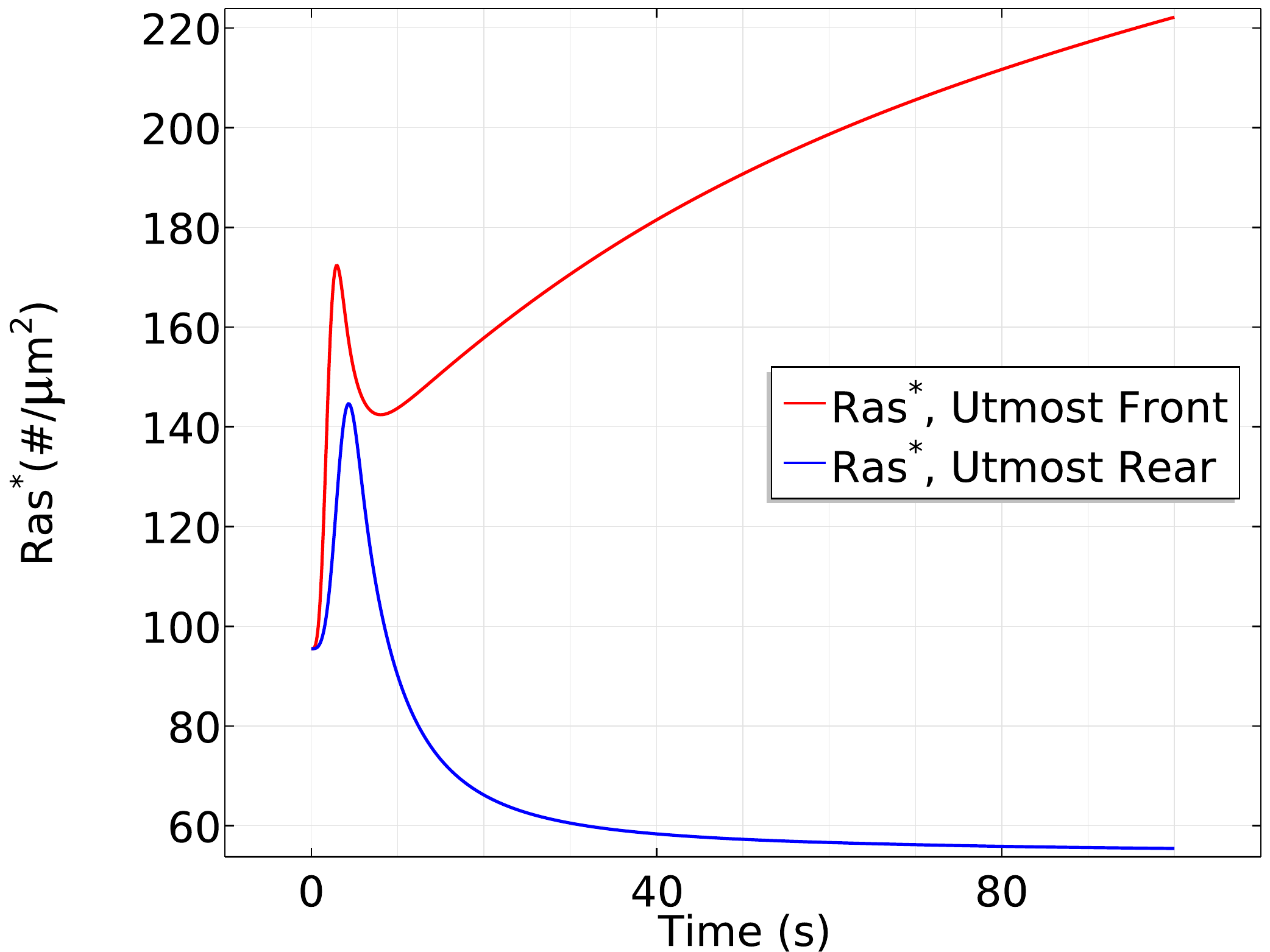}\includegraphics[width=6cm] {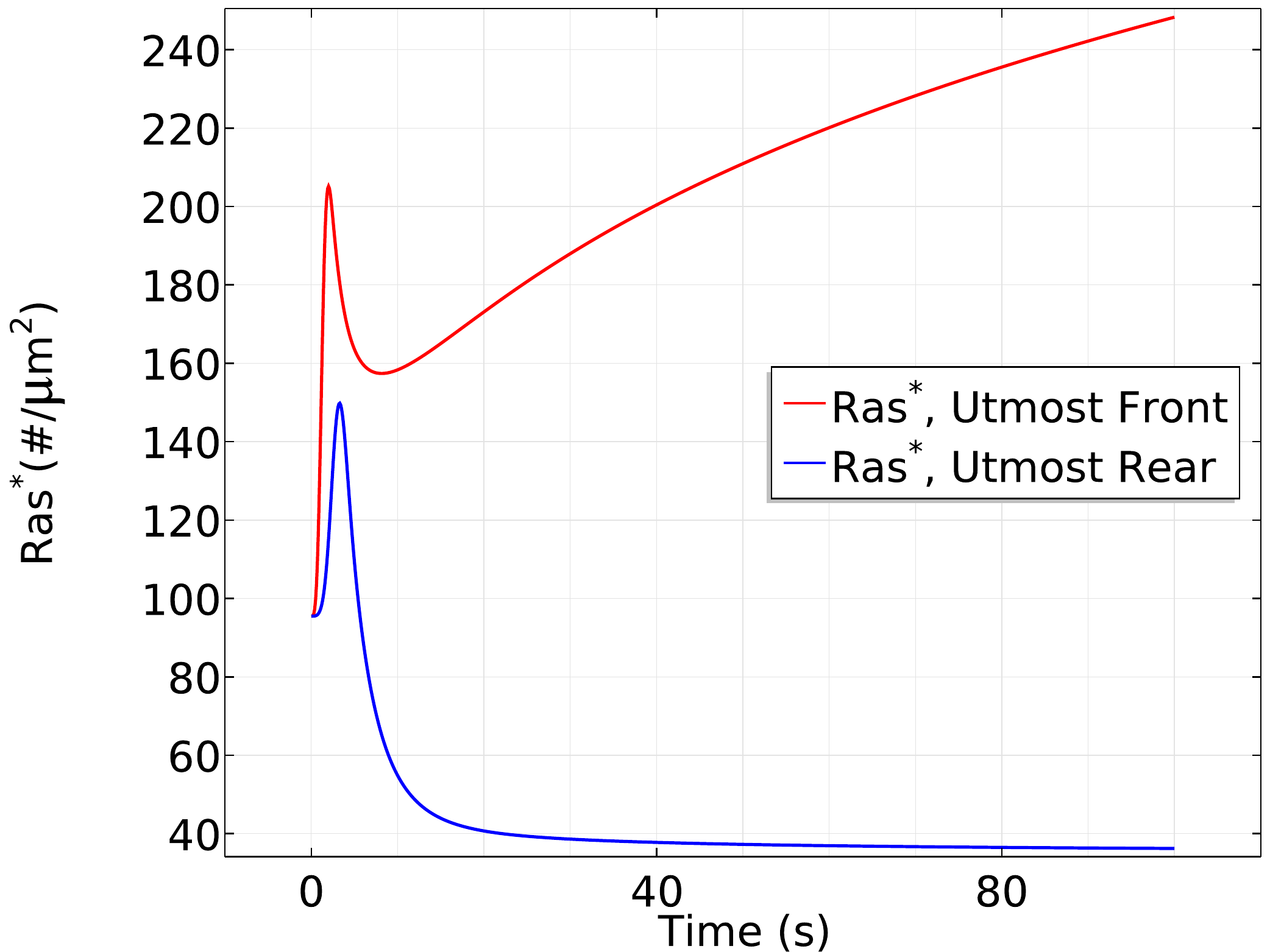}\includegraphics[width=6cm] {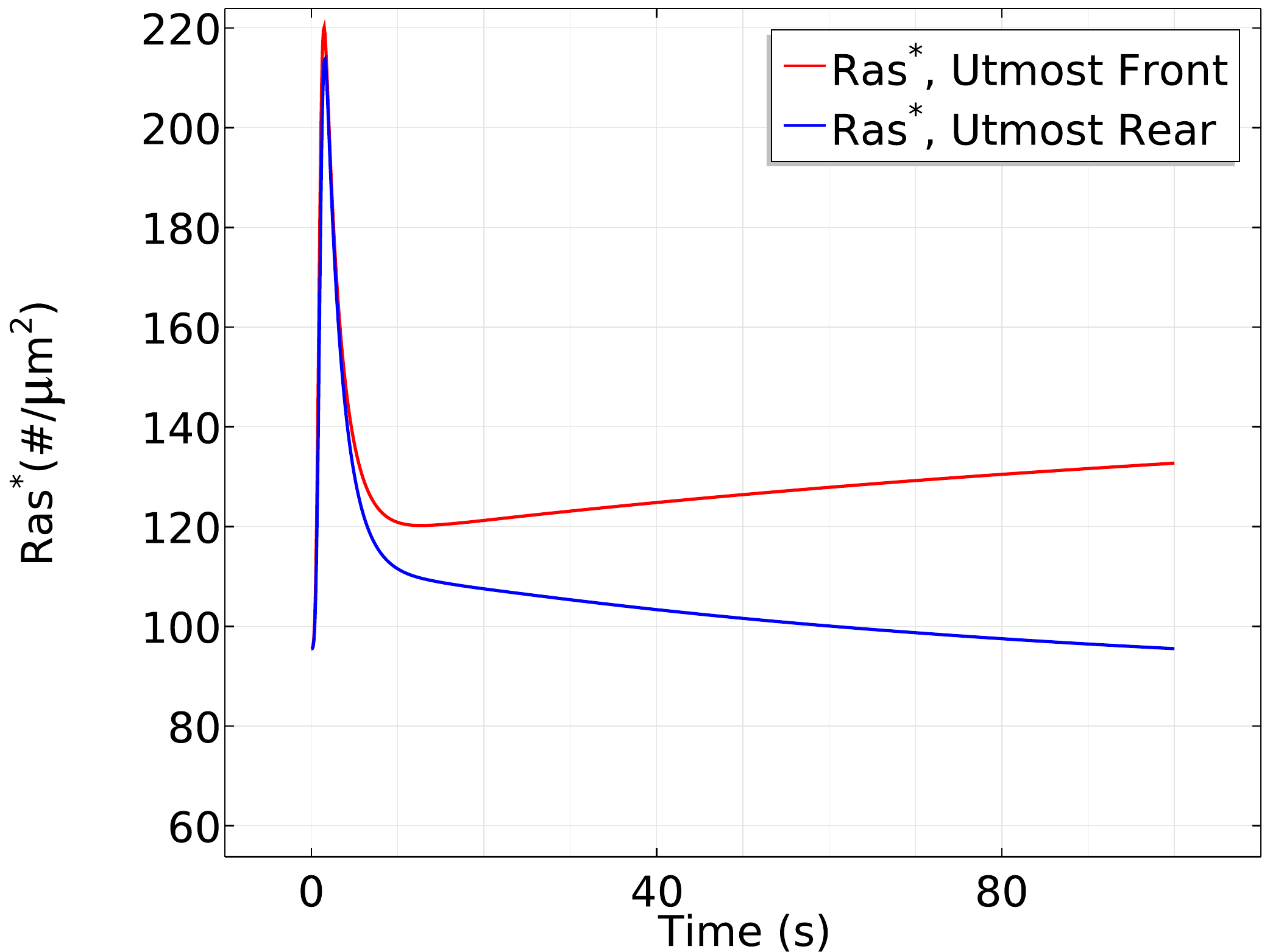}}
\caption{\textbf{time course of $Ras^*$ at front and rear half (top) and at $x_f$ and $x_r$ (bottom) in various gradient in Mode 2.}\\
\emph{Left}: $C_f=10$ nM and $C_r=1$ nM. \emph{Center}: $C_f=50$ nM and $C_r=0$ nM. \emph{Right}: $C_f=175$ nM and $C_r=125$ nM.}
\label{mode2halfhalf}
\end{figure}

\begin{figure}[H]
\centerline{\includegraphics[width=7cm]{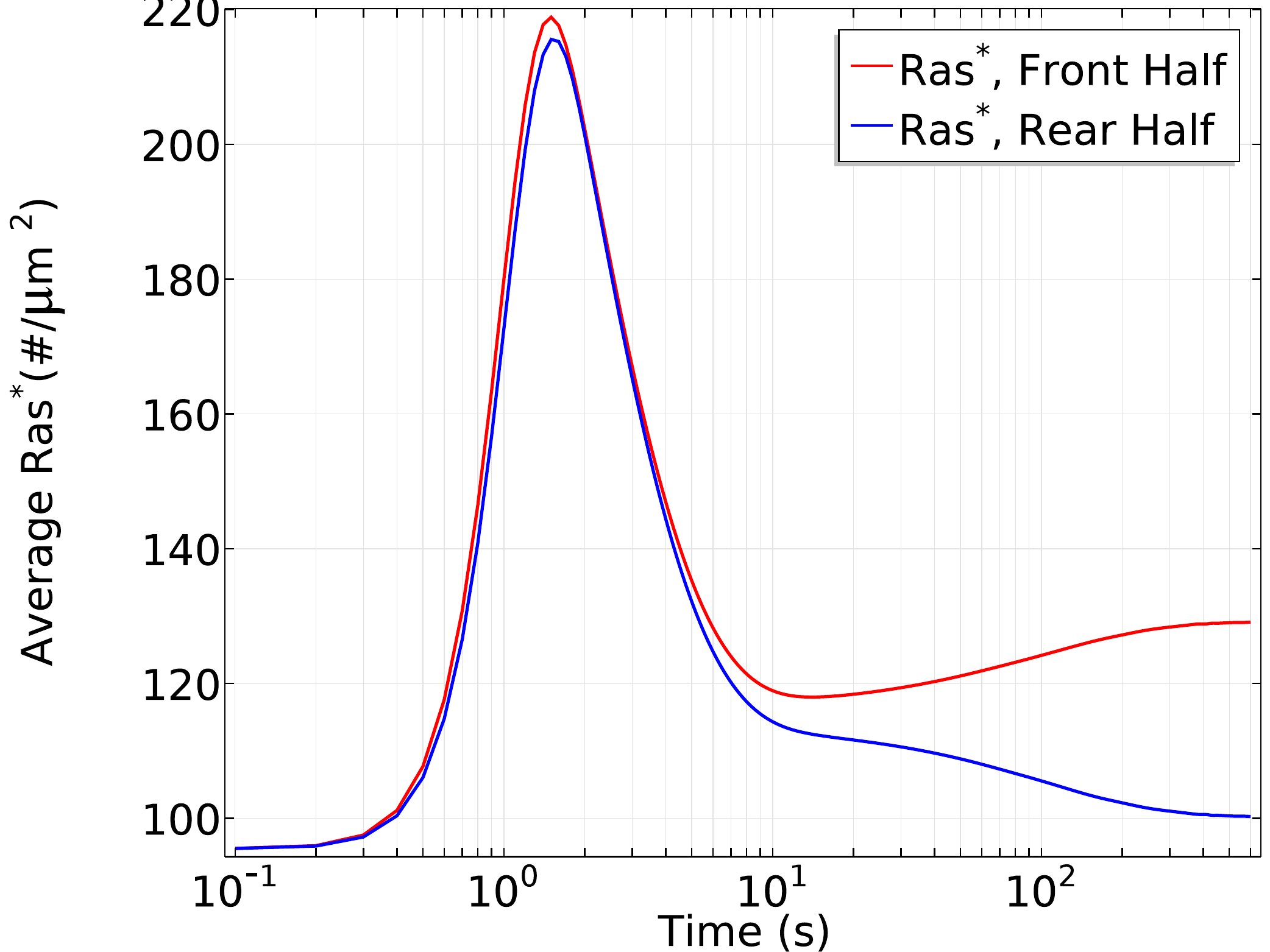}\includegraphics[width=7cm]{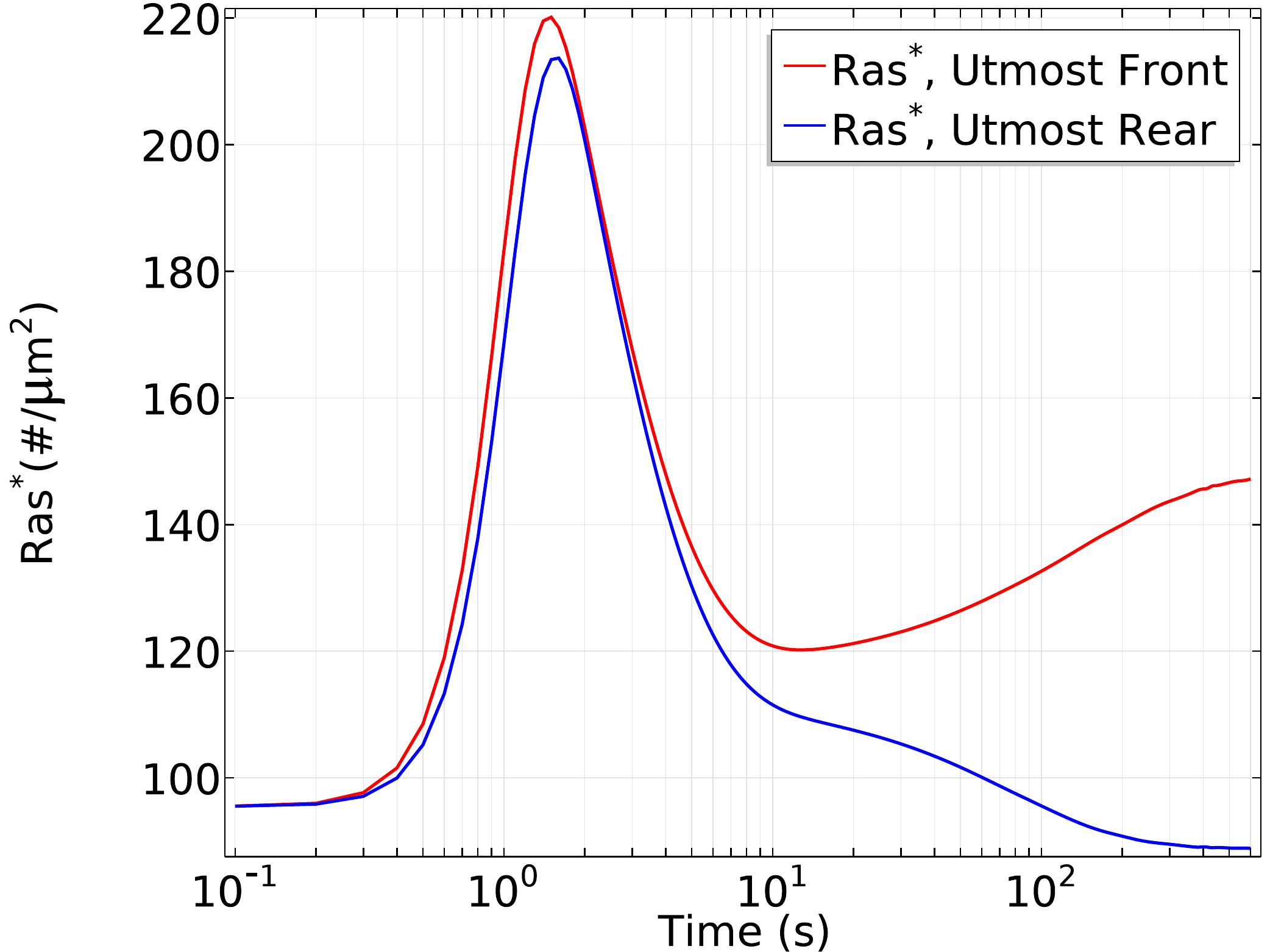}}
\caption{\textbf{Time course of $Ras^*$ at front and rear half (left) and at $x_f$ and $x_r$ (right) in a cAMP gradient in Mode 2 defined by $C_f=175$ nM and $C_r=125$ nM.}}
\label{mode2backhigh}
\end{figure}

We tested three gradients in Mode 2 and the simulation results are illustrated in Fig.~\ref{mode2halfhalf} and Fig.~\ref{mode2backhigh}.
Fig.~\ref{mode2halfhalf} and Fig.~\ref{mode2backhigh} demonstrate that Mode 2 still capture the basic characteristics of Ras activation, almost the same as Mode 1 except the magnitudes are slightly changed. This suggests that the robustness of the network and $G_{\beta\gamma}$ activation is not an essential step.

\begin{figure}[H]
\centerline{\includegraphics[width=6cm] {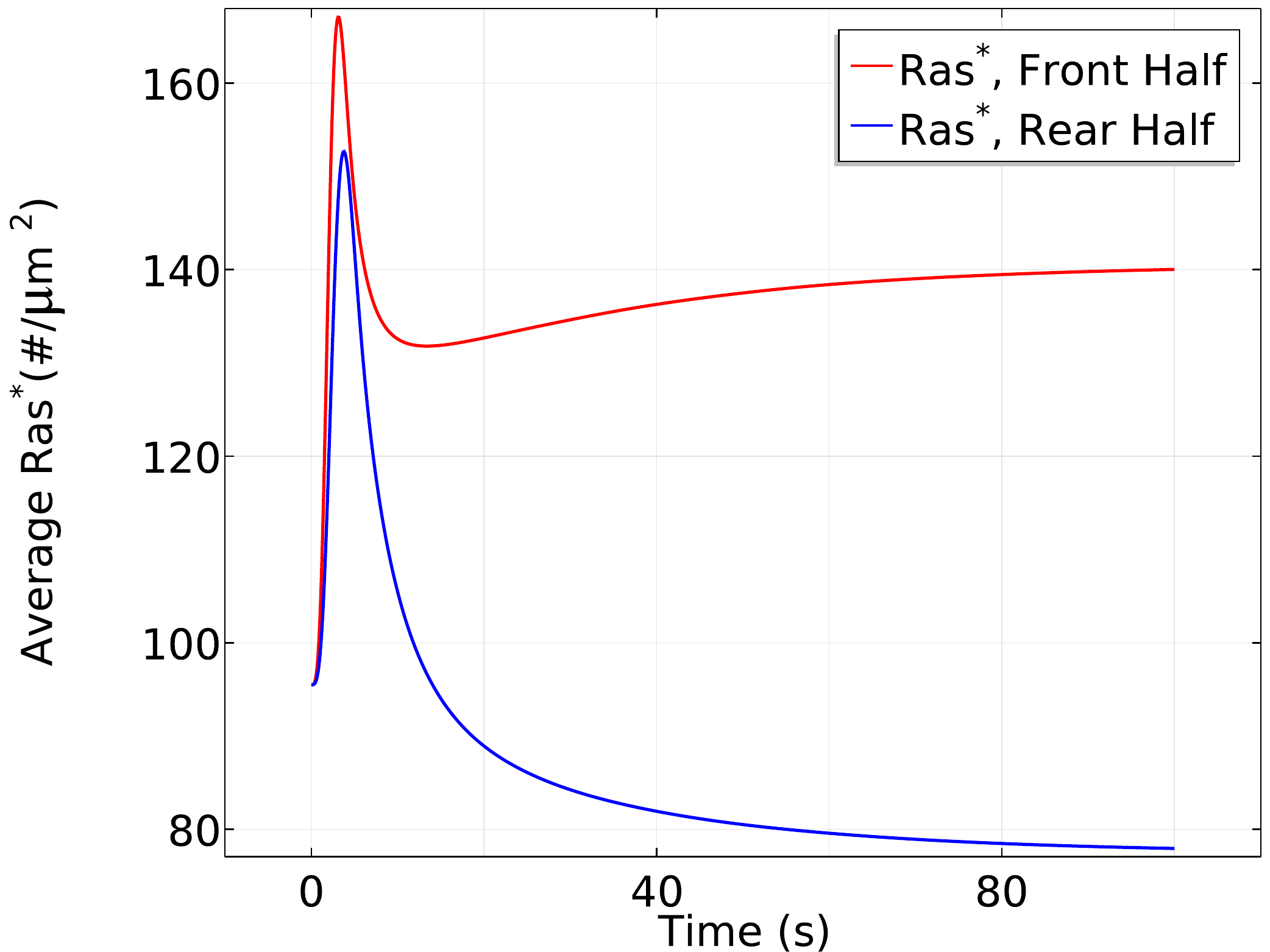}\includegraphics[width=6cm] {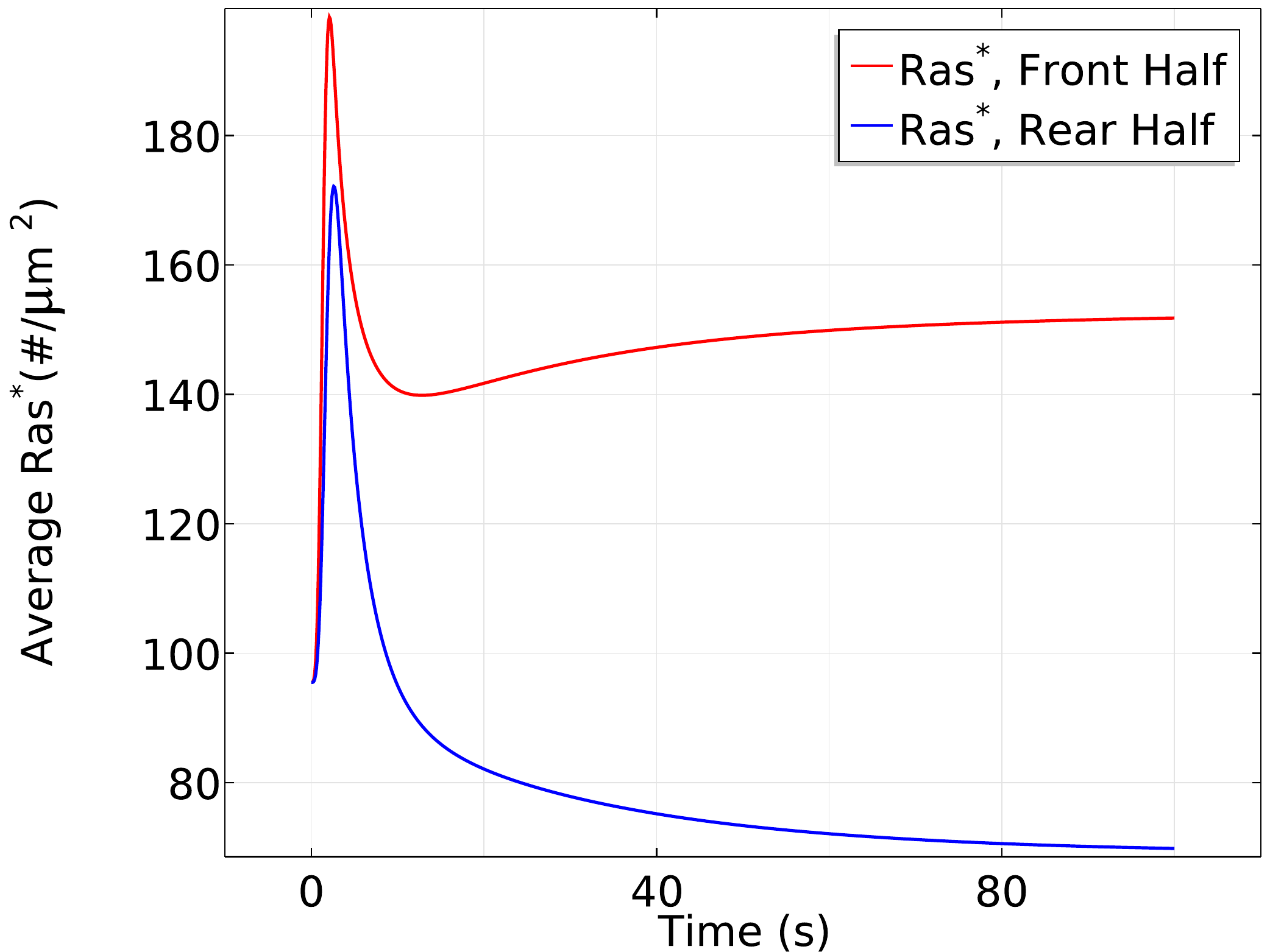}\includegraphics[width=6cm] {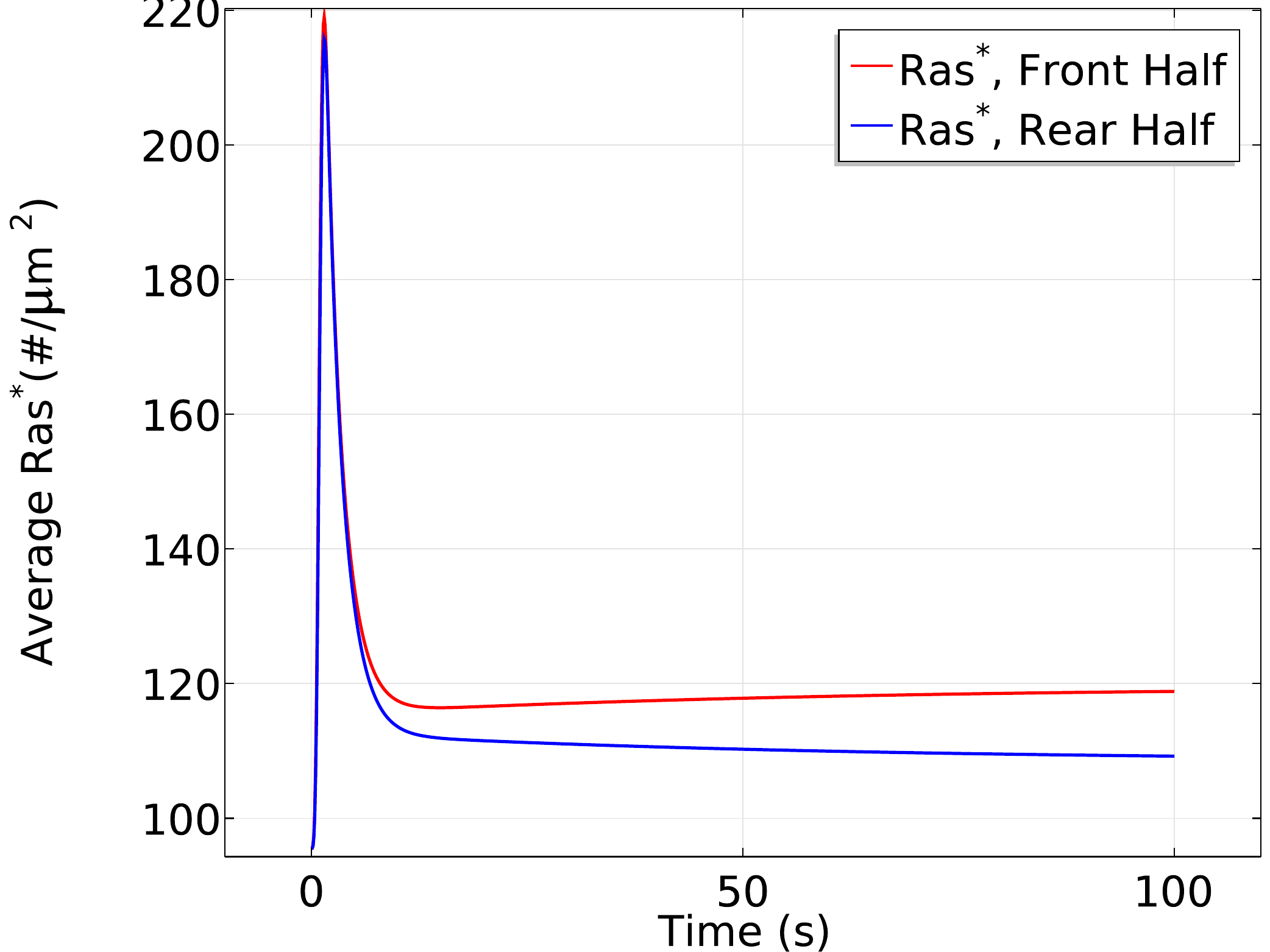}}
\centerline{\includegraphics[width=6cm] {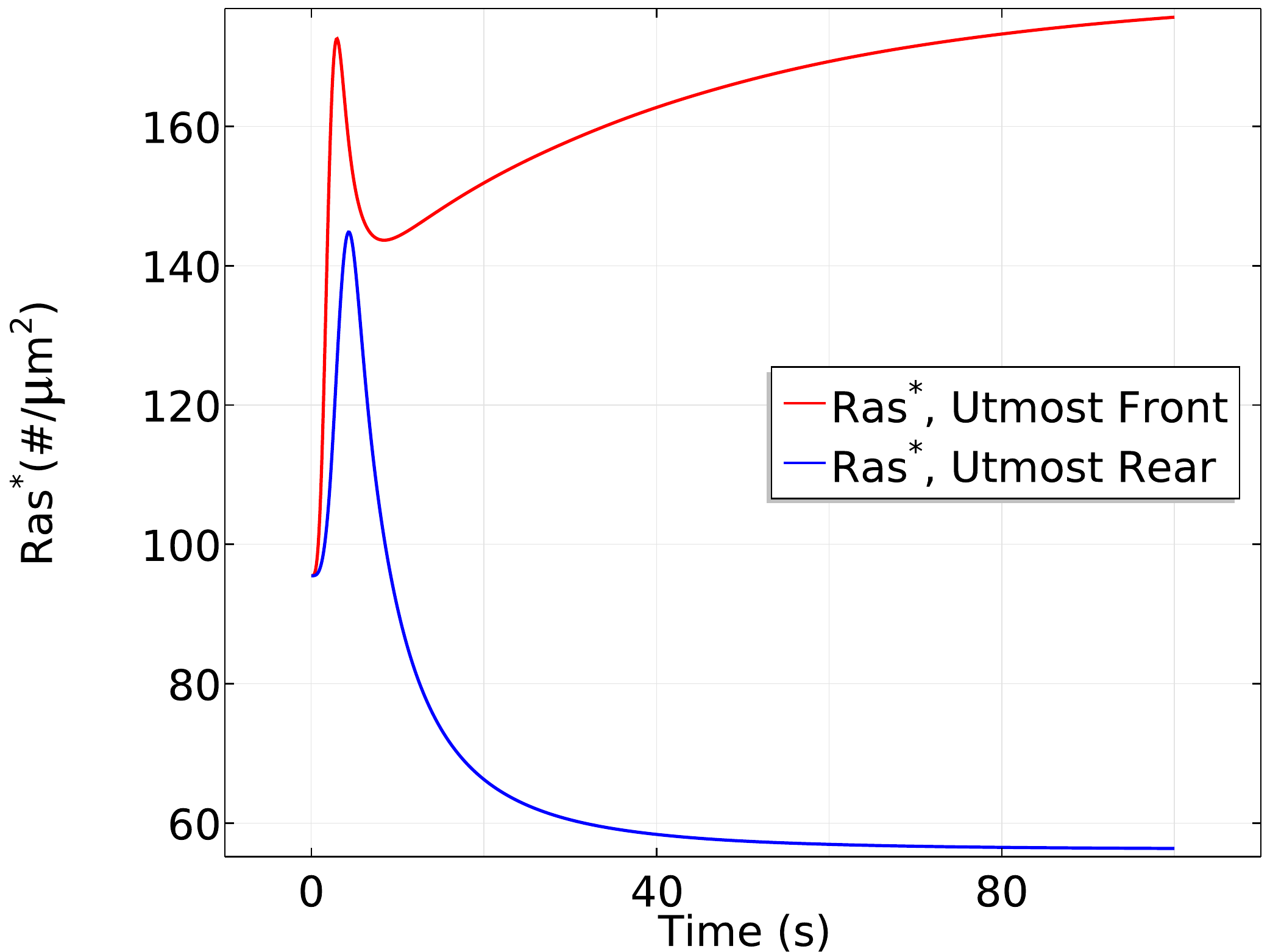}\includegraphics[width=6cm] {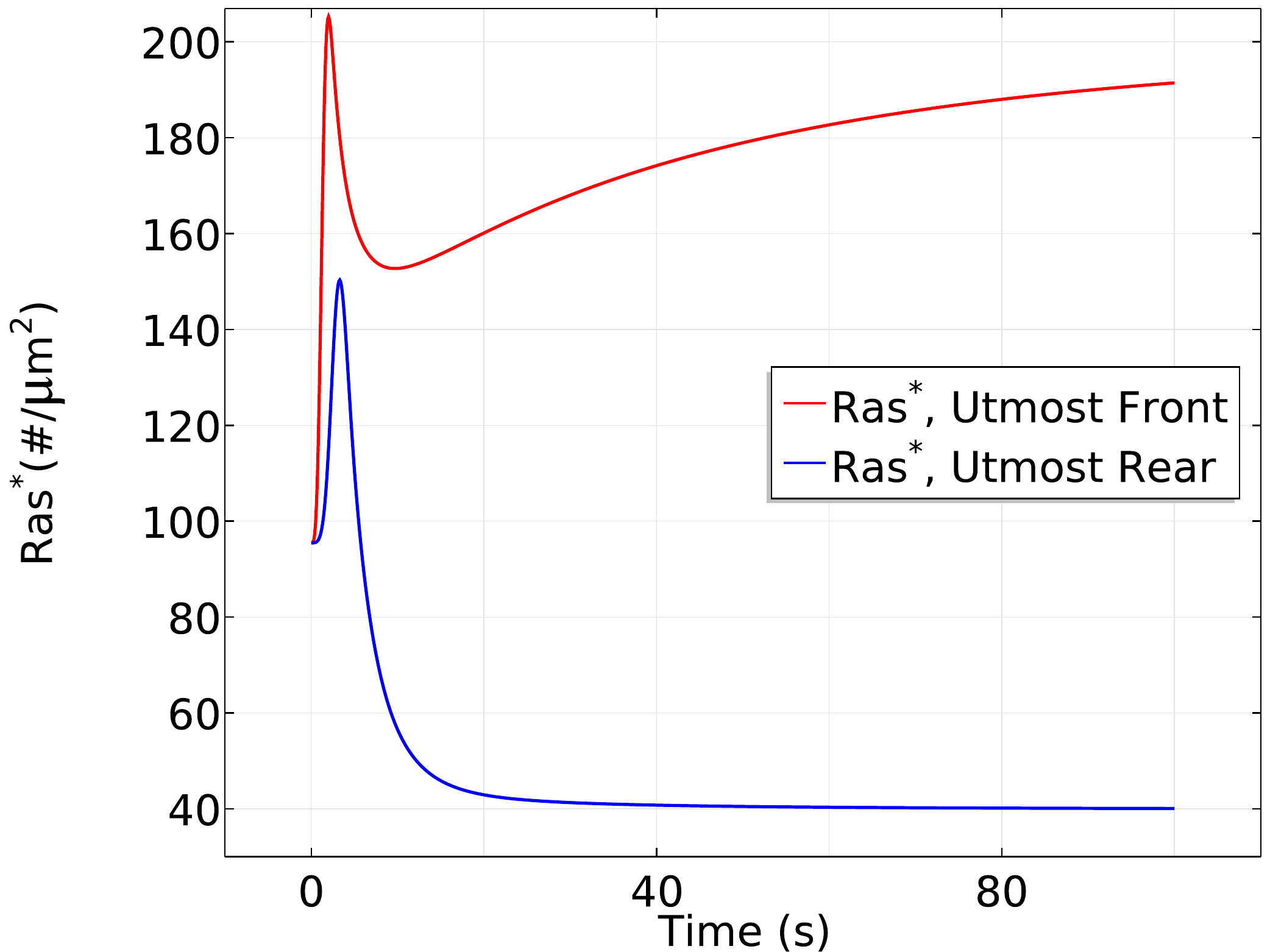}\includegraphics[width=6cm] {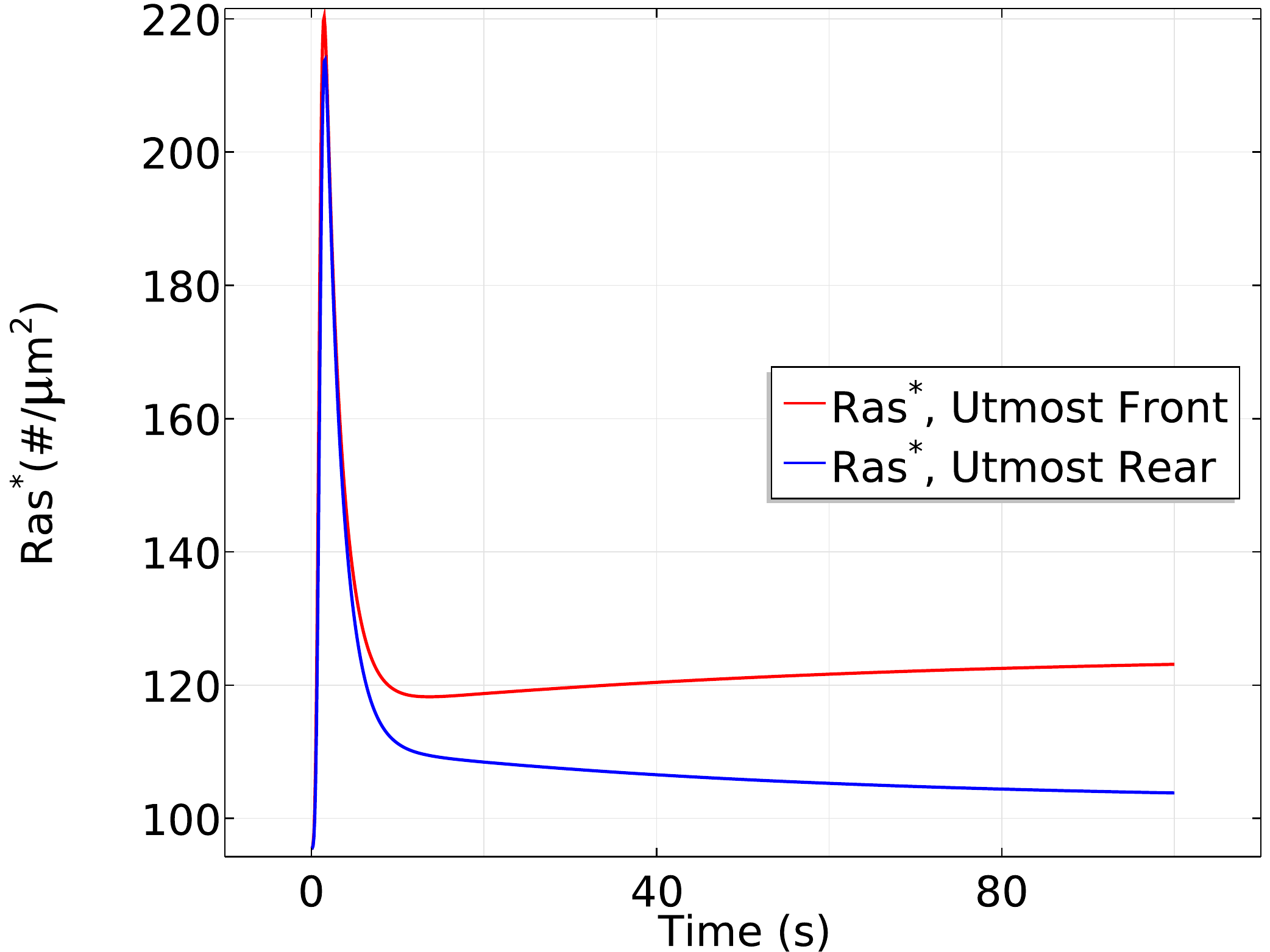}}
\caption{\textbf{time course of $Ras^*$ at front and rear half (top) and at $x_f$ and $x_r$ (bottom) in various gradient in Mode 3.}\\
\emph{Left}: $C_f=10$ nM and $C_r=1$ nM. \emph{Center}: $C_f=50$ nM and $C_r=0$ nM. \emph{Right}: $C_f=175$ nM and $C_r=125$ nM.}
\label{mode3halfhalf}
\end{figure}

\begin{figure}[H]
\centerline{\includegraphics[width=7cm]{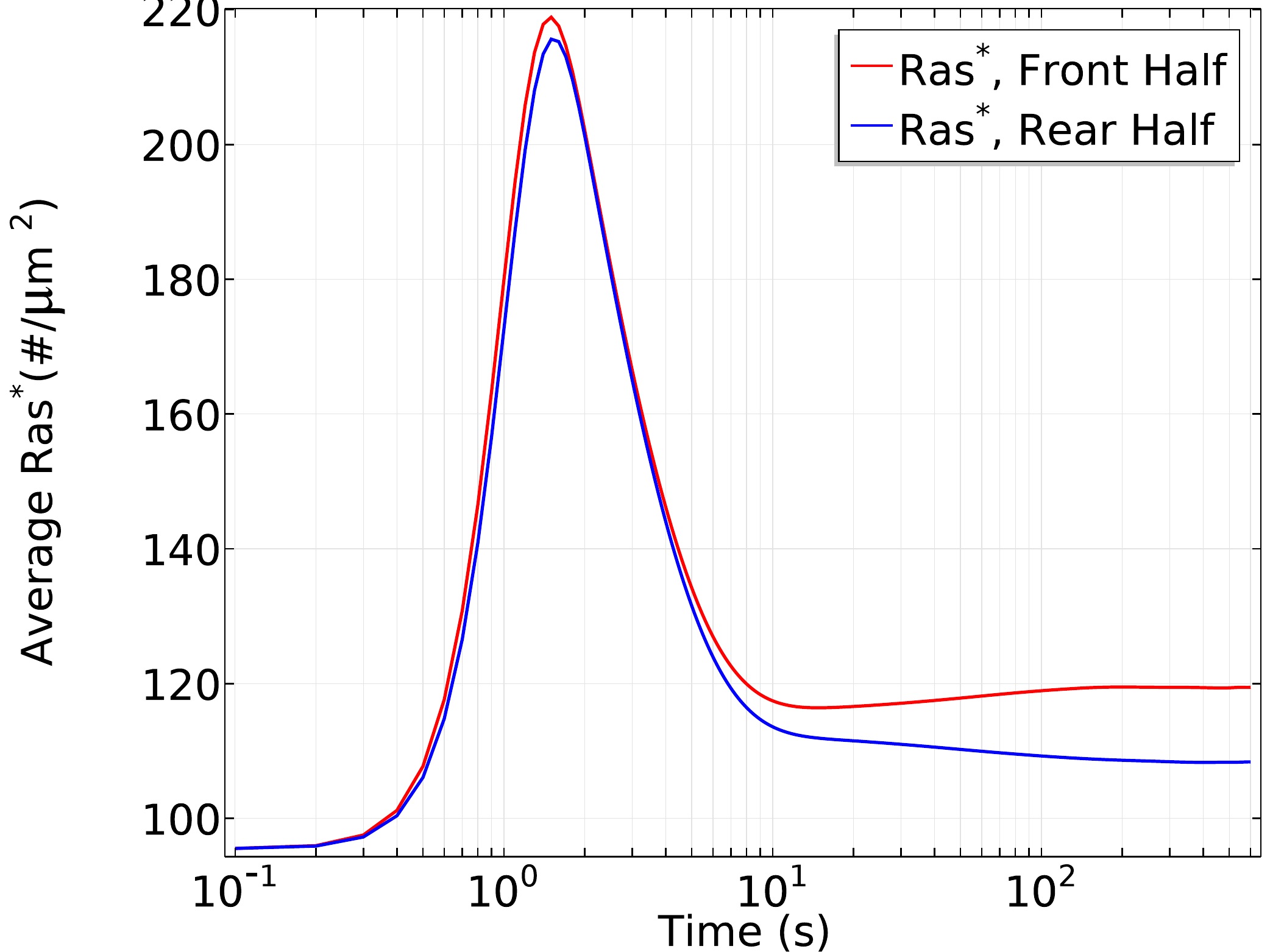}\includegraphics[width=7cm]{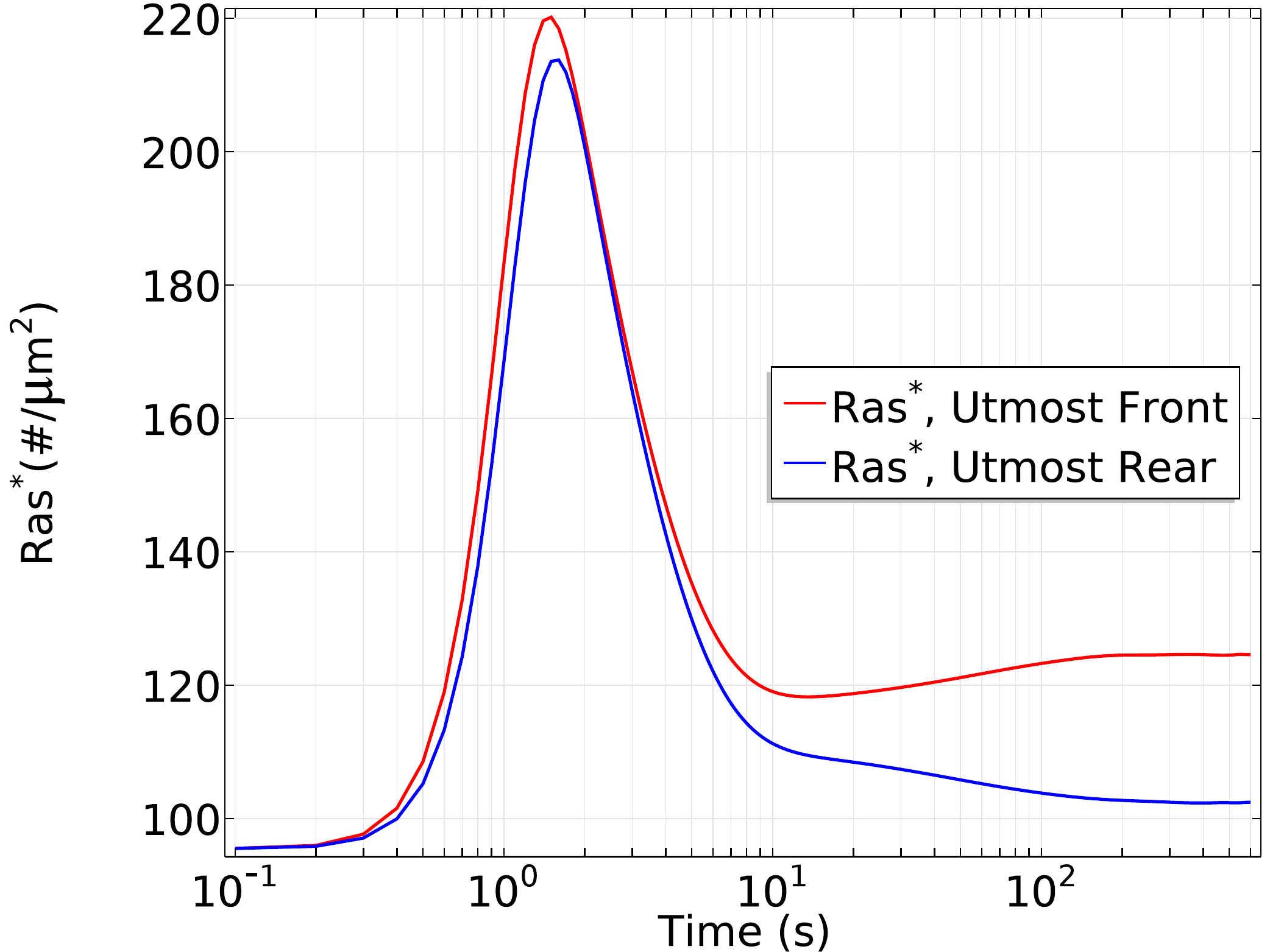}}
\caption{\textbf{Time course of $Ras^*$ at front and rear half (left) and at $x_f$ and $x_r$ (right) in a cAMP gradient in Mode 3 defined by $C_f=175$ nM and $C_r=125$ nM.}}
\label{mode3backhigh}
\end{figure}

Next, we test the possibility that Membrane recruitment of Ric8 is promoted by $G\alpha$ (Mode 3). The results are illustrated in Fig.~\ref{mode3halfhalf} and Fig.~\ref{mode3backhigh}. These plots suggest that the cell is still able to sense direction and exhibit biphasic responses under various cAMP gradients. They differ from the plots in Mode 1 and Mode 2 in the sense the point Ras activity equilibrates quicker and the magnitudes of the symmetry breaking are smaller.
\linenumbers
\subsection*{Reaction Rates}
The reactions considered in signal transdution are mostly bimolecular reactions:
$$
\cR: A+B\rightarrow C,
$$
and we model the reaction rate as
$$
\rxn = k A \cdot B ,
$$
where $k$ is the reaction rate and $ A ,  B $ represent the concentrations of A and B.

Reactions on the membrane and membrane-cytosol translocation are expressed in
the form
\begin{equation}
\label{reaction1}
A+B\rightarrow A+B^*.
\end{equation}
To model this kind of reactions in a realistic way, we first write the full
dynamics
$$
A+B\xrightleftharpoons[k_{-1}]{k_{1}}\overline{A\cdot B} \xrightarrow  {k_2}
\overline{A \cdot B^*} \xrightarrow  {k_3} A+B^*,
$$
then the full dynamics can be described by
\begin{eqnarray}
&\frac{dA}{dt}& = -k_1 A B+k_{-1} \overline{A\cdot B}+k_3 \overline{A\cdot
    B^*}\\ &\frac{d\overline{A\cdot B}}{dt}& = k_1 A B - k_{-1} \overline{A\cdot
    B} -k_2 \overline{A\cdot B}\\ &\frac{d\overline{A\cdot B^*}}{dt}&=
  k_2\overline{A\cdot B} -k_3 \overline{A\cdot B^*}\\ &\frac{dB^*}{dt}&=
  k_3\overline{A\cdot B^*}.
\end{eqnarray}
Now we assume that fast relaxation to a steady state for the intermediate enzyme-substrate
complexes is achieved,
$$
\frac{d\overline{A\cdot B}}{dt} = \frac{d\overline{A\cdot B^*}}{dt}=0.
$$
Hence
$$
k_1 A B - k_{-1} \overline{A\cdot B} =k_2 \overline{A\cdot B}=k_3
\overline{A\cdot B^*}.
$$
Therefore,
$$
\overline{A\cdot B} = \frac{k_1}{k_{-1}+k_2} AB.
$$
Then we have
\begin{eqnarray*}
\frac{dA}{dt} &=& -k_1 A B+k_{-1} \overline{A\cdot B}+k_3 \overline{A\cdot
  B^*}\\ &=& -k_2 \overline{A\cdot B}+k_3 \overline{A\cdot B^*}\\ &=& -k_3
\overline{A\cdot B^*}+k_3 \overline{A\cdot B^*}\\ &=& 0,
\end{eqnarray*}
and
\begin{eqnarray}
\frac{dB^*}{dt} &=& k_3 \overline{A\cdot B^*}\\ &=& k_2 \overline{A\cdot
  B}\\ &=& \frac{k_1k_2}{k_{-1}+k_2} AB.
\end{eqnarray}
Denote
$$
K = \frac{k_1k_2}{k_{-1}+k_2},
$$
we obtain the reaction rate of (\ref{reaction1})
\begin{equation}
\frac{dB^*}{dt} = K AB.
\end{equation}
\subsection*{Parameter estimation}
\label{parameter}
 We first show that a spatially lumped model can be derived by mean
 approximation of the generic reaction diffusion system. The spatially lumped
 model will be used to analyse the adaptation of Ras activity under uniform
 stimulation.

Define the mean concentration $\overline{C}$ of a given species $A$ in the
cytosol $\Omega$ to be
$$
\overline{C}(t) = \frac{1}{|\Omega|} \int_{\Omega} C(t,x)dx,
$$
where $|\Omega|$ is the volume of the cytosol, or the volume of the cell.

Integrating both sides of the reaction diffusion equation,
we obtain by the divergence theorem,
\begin{eqnarray}
\label{lumpedreactiondiffusion}
\frac{d\overline{C}}{dt}& =& \frac{1}{|\Omega|} \int_{\Omega}\frac{\partial
  C(t,x)}{\partial t}dx\nonumber\\ &=& \frac{1}{|\Omega|} \int_{\Omega}
\nabla\cdot (D\nabla C)dx
+\frac{1}{|\Omega|}\int_{\Omega}\sum_{i}s^i\rxn^idx\nonumber\\ &=&
\frac{1}{|\Omega|} \int_{\partial \Omega} D\nabla C \cdot n ds +\sum_{i}s^i
\frac{1}{|\Omega|}\int_{\Omega} \rxn^i dx\nonumber\\ &=& \frac{1}{|\Omega|}
\int_{\partial \Omega} D\frac{\partial C}{\partial n}
ds+\sum_{i}s^i\overline{\rxn^i},
\end{eqnarray}
in which the average reaction rates are defined as
$$
\overline{\rxn^i} = \frac{1}{|\Omega|}\int_{\Omega} \rxn^i dx.
$$
Substitute the boundary conditions into
(\ref{lumpedreactiondiffusion}), we have
\begin{equation}
\label{lumpedcytosol}
\frac{d\overline{C}}{dt} = \frac{1}{|\Omega|} \int_{\partial \Omega}
\left(-j^{+}+j^{-}\right)ds+\sum_{i}s^i\overline{\rxn^i}.
\end{equation}

If we assume that $j^{+}$ and $j^{-}$ are space invariant, (\ref{lumpedcytosol})
can be simplified as
\begin{equation}
\frac{d\overline{C}}{dt} = \frac{|\partial \Omega|}{|\Omega|}
\left(-j^{+}+j^{-}\right)+\sum_{i}s^i\overline{\rxn^i},
\end{equation}
where $|\partial \Omega|$ is the surface area of the cell membrane. If the cell
has a spherical shape, then
$$
\frac{|\partial \Omega|}{|\Omega|} = \frac{r}{3}.
$$
Since the membrane diffusion is small compared to  diffusion in the  cytosol, we omit
the membrane diffusion for simplicity, and similarly, define the mean
concentration $\overline{C_m}$ of a given species $A$ on the membrane $\partial
\Omega$ to be
$$
\overline{C_m}(t) = \frac{1}{|\partial \Omega|} \int_{\partial \Omega}
C_m(t,x)ds,
$$
and integrate both sides of the translocation reaction equation, we obtain
\begin{eqnarray}
\label{lumpedmembrane}
\frac{d \overline{C_m}(t)}{dt} &=&\frac{1}{|\partial \Omega|} \int_{\partial
  \Omega} \frac{\partial C_m(t,x)}{\partial t} ds\nonumber\\ &=&
-\kappa\frac{1}{|\partial \Omega|}\int_{\partial
  \Omega}\left(j^{+}-j^{-}\right)ds+\sum_{i}s_m^i\frac{1}{|\partial
  \Omega|}\int_{\partial \Omega} \rxn_m^ids\nonumber\\ &=& -\kappa
\left(j^{+}-j^{-}\right) +\sum_{i}s_m^i\overline{\rxn_m}^i,
\end{eqnarray}
where
$$
\overline{\rxn_m}^i = \frac{1}{|\partial \Omega|}\int_{\partial \Omega}
\rxn_m^ids.
$$
If we assume the parameters and the concentrations on the membrane are spatially
invariant, the parameters in the spatially distributed model and the spatially
lumped model are identical in the sense that
\begin{eqnarray*}
\overline{\rxn_m}^i &=& \frac{1}{|\partial \Omega|}\int_{\partial \Omega}
\rxn_m^ids\\ &=& \frac{1}{|\partial \Omega|}\int_{\partial \Omega} k AB ds\\ &=&
kAB.
\end{eqnarray*}
Through mean approximation, we obtain a spatially lumped model consisting of
equations in the form (\ref{lumpedcytosol}) and (\ref{lumpedmembrane}).

Now we explain how we the parameters are estimated using the spatially lumped
model and steady state analysis (SSA). The spatially lumped model for the G
protein module is given by
\begin{align}
\frac{d  G_{\alpha\beta\gamma,m} }{dt} &=
\tr_2 G_{\alpha\beta\gamma},c -\tr_1
        G_{\alpha\beta\gamma,m} -\rk_2 G_{\alpha\beta\gamma,m} R^*+\rk_7 G_{\alpha} \cdot \Gbgm \label{1}\\ \frac{d G_{\alpha\beta\gamma},c }{dt}&=\frac{3}{r}(-\tr_2 G_{\alpha\beta\gamma},c +\tr_1
        G_{\alpha\beta\gamma,m} )\label{2}\\ \frac{d G^*_{\alpha} }{dt}&=\rk_2 G_{\alpha\beta\gamma,m} R^*-\rk_3G^*_{\alpha}+\rk_5 Ric8^*
        G_{\alpha} \label{3}\\ \frac{d G_{\alpha} }{dt}&=\rk_3G^*_{\alpha}-\rk_7 G_{\alpha} \cdot \Gbgm -\rk_5 Ric8^*
        G_{\alpha} \label{4}\\ \frac{d \Gbgm }{dt}&=
       -\tr_3 \Gbgm +\tr_4 \Gbgc  -\rk_7 G_{\alpha} \cdot \Gbgm +\rk_2 G_{\alpha\beta\gamma,m} R^*\label{5}\\ \frac{d
          \Gbgc  }{dt}&=
       \frac{3}{r}(\tr_3 \Gbgm -\tr_4 \Gbgc  )\label{6}\\ \frac{d Ric8_m }{dt}&=
       -\tr_5 Ric8_m +\tr_6 Ric8_c -\rk_4 Ric8_m \cdot \Gbgm +\rk_6 Ric8^* \nonumber\\ &+\tr_7\delta
        G^*_{\alpha} \cdot Ric8_c \label{7}\\ \frac{d Ric8_c }{dt}&=
       \frac{3}{r}(\tr_5 Ric8_m -\tr_6 Ric8_c -\tr_7\delta
             G^*_{\alpha} \cdot Ric8_c )\label{8}\\ \frac{d Ric8^* }{dt}&=\rk_4 Ric8_m \cdot \Gbgm -\rk_6 Ric8^* \label{9}.
\end{align}
At steady state, from (\ref{2}), we have
$$
\frac{\tr_2}{\tr_1} = \frac{ G_{\alpha\beta\gamma,m} }{\delta
   G_{\alpha\beta\gamma},c }.
$$
It is reported in \cite{elzie2009} that roughly 30\% of the heterotrimeric G
protein is in the cytosol, hence
$$
\frac{G_{\alpha\beta\gamma,m} S}{G_{\alpha\beta\gamma},c V} = \frac{7}{3}.
$$
This is true at all stimulus level. When no stimulus presents, we have
$$
G_{\alpha\beta\gamma,m}^t = 0.7G_{\alpha\beta\gamma}^t/S, G_{\alpha\beta\gamma}\beta_c^t
= 0.3G_{\alpha\beta\gamma}^t/V.
$$
Also,
$$
\tr_2 = \frac{7V}{3S\delta}\tr_1, S=4\pi r^2, V = 4/3\pi r^3.
$$
\cite{elzie2009} measures the recovery rate for the G proteins in fluorescence
recovery after photobleaching (FRAP) was independent of the amount of bleached
area with a half-time of approximately 5 seconds. Hence we estimate $\tr_1=
1s^{-1}$, which value is also assigned for $\tr_3$, $\tr_5$, $\tr_9$ and $\tr_{11}$.

It is reported in \cite{janetopoulos}, half of the G protein dissociates at cAMP
concentration $10 nM$. Assume $p$ represents the ratio of quantities of
$\Gbgc $ and $\Gbgm$ at this cAMP concentration level, then by
(\ref{6}), we have
$$
\frac{\tr_4}{\tr_3} = \frac{ \Gbgm }{ \Gbgc  } =
\frac{V}{p\delta S}.
$$
We speculate that G$_{\beta\gamma,m}$ dissociates from the membrane with the same
rate of G$\alpha_{\beta\gamma,m}$, $\tr_3=\tr_1$, and then $\tr_4$ can be
calculated. In the numerical simulations we assign $p=3/7$.

From conservation of G protein, at $10 nM$ cAMP concentration, we have
\begin{equation}
G_{\alpha\beta\gamma,m}S+G_{\alpha\beta\gamma},cV+\Gbgm S+\Gbgc V=
G_{\alpha\beta\gamma}^t,
\end{equation}
and
\begin{equation}
G_{\alpha\beta\gamma,m}S+G_{\alpha\beta\gamma},cV+G^*_{\alpha}S+G_{\alpha}S=G_{\alpha\beta\gamma}^t.
\end{equation}
Note that
$$
G_{\alpha\beta\gamma,m}S+G_{\alpha\beta\gamma},cV = \frac{1}{2}G_{\alpha\beta\gamma}^t,
$$
we have
\begin{equation}
\label{im}
\Gbgm S+\Gbgc V=G^*_{\alpha}S+G_{\alpha}S=\frac{1}{2}G_{\alpha\beta\gamma}^t.
\end{equation}
From (\ref{9}) and (\ref{im}), we obtain
\begin{equation}
\label{riceq1}
\Gbg=\frac{\frac{1}{2}G_{\alpha\beta\gamma}^t}{S+pV}=\frac{\rk_6
  Ric8^*}{\rk_4Ric8_m},
\end{equation}
which leads to
\begin{equation}
\label{riceq2}
Ric8^* = \alpha Ric8_m, \alpha =
\frac{\frac{1}{2}G_{\alpha\beta\gamma}^t}{S+pV}\cdot \frac{\rk_4}{\rk_6}.
\end{equation}
Also, from (\ref{im}),
\begin{equation}
\label{galphaeq}
G^*_{\alpha}+G_{\alpha}=\frac{\frac{1}{2}G_{\alpha\beta\gamma}^t}{S}.
\end{equation}
Moreover, from (\ref{8}), we have
\begin{equation}
\label{gaaeq1}
G^*_{\alpha} = \frac{\tr_5 Ric8_m-\tr_6\delta Ric8_c}{\tr_7\delta
  Ric8_c}.
\end{equation}
From (\ref{gaaeq1}) and (\ref{3})
\begin{equation}
\label{gameq}
G_{\alpha} = \frac{\rk_3 \frac{\tr_5 Ric8_m-\tr_6\delta
    Ric8_c}{\tr_7\delta Ric8_c}-\rk_2 G_{\alpha\beta\gamma,m} R^*}{\rk_5 Ric8_a}.
\end{equation}
Substitute (\ref{gaaeq1}) and (\ref{gameq}) into (\ref{galphaeq}), we obtain a
equation consisting of $Ric8_c$, $Ric_m$ and $Ric8^*$,
\begin{equation}
\label{master1}
\frac{\tr_5 Ric8_m-\tr_6\delta Ric8_c}{\tr_7\delta Ric8_c}+\frac{\rk_3
  \frac{\tr_5 Ric8_m-\tr_6\delta Ric8_c}{\tr_7\delta Ric8_c}-\rk_2
  G_{\alpha\beta\gamma,m} R^*}{\rk_5 Ric8_a} = \frac{G_{{\alpha\beta\gamma}_t}}{2S}.
\end{equation}
By conservation of total Ric8 and (\ref{riceq2}), we have
$$
Ric8_cV+(Ric8_m+\alpha Ric8_m)S= Ric8_t,
$$
from which we have
\begin{equation}
\label{master2}
Ric8_c = \beta +\gamma Ric8_m, \beta = \frac{Ric8_t}{V}, \gamma =
-\frac{(1+\alpha)S}{V}<0.
\end{equation}
Substitute (\ref{master2}) and (\ref{riceq2}) into (\ref{master1}), we obtain a
equation of $Ric8_m$, which can be simplified into a quadratic equation
\begin{equation}
\label{master3}
 a\cdot (Ric8_m)^2+ b\cdot Ric8_m+c=0,
 \end{equation}
 where
 $$
a = \alpha(\rk_5\tr_5-\rk_5\tr_6\delta\gamma-C_2C4\gamma),
 $$
 $$
b = -\alpha \rk_5\tr_6\delta\beta+\rk_3\tr_5-C_3\gamma-C_2C_4\alpha\beta,
c=- C_3\beta,
 $$
 $$
C_1 = \rk_2 G_{\alpha\beta\gamma,m} R^*, C_2=\frac{G_{{\alpha\beta\gamma}_t}}{2S},
 $$
 and
 $$
C_3 = \rk_3\tr_6\delta+C_1\tr_7\delta, C_4 = \rk_5 \tr_7\delta.
 $$
 Finally, by solving (\ref{master3}), we have
 $$
Ric8_m = \frac{-b+\sqrt{b^2-4ac}}{2a}.
 $$
Then we can calculate backward to get $G_{\alpha}$, and estimate
$$
\rk_7 = \frac{C_1}{G_{\alpha} \Gbgm}.
$$
\subsection*{Imperfect adaptation}
\label{imperfect}
The spatially lumped model for the RasGTPase module is given by
\begin{align}
\frac{d RasGEF_c }{dt}&=\frac{3}{r}(-\tr_9\delta  RasGEF_c +\tr_8
      RasGEF_m -\delta \tr_{10}
      G^*_{\alpha} \cdot RasGEF_c )\\ \frac{d RasGAP_c }{dt}&=
     \frac{3}{r}(-\tr_{12}\delta
           RasGAP_c +\tr_{11} RasGAP_m )\label{gapeq}\\ \frac{d RasGEF_m }{dt}&=\tr_9 RasGEF_c -\tr_{8} RasGEF_m \nonumber\\ &+\tr_{10} G^*_{\alpha} \cdot RasGEF_c -\rk_8 \Gbgm \cdot RasGEF_m +\rk_9 RasGEF^* \\ \frac{d RasGAP_m }{dt}&=\tr_{12} RasGAP_c -\tr_{11} RasGAP_m -\rk_{10} \Gbgm \cdot RasGAP_m \nonumber\\ &+\rk_{11} RasGAP* \\ \frac{d RasGEF^* }{dt}&=
          \rk_8 \Gbgm \cdot RasGEF_m -\rk_9 RasGEF^* \\ \frac{d RasGAP^* }{dt}&=
          \rk_{10} \Gbgm \cdot RasGAP_m -\rk_{11} RasGAP^* \\ \frac{d Ras^* }{dt}
          &=
          \rk_{12} RasGEF^* \cdot Ras -\rk_{13} RasGAP^* \cdot Ras^* +\rk_{14} Ras -\rk_{15} Ras^* .
\end{align}
At steady states, we have
$$
 RasGEF^* =\frac{\rk_8 \Gbgm \cdot RasGEF_m }{\rk_9},
 RasGEF_c =\frac{\tr_8 RasGEF_m }{\tr_9\delta+\tr_{10} G^*_{\alpha} }.
$$
From the conservation law for RasGEF,
$$
 RasGEF_c V+ RasGEF_m S+ RasGEF^* S=RasGEF^t,
$$
we obtain
\begin{align}
 RasGEF_m  &= \frac{RasGEF^t}{S}\times\nonumber\\ &\frac{\rk_9
  (\tr_9\delta+\tr_{10} G^*_{\alpha} )}{(r/3\tr_8+\tr_9\delta)\rk_9+\tr_{10}\rk_9 G^*_{\alpha} +\tr_9\delta
  \rk_8 \Gbgm +\tr_{10}\rk_8 \Gbgm \cdot G^*_{\alpha} }.
\end{align}
Similarly, from steady states and conservation of RasGAP, we have
$$
 RasGAP_m  =
\frac{RasGAP^t}{S}\frac{\rk_{11}\tr_{12}\delta}{\rk_{11}(r/3\tr_{11}+\tr_{12}\delta)+\rk_{10}\tr_{12}\delta
   \Gbgm }.
$$
Then
$$
f =\frac{ Ras^* }{ Ras^* } = \frac{\rk_{12}  RasGEF^* }{\rk_{13} RasGAP* } =
\frac{\rk_{12}}{\rk_{13}}\frac{\rk_8}{\rk_9}\frac{\rk_{11}}{\rk_{10}}
\frac{ RasGEF_m }{ RasGAP_m }.
$$
We impose
$$\tr_{12}=\tr_{9}, \tr_{11}=\tr_8,$$
and
$$
\frac{\rk_9}{\rk_{11}}=\frac{\rk_{8}}{\rk_{10}}=\theta,
$$
then there exists perfect adaptation when $\tr_{10}=0$. But when
$\tr_{10}\neq 0$,
$$
\frac{\partial f}{\partial \Gbgm }\neq 0, \frac{\partial
  f}{\partial G^*_{\alpha} }\neq 0,
$$
which means adaptation can not be perfect.

To determine the values of $\tr_{11} (\tr_{8})$ and $\tr_{12}
(\tr_{9})$, we use steady state of equation (\ref{gapeq})
$$
\frac{\tr_{12}}{\tr_{11}} = \frac{ RasGAP_m }{\delta  RasGAP_c }.
$$
If a partition of $RasGAP_c$ and $RasGAP_m$ is determine, we can calculate
$\tr_{12}(\tr_8)$ based on $\tr_{11}(\tr_9)$. In the numerical
simulation, we choose the ratio as
$$
\frac{ RasGAP_c V}{ RasGAP_m S}=\frac{3}{7},
$$
the same as G2 partition without further information. The simulation results do not change significantly by varying this ratio.

We assign $\tr_{11}=1s^{-1}$ based on a measurement of PTEN dissociation
rate in \cite{vazquez2006}. We speculate that PTEN and RasGEF and RasGAP share
similar time constants, since no explicit values for RasGEF and RasGAP dissociation are available.

To ensure Ras is activated when G$_{\beta\gamma}$ both activates $RasGEF$ and
$RasGAP$, we require
$$
\theta>1.
$$
In the simulation, we choose $\theta=4$. Varying this value would only change
the peak value of Ras activation.

Note that when $\tr_{10}=0$, perfect adaptation gives us
$$
\frac{  Ras^*  }{ Ras }=\frac{\rk_{12}}{\rk_{13}}\frac{RasGEF^t}{RasGAP^t}.
$$
In simulation we assign the ratio of $Ras^*$ and $Ras$ as $1:9$, whose value
does not alter the system behaviors.

\bibliography{SignalTransductionref,newrefs}
\end{document}